%% file: draftMain.tex
\documentclass[prb,twocolumn, superscriptaddress]{revtex4}

\usepackage{times}
\usepackage{yfonts}
\usepackage{amsmath}
\usepackage{amsfonts}
\usepackage{dsfont}
\usepackage{amssymb}
\usepackage{graphicx}
\usepackage{hyperref}
\usepackage{tikz}
\usepackage{color}

\usepackage{graphicx,bm,amssymb,amsmath}
\usepackage[caption=false]{subfig}
\usepackage{subfig}
\usepackage{ulem}

\usepackage{dsfont}
\usepackage{epsfig}
\usepackage{verbatim}
\usepackage{yhmath}
\usepackage{bbm}
\usepackage{slashed}
\usepackage{tikz}

\usetikzlibrary{decorations.pathreplacing,decorations.markings}
\tikzset{middlearrow/.style={
        decoration={markings,
            mark= at position 0.5 with {\arrow{#1}} ,
        },
        postaction={decorate}
    }
}

\pdfoutput=1

\newcommand{\ket}[1]{|#1\rangle}

\newcommand{\mb}[1]{\mathbf{#1}}
\newcommand{\coho}[1]{\textswab{#1}}
\newcommand{\cohosub}[1]{\scalebox{0.7}{\textswab{#1}}}

\def\R{\mb{r}}
\def\r{\mb{r}}
\newcommand{\ra}[1]{ {}^{\R}{#1}}
\def\Z{\mathcal{Z}} % partition function
\def\Tet{\text{Tet}} % tetrahedron function
\def\D{\mathrm{D}} % quantum double/Drinfeld center
\def\V{\mathcal{V}} % TQFT vector space

\def\G{H}

\def\CZ{\mathbf{CZ}}
\def\vac{\text{vac}}

\def\bfg{{\bf g}}
\def\sgn{\text{sgn}}

\makeatletter
\def\l@subsubsection#1#2{}
\makeatother

\definecolor{dualblue}{RGB}{3,101,192}

\newcommand{\parsa}[1]{\textcolor{cyan}{(Parsa: #1)}}

\begin{document}

\title{Reflection and time reversal symmetry enriched topological phases of matter: \\ path integrals, non-orientable manifolds, and anomalies}

\author{Maissam Barkeshli}
\affiliation{Department of Physics, Condensed Matter Theory Center, University of Maryland, College Park, Maryland 20742, USA}
\affiliation{Joint Quantum Institute, University of Maryland, College Park, Maryland 20742, USA}
\affiliation{Kavli Institute for Theoretical Physics, University of California, Santa Barbara, California, 93106, USA}
\affiliation{Station Q, Microsoft Research, Santa Barbara, California 93106-6105, USA}
\author{Parsa Bonderson}
\affiliation{Station Q, Microsoft Research, Santa Barbara, California 93106-6105, USA}
\author{Meng Cheng}
\affiliation{Department of Physics, Yale University, New Haven, CT 06511-8499, USA}
\affiliation{Station Q, Microsoft Research, Santa Barbara, California 93106-6105, USA}
\author{Chao-Ming Jian}
\affiliation{Kavli Institute for Theoretical Physics, University of California, Santa Barbara, California, 93106, USA}
\affiliation{Station Q, Microsoft Research, Santa Barbara, California 93106-6105, USA}
\author{Kevin Walker}
\affiliation{Station Q, Microsoft Research, Santa Barbara, California 93106-6105, USA}
%\date{\today}

\begin{abstract}
We study symmetry-enriched topological (SET) phases in 2+1 space-time dimensions with
spatial reflection and/or time-reversal symmetries. We provide a systematic construction of a wide class
of reflection and time-reversal SET phases in terms of a topological path integral defined on general space-time manifolds.
An important distinguishing feature of different topological phases with reflection and/or time-reversal symmetry is the value of the path integral
on non-orientable space-time manifolds. We derive a simple general formula for the path integral on the manifold $\Sigma^2 \times S^1$,
where $\Sigma^2$ is a two-dimensional non-orientable surface and $S^1$ is a circle.
This also gives an expression for the ground state degeneracy of the SET on the surface $\Sigma^2$ that depends on the reflection symmetry fractionalization class,
generalizing the Verlinde formula for ground state degeneracy on orientable surfaces.
Consistency of the action of the mapping class group on non-orientable manifolds leads us to a constraint that can detect when
a time-reversal or reflection SET phase is anomalous in (2+1)D and, thus, can only exist at the surface of a (3+1)D
symmetry protected topological (SPT) state. Given a (2+1)D reflection and/or time-reversal SET phase, we further derive a general formula that
determines which (3+1)D reflection and/or time-reversal SPT phase hosts the (2+1)D SET phase as its surface termination. A number of explicit examples
are studied in detail.
\end{abstract}

\maketitle

\tableofcontents

\input{intro}

\input{1dSPT}

\input{symfrac}

\input{state-sum2d}

\input{state-sum3d}

\input{gsd}

\input{anomaly}

\input{example}

\acknowledgements
We thank R. Bela Bauer, Shawn X. Cui, Michael H. Freedman, Max A. Metlitski, Chenjie Wang, Zhenghan Wang, and M. Michael
P. Zaletel for useful discussions.  MC would like to especially thank Chenjie Wang for sharing unpublished results.
MB is supported by startup funds from the University of Maryland and NSF-JQI-PFC. CMJ is
supported by a fellowship from the Gordon and Betty
Moore Foundation (Grant 4304). This research was supported
in part by the National Science Foundation under Grant No. NSF PHY-1125915 (KITP).
KW thanks the Aspen Center for Physics (NSF grant PHY-1066293) for providing a pleasant and productive work environment.

\emph{Note added}: The main results of this paper were presented by one of the authors at the Fall 2016 KITP
conference on \it Topological Quantum Matter.\rm\cite{barkeshli2016kitp} As this manuscript was being prepared, several related papers
also appeared.\cite{wang2016,tachikawa2016a,tachikawa2016b,bhardwaj2016}

\appendix

\input{parity-eigenvalue}

\input{1dTQFT}

\input{4dPF}

\input{string}

\input{dgRP2}

\input{DS3}

\input{data}

\bibliographystyle{apsrev}
%\bibliography{TI,corr}
\bibliography{TI}

\end{document}

%% file: intro.tex
\section{Introduction}

A fundamental question in condensed matter physics is to understand the possible quantum phases of matter.
The last few years have seen remarkable advances in the systematic understanding of this question
for gapped quantum systems -- where the system in the thermodynamic limit has a non-vanishing
bulk energy gap separating the ground state(s) of the quantum many-body system from the excited states.

In general, we can consider a gapped quantum system with symmetries described by the group $G$. Two quantum states
belong to different phases of matter if continuously tuning from one to the other while preserving
the symmetry $G$ necessarily passes through a quantum phase transition.

For one-dimensional quantum systems, the characterization and classification of gapped quantum
many-body states is believed to be complete and is related to projective representations of $G$, which
physically characterize the zero energy edge states on one-dimensional systems with boundary and
also the entanglement spectra of reduced density matrices.~\cite{pollmann2010,chen2011,fidkowski2011,schuch2011}
This advance was enabled in large part by developments in using matrix product states for describing
ground states of gapped, local Hamiltonians.~\cite{fannes1992,Verstraete2006,Hastings1D}

For higher-dimensional quantum systems, however, the understanding of how to characterize and classify
general gapped quantum states is still in progress. Remarkably, even in the absence of any symmetries,
i.e. when $G$ is trivial, gapped quantum systems can still form distinct phases of matter exhibiting topological
order.~\cite{wen04,nayak2008} Topologically ordered phases are distinguished by a number of exotic properties, including
quasiparticles with fractional or non-Abelian braiding statistics, robust topological ground state
degeneracies, various patterns of long-range quantum entanglement, and protected gapless edge modes.

In two-dimensional quantum systems (in (2+1)D space-time), well-known topologically ordered phases
include fractional quantum Hall (FQH) states and quantum spin liquids (QSL). Topological phases
in (2+1)D are believed to be mathematically characterized completely by two objects: (1) the chiral central charge
$c_{-}$ of the phase, which dictates the chiral thermal transport of the gapless edge modes,
and (2) an algebraic theory, known as a unitary modular tensor category (UMTC) $\mathcal{B}$, which specifies
the fusion rules, braiding statistics, and other universal properties of the emergent quasiparticle
excitations.~\footnote{See e.g. Refs.~\onlinecite{kitaev2006,Bonderson07b} for a discussion written for physicists of
UMTCs as applied to anyons and topological phases.}

Given a general (2+1)D topological phase described by the pair $(\mathcal{B}, c_-)$,
we wish to understand the distinct possible quantum states that can be formed in the presence of a
symmetry $G$. When $\mathcal{B}$ is trivial and $c_-=0$, the resulting class of states are referred to as
symmetry-protected topological (SPT) states~\cite{XieScience2012, chen2013, lu2012, kapustin2014,senthil2015}.
More generally, when $\mathcal{B}$ is trivial (but $c_-$ is not necessarily 0), the resulting class of states are called
``invertible'' phases.~\cite{freed2014, freed2016}
When $\mathcal{B}$ is nontrivial, the resulting class of states are referred to as symmetry-enriched topological (SET)
phases~\cite{wen2002psg, wen04, LevinPRB2012, essin2013, mesaros2013, lu2013, barkeshli2014SDG, Tarantino_SET, Chen2014, Lan2015, Lan2016},
because the system is still highly nontrivial even when the symmetry group $G$ is broken by perturbations
to the Hamiltonian of the system.

In order to make progress, it is useful to distinguish between various types of symmetry groups $G$.
For example, group elements of $G$ can act on the physical quantum states in either a unitary or anti-unitary
manner, and $G$ can include either on-site (internal) symmetries, spatial translation symmetries, or
point-group symmetries (such as rotations and reflections).

When the symmetry $G$ is represented on the quantum states in a unitary and on-site manner,
it has been shown how (2+1)D SET phases can be characterized, in general, by studying the algebraic
properties of symmetry defects associated with group elements of $G$.~\cite{barkeshli2014SDG}
These algebraic properties include fusion rules of defects and a generalized notion of braiding
transformations, which collectively give rise to a mathematical object known
as a $G$-crossed braided tensor category (BTC).~\cite{turaev2000,kirillov2004,turaev2010book,ENO2009,barkeshli2014SDG}

The understanding of (2+1)D SET phases in terms of $G$-crossed BTC has recently also been extended
to the case where $G$ includes lattice translational symmetries, which have led to powerful microscopic
constraints on the types of allowed patterns of symmetry fractionalization that can occur, given
the quantum numbers of the microscopic system within each unit cell.~\cite{cheng2015}

When $G$ contains time-reversal and/or reflection symmetries, there has also been recent progress
in the study of (2+1)D SPT and SET phase~\cite{LevinPRB2012, lu2012, hsieh2014, Hsieh2014b, kapustin2014, essin2013, zaletel2015, YangPRL2015,hermele2016, cheng2016, song2016},
yet the understanding for such SET phases is still incomplete.
It is not clear to what extent the $G$-crossed BTC framework is applicable to the case where
$G$ contains time-reversal and/or reflection symmetries. This is because we cannot, in the same way, create
defects of time-reversal or reflection symmetry and discuss their fusion and braiding properties.

In this paper, we study (2+1)D SET phases where the symmetry group $G$ contains time-reversal and/or reflection symmetries by
developing a different approach, where we utilize the topological path integrals (partition functions) that describe the long-wavelength
universal properties of the SET phases. This leads us to a systematic construction of a wide class of bosonic time-reversal and/or spatial reflection
SET phases (where the microscopic degrees of freedom are bosons, as opposed to fermions) through an
exact state sum construction of the topological path integral. Moreover, this allows us to define
topological path integrals on non-orientable space-time manifolds. For SPT states, the topological path integral
on non-orientable space-time manifolds has been used to distinguish different time-reversal and/or reflection SPT
states.~\cite{kapustin2014,hsieh2014,cho2015,metlitski2015,witten2016} We demonstrate how to compute
path integrals of (2+1)D time-reversal and/or reflection SET phases on non-orientable space-time manifolds.

Over the past few years, it was discovered that SET phases with certain patterns of symmetry fractionalization are
``anomalous,'' in the sense that they cannot exist in purely (2+1)D systems, but can exist at the (2+1)D surface
of a (3+1)D SPT state.~\cite{vishwanath2013, wang2013, MetlitskiPRB2013, Chen2014, Bonderson13d, chen2014b,
YangPRL2015,hermele2016, song2016, wang2013b, metlitski2014, MetlitskiPRB2015, Fidkowski13,
SW1, ChoPRB2014, kapustin2014b} For the case of unitary on-site and translation symmetries, there is a systematic procedure
to determine whether a given pattern of symmetry fractionalization is anomalous, by systematically solving the consistency
equations for the algebraic theory ($G$-crossed BTC) of defects.~\cite{barkeshli2014SDG,cheng2015} (For other related approaches, see also Ref.~\onlinecite{Chen2014, WangPRX2016, cui2016}). However, in the case of time-reversal SET phases, while some partial results have been obtained~\cite{cheng2016,hermele2016}, there has not been a general understanding of whether and why certain time-reversal and/or reflection SET phases are anomalous and, if so, which (3+1)D SPT states (if any) can host them on their surface.
Our study of time-reversal and/or reflection SET phases in terms of topological path integrals provides an answer to this open question.

\subsection{Summary of Main Results}

We now provide an overview of our main results.

\subsubsection{Topological path integrals and systematic construction of SET phases}

We present a systematic construction of a wide class of bosonic time-reversal and/or reflection SET phases,
where, for each SET phase, we provide a topologically invariant state sum for the path integral on general (2+1)D
space-time manifolds. The construction is a hybrid of the well-known Turaev-Viro-Barrett-Westbury (TVBW)
construction of (2+1)D topological quantum field theories (TQFTs) from fusion categories~\cite{turaev1992,barrett1996}, and the bosonic SPT state sum construction
discussed in Ref.~\onlinecite{chen2013}.

The TVBW path integral construction takes a unitary fusion category $\mathcal{C}$ as its input.~\footnote{More generally, it can take a spherical fusion category as its input, though
for the purpose of describe topological quantum phases of matter, we restrict to the unitary case.} The quasiparticles (anyons) of the
resulting TQFT are described by a UMTC given by the quantum double $\D(\mathcal{C})$.~\cite{turaev1994, Roberts1995, levin2005,turaev2010,kirillov2010,balsam2010,balsam2010b}

In the SET generalization that we introduce, when $G$ corresponds to unitary, space-time orientation-preserving symmetries,
the input data into the path integral is a unitary $G$-graded fusion category $\mathcal{C}_G$
(see Ref.~\onlinecite{barkeshli2014SDG} for a discussion of $G$-graded fusion categories written for physicists):
\begin{align}
\mathcal{C}_G = \bigoplus_{{\bf g} \in G} \mathcal{C}_{\bf g} ,
\end{align}
where objects $a_{\bf g} \in \mathcal{C}_{\bf g}$ have a $G$-grading. The intrinsic topological order associated with the TQFT resulting from the path integral construction is given by
the quantum double $\D(\mathcal{C}_{\bf 0})$, where ${\bf 0}$ is the identity element in $G$. The additional structure from the
$G$-grading allows us to describe a general class of SET phases with $G$ symmetry. Different SET phases having the same intrinsic topological order
$\D(\mathcal{C}_{\bf 0})$ correspond to distinct $G$-graded extensions $\mathcal{C}_G$ of the same $\mathcal{C}_{\bf 0}$.
This construction is closely related to recent exactly solvable Hamiltonians presented in Ref.~\onlinecite{heinrich2016,cheng2016}.
It appears to also be related to the construction of Ref.~\onlinecite{turaev2012}, though the precise relation is unclear to us.

When $G$ contains time-reversal and/or spatial reflection symmetries, the input data forms a type of
anti-unitary extension of $\mathcal{C}_{\bf 0}$, which technically is not a $G$-graded fusion category, though it has similar structure.
This anti-unitary extension has to our knowledge not been discussed in the mathematics literature before.
A special case was briefly discussed in the physics literature, in Ref.~\onlinecite{cheng2016}, in
terms of a Hamiltonian construction. In this paper, we identify the appropriate mathematical structure in terms of $G$-equivariant
2-categories with $G$ actions, which we define in Sec.~\ref{2catSec}.

\it We thus conclude that a wide class of time-reversal and/or reflection SET phases can be systematically constructed
in terms of such $G$-equivariant 2-categories with $G$ actions. \rm We conjecture that this construction is
capable of realizing a universal long-wavelength description of all possible (2+1)D time-reversal and/or spatial reflection SET phases
that admit gapped boundaries.

\subsubsection{Non-orientable manifolds}

An important property of a topological phase of matter is the topological ground state degeneracy on an oriented genus $g$
surface. In the effective TQFT description, this is given by the value of the path integral $\Z(\Sigma^2 \times S^1)$,
where $\Sigma^2$ is the genus $g$ surface, and $S^1$ is a circle representing the time direction. Thus,
\begin{align}
\Z(\Sigma^2 \times S^1) = \text{dim } \V(\Sigma^2),
\end{align}
where $\V(\Sigma^2)$ is the Hilbert space of the TQFT on $\Sigma^2$, and
its dimension $\text{dim } \V(\Sigma^2)$ is therefore the topologically protected ground state degeneracy
of a system defined on $\Sigma^2$ in the corresponding topological phase. For closed orientable manifolds, the ground state degeneracy is given by the
well-known formula~\cite{witten1989,moore1989b}
\begin{align}
\label{gsdOrientedEq}
\text{dim }\V(\Sigma^2) = \sum_{a \in \mathcal{B}} S_{0a}^{\chi(\Sigma^2)} ,
\end{align}
where the sum is over the set of topologically distinct quasiparticle types (anyons) described by $\mathcal{B}$ and $\chi(\Sigma^2) = 2 - 2g$ is the Euler
characteristic. The quantity $S_{0a} = d_a/ \mathcal{D}$, where $S$ is the modular $S$-matrix of $\mathcal{B}$,
$d_a$ is the quantum dimension of the anyon $a$, and $\mathcal{D} = \sqrt{\sum_{a\in \mathcal{B}} d_a^2}$ is the total quantum dimension.

In the presence of reflection symmetry, one can also define the system on a non-orientable surface by picking a branch cut
along which the state is twisted by the action of reflection. Previous studies of the ground state degeneracy of topological phases on non-orientable surfaces are limited.
Ref.~\onlinecite{freedman2001} defined Kitaev's $\mathbb{Z}_2$ toric code model~\cite{kitaev2003} on the real projective plane 
$\mathbb{RP}^2$, and found that for a particular triangulation it can be viewed as Shor's 9-qubit quantum error correcting code.\cite{shor1995} 
Ref.~\onlinecite{freedman2016} studied the doubled semion model~\cite{freedman2004} on $\mathbb{RP}^2$ (while generalizing it to higher dimensions), and found 
its ground state degeneracy on $\mathbb{RP}^2$ to be distinct from that of the toric code. Ref.~\onlinecite{chan2016} provided a 
prescription for computing the ground state degeneracy on non-orientable manifolds for Abelian (2+1)D topological
phases, with the further restrictions that reflection can only permute anyon types with trivial mutual braiding statistics and that the reflection 
symmetry fractionalization quantum numbers (of invariant anyons) are trivial. In any case, a simple formula has not been presented to date 
for the topological ground state degeneracy of an arbitrary reflection-symmetric topological state on a non-orientable manifold.

In this paper, we demonstrate that, for a general reflection SET phase on closed non-orientable surfaces $\Sigma^2$, the ground state degeneracy
is given by the following simple formula
\begin{align}
\label{gsdUnorientedREq}
\Z(\Sigma^2 \times S^1) = \text{dim }\V(\Sigma^2) = \sum_{a \in \mathcal{B}^{{\bf rc}}} (\eta_a^{\bf r} S_{0a}) ^{\chi(\Sigma^2)} .
\end{align}
In this expression, we have an action of reflection on anyon types ${\bf r}$: $a \rightarrow \,^{\bf r}a$, where $a$ transforms into $\,^{\bf r}a$
as it crosses an orientation-reversing branch sheet in $\Sigma^2 \times S^1$. We define
\begin{align}
\mathcal{B}^{\bf rc} = \{a \in \mathcal{B} | \,^{\bf r}a = \bar{a} \},
\end{align}
to be the set of topological charges (anyon types) that are invariant under reflection followed by topological charge conjugation, which we sometimes denote as $\,^{\bf c}a \equiv \bar{a}$. The quantity $\eta_a^{\bf r} = \pm 1$ is an invariant associated with the reflection symmetry fractionalization properties of $a$,
and will be defined and discussed in Sec.~\ref{1dSPT}-\ref{symfrac}. At this point, we just note that ${\bf r}$ and $\{\eta^{\bf r}_a\}$
are physical characteristics of a reflection SET phase.

It is notable that the reflection symmetry fractionalization class, determined by ${\bf r}$ and $\{\eta^{\bf r}_a\}$,
plays a role in determining the ground state degeneracy of the topological phase on a closed surface.
One may understand this by thinking of the branch sheets used to construct the non-orientable manifold as
reflection symmetry ``defects.'' Previous studies of surfaces containing defect branch lines associated with
unitary on-site symmetries also found that the symmetry fractionalization class affects the topological ground state
degeneracy.~\cite{barkeshli2014SDG}

Since our description is in terms of a Euclidean quantum field theory, time and space are treated on equal footing. We can thus
directly carry over the previous discussion to the case of time-reversal symmetry. Here, we interpret
the $S^1$ as purely spatial, and there is a branch cut in $\Sigma^2$ along which the direction of the imaginary time is reversed.
In this way, a similar expression as Eq.~(\ref{gsdUnorientedREq}) for time-reversal symmetry is given by
\begin{align}
\label{gsdUnorientedTEq}
\Z(\Sigma^2 \times S^1) =  \sum_{a \in \mathcal{B}^{\bf T} } (\eta_a^{\bf T} S_{0a}) ^{\chi(\Sigma^2)} .
\end{align}
Notice that the action of ${\bf rc}$ on anyon types is replaced by the action of time-reversal ${\bf T}$. Thus, the sum is over ${\bf T}$-invariant topological charges, i.e. $a \in \mathcal{B}$ such that $a = \,^{\bf T}a$. The quantity $\eta^{\bf T}_a = \pm 1$ is an invariant for ${\bf T}$-invariant topological charges
and it describes whether a quasiparticle of topological charge $a$ carries a local Kramers degeneracy associated with the time-reversal symmetry.
The replacement of ${\bf rc}$ with ${\bf T}$ can be viewed to be a consequence of CPT invariance of quantum field theory.

Oriented surfaces can be obtained by gluing together 3-punctured spheres. In the TQFT, a
three-punctured sphere labeled by topological charges $a$, $b$, $c$, gives rise to a Hilbert space with dimension
$N_{ab}^{\bar{c}}$. Eq.~(\ref{gsdOrientedEq}) can be obtained by gluing together 3-punctured spheres, summing
over topological charges at the punctures that are glued together, and using the Verlinde formula
$N_{ab}^c = \sum\limits_{x \in \mathcal{B}} \frac{S_{ax} S_{bx} S_{cx}^*}{S_{0x}}$ relating the fusion rules to the modular $S$ matrix.

Non-orientable surfaces can be obtained by gluing together 3-punctured spheres together with M\"obius bands, 
i.e. one-punctured real projective planes. We derive a formula
\begin{align}
\label{Maformula}
M_a = \sum_{x \in \mathcal{B}^{\bf rc}} \eta^{\bf r}_x S_{xa}
\end{align}
for the ground state degeneracy of a topological phase on a M\"obius band whose boundary carries topological
charge $a$. Again, this can be interpreted as
\begin{align}
M_a = \Z_a(\mathbb{RP}^2 \times S^1),
\end{align}
where $\Z_a(\mathbb{RP}^2 \times S^1)$ is the path integral on the manifold $\mathbb{RP}^2 \times S^1$ containing a loop of topological charge $a$ encircling $S^1$ at a given point on $\mathbb{RP}^2$. As before, if the branch cut of $\mathbb{RP}^2$ is associated with time-reversal symmetry, the expression is modified to
\begin{align}
\Z_a(\mathbb{RP}^2 \times S^1) = \sum_{x \in \mathcal{B}^{\bf T}} \eta^{\bf T}_x S_{xa}
.
\end{align}

We note that the existence of the Hilbert space on $\mathbb{RP}^2$ (possibly with punctures) marks a departure
from some usual notions of TQFTs. Any oriented surface $\Sigma^2$ can bound a 3-manifold $M^3$, and
thus states on $\Sigma^2$ can be obtained from the path integral on $M^3$. However, non-orientable surfaces
with odd Euler characteristic, such as $\mathbb{RP}^2$, do not bound any 3-manifold. Therefore, states on
$\mathbb{RP}^2$ cannot be obtained by evaluation of a path integral on any 3-manifold. Nevertheless,
the Hilbert space can be defined for theories with reflection symmetry through a Hamiltonian construction,
which we demonstrate for a class of examples in Sec.~\ref{exampleSec} .

\subsubsection{Anomaly detection}

The study of time-reversal and reflection SET phases on non-orientable manifolds leads naturally to an understanding
of anomalous SET phases, as we discuss in Sec.~\ref{anomalySec}. We start by considering a Dehn twist along the
boundary of a M\"obius band (i.e. around a puncture on $\mathbb{RP}^2$ with a puncture). If the puncture carries 
topological charge $a$, then the Dehn twist yields a phase $\theta_a$ corresponding to the topological twist of $a$.
On the other hand, this Dehn twist is actually a trivial operation in the mapping class group of the M\"obius band.
This leads us to the requirement that
\begin{align}
\label{dehnAnomalyEq}
M_a > 0  \quad \Rightarrow \quad \theta_a = 1 ,
\end{align}
which must be satisfied by all (2+1)D time-reversal and/or reflection SET phases. Failure to satisfy the above condition indicates that the
SET is anomalous. Indeed, one can verify for various known anomalous SET phases, such as the eTmT
state~\cite{vishwanath2013,wang2013}, that Eq.~(\ref{dehnAnomalyEq}) is violated.

Eq.~(\ref{dehnAnomalyEq}) corresponds to whether a given SET phase has a ``Dehn twist'' anomaly. Given a
time-reversal and/or reflection SET phase, one can further ask whether a (3+1)D SPT state can host the (2+1)D SET phase
as a possible surface termination and, if so, how to determine the SPT state. (3+1)D SPT states with time-reversal or reflection symmetry
are believed to have a $\mathbb{Z}_2 \times \mathbb{Z}_2$ classification~\cite{vishwanath2013, kapustin2014},
and are completely characterized by the value of their topological path integrals on $\mathbb{RP}^4$ and $\mathbb{CP}^2$:~\cite{kapustin2014} 
$\mathcal{Z}(\mathbb{RP}^4) = \pm 1$ and $\mathcal{Z}(\mathbb{CP}^2) = \pm 1$. 
Another main result of our paper is to provide an explicit connection between the properties of anomalous 
SET phases and the path integrals on $\mathbb{RP}^4$ and $\mathbb{CP}^2$.

More concretely, given $\mathcal{B}, \R$, and $\{ \eta^{\bf r}_a\}$ as (part of) the data that characterizes the SET phase,
we can define a (3+1)D TQFT. This (3+1)D TQFT corresponds to a (3+1)D SPT state whose (2+1)D boundary
hosts an SET phase. The topological order of this SET phase is described by $\mathcal{B}$, and its reflection symmetry fractionalization is described by
$\{ \eta^{\bf r}_a\}$. In Sec.~\ref{anomalySec}, we compute the path integral of this (3+1)D SPT state on $\mathbb{RP}^4$ and $\mathbb{CP}^2$,
and find
\begin{align}
\Z(\mathbb{RP}^4) &= \frac{1}{\mathcal{D}} \sum_{a \in \mathcal{B}^{\bf rc}} d_a \theta_a \eta^{\bf r}_a ,
\nonumber \\
\Z(\mathbb{CP}^2 ) &= \frac{1}{\mathcal{D}} \sum_{a \in \mathcal{B}} d_a^2 \theta_a = e^{i \frac{2\pi }{8}c_{-}}.
\end{align}
As previously discussed, for time-reversal symmetry the expression is modified to
\begin{align}
\Z(\mathbb{RP}^4) &= \frac{1}{\mathcal{D}} \sum_{a \in \mathcal{B}^{\bf T}} d_a \theta_a \eta^{\bf T}_a .
\end{align}

If $\Z(\mathbb{RP}^4) = \Z(\mathbb{CP}^2 ) = 1$, then the bulk (3+1)D SPT state is trivial,
which indicates that the given (2+1)D SET phase is non-anomalous and can exist in purely (2+1)D. On the other hand, as discussed in
Ref.~\onlinecite{kapustin2014}, the (3+1)D SPT state with $\Z(\mathbb{RP}^4) = -1$ and $\Z(\mathbb{CP}^2 ) = 1$
is the (3+1)D time-reversal and/or reflection SPT state within the group cohomology construction of Ref.~\onlinecite{chen2013}.
The (3+1)D SPT states with $\Z(\mathbb{CP}^2 ) = -1$ include the ``beyond group cohomology'' state first discussed in Ref.~\onlinecite{vishwanath2013}.

Finally, we find that the Dehn twist anomaly can be converted into a quantity that takes values $\pm 1$
depending on whether there is an anomaly, and which can be related to $\Z(\mathbb{RP}^4)$ and
$\Z(\mathbb{CP}^2 )$. In particular, we derive the identity
\begin{align}
\frac{1}{ |\mathcal{B}^{\bf rc}|} \sum_{a \in \mathcal{B}} \theta_a M_{a}^{2}
	= \Z(\mathbb{RP}^4) \Z(\mathbb{CP}^2) ,
\end{align}
where $|\mathcal{B}^{\bf rc}|$ is the number of topological charges in $\mathcal{B}^{\bf rc}$.

\subsubsection{Examples}

In Sec.~\ref{exampleSec}, we study a number of explicit examples illustrating various aspects of the general theoretical discussion.

We provide a general discussion of reflection and time-reversal symmetry in the Kitaev quantum
double Hamiltonians based on a finite group $\G$.~\cite{kitaev2003} The quasiparticles of the corresponding topological phase are described by a UMTC given by the
quantum double $\D(\G)$. The quasiparticle types are labeled by two objects, $(C, \pi)$, where $C$ is a conjugacy class
and $\pi$ is an irreducible representation of the centralizer of $C$. We demonstrate that, in these models, the sets of ${\bf rc}$-invariant topological charges and ${\bf T}$-invariant topological charges are equal, i.e. $\mathcal{B}^{\bf r c} = \mathcal{B}^{\bf T}$. Furthermore, we demonstrate that the symmetry fractionalization quantum
numbers
\begin{align}
\eta^{\bf r}_{(C, \pi)} = \eta^{\bf T}_{(C, \pi)} = \nu_\pi,
\end{align}
where $\nu_\pi$ is the Frobenius-Schur indicator of the representation $\pi$.
We show that a simple example where $\eta^{\bf r}_{a} = \eta^{\bf T}_{a}$ are nontrivial for these models arises in the
case where $\G = Q_8$, the quaternion group with 8 elements. $Q_8$ can be thought of as the subgroup of
SU$(2)$ generated by $\pi$ spin rotations.

The topological order associated with $\D(\G)$ can also be obtained from a Levin-Wen Hamiltonian,\cite{levin2005} where the input
is the fusion category formed by the irreducible representations of $\G$, denoted $\text{Rep}(\G)$. We demonstrate via a computation of $\mathcal{Z}(\mathbb{RP}^2 \times S^1)$ that, for the case of $\G = Q_8$, the Kitaev quantum double model and Levin-Wen model realize distinct reflection and time-reversal SET phases,
even though both have the same intrinsic topological order.

We further focus more specifically on the Kitaev $\mathbb{Z}_N$ toric code model and analyze the model on a lattice discretization
of $\mathbb{RP}^2$. We use this to compute the ground state degeneracies and verify our general formulae. We systematically
solve for $\mathbb{Z}_2$ extensions of the fusion category associated with $\mathbb{Z}_N$, which classifies the different
possible time-reversal and reflection SET phases given by our state sum construction.

We then provide a solvable Hamiltonian realizing a version of the $\mathbb{Z}_2$ toric code state with nontrivial
$\eta^{\bf r}_e$ for the ``electric'' quasiparticle (labeled $e$). This is closely related to a class of 
time-reversal symmetric Hamiltonians presented in Ref.~\onlinecite{zion2016}. We study our Hamiltonian on a discretization of $\mathbb{RP}^2$. This allows us to make precise an argument presented
in Sec.~\ref{sec:loop-gas} for Eq.~(\ref{Maformula}) based on loop gases. This loop gas argument provides a simple physical picture for how to
understand Eq.~(\ref{Maformula}) in the case of Abelian topological phases.

We study a number of anomalous SET phases within our framework. We discuss how certain choices of symmetry fractionalization
quantum numbers $\{\eta^{\bf r}_a\}$ and $\{\eta^{\bf T}_a\}$ for the $\mathbb{Z}_N$ toric code
are anomalous. This provides a class of anomalous SET phases for $\mathbb{Z}_N$, which generalizes the anomalous eTmT state known to
arise in the case $N = 2$. We explicitly compute $\mathcal{Z}(\mathbb{RP}^4) = -1$ and $\mathcal{Z}(\mathbb{CP}^2) = 1$
for the eTmT state and its $\mathbb{Z}_N$ generalizations.

We demonstrate within our framework that the efmf state (the anyon content of which is
equivalent to that of SO$(8)_1$ Chern-Simons theory), studied in Ref.~\onlinecite{vishwanath2013} and \onlinecite{wang2013}, has
$\mathcal{Z}(\mathbb{CP}^2) = \mathcal{Z}(\mathbb{RP}^4) = -1$.

We find that the gauged T-Pfaffian$_{\pm}$ state, discussed in Refs.~\onlinecite{Bonderson13d} and \onlinecite{chen2014b}, has
$\mathcal{Z}(\mathbb{RP}^4) = \pm 1$ and $\mathcal{Z}(\mathbb{CP}^2) = 1$, agreeing with the analysis of Ref.~\onlinecite{metlitski2015}.

Finally, using our framework, we discover an intriguing new anomalous SET phase associated with the quantum double $\D(S_3)$,
where $S_3$ is the permutation group on three elements. The topological charges of $\D(S_3)$ can be labeled $\{ A, B, C , \cdots, H\}$, with $A$ being the vacuum topological charge (often denoted $0$). We can consider a certain action of time-reversal symmetry that permutes the topological charges as follows:
\begin{align}
\,^{\bf T}C = F, \qquad ^{\bf T} G= H.
\end{align}
The permutation of $C$ and $F$ here is unconventional, in the sense that time-reversal acting on a realization of $\D(S_3)$ by Kitaev's quantum double model would only interchange $G$ and $H$. We show that, through a highly subtle constraint on the properties of time-reversal symmetry fractionalization,
such a permutation of topological charges under time-reversal symmetry fixes the value of $\eta^\mb{T}_{B}$ for the unique nontrivial Abelian anyon $B$ to
\begin{align}
\eta^{\bf T}_{B} = -1.
\end{align}
This must be satisfied even if the corresponding SET phase exists at the (2+1)D surface of a (3+1)D SPT state. We further show that the
case where $\eta^{\bf T}_D = -1$ is anomalous and the corresponding SET phase must therefore exist at the surface
of a (3+1)D SPT state, while the case where $\eta^{\bf T}_D = 1$ is non-anomalous and can exist in purely (2+1)D. In fact, we can construct this non-anomalous (2+1)D SET phase with our state sum model. We note that the above results hold also when time-reversal is replaced by spatial reflection symmetry.

%% file: 1dSPT.tex
\section{Review of (1+1)D spatial reflection or time-reversal SPT states}
\label{1dSPT}

In (1+1)D, SPT states with symmetry group $G$ are classified by the second cohomology group
$\mathcal{H}^{2}_{\sigma}(G, \text{U}(1))$~\cite{pollmann2010,ChenPRB2011a, ChenPRB2011b, fidkowski2011,schuch2011,kapustin2015}.
This classifies the projective representations of $G$, which for
on-site symmetries physically corresponds to the zero-energy edge states of the system.
For time-reversal symmetry, the symmetry group is $G=\mathbb{Z}_2$ and the nontrivial group element, which we denote as ${\bf T}$, acts on the U$(1)$ elements with complex conjugation. This is encoded by the symmetry action $\sigma$, which we generally define to be
\begin{equation}
\label{eq:G_action_sigma}
\sigma({\bf g}) = \left\{
\begin{array}{lll}
\openone & \quad  & \text{if } {\bf g} \text{ is space-time parity even} \\
\ast  & \quad  & \text{if } {\bf g} \text{  is space-time parity odd}
\end{array}
\right.
\end{equation}
In other words, $\sigma({\bf g})$ acts as complex conjugation when ${\bf g}$ is a space-reflecting or time-reversing symmetry, and otherwise acts trivially.
The time-reversal symmetry group is sometimes denoted as $\mathbb{Z}_2^{\bf T}$ to convey the distinction of its group action. We use similar notation with ${\bf r}$ and $\mathbb{Z}_2^{\bf r}$ for spatial reflection symmetry. We further use the notation $\mathbb{Z}_2^{\bf T} = \{ {\bf 0}, {\bf T}\}$ and $\mathbb{Z}_2^{\bf r} = \{ {\bf 0}, {\bf r}\}$, where ${\bf 0}$
indicates the identity element, and therefore ${\bf T}^2 = {\bf 0}$, ${\bf r}^2 = 0$. 

We note that in general one could consider situations where the space-time parity odd symmetries of a system are
not order 2, but rather order 4 or higher. For example, a spin system may not be invariant under reflection symmetry,
but only under reflection symmetry combined with a $\pi/2$ spin rotation; in this case, one has to apply the symmetry
four times to reach the identity. Unless stated otherwise, the spatial reflection ${\bf r}$ and time-reversal ${\bf T}$ symmetries
discussed in this paper are order two.

As was shown in Refs.~\onlinecite{pollmann2010, ChenPRB2011b, kapustin2015}, the group cohomology
classification for reflection is the same as for time-reversal, in the sense that the
$G=\mathbb{Z}_2^{\bf r}$ reflection symmetry also acts on the U$(1)$ elements with the group action $\sigma$. Therefore, $\mathcal{H}^2_{\sigma}(G, \text{U}(1)) = \mathbb{Z}_2$ gives the classification of (1+1)D reflection SPT states and time-reversal SPT states into ``trivial'' and ``nontrivial'' phases.

In this paper, we will utilize two important features of (1+1)D time-reversal or reflection symmetric SPT states:
the value of the topological path integral on the real projective plane $\mathbb{RP}^2$, and the eigenvalue of
the reflection operator on (1+1)D systems with boundaries. In this section, we provide a brief discussion of these
two features of (1+1)D SPT states.
%While the results of this section are mainly review, to our knowledge the specific
%computation of the topological path integral on $\mathbb{RP}^2$ is new.

\subsection{Path integrals on $\mathbb{RP}^2$}
\label{sec:1dspt-pi}

As described in Refs.~\onlinecite{chen2013,kapustin2014}, SPT states are generally characterized
by a purely topological path integral. The two topologically distinct (1+1)D space-time reflection invariant SPT
states can be distinguished by the value of their topological path integral on the real projective
plane $\mathbb{RP}^2$. In particular, the path integral is~\cite{kapustin2014}
\begin{align}
\label{1dRP2sum}
\mathcal{Z}(\mathbb{RP}^2) = \left\{ \begin{array}{rl}
        1 & \mbox{ for a trivial SPT state}\\
        -1 & \mbox{ for a nontrivial SPT state}\end{array} \right.
.
\end{align}
In the following, we derive this result via an explicit computation of the topological path integral by extending the
approach of Ref.~\onlinecite{chen2013} to the case of non-orientable manifolds. A computation of
$\mathcal{Z}(\mathbb{RP}^2)$ using matrix product states can also be found in Ref.~\onlinecite{ShiozakiMPS}.

First, the space-time manifold $\Sigma^2$ is discretized into a triangulation. (A triangulation technically requires that two different triangles be glued together along only one edge, whereas more general
cell decompositions are referred to as ``cellulations.'' The arguments used in this paper to establish topological state sums require the use of triangulations of the manifold, but more general arguments may be used to ensure the same results hold for general cellulations.) We denote the set of $n$-cells of the triangulation by $\mathcal{I}_{n}$ and provide an ordering to the elements of $\mathcal{I}_{0}$, i.e. we assign the labels $j=0,1,\ldots,N_{v}-1$ to the vertices (0-cells) of the triangulation, where $N_{v}$ is the number of vertices in the triangulation. The ordering of vertices provides a direction to each edge (1-cell), pointing from the lower ordered endpoint of the edge to the higher one, and an orientation to each triangle (2-cell) $\Delta^2\in \mathcal{I}_{2}$. The ordering of vertices defines a ``branching structure'' on the triangulation. A branching
structure can be thought of as an orientation of all edges (1-simplices) of the triangulation such that oriented edges of each triangle (2-simplex) do not form a closed loop for any triangle. For general cellulations, a global ordering of the vertices does not imply a branching structure, while the state sum construction applied to more general cellulations requires the branching structure.

\begin{figure}[t!]
\includegraphics[width=1.8in]{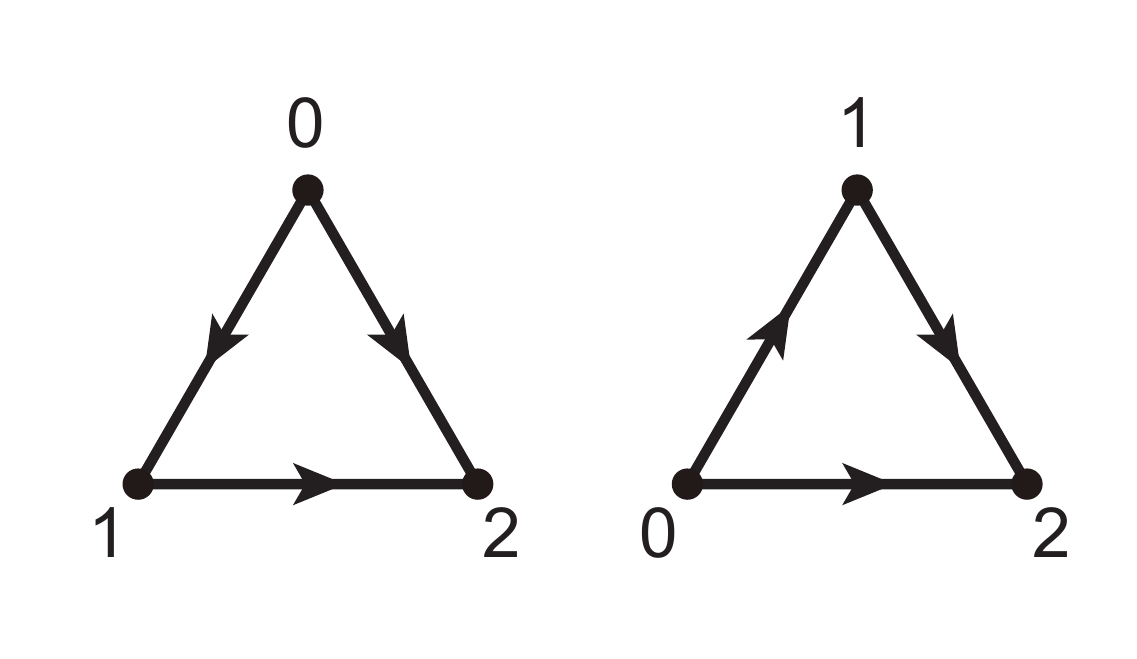}
\caption{Two triangles with opposite orientations. A triangle with positive orientation (right) is assigned $s(\Delta^2)=\openone$. A triangle with negative orientation (left) is assigned $s(\Delta^2)=\ast$.
}
\label{triangles2}
\end{figure}

We can define the relative orientation $s(\Delta^2)$ of triangles by thinking of the surface as being constructed from an oriented surface with boundary, e.g. a disk, by gluing together segments of the boundary, possibly with twists that spoil the global orientation. Then we use an appropriate triangulation of the oriented surface with boundary to define a triangulation of the surface in question, by identifying vertices and edges appropriately. The orientation of the oriented surface then provides a reference orientation for the 2-cells of the triangulation. When the order of the vertices of a triangle $\Delta^2$ agree with the reference orientation, we set $s(\Delta^2)=1$; when the order of the vertices of a triangle $\Delta^2$ is opposite that of the reference orientation, we set $s(\Delta^2)=\ast$. Fig.~\ref{triangles2} shows two triangles with oppositely ordered vertices and hence opposite orientations.

We next assign a group element ${\bf g}_{j} \in G$ to the $j$th vertex of the triangulation. Then, for each configuration $\{{\bf g}_j\}$ on the vertices, we associate a U$(1)$ phase to each triangle $\Delta^2\in \mathcal{I}_{2}$, given by
\begin{equation}
\mathcal{Z}({\Delta^2} ; \{{\bf g}_j\}) = \nu_2({\bf g}_{j_{0}}, {\bf g}_{j_{1}}, {\bf g}_{j_{2}})^{s(\Delta^2)}
,
\end{equation}
where the vertices $j_0,j_1$ and $j_2$ of $\Delta^2$ have been ordered to respect the vertex numbering order, i.e. ${j_{0}}< {j_{1}} < {j_{2}}$.
The path integral is then given by
\begin{align}
\mathcal{Z}(\Sigma^2) = \frac{1}{|G|^{N_v}} \sum_{\{{\bf g}_j\}} \prod_{\Delta^2 \in \mathcal{I}_{2}} \mathcal{Z}({\Delta^2} ; \{{\bf g}_j\}) ,
\end{align}
where $|G|$ is the number of group elements in $G$.

\begin{figure}[t!]
\includegraphics[width=1.8in]{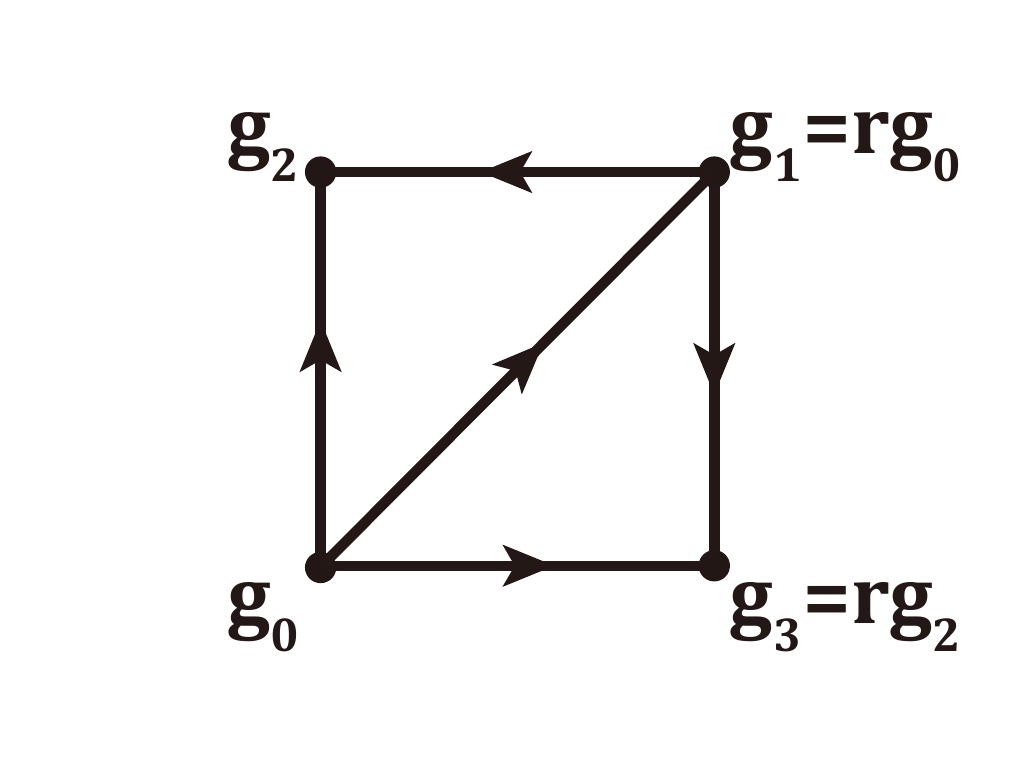}
\caption{
A cellulation of $\mathbb{RP}^2$ with two vertices labeled by group elements. $\mathbb{RP}^2$ is obtained from a square by identifying antipodal points along the boundary, i.e. the top edge is identified with the bottom edge in opposite directions and the left edge is identified with the right edge in opposite directions. When vertices of the triangulation of the square are identified, their corresponding group labels are equated through multiplication by ${\bf r}$, which reflects the orientation reversal involved in identifying antipodal points of the boundary to form $\mathbb{RP}^2$. The relative orientation of the triangles of the cellulation of $\mathbb{RP}^2$ is induced from triangulation of the square.
}
\label{rp2Fig}
\end{figure}

The amplitudes $\nu_2$ satisfy the symmetry property~\cite{chen2013}
\begin{align}
\nu_2({\bf h g}_{j_0}, {\bf h g}_{j_1}, {\bf h g}_{j_2}) = \nu_2({\bf g}_{j_0}, {\bf g}_{j_1}, {\bf g}_{j_2})^{\sigma({\bf h})},
\end{align}
for any ${\bf h}\in G$. The phases $\nu_2$ can be shown to correspond to a U$(1)$ $2$-cocycle with action $\sigma$. Moreover, there is a gauge freedom that physically equates values of $\nu_2$ that differ by a U$(1)$ $2$-coboundary (with action $\sigma$). Thus, a given $\nu_2$ represents a cohomology class in $\mathcal{H}^{2}_{\sigma}(G,\text{U}(1))$.

In order to compute the topological path integral on $\mathbb{RP}^2$, we use a cellulation, rather than a triangulation, since it allows for a more efficient discretization of the manifold. (Cellulations can generally provide more efficient discretizations of manifolds than triangulations, so we will use them when performing explicit computations.) Recall that $\mathbb{RP}^2$ can be viewed topologically as a disk with antipodal points along the boundary of the disk identified. As shown in Fig.~\ref{rp2Fig}, we start with the triangulation of a disk (topologically equivalent to a square) using four vertices, five edges, and two triangles. The integer labeling of vertices defines the relative orientation of the triangles. Identifying antipodal points of the boundary of the disk so that opposite edges of the square are glued together with an orientation reversal yields $\mathbb{RP}^2$ with a cellulation that contains two vertices, three edges, and two triangles with an induced relative orientation. When identifying vertices along the boundary of the disk, the corresponding group elements are equated through multiplication by ${\bf r}$. This reflects the fact that we are placing an ``orientation reversing branch cut in space-time'' and that the group elements at each vertex keep track of the local orientation of space-time. Thus, the state sum is given by
\begin{widetext}
\begin{eqnarray}
\mathcal{Z}(\mathbb{RP}^2) &=& \frac{1}{2^2} \sum_{{\bf g}_0, {\bf g}_1, {\bf g}_2, {\bf g}_3= {\bf 0}, {\bf r}} \nu_2({\bf g}_0,  {\bf g}_1, {\bf g}_2)^{\ast} \nu_2 ({\bf g}_0, {\bf g}_1, {\bf g}_3) \delta_{{\bf g}_1 , {\bf r g}_0} \delta_{{\bf g}_3 , {\bf r g}_2} \notag \\
&=& \frac{1}{4} \sum_{{\bf g}_0, {\bf g}_2 = {\bf 0}, {\bf r}} \nu_2({\bf g}_0,  {\bf r g}_0, {\bf g}_2)^{\ast} \nu_2 ({\bf g}_0, {\bf r g}_0, {\bf r g}_2)
.
\label{eq:RP2_statesum}
\end{eqnarray}
\end{widetext}
In the trivial $\mathbb{Z}_2^{\bf r}$ SPT state, we can choose a gauge in which $\nu_2({\bf g}, {\bf h}, {\bf k}) = 1$ for all ${\bf g}, {\bf h}, {\bf k} \in \mathbb{Z}_2^{\bf r}$. For the nontrivial
$\mathbb{Z}_2^{\bf r}$ SPT state, we can choose a gauge in which $\nu_2({\bf g}, {\bf h}, {\bf k}) = 1$ except for
$\nu_2({\bf r}, {\bf 0}, {\bf r}) = \nu_2({\bf 0}, {\bf r}, {\bf 0}) = -1$.~\cite{chen2013} Using these values of $\nu_2$ in Eq.~(\ref{eq:RP2_statesum}), we find Eq.~(\ref{1dRP2sum}).

\subsection{Reflection eigenvalue on systems with boundary}
\label{1dSPT_reflection_boundary}

\begin{figure}[t!]
\includegraphics[width=3.0in]{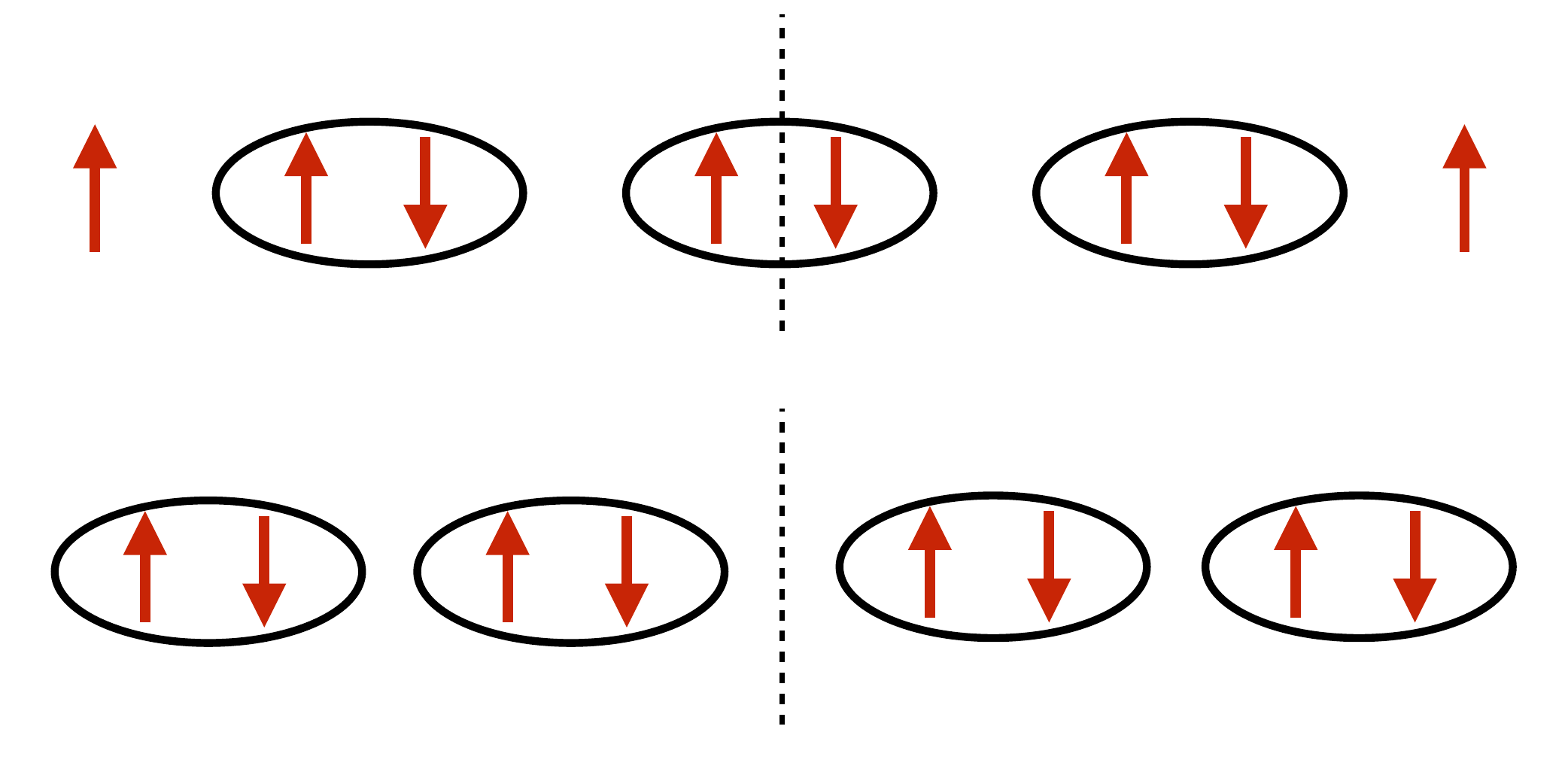}
\caption{(1+1)D SPT states. Ellipses enclosing neighboring spins indicate that the enclosed pair of spins form a spin singlet state:
$\frac{1}{\sqrt{2}} ( |\uparrow \downarrow\rangle - |\downarrow \uparrow\rangle)$.
Top: The ``nontrivial'' (1+1)D SPT state, analogous to the AKLT state describing the Haldane chain. Bottom: The ``trivial'' (1+1)D SPT state. Bond-centered
reflection corresponds to a reflection about the dashed line.}
\label{fig:aklt}
\end{figure}

An important property of reflection SPTs, which we will utilize in subsequent sections, are the
reflection matrix elements on (1+1)D systems with boundary.~\cite{pollmann2012, PollmannPRB2012}
As a simple example, let us consider a simple limit of the Haldane chain, which can be considered to correspond to the case where
neighboring spin-1/2 degrees of freedom pair into a spin singlet (see Fig.~\ref{fig:aklt}). We take the reflection symmetry to correspond
to a site-centered reflection. On a chain with an even number of sites, the two distinct SPT states correspond to whether there is a
singlet across the reflection axis. In this limiting case, there also appears to be unpaired, dangling spin-1/2 edge modes, though these are not protected by symmetry, as they can be gapped out without breaking the reflection symmetry, e.g. by turning on the same Zeeman fields locally on the two ends. For the
nontrivial SPT phase, clearly the trace over all ground states of a bond-centered reflection gives $-1$, that is
\begin{align}
\frac{1}{2}\sum_{\alpha,\beta = \uparrow, \downarrow} \langle \Psi_{0;\alpha\beta} | R_{\R} |\Psi_{0;\alpha\beta} \rangle = -1 ,
\end{align}
where $R_{\R}$ is the operator representing the action of $\R$ on the physical Hilbert space and  $|\Psi_{0; \alpha \beta}\rangle$ for $\alpha,\beta = \uparrow, \downarrow$
corresponds to the ground state of the nontrivial SPT state, with $\alpha,\beta$ describing the spin state
of the edge zero modes. On the other hand, the trivial SPT state has a unique ground state
on the open chain with a $+1$ eigenvalue for $R_{\R}$.

The above discussion can be made more general. Let us consider a reflection symmetric Hamiltonian
$H$, acting in a Hilbert space $\mathcal{H}$ which consists of an open chain with $N_s$ sites.
(As usual we consider the limit where $N_s$ is large compared to the correlation length, in units of the lattice spacing,
of the ground state). Let $\mathcal{H}_0$ denote the space of ground states of $H$. In general, even when $H$ corresponds to a
gapped SPT phase, $\mathcal{H}_0$ can be multi-dimensional due to the possible existence of boundary zero modes.
Let us define $\Pi_{\mathcal{H}_0}$ to be the projector onto $\mathcal{H}_0$.
Then, a fundamental property of (1+1)D reflection SPT states is the relation
\begin{align}
\label{Rvalue}
\frac{ \text{Tr}[ \Pi_{\mathcal{H}_0}  R_{\R} ]} { | \text{Tr}[ \Pi_{\mathcal{H}_0} R_{\R} ] | } = \pm 1 .
\end{align}
The value $\pm 1$ in the RHS of the above equation is a topological invariant characterizing the reflection SPT class.
In Appendix \ref{ReigenvalueAppendix}, we will provide the justification of this in general using the framework of
matrix product states.
%Let us consider a reflection symmetric Hamiltonian $H[\lambda]$, acting in a Hilbert space $\mathcal{H}$. We consider $H[\lambda]$ such that
%as the real parameter $\lambda$ is tuned, $H[\lambda]$ tunes between the two possible reflection SPT phases.
%Let $\mathcal{H}_0[\lambda]$ denote the space of ground states of $H[\lambda]$ on an open chain with
%$N_s$ sites. In general, even when $H[\lambda]$ corresponds to a gapped SPT phase, $\mathcal{H}_0[\lambda]$
%can be multi-dimensional due to the existence of boundary zero modes. Let us define $\Pi_{\mathcal{H}_0[\lambda]}$ to be the projector onto the ground state subspace of $H[\lambda]$.
%Then, a fundamental property of (1+1)D reflection SPT states is the relation
%\begin{align}
%\label{Rvalue}
%\frac{ \text{Tr}[ \Pi_{\mathcal{H}_0[\lambda_1]}  R_{\R} ]} { | \text{Tr}[ \Pi_{\mathcal{H}_0[\lambda_1]} R_{\R} ] | }
%\left(\frac{\text{Tr}[ \Pi_{\mathcal{H}_0[\lambda_2]} R_{\R} ] }{|\text{Tr}[ \Pi_{\mathcal{H}_0[\lambda_2]} R_{\R} ] |}\right)^{-1} = -1
%,
%\end{align}
%where $\lambda_1$ and $\lambda_2$ are such that the ground states of $H[\lambda_1]$ and $H[\lambda_2]$ respectively correspond to two
%distinct reflection SPT phases.

%% file: symfrac.tex
\section{Symmetry fractionalization for reflection and time-reversal symmetries}
\label{symfrac}

In this section, we discuss aspects of symmetry fractionalization, with a particular emphasis on
time-reversal and reflection symmetry. Sections~\ref{topsym} - \ref{globalsym} provide
a brief review of the discussion presented in Ref.~\onlinecite{barkeshli2014SDG}, together with an extension of the machinery to space reflecting symmetries.

\subsection{Topological symmetry and braided auto-equivalence}
\label{topsym}

Let $\mathcal{B}$ be a unitary modular tensor category (UMTC). An important property of $\mathcal{B}$ is the group of ``topological symmetries,''
which are sometimes referred to as ``braided auto-equivalences'' in the mathematical literature. They are associated
with the internal symmetries of the emergent TQFT described $\mathcal{B}$, irrespective of any microscopic global symmetries of a quantum system in which the TQFT emerges.

The topological symmetries consists of the invertible maps
\begin{align}
\varphi: \mathcal{B} \rightarrow \mathcal{B}
.
\end{align}
Importantly, the different $\varphi$, modulo equivalences known as natural transformations,
form a group, which we denote as $\text{Aut}(\mathcal{B})$.

The symmetry maps can be classified according to a $\mathbb{Z}_2 \times \mathbb{Z}_2$ grading defined by the homomorphisms
\begin{equation}
q\left( \varphi \right) = \left\{
\begin{array}{lll}
0  & & \text{ if $\varphi$ is not time-reversing} \\
1  & & \text{ if $\varphi$ is time-reversing}
\end{array}
\right.
,
\end{equation}
\begin{equation}
p\left( \varphi \right) = \left\{
\begin{array}{lll}
0  & & \text{ if $\varphi$ is spatial parity even} \\
1  & & \text{ if $\varphi$ is spatial parity odd}
\end{array}
\right.
,
\end{equation}
from $\text{Aut}(\mathcal{B})$ to  $\mathbb{Z}_2$, where spatial parity even or odd means the map involves a reflection in an even or
odd number of spatial directions, respectively. We note that $\varphi$ is unitary iff $q\left( \varphi \right)=0$ and it is anti-unitary
iff $q\left( \varphi \right)=1$. In this way, we can write the topological symmetry group as
\begin{align}
\text{Aut}(\mathcal{B}) = \bigsqcup_{q,p=0,1} \text{Aut}_{q,p}(\mathcal{B}) .
\end{align}
The $\text{Aut}_{0,0}(\mathcal{B})$ is the subgroup corresponding to the topological symmetries that are unitary and space-time parity even, which is referred to in the mathematics literature as the group of ``braided auto-equivalences'' (where it is instead denoted as $\text{Aut}(\mathcal{B})$). The generalization involving reflection and time-reversal symmetries appears to be beyond what has been considered in the mathematics literature to date.

It is convenient to also define
\begin{equation}
\sigma(\varphi) = \left\{
\begin{array}{lll}
\openone & \quad  & \text{if } \varphi \text{ is space-time parity even} \\
\ast  & \quad  & \text{if } \varphi \text{  is space-time parity odd}
\end{array}
\right.
,
\end{equation}
where space-time parity even or odd means the symmetry involves a reflection of an even or odd number of dimensions of space-time, respectively;
that is, $\sigma(\varphi) = \ast^{q\left( \varphi \right)+p\left( \varphi \right)}$. We can think of $\sigma$ as another $\mathbb{Z}_2$ grading
where $\openone$ is the identity element of $\mathbb{Z}_2$ and $\ast$ is the nontrivial element of $\mathbb{Z}_2$.

The topological symmetries may permute the topological charges while mapping $\mathcal{B}$ to itself. Denoting this permutation of topological charges as
\begin{equation}
\varphi(a) = a' \in \mathcal{B},
\end{equation}
the invariant quantities of $\mathcal{B}$ must satisfy
\begin{align}
\label{topsymEq}
N_{a' b'}^{c'} &= N_{ab}^c,
\nonumber \\
S_{a'b'} &= S_{ab}^{\sigma(\varphi)},
\nonumber \\
\theta_{a'} &= \theta_a^{\sigma(\varphi) } ,
\end{align}
where $N_{ab}^c$, $S$, and $\theta_a$ are the fusion coefficients, modular $S$-matrix, and topological twist factors of the UMTC $\mathcal{B}$.
Notice, in particular, that these equations above require $\varphi$ to leave the vacuum invariant $0' = 0$ and to commute with topological charge conjugation, namely $\varphi(\bar{a}) = \overline{\varphi(a)}$ for all topological charges $a\in \mathcal{B}$.

The maps $\varphi$ also have an action on the fusion and splitting state spaces of $\mathcal{B}$, as well as the $F$-symbols and $R$-symbols of $\mathcal{B}$. This is described in detail for $p=0$ in Ref.~\onlinecite{barkeshli2014SDG}. The case of $p=1$ was briefly discussed in Ref.~\onlinecite{barkeshli2014SDG}, but not developed in detail, because of the complication involved in the positions of quasiparticles not being fixed under such symmetries. In the next subsection, we will develop the $p=1$ case further, but only for the situation needed for our purposes.

\subsection{Global symmetry and fractionalization}
\label{globalsym}

Let us now suppose that the microscopic Hamiltonian of the system in question has a global symmetry $G$. The global
symmetry acts on the quasiparticles through the action of a group homomorphism
\begin{align}
[\rho]: G \rightarrow \text{Aut}(\mathcal{B}) .
\end{align}
We will use the notation $[\rho_{\bf g}] \in \text{Aut}(\mathcal{B})$ for a specific element ${\bf g}\in G$. Note that
the brackets indicate the equivalence class of symmetry maps related by natural isomophisms (which do not permute
topological charge values), so $\rho_{\bf g}$ is a representative symmetry map of the equivalence class $[\rho_{\bf g}]$.\cite{barkeshli2014SDG}
We will frequently use the shorthand notation
\begin{equation}
^{\bf g}a \equiv \rho_{\bf g}(a)
.
\end{equation}

We similarly can define a $\mathbb{Z}_2 \times \mathbb{Z}_2$ grading on $G$, given (using the same notation) by
\begin{equation}
q\left( {\bf g} \right) = \left\{
\begin{array}{lll}
0  & & \text{ if ${\bf g}$ is not time-reversing} \\
1  & & \text{ if ${\bf g}$ is time-reversing}
\end{array}
\right.
,
\end{equation}
\begin{equation}
p\left( {\bf g} \right) = \left\{
\begin{array}{lll}
0  & & \text{ if ${\bf g}$ is spatial parity even} \\
1  & & \text{ if ${\bf g}$ is spatial parity odd}
\end{array}
\right.
,
\end{equation}
and require that the action respect the grading, i.e. $q\left( \rho_{\bf g} \right)= q\left( {\bf g} \right)$ and $p\left( \rho_{\bf g} \right)= p\left( {\bf g} \right)$.

The UMTC $\mathcal{B}$ is equipped with fusion spaces $V_{ab}^c$ and their dual (``splitting'') spaces $V_c^{ab}$, respectively,
such that $\text{dim } V_{ab}^c = \text{dim } V_{c}^{ab}  = N_{ab}^c$. One can pick basis states
$|a, b; c,\mu \rangle \in V_{c}^{ab} $, with $\mu = 1, \cdots, N_{ab}^c $. A basis transformation
$|a, b;c, \mu \rangle \rightarrow \sum_{\nu} [\Gamma^{ab}_c]_{\mu\nu} |a, b; c, \nu \rangle$ in this space is referred
to as a ``vertex basis gauge transformation.'' As discussed in Ref.~\onlinecite{barkeshli2014SDG},
the symmetry action $\rho_{\bf g}$ acts on these fusion as
\begin{align}
\rho_{\bf g} | a, b; c \rangle = U_{\bf g}(\,^{\bf g}a, \,^{\bf g}b; \,^{\bf g}c) | \,^{\bf g}a, \,^{\bf g}b; \,^{\bf g}c \rangle,
\end{align}
where we have left the additional index $\mu$ implicit. $U_{\bf g}(\,^{\bf g}a, \,^{\bf g}b; \,^{\bf g}c)$ can be chosen
to keep the $F$ and $R$ symbols of $\mathcal{B}$ invariant. The action of $\rho_{\bf g}$ can be modifed by a
``symmetry action gauge transformation,'' which takes
$U_{\bf g}(a, b; c) \rightarrow U_{\bf g}(a, b;c) \frac{\gamma_a({\bf g}) \gamma_b({\bf g})}{\gamma_c({\bf g})}$,
where $\gamma_a({\bf g})$ are $U(1)$ phases. This amounts to choosing a different representative $\check{\rho}_{\bf g}$ in
the same equivalence class: $[\check{\rho}_{\bf g}] = [\rho_{\bf g}]$.

In Ref.~\onlinecite{barkeshli2014SDG} it was further shown that the different types of symmetry fractionalization
were parameterized by elements of the second group cohomology
$\mathcal{H}^2_{[\rho]}(G, \mathcal{A})$, provided a certain obstruction class for $[\rho]$ vanishes.
In particular, different patterns of symmetry fractionalization can be related to each other through
the operation of elements of $\mathcal{H}^2_{[\rho]}(G, \mathcal{A})$. Here, $\mathcal{A}$ is an Abelian group
whose elements are the subset of Abelian topological charges of $\mathcal{B}$ with group multiplication given
by their fusion rules in $\mathcal{B}$. The action $[\rho]$ on an $\mathcal{A}$-cochain $\coho{w}$ is given by
$[\rho]: \coho{w} \mapsto \rho_{\bf g}(\coho{w})$ for $p({\bf g})=0$ and $[\rho]: \coho{w} \mapsto \rho_{\bf g}(\bar{\coho{w}})$ for $p({\bf g})=1$.~\cite{barkeshli2014SDG}

In the case where ${\bf T} \in G$ corresponds to a time-reversal symmetry, i.e. ${\bf T}^2 = {\bf 0}$, $q({\bf T}) = 1$, and $p({\bf T}) = 0$,
Refs.~\onlinecite{Bonderson13d} and \onlinecite{barkeshli2014SDG} defined the local projective phase $\eta_a({\bf T}, {\bf T})$ associated with topological charge $a$, which characterizes the symmetry fractionalization class. Moreover, it was shown that
when a topological charge $a$ is invariant under ${\bf T}$, i.e. $^{\bf T}a = a$, this phase is a gauge invariant quantity that takes the values
\begin{align}
\label{tsymfrac}
\eta_a^\mb{T} \equiv \eta_a({\bf T}, {\bf T}) = \pm 1 ,
\end{align}
which can be interpreted as the ``local ${\bf T}^2$'' value of $a$. As such, $\eta_a^\mb{T}$ indicates whether quasiparticles with ${\bf T}$-invariant topological
charge $a$ carry a local Kramers degeneracy associated with the anti-unitary action of $\mb{T}$. Note that we always have $\eta_a^\mb{T}=1$

\subsubsection{Reflection symmetry action and fractionalization}

\begin{figure}[t!]
	\centering
	\includegraphics[width=0.5\columnwidth]{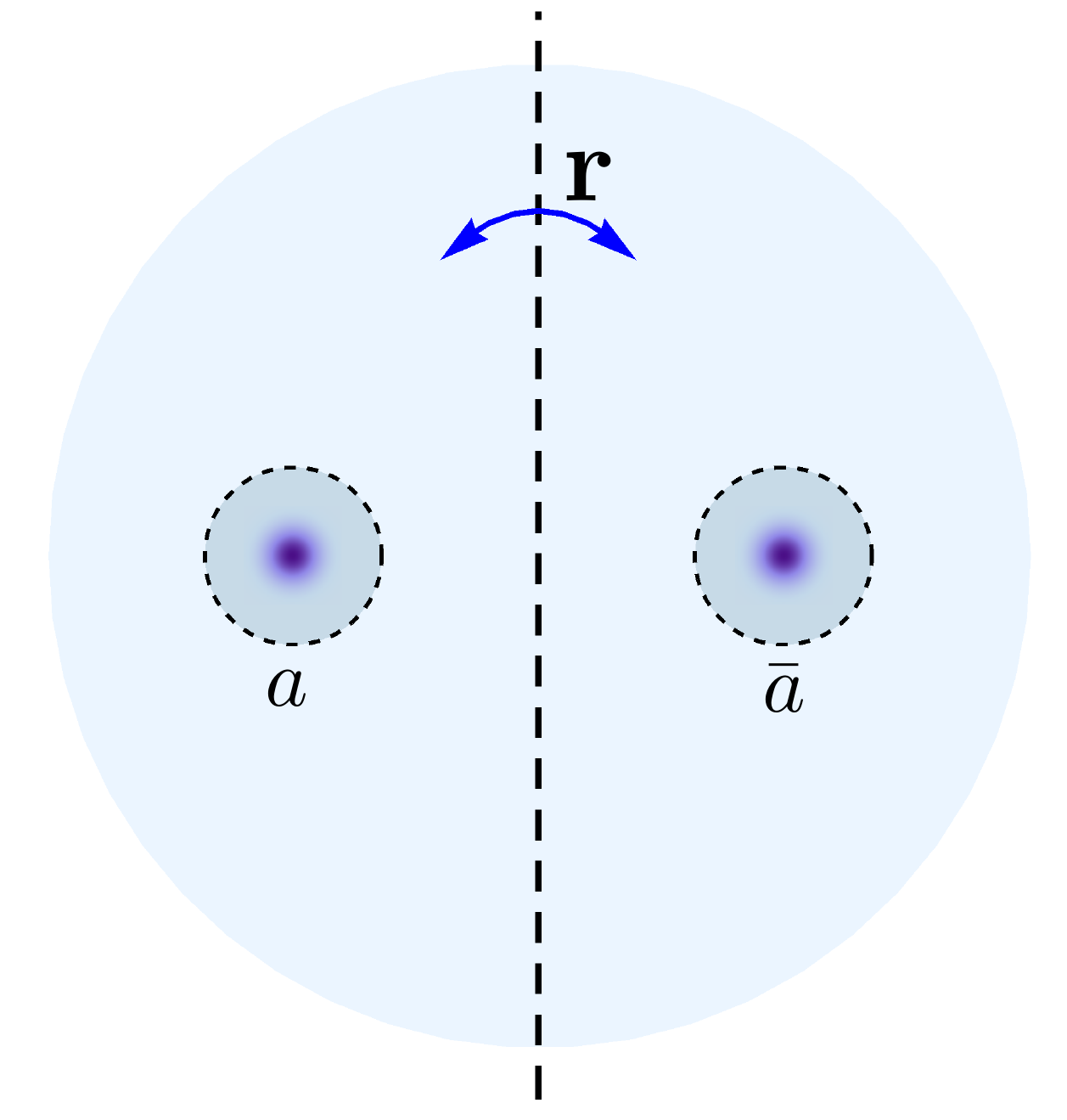}
	\caption{A pair of anyons $a$ and $\bar{a}$ created from vacuum, located in regions that are related by the reflection ${\bf r}$.}
	\label{fig:anyon-pair}
\end{figure}

In the case where ${\bf r} \in G$ corresponds to a spatial reflection symmetry, i.e. ${\bf r}^2 = {\bf 0}$, $q({\bf r}) = 0$,
and $p({\bf r}) = 1$, we restrict our attention to a state with two quasiparticles of topological charge $a$ and $\bar{a}$
created from vacuum. Moreover, we focus on a configuration in which the reflection symmetry interchanges the regions
$\mathcal{R}_{1}$ and $\mathcal{R}_{2}$ in which the quasiparticles are respectively localized, as indicated in Fig.~\ref{fig:anyon-pair}.

In this case, the symmetry action on the topological state space is
\begin{equation}
\rho_{\bf r} \left| a , \bar{a} ; 0 \right\rangle = U_{\bf r} (\,^{\bf r}\bar{a},\,^{\bf r}a ;0)  \left| \,^{\bf r}\bar{a},\,^{\bf r}a ;0 \right\rangle
,
\end{equation}
where $U_{\bf r} (\,^{\bf r}\bar{a},\,^{\bf r}a ;0)$ here is a phase factor associated with a vertex transformation of the topological state space, which is chosen such that the braiding $R$-symbol satisfies
\begin{equation}
\label{eq:reflection_R}
U_{\bf r} (\,^{\bf r}\bar{a} , \,^{\bf r}a ;0) \left[ R^{\,^{\bf r}a \,^{\bf r}\bar{a}}_{0} \right] U_{\bf r} (\,^{\bf r}a , \,^{\bf r}\bar{a} ;0)^{-1} =  \left[ R^{a \bar{a}}_{0} \right]^{-1}
.
\end{equation}

The transformation of $U_{\bf r}$ under a vertex basis gauge transformation is
\begin{equation}
\label{eq:vertex_gauge_U_r}
\widetilde{U}_{\bf r} (\,^{\bf r}\bar{a},\,^{\bf r}a ;0)  = \Gamma^{a \bar{a}}_{0}   U_{\bf r} (\,^{\bf r}\bar{a},\,^{\bf r}a ;0) \left[ \Gamma^{\,^{\bf r}\bar{a} \,^{\bf r}a}_{0} \right]^{-1}
.
\end{equation}

In the physical Hilbert space, we write the states for a system with a quasiparticle pair produced from vacuum as
$\left| \Psi_{ a , \bar{a} ; 0}^{\alpha, \beta} \right\rangle$. These can be thought of as ground states of a Hamiltonian with
local pinning potentials that trap quasiparticles carrying topological charges $a$ and $\overline{a}$ in two well-separated regions $\mathcal{R}_1$ and
$\mathcal{R}_2$, respectively. Due to the symmetry of the system (and details of the pinning potentials), the ground state space of the Hamiltonian with pinning potentials may be degenerate. Such a degeneracy will be localized in the regions $\mathcal{R}_1$ and $\mathcal{R}_2$ containing the quasiparticles. We denote the corresponding local state spaces as $\mathcal{H}_a$ and $\mathcal{H}_{\bar{a}}$, respectively, and label their respective basis states by $\alpha$ and $\beta$. (We note that, while we impose symmetry on the Hamiltonian, the process that pair creates the quasiparticles is not required to respect the symmetry, so the states in the localized state spaces are not required to be correlated with each other.) For example, $\alpha,\beta = \uparrow, \downarrow$ could label spin states if quasiparticles with topological charge $a$ carry a local spin-$1/2$ degree of freedom.

The global symmetry ${\bf r}$ acts on the physical Hilbert space as follows
\begin{eqnarray}
R_{\bf r} \left| \Psi_{ a , \bar{a} ; 0}^{\alpha, \beta} \right\rangle &=& U_{\bf r}^{(1)} U_{\bf r}^{(2)} U_{\bf r} (\,^{\bf r}\bar{a},\,^{\bf r}a ;0) \left| \Psi_{ \,^{\bf r}\bar{a},\,^{\bf r}a ; 0}^{\beta, \alpha} \right\rangle \notag \\
&=& U_{\bf r}^{(1)} U_{\bf r}^{(2)} \rho_{\bf r} \left| \Psi_{ a , \bar{a} ; 0}^{\alpha, \beta} \right\rangle
,
\end{eqnarray}
where $U_{\bf r}^{(j)}$ is an operator whose nontrivial action is localized in the region $\mathcal{R}_{j}$, i.e. they act on the local state
spaces labeled by $\alpha$ and $\beta$, and $\rho_{\bf r} $ here is an operator on the physical Hilbert space that has the same
action on the topological quantum numbers as does the symmetry action upon the topological state space. We emphasize that
$\rho_{\bf r}$ also includes the nonlocal action on the quasiparticles responsible for interchanging their locations, along with
their corresponding local state spaces (as indicated by the interchange of the labels $\alpha$ and $\beta$). This form of the symmetry action
is expected when the symmetry acts on the physical Hilbert space in a locality preserving fashion as defined in Ref. \onlinecite{barkeshli2014SDG}.

The projective phase factors $\eta_{a}({\bf r},{\bf r})$ associated with symmetry fractionalization are defined by the action of the
localized operators of the regions containing the quasiparticles
\begin{eqnarray}
\label{eq:eta_reflection}
\eta_{a} ({\bf r},{\bf r}) \left| \Psi_{ a , \bar{a} ; 0}^{\alpha,\beta} \right\rangle &=& U_{\bf r}^{(1)} \rho_{\bf r} U_{\bf r}^{(2)} \rho_{\bf r}^{-1} \left| \Psi_{ a , \bar{a} ; 0}^{\alpha,\beta} \right\rangle  \\
\eta_{\bar{a}} ({\bf r},{\bf r}) \left| \Psi_{ a , \bar{a} ; 0}^{\alpha,\beta} \right\rangle &=& U_{\bf r}^{(2)} \rho_{\bf r} U_{\bf r}^{(1)} \rho_{\bf r}^{-1} \left| \Psi_{ a , \bar{a} ; 0}^{\alpha,\beta} \right\rangle
.
\end{eqnarray}
(Note that $\rho_{\bf r} U_{\bf r}^{(2)} \rho_{\bf r}^{-1}$ is an operator whose nontrivial action is localized in region $\mathcal{R}_{1}$, and $\rho_{\bf r} U_{\bf r}^{(1)} \rho_{\bf r}^{-1}$ is similarly localized in $\mathcal{R}_{2}$.)
Associativity of the group action leads to the constraint
\begin{equation}
\label{eq:reflection_eta_assoc}
\eta_{\,^{\bf r}a} ({\bf r},{\bf r}) = \eta_{a} ({\bf r},{\bf r})
.
\end{equation}
We define the quantity
\begin{equation}
{\eta}_{a}^\R \equiv \eta_{a} ({\bf r},{\bf r}) U_{\bf r} (a,\bar{a} ;0)
.
\end{equation}
Note that we always have ${\eta}_{0}^\R=1$

We now consider the case when $^{\bf r}a = \bar{a}$. Eq.~(\ref{topsymEq}) indicates that $\theta_{\bar{a}}=\theta_{a}^{\ast}=\theta_{\bar{a}}^{\ast}$, so that $\theta_{a}= \pm 1$. Combining this with Eq.~(\ref{eq:reflection_R}) and the ribbon identity, we find
\begin{equation}
\frac{ U_{\bf r} (a,\bar{a} ;0)}{ U_{\bf r} (\bar{a},a ;0)} = R^{a \bar{a}}_{0} R^{ \bar{a} a}_{0} = \frac{\theta_{0}}{\theta_{a}^{2}}=1
,
\end{equation}
so $U_{\bf r} (a,\bar{a} ;0)= U_{\bf r} (\bar{a},a ;0)$. From Eq.~(\ref{eq:reflection_eta_assoc}), it follows that $\eta_{a} ({\bf r},{\bf r})=\eta_{\bar{a}} ({\bf r},{\bf r})$, and, thus, ${\eta}_{a}^\R={\eta}_{\bar{a}}^\R$. Since $R_{\bf r}^{2} = \openone$, it follows that
\begin{widetext}
\begin{align}
\left| \Psi_{ a , \bar{a} ; 0}^{\alpha,\beta} \right\rangle &=  R_{\bf r}^{2} \left| \Psi_{ a , \bar{a} ; 0}^{\alpha,\beta} \right\rangle
= U_{\bf r}^{(1)} \rho_{\bf r} U_{\bf r}^{(2)} \rho_{\bf r}^{-1} U_{\bf r}^{(2)} \rho_{\bf r} U_{\bf r}^{(1)} \rho_{\bf r}^{-1} U_{\bf r} (a,\bar{a} ;0) U_{\bf r} (a,\bar{a} ;0) \left| \Psi_{ a , \bar{a} ; 0}^{\alpha,\beta} \right\rangle
\nonumber \\
&= {\eta}_{a}^\R {\eta}_{\bar{a}}^\R  \left| \Psi_{ a , \bar{a} ; 0}^{\alpha,\beta} \right\rangle
= \left({\eta}_{a}^\R \right)^2  \left| \Psi_{ a , \bar{a} ; 0}^{\alpha,\beta} \right\rangle
,
\end{align}
\end{widetext}
%\begin{eqnarray}
%\left| \Psi_{ a , \bar{a} ; 0}^{\alpha,\beta} \right\rangle &=&  R_{\bf r}^{2} \left| \Psi_{ a , \bar{a} ; 0}^{\alpha,\beta} \right\rangle \notag \\
%&=& U_{\bf r}^{(1)} \rho_{\bf r} U_{\bf r}^{(2)} \rho_{\bf r}^{-1} U_{\bf r}^{(2)} \rho_{\bf r} U_{\bf r}^{(1)} \rho_{\bf r}^{-1} U_{\bf r} (a,\bar{a} ;0) U_{\bf r} (a,\bar{a} ;0) \left| \Psi_{ a , \bar{a} ; 0}^{\alpha,\beta} \right\rangle \notag \\
%&=& {\eta}_{a}^\R {\eta}_{\bar{a}}^\R  \left| \Psi_{ a , \bar{a} ; 0}^{\alpha,\beta} \right\rangle
%,
%\end{eqnarray}
so we find that
\begin{equation}
{\eta}_{a}^\R = \pm 1
.
\end{equation}

Moreover, ${\eta}_{a}^\R$ is an invariant quantity when $^{\bf r}a = \bar{a}$. In particular, vertex basis transformations described in Eq.~(\ref{eq:vertex_gauge_U_r}) give $\widetilde{U}_{\bf r} (a,\bar{a} ;0) = U_{\bf r} (a,\bar{a} ;0)$ and do not change $\eta_{a} ({\bf r},{\bf r})$. Additionally, symmetry action gauge transformations\cite{barkeshli2014SDG} %(which modify the symmetry action by a natural transformation)
give $\check{U}_{\bf r} (a,\bar{a} ;0) = \frac{\gamma_{a}({\bf r}) \gamma_{\bar{a}}({\bf r})}{\gamma_{0}({\bf r})} U_{\bf r} (a,\bar{a} ;0)$ and $\check{\eta}_{a}({\bf r},{\bf r})= \frac{\gamma_{a}({\bf 0})}{\gamma_{\bar{a}}({\bf r}) \gamma_{a}({\bf r})} \eta_{a}({\bf r},{\bf r})$. Since we always require $\gamma_{0}({\bf r}) = \gamma_{a}({\bf 0}) =1$, this transformation cancels out to leave $\check{\eta}_{a}^\R =\eta_{a}^\R$.

Finally, when $^{\bf r}a = \bar{a}$, we also can see that
\begin{widetext}
\begin{eqnarray}
\sum_{\alpha,\beta}  \left\langle \Psi_{ a , \bar{a} ; 0}^{\alpha,\beta} \right| R_{\bf r} \left| \Psi_{ a , \bar{a} ; 0}^{\alpha,\beta} \right\rangle
&=& \sum_{\alpha,\beta}  \left\langle \Psi_{ a , \bar{a} ; 0}^{\alpha,\beta} \right| U_{\bf r}^{(1)} U_{\bf r}^{(2)} U_{\bf r} (a,\bar{a} ;0) \left| \Psi_{ a , \bar{a} ; 0}^{\beta , \alpha} \right\rangle
= \sum_{\alpha,\beta}  \left[ U_{\bf r}^{(1)} \right]_{\alpha \beta} \left[U_{\bf r}^{(2)}\right]_{\beta \alpha} U_{\bf r} (a,\bar{a} ;0) \notag
\end{eqnarray}
\begin{eqnarray}
&=& \sum_{\alpha,\beta,\beta'}  \left[ U_{\bf r}^{(1)} \right]_{\alpha \beta'}  U_{\bf r} (a,\bar{a} ;0) \left[U_{\bf r}^{(2)}\right]_{\beta' \beta} \delta_{\beta \alpha}
= \sum_{\alpha,\beta}  \left\langle \Psi_{ a , \bar{a} ; 0}^{\alpha,\beta} \right| U_{\bf r}^{(1)} \rho_{\bf r} U_{\bf r}^{(2)} \left| \Psi_{ a , \bar{a} ; 0}^{\alpha, \beta} \right\rangle \notag
\end{eqnarray}
\begin{eqnarray}
&=& \sum_{\alpha,\beta}  \left\langle \Psi_{ a , \bar{a} ; 0}^{\alpha,\beta} \right| U_{\bf r}^{(1)} \rho_{\bf r} U_{\bf r}^{(2)} \rho_{\bf r}^{-1} U_{\bf r} (a,\bar{a} ;0) \left| \Psi_{ a , \bar{a} ; 0}^{\beta , \alpha} \right\rangle
= \eta_{a}^{\bf r} \sum_{\alpha,\beta}  \left\langle \Psi_{ a , \bar{a} ; 0}^{\alpha,\beta} \right| \left. \Psi_{ a , \bar{a} ; 0}^{\beta , \alpha} \right\rangle \notag \\
&=& \eta_{a}^{\bf r} \text{dim}[\mathcal{H}_{a}]
,
\end{eqnarray}
\end{widetext}
where recall that we assumed $\alpha$ and $\beta$ to be labels of orthonormal basis states of the local state spaces
$\mathcal{H}_{a}$ and $\mathcal{H}_{\bar{a}}$, which have the same dimension. In other words,
\begin{equation}
\label{eq:reflection_SET_eta}
\frac{\sum\limits_{\alpha,\beta}  \left\langle \Psi_{ a , \bar{a} ; 0}^{\alpha,\beta} \right| R_{\bf r} \left| \Psi_{ a , \bar{a} ; 0}^{\alpha,\beta} \right\rangle}{\left| \sum\limits_{\alpha,\beta}  \left\langle \Psi_{ a , \bar{a} ; 0}^{\alpha,\beta} \right| R_{\bf r} \left| \Psi_{ a , \bar{a} ; 0}^{\alpha,\beta} \right\rangle \right| } = {\eta}_{a}^\R
.
\end{equation}

\subsection{(1+1)D SPT invariant from dimensional reduction}

An alternative perspective on the above symmetry fractionalization
quantum numbers $\eta_a^{\bf r}$ and $\eta_{a}^{\bf T}$ is
through the notion of dimensional reduction and asking whether the dimensionally reduced
system forms a (1+1)D SPT state. This point of view was used to detect symmetry fractionalization in
$\mathbb{Z}_2$ quantum spin liquids in Ref.~\onlinecite{zaletel2015}.

More specifically, we consider the state of a (2+1)D topological phase on a cylinder in which there is a definite
topological charge $a$ and $\overline{a}$ localized on the two ends of the cylinder, respectively. This
configuration can be thought of as having an anyon ``flux'' of $a$ threading the cylinder.
The ground state(s) of the system in such a sector $|\Psi^{\alpha, \beta}_{a, \overline{a}} \rangle$
form a ground state Hilbert space, which we label as $\V_{a,\overline{a}}$. The indices $\alpha, \beta$
are related to local degrees of freedom associated with the two ends of the cylinder.

Next, viewing the circumference of the cylinder as an internal degree of freedom, we can thus
view $|\Psi_{a,\overline{a}}^{\alpha,\beta}\rangle$ as the state of a (1+1)D system.

Let us first consider the case where $G = \mathbb{Z}_2^{\bf T}$ corresponds to an anti-unitary time-reversal symmetry.
A ground state of the 1D system with boundary transforms under time-reversal as
\begin{align}
\label{eq:Ttrans_1D}
R_{\bf{T}} |\Psi \rangle = U^{(1)}_{\bf{T}} U^{(2)}_{\bf{T}}|\Psi \rangle ,
\end{align}
where $U^{(1)}_{\bf T}$ and $U^{(2)}_{\bf T}$ are unitary operators whose nontrivial action is localized near the two endpoints of the system, respectively.
Applying $R_{\mathbf{T}}$ again, we find
\begin{equation}
|\Psi \rangle = R_{\bf{T}}^2|\Psi \rangle = U^{(1)}_{\bf{T}}{U^{(1)}_{\bf{T}}}^* U^{(2)}_{\bf{T}}{U^{(2)}_{\bf{T}}}^* |\Psi \rangle.
\end{equation}
Since $U^{(j)}_{\bf T}$ are local unitary operators, we must have
\begin{equation}
U^{(1)}_{\bf{T}}{U^{(1)}_{\bf{T}}}^* |\Psi \rangle = U^{(2)}_{\bf{T}}{U^{(2)}_{\bf{T}}}^* |\Psi \rangle = \pm |\Psi \rangle.
\end{equation}
The $\pm$ sign here is an invariant which indicates the absence or presence of a local Kramers degeneracy at each endpoint, and which distinguishes $|\Psi \rangle$ as belonging to a trivial or non-trivial time-reversal SPT phase.

Assuming that time-reversal does not permute the topological charge value $a$ to another topological charge type, i.e. $\,^{\bf T}a \equiv \rho_{\bf{T}} (a) = a$,
we can view the (2+1)D state $|\Psi_{a,\overline{a}}\rangle \in \mathcal{V}_{a \overline{a}} $ on the cylinder as a (1+1)D time-reversal SPT state on an interval through dimensional reduction.~\footnote{We note that the one can always choose a gauge such that the transformation of the (2+1)D state $|\Psi_{a,\overline{a}}\rangle $ under $R_{\bf T}$ takes the form in Eq.~(\ref{eq:Ttrans_1D}).}
Since nontrivial (1+1)D time-reversal SPT states possess a local Kramers degeneracy at each end of a 1D system with boundary, it follows that the (1+1)D SPT invariant is equal to the quantity
\begin{align}
{\eta}_a^{\bf T} =\pm 1 ,
\end{align}
introduced in the previous section, with the $\pm 1$ indicating whether the dimensional reduction of $|\Psi_{a,\overline{a}}\rangle$ corresponds to a trivial or nontrivial (1+1)D time-reversal SPT state.

\begin{figure}
\includegraphics[width=3.5in]{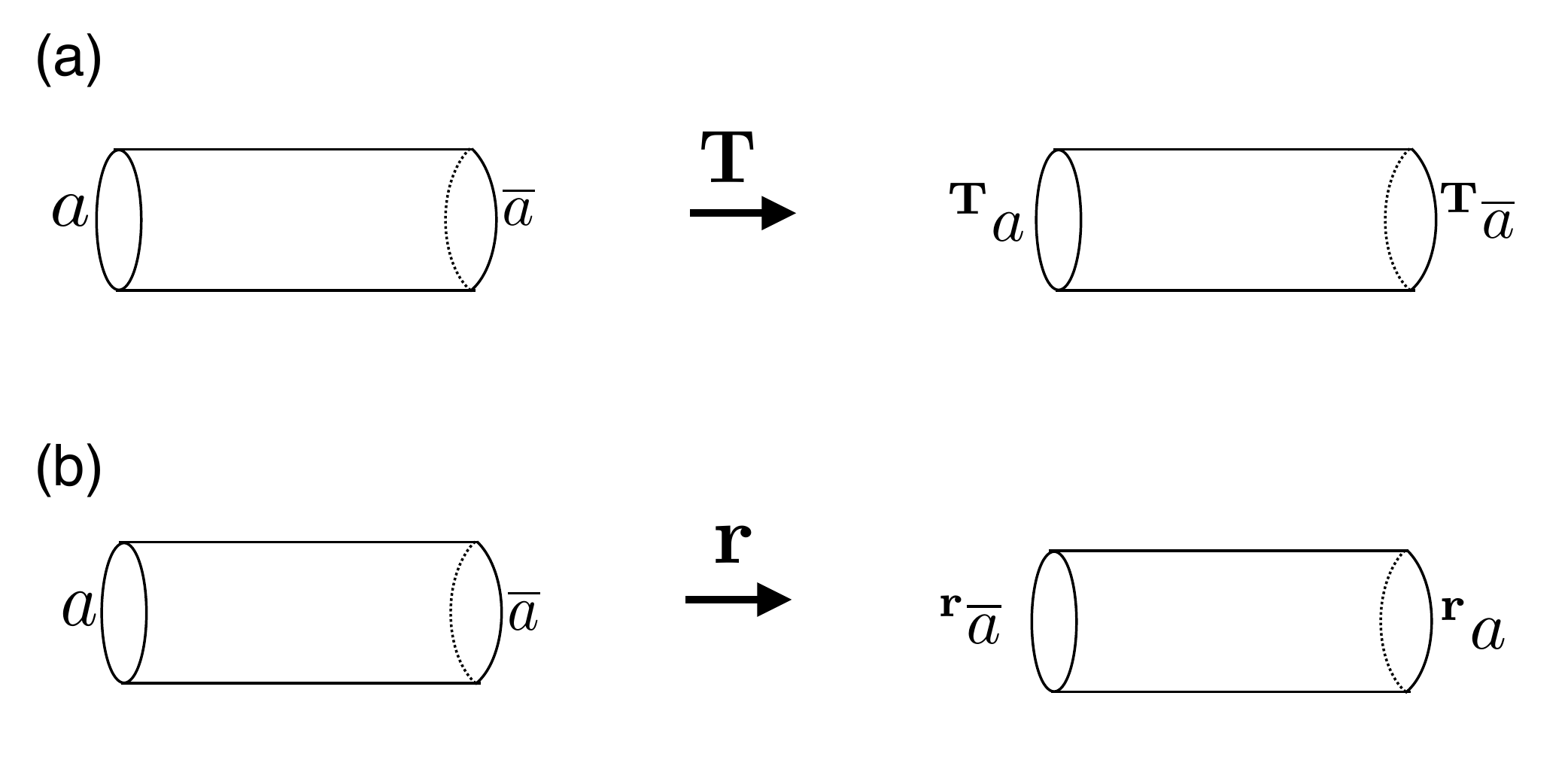}
\caption{(a) The left and right boundaries of the cylinder have topological charge $a$ and $\overline{a}$ respectively.
Under the action of time reversal, ${\bf T}$, these get permuted to $\rho_{\bf T}(a) \equiv \,^{\bf{T}}a$ and $\rho_{\bf T}(\overline{a}) \equiv \,^{\bf{T}}\overline{a}$,
respectively. (b) Reflection ${\R}$ takes the coordinate along the cylinder $x \rightarrow -x$, so that
the topological charges at the boundary get permuted to $\rho_{\R}(\overline{a}) \equiv \,\ra{a}$ and
$\rho_{\bf T}(\overline{a}) \equiv \,^{\bf{T}}\overline{a}$, respectively.
%(c) A line defect (red) at $x = 0$ wrapping around the circumference of the cylinder, such that an anyon $a$ crossing the line defect (from left to right) transforms into $\varphi(a)$.
%In this case, the two boundaries of the cylinder have topological charge $a$ and $\varphi(\overline{a})$, which transform under
%under ${\R}$ to $\,^{\R}\varphi(\overline{a})$ and $^{\R} a$, respectively.
}
\label{cylFig}
\end{figure}

Let us now consider the case where $G = \mathbb{Z}_2^{\bf r}$ corresponds to a spatial
reflection symmetry. As discussed in Sec.~\ref{1dSPT_reflection_boundary}, the (1+1)D SPT invariant for a state on a system with boundary is given by Eq.~(\ref{Rvalue}).
For the (2+1)D states on the cylinder, we consider the action of reflection to reverse the coordinate along the cylinder.
In this case, we have:
\begin{align}
R_{\R} : \mathcal{V}_{a,\overline{a}} \rightarrow \mathcal{V}_{ \ra{\bar{a}}, \ra{a}}
\end{align}
When $\ra{a} = \overline{a}$, we can view the (2+1)D state $|\Psi_{a,\overline{a}}\rangle \in \mathcal{V}_{a \overline{a}} $ on the cylinder as a (1+1)D reflection SPT state on an interval through dimensional reduction.
Comparing Eq.~(\ref{Rvalue}) and Eq.~(\ref{eq:reflection_SET_eta}), we see that the corresponding (1+1)D SPT invariant is equal to the quantity
\begin{align}
{\eta}_a^\R = \pm 1
\end{align}
defined in the previous section.

In general, one can consider other configurations of the system. For example, instead of considering anyons $a$ and $\bar{a}$ on the left
and right ends of the cylinder, one can consider multi-anyon configurations, with $n$ anyons $a_1, \ldots, a_n$ arranged on a line along the
cylinder. Alternatively, one can consider configurations of the system with a line defect~\cite{kitaev2012,barkeshli2013defect2,barkeshli2014SDG}
associated with $\varphi \in \text{Aut}_{0,0}(\mathcal{B})$. In these situations, one can further define an invariant associated with whether the state,
viewed as a (1+1)D reflection SPT state, is trivial or nontrivial. Presumably these invariants are determined by the set $\{\eta_a^{\bf r} \}$ defined above,
though we leave a systematic study for future work.

%% file: state-sum2d.tex
\section{Path integral state-sums for (2+1)D TQFTs}
\label{statesumSec}

One way to characterize a topological phase of matter is to specify a fully topological effective action that captures the universal, long
wavelength properties of the phase. Such an effective topological action is called a topological quantum field theory.~\cite{atiyah}
This allows one to compute the path integral on closed space-time manifolds, and associates a (potentially multi-dimensional) topological
state space to manifolds with boundaries. In many cases, the TQFTs can be defined as a state-sum model on triangulated space-time manifolds,
yielding fixed-point wavefunctions having a correlation length of zero. Before developing our general SET state-sum construction, we
briefly review two limiting cases: the case of SPT phases, with no intrinsic topological order, and the case of topological
phases with no symmetries.

\subsection{Topological state-sums for SPT phases}

In this section, we briefly review the state-sum approach to describing SPT phases, which was developed in Ref.~\onlinecite{chen2013}. This approach describes
a large class of bosonic SPT phases in any dimension, but it misses certain SPT phases with space-time symmetries for dimensions greater than (2+1)D.~\cite{kapustin2014}
A notable example of this is one of the bosonic SPT phases in (3+1)D with time-reversal symmetry.~\cite{vishwanath2013}

In order to construct a Euclidean path integral for SPT phases with symmetry group $G$ on a ($d$+1)D space-time manifold $M^{d+1}$, we proceed with a straightforward generalization of the construction of Sec.~\ref{sec:1dspt-pi}.

First, the space-time manifold $M^{d+1}$ is discretized into a triangulation. We denote the set of $n$-simplexes of the triangulation by $\mathcal{I}_{n}$ and provide an ordering to the elements of $\mathcal{I}_{0}$, i.e. we assign the labels $j=0,1,\ldots,N_{v}-1$ to the vertices (0-simplexes) of the triangulation, where $N_{v}$ is the number of vertices in the triangulation. The ordering of vertices provides a direction to each edge (1-simplex), pointing from the lower ordered endpoint of the edge to the higher one, and an orientation to each ($d$+1)-simplex $\Delta^{d+1} \in \mathcal{I}_{d+1}$. The ordering of vertices defines a ``branching structure'' on the triangulation.

We can define the relative orientation $s(\Delta^{d+1})$ of ($d$+1)-simplexes by thinking of the manifold as being constructed from an oriented manifold with boundary by gluing together regions of the boundary, possibly with twists that spoil the global orientation. Then we use an appropriate triangulation of the oriented surface with boundary to define a triangulation of the surface in question, by identifying vertices and edges appropriately. The orientation of the oriented surface then provides a reference orientation for the ($d$+1)-simplexes of the triangulation. When the order of the vertices of a ($d$+1)-simplex $\Delta^{d+1}$ agree with the reference orientation, we set $s(\Delta^{d+1})=1$; when the order of the vertices of a ($d$+1)-simplex $\Delta^{d+1}$ is opposite that of the reference orientation, we set $s(\Delta^{d+1})=\ast$. Fig.~\ref{tetraFig} shows two 3-simplexes with oppositely ordered vertices and hence opposite orientations.

We next assign a group element ${\bf g}_{j} \in G$ to the $j$th vertex of the triangulation. Then, for each configuration $\{{\bf g}_j\}$ on the vertices, we associate a U$(1)$ phase to each $(d+1)$-simplex $\Delta^{d+1} \in \mathcal{I}_{d+1}$, given by
\begin{equation}
\Z(\Delta^{d+1}; \{ {\bf g}_j \}) = \nu_{d+1}({\bf g}_{j_0}, {\bf g}_{j_1}, \ldots, {\bf g}_{j_{d+1}})^{s(\Delta^{d+1})}
,
\end{equation}
where the vertices $j_0,j_1, \ldots , j_{d+1}$ of $\Delta^{d+1}$ have been ordered to respect the vertex numbering order, i.e. ${j_{0}}< {j_{1}}<\ldots < {j_{d+1}}$.
The path integral for a closed manifold $M^{d+1}$ is then given by
\begin{align}
\Z(M^{d+1}) = \frac{1}{|G|^{N_v}} \sum_{\{ {\bf g}_j \}} \prod_{ \Delta^{d+1} \in \mathcal{I}_{d+1} } \Z(\Delta^{d+1}; \{ {\bf g}_j \})
.
\end{align}

On a manifold with boundary, one can generalize the above sum by fixing the group elements on the boundary vertices and summing over the
group elements at the internal vertices, i.e. those in $\text{int}(M) = M \setminus \partial M$. This yields a wavefunction amplitude for each configuration of group elements on the boundary vertices
\begin{widetext}
\begin{align}
\Psi(\{{\bf g}_j\}_{\partial M^{d+1}}) = \frac{1}{|G|^{N_{v;\text{int}(M)}}} \sum_{\{{\bf g}_j \}_{\text{int}(M)}} \prod_{\Delta^{d+1} \in \mathcal{I}_{d+1}} \Z(\Delta^{d+1}; \{ {\bf g}_j \})
,
\end{align}
\end{widetext}
where the sum is over group elements of the vertices in the interior of the manifold, and $N_{v;\text{int}(M)}$ is the number of vertices in the interior.

In order for $\Z(M^{d+1})$ to be a topological invariant of closed ($d$+1)D manifolds, it must be independent of the specific
choice of triangulation and also invariant under reordering the vertices, i.e. a change of the branching structure.
It is known that all triangulations can be related to each other via a finite series of moves, known as Pachner moves.
Imposing that the state sum be invariant under these Pachner moves leads to the requirement~\cite{chen2013}
\begin{align}
\label{cocycleEq}
\prod_{n = 0}^{d+2} \nu_{d+1}^{(-1)^n} ({\bf g}_0, \ldots, {\bf g}_{n-1}, {\bf g}_{n+1}, \ldots, {\bf g}_{d+1})  = 1.
\end{align}

$\Z(M^{d+1})$ is also invariant under reordering the vertices, i.e. a change of the branching structure, though this is more nontrivial
to verify.

In order to describe an SPT state, we require the ground state wavefunction $\Psi(\{{\bf g}_j\})$ to be invariant under a symmetry transformation
\begin{align}
\Psi(\{{\bf h} {\bf g}_j\}) = \Psi(\{{\bf g}_j\})^{\sigma({\bf h})},
\end{align}
for any ${\bf h} \in G$, where $\sigma({\bf g})$ was defined in Eq.~(\ref{eq:G_action_sigma}), i.e. it acts as complex conjugation iff ${\bf g}$ is a space-time orientation reversing symmetry, e.g. reflection or time-reversal symmetry. One can show that the symmetry invariance of the wavefunction implies the condition
\begin{widetext}
\begin{align}
\label{symm1}
 \nu_{d+1}({\bf h} {\bf g}_{j_0}, {\bf h} {\bf g}_{j_1}, \ldots, {\bf h} {\bf g}_{j_{d+1}}) =  \nu_{d+1} ({\bf g}_{j_0}, {\bf g}_{j_1}, \ldots, {\bf g}_{j_{d+1}})^{\sigma({\bf h})} .
\end{align}
\end{widetext}

Using Eq.~(\ref{symm1}), we see that $\nu_{d+1}$ can actually be interpreted as depending on only $d+1$ group elements, e.g. ${\bf g}_{j_0}^{-1}{\bf g}_{j_1}, \ldots, {\bf g}_{j_0}^{-1}{\bf g}_{j_{d+1}}$. As such, we can think of it as a map $\nu_{d+1}: G^{d+1} \rightarrow \text{U}(1)$, which is to say it is a $d+1$ cochain. Thus Eq.~(\ref{cocycleEq}) can be viewed as the $d+1$-cocycle condition.

Finally, we note that transforming phases $\nu_{d+1}$ by
\begin{multline}
\nu'_{d+1} ({\bf g}_{j_0}, {\bf g}_{j_1}, \ldots, {\bf g}_{j_{d+1}}) =
\nu_{d+1} ({\bf g}_{j_0}, {\bf g}_{j_1}, \ldots, {\bf g}_{j_{d+1}}) \\
\times \prod_{n = 0}^{d+1} \mu_d({\bf g}_{j_0}, \ldots, {\bf g}_{j_n} {\bf g}_{j_{n+1}}, \ldots, {\bf g}_{j_{d+1}})^{(-1)^n},
\label{coboundEq}
\end{multline}
where $\mu_d ({\bf g}_{0}, \ldots, {\bf g}_{d})$ are U$(1)$ phases, leaves the path integral invariant $\Z'(M^{d+1})=\Z(M^{d+1})$ on closed manifolds.

Under such transformations, the wavefunction transforms as
\begin{align}
\Psi'(\{ {\bf g}_j \}) = \Psi(\{{\bf g}_j\}) \prod_{\Delta^{d} \in \mathcal{J}_{d} } \mu_d ({\bf g}_{j_{0}}, \ldots, {\bf g}_{j_{d}})^{s(\Delta^{d})}
,
\end{align}
where $\mathcal{J}_{d} \subset \mathcal{I}_d$ is the subset of $d$-simplexes that provide a triangulation of $\partial M^{d+1}$.
For the wavefunction to remain symmetric under $G$, we require that $\mu_d$ be $G$-symmetric, i.e. that
\begin{align}
\label{symm2}
\mu_{d}({\bf h} {\bf g}_{j_0}, {\bf h} {\bf g}_{j_1}, \ldots, {\bf h} {\bf g}_{j_{d}}) =  \mu_{d}({\bf g}_{j_0}, {\bf g}_{j_1}, \ldots, {\bf g}_{j_{d}})^{\sigma({\bf h})}
.
\end{align}

Putting together Eqs.~(\ref{cocycleEq}), (\ref{symm1}), (\ref{coboundEq}), and (\ref{symm2}), implies that the
set of possible effective actions are classified by the cohomology group $\mathcal{H}^{d+1}_{\sigma}(G; \text{U}(1))$.\cite{chen2013}
If $G$ has space-time orientation-reversing elements, then the corresponding elements act on the U$(1)$ coefficients by
complex conjugation, as indicated by $\sigma$. However, it has not been argued that distinct elements of $\mathcal{H}^{d+1}(G, \text{U}(1))$ are truly
different phases of matter. This requires more detailed considerations of the Hamiltonian for which $\Psi(\{g_i\})$
is the ground state wavefunction.~\cite{levin2012,chen2013, ElsePRB2014}

\subsection{Turaev-Viro-Barrett-Westbury construction}
\label{sec:TVBW}

In this section, we will review a construction due to Barrett and Westbury~\cite{barrett1996}, which generalized earlier constructions
by Turaev and Viro~\cite{turaev1992}, Dijkgraaf and Witten~\cite{dijkgraaf1989}, and others~\cite{ponzano1968,kogut1979}.
Since then, this construction has been put within an even more general framework.~\cite{walker2006}
The input to the construction is a spherical fusion category, $\mathcal{C}$,
although for the purposes of describing topological phases of matter, we further require that $\mathcal{C}$ be a unitary fusion
category (unitarity implies sphericity).~\footnote{Here by unitary, we mean that the $F$ matrices are all unitary and the quantum dimensions are positive. }  For a review of unitary fusion categories written for physicists, see Refs.~\onlinecite{kitaev2006,Bonderson07b,barkeshli2014SDG}. Given an arbitrary $\mathcal{C}$,
Barrett and Westbury defined a topological invariant $\Z(M^3)$, where $M^3$ is a
3D manifold, and where $\Z(M^3)$ corresponds to the path integral of a (2+1)D TQFT.

\subsubsection{Definition}

\begin{figure}[t!]
\includegraphics[width=3.3in]{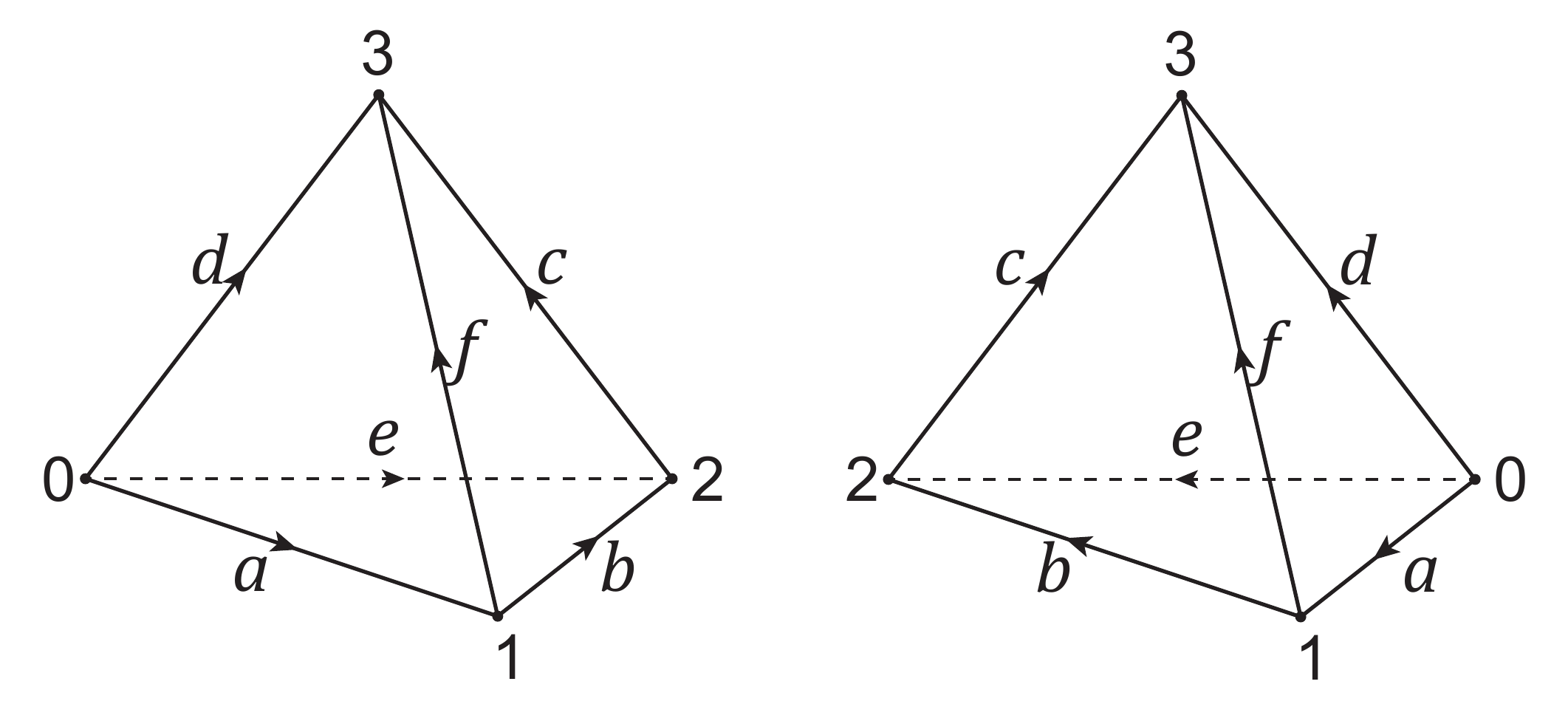}
\caption{Two tetrahedra with opposite orientations. A tetrahedron with positive orientation (left) is assigned $s(\Delta^3)=\openone$ and a tetrahedron with negative orientation (right) is assigned $s(\Delta^3)=\ast$. Edges of the tetrahedra are labeled by topological charges of the fusion category $\mathcal{C}$.}
\label{tetraFig}
\end{figure}

To define $\Z(M^3)$, we pick a triangulation of $M^3$. As before, we denote the set of $n$-simplexes of the triangulation by $\mathcal{I}_{n}$ and provide an ordering to the elements of $\mathcal{I}_{0}$, assigning the labels $j=0,1,\ldots,N_{v}-1$ to the vertices (0-simplexes) of the triangulation, where $N_{v}$ is the number of vertices in the triangulation. Again, this provides a direction to each edge (1-simplex), pointing from the lower ordered endpoint of the edge to the higher one, a branching structure, and an orientation to each 3-simplex $\Delta^{3} \in \mathcal{I}_{3}$ (see Fig.~\ref{tetraFig}).

\begin{figure}[t!]
\includegraphics[width=3in]{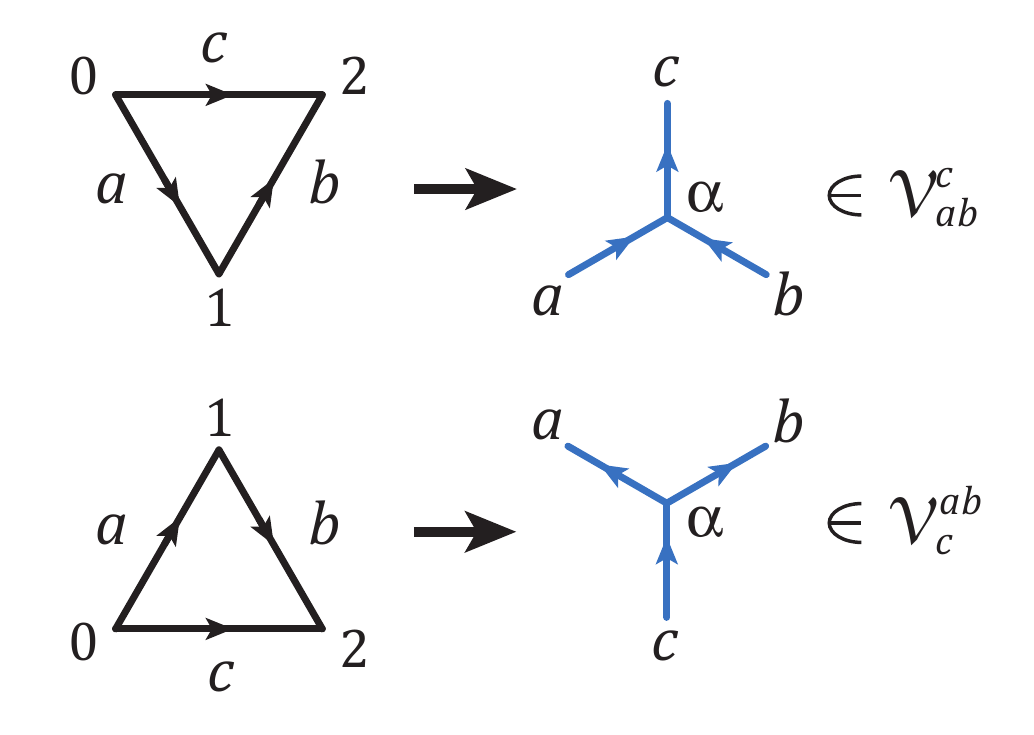}
\caption{Triangles are assigned states in vector spaces $V_{a,b;c}$, which are mapped to fusion or splitting spaces of $\mathcal{C}$ for the dual diagrams.}
\label{fig:vertex_states}
\end{figure}

In this case, we assign a topological charge label $a_{E} \in \mathcal{C}$ to each edge $E \in \mathcal{I}_{1}$ of the triangulation. For a given triangle $\Delta^{2} \in \mathcal{I}_{2}$, we label its vertices $j_{0}$, $j_{1}$, and $j_{2}$, such that $j_{0}<j_{1}<j_{2}$, and denote the corresponding edges connecting these vertices as $E_{01}$, $E_{12}$, and $E_{02}$. When these edges of the triangle are labeled by the corresponding topological charges $a$, $b$, and $c$, respectively, we assign a vector space $V_{a,b;c}$ to the triangle $\Delta^{2}$, such that $\text{dim } V_{a,b;c} = N_{ab}^c$, the fusion coefficients of $\mathcal{C}$. We label an orthogonal basis of $V_{a,b;c} $ by $\alpha = 1,\ldots, N_{ab}^c$. We then assign a state label $\alpha_{\Delta^2}$ of the corresponding vector space to each triangle $\Delta^2 \in \mathcal{I}_2$ in the triangulation (see Fig.~\ref{fig:vertex_states}). The configuration of the topological charge labels on all the edges and corresponding (basis) states on all the triangles of the entire triangulation will be collectively denoted as $\ell$.

Next, consider a tetrahedron $\Delta^{3} \in \mathcal{I}_{3}$ in the triangulation for a configuration $\ell$ such that the edges $E_{01},E_{12},E_{23},E_{03},E_{02},E_{13}$ of $\Delta^{3}$ are respectively labeled by the topological charges $a,b,c,d,e,f$, as shown in Fig.~\ref{tetraFig}, and where the faces $F_{012}, F_{023}, F_{123}, F_{013}$ of $\Delta^{3}$ are respectively labeled by the corresponding basis states $\alpha, \beta, \mu, \nu$. We note that $\alpha = 1, \ldots, N_{ab}^e$; $\beta =1,\ldots, N_{ec}^d$; $\mu = 1, \ldots, N_{bc}^f$; and $\nu = 1, \ldots, N_{af}^d$. We define an amplitude for the tetrahedron in this configuration to be
\begin{equation}
\label{eq:Tet}
\Tet(\Delta^3; \ell) = \left[ F^{abc}_d \right]^{s(\Delta^3)}_{(e,\alpha, \beta) (f,\mu,\nu)}\sqrt{d_a d_b d_c d_d}
.
\end{equation}

\begin{figure}[t!]
\includegraphics[width=3.3in]{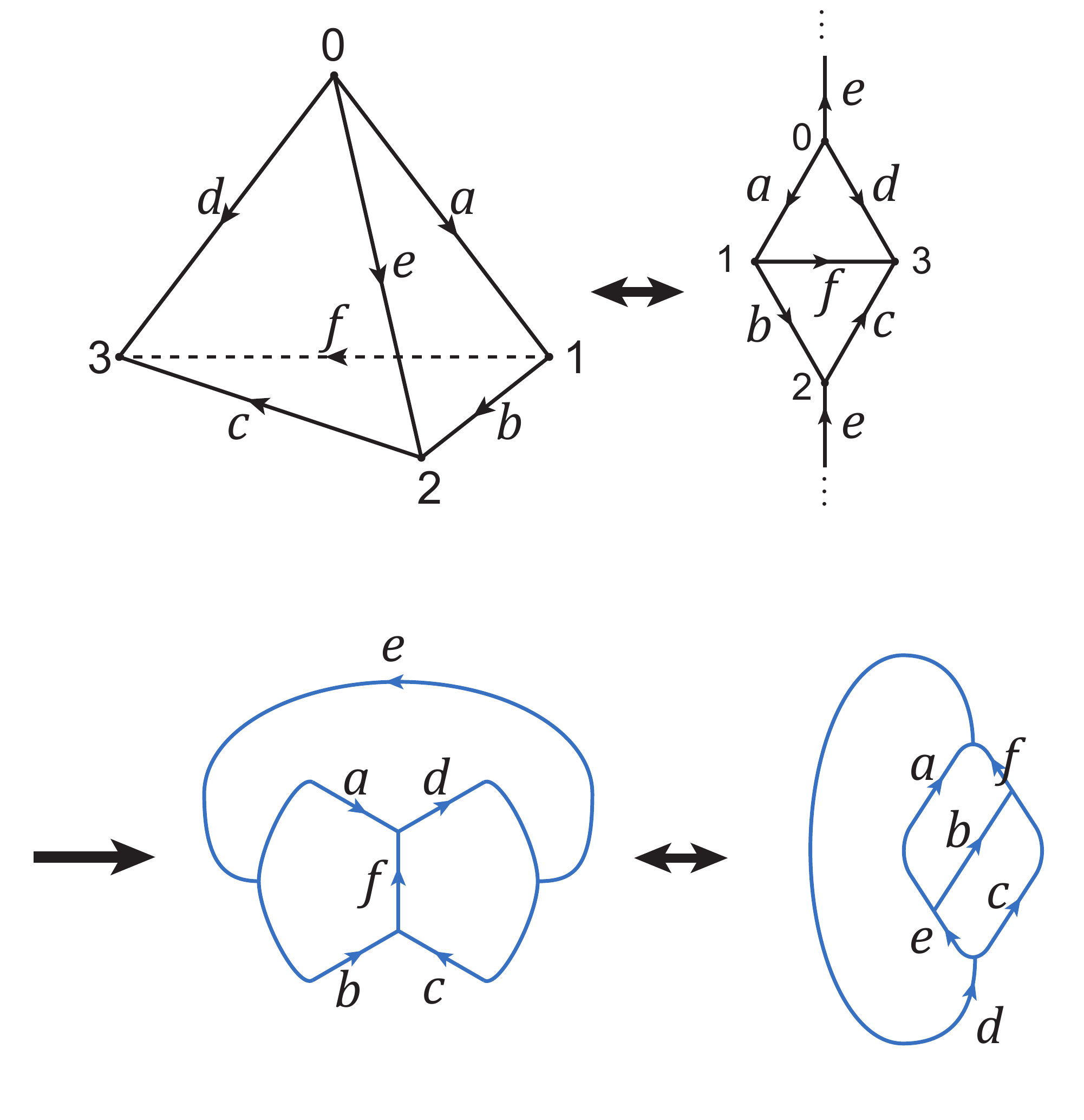}
\caption{(Top) Positively oriented tetrahedron with edge labels as shown can be represented in a planar way. (Bottom) In order to define an amplitude for the tetrahedron, we draw the dual graph (blue), which we represent using a fixed convention as shown on the right diagram. This diagram is evaluated using the diagrammatic calculus of the fusion category.}
\label{graph}
\end{figure}

\begin{figure}[t!]
\includegraphics[width=3in]{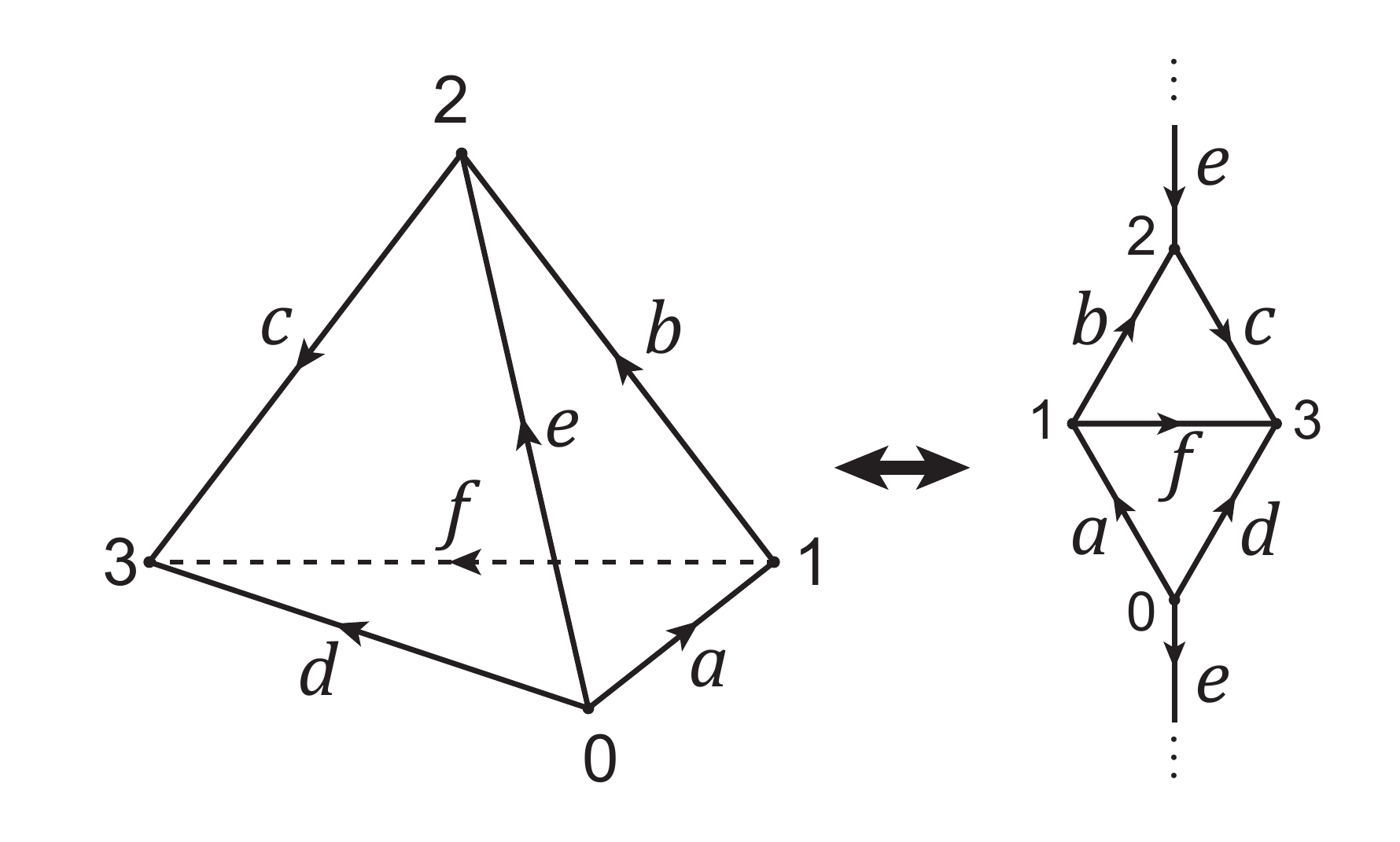}
\caption{Negatively oriented tetrahedron can be represented in a planar way using a fixed convention to define the tetrahedron amplitude using the dual diagram.}
\label{graph_reversed}
\end{figure}

\begin{figure}[t!]
\includegraphics[width=2.5in]{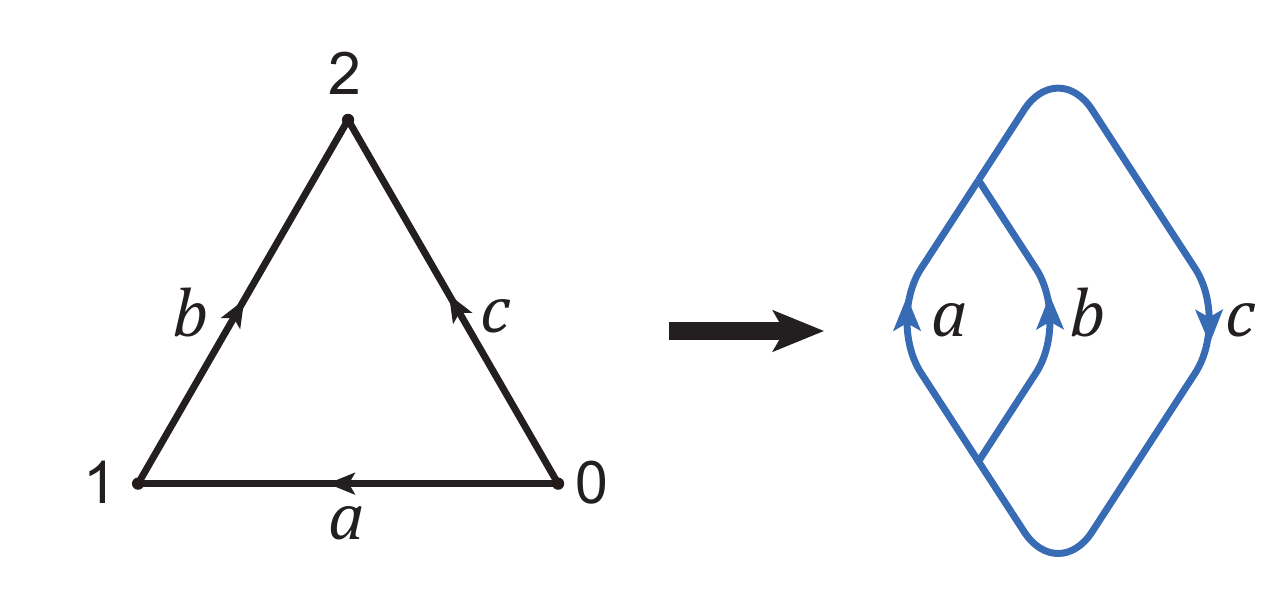}
\caption{``Theta diagram'' (right) associated to the triangle (left) with labels $a$, $b$, and $c$. }
\label{theta}
\end{figure}

We can heuristically think of this definition for $\Tet(\Delta^3; \ell)$ as being obtained in the following manner, which is shown schematically in Fig.~\ref{graph}.
First, we draw the tetrahedron in a planar representation using fixed convention depending on its orientation (as shown in Figs.~\ref{graph} and \ref{graph_reversed}) for where we place the ordered vertices, as shown. We draw dual diagrams on the surface of a tetrahedron such that there is a trivalent vertex at the center of each face and each line in the dual diagram is labeled by the topological charge value of the edge over which it crosses. The direction assigned to a line of the dual diagram is such that the edge it crosses over is oriented towards its right. The state vector of a triangle is assigned to the corresponding vertex of the dual diagram. This is equivalent to saying that the orientation of the triangle determines whether its vector space is mapped to the corresponding fusion space or splitting space of $\mathcal{C}$ when considering the dual diagrams, as shown in Fig.~\ref{fig:vertex_states}. Since a triangle's orientation is specified with respect to a given tetrahedron, it will show up with opposite orientations for the two different tetrahedra that share it as one of their faces, and so it will give a splitting space for one tetrahedron and a fusion space for the other.

Next, we draw the dual diagram in planar form by orienting all its arrows upward (with one line looping around from the top of the diagram back to the bottom). Finally, the dual diagram is evaluated using the diagrammatic calculus of the fusion category $\mathcal{C}$ and the result is the amplitude $\Tet(\Delta^3; \ell)$. The precise value of this amplitude, thus, depends on the normalization convention of the diagrammatic inner products. Using the conventions of Ref.~\onlinecite{barkeshli2014SDG}, we find Eq.~(\ref{eq:Tet}). Regardless of the normalization convention, the evaluation of the diagram will be proportional to $\left[ F^{abc}_d \right]^{s(\Delta^3)}_{(e,\alpha, \beta) (f,\mu,\nu)}$.

To each triangle $\Delta^2$, we similarly assign an amplitude $\Theta(\Delta^2; \ell)^{-1}$, which can be thought of as accounting for the normalization of states in the state spaces assigned to triangles. The factor $\Theta(\Delta^2; \ell)$ for a configuration $\ell$ in which the edges $E_{01}$, $E_{12}$, and $E_{02}$ of $\Delta^2$ are labeled by the corresponding topological charges $a$, $b$, and $c$ is given by evaluating the ``Theta diagram,'' which is the diagrammatic inner product of a basis state in $V_{a,b;c}$ with itself, as shown in Fig.~\ref{theta}. Using the conventions of Ref.~\onlinecite{barkeshli2014SDG}, this diagram evaluates to
\begin{equation}
\Theta(\Delta^2; \ell) = \sqrt{d_a d_b d_c}
.
\end{equation}

Finally, we assign a weight $d(\Delta^1; \ell)$ to each edge $\Delta^1$, where we take
\begin{equation}
d(\Delta^1; \ell) = d_a
,
\end{equation}
when the configuration $\ell$ is such that $a$ is the topological charge value labeling $\Delta^1$.

Putting these all together, the path integral for the manifold $M^3$ is given by
\begin{widetext}
\begin{align}
\Z(M^3) = \mathcal{D}^{-2N_v} \sum_{\ell} \prod\limits_{\Delta^3 \in \mathcal{I}_{3}} \Tet(\Delta^3;\ell) \prod\limits_{\Delta^2 \in \mathcal{I}_{2}} \Theta(\Delta^2; \ell)^{-1}  \prod\limits_{\Delta^1 \in \mathcal{I}_{1}} d(\Delta^1; \ell)  ,
\end{align}
\end{widetext}
where the sum is over all configurations $\ell$ of the topological charge labels on the edges of the triangulation and (basis) states on the triangles of the triangulation.
The quantity $\mathcal{D}^2 = \sum\limits_{a \in \mathcal{C}} d_a^2$  is the total quantum dimension squared of the fusion category.

Since each triangle $\Delta^2$ of the triangulation of a closed manifold is the face of exactly two tetrahedra, we can absorb a factor of $\Theta(\Delta^2, l)^{-1/2}$ into $\Tet(\Delta^3;\ell)$ by for each of the four faces of a tetrahedron $\Delta^3$, defining
\begin{eqnarray}
\widetilde{\Tet}(\Delta^3;\ell) &=& \frac{ \Tet(\Delta^3; \ell) }{ \sqrt{\Theta(F_{012};\ell) \Theta(F_{013};\ell) \Theta(F_{023};\ell) \Theta(F_{123};\ell)}} \notag \\
&=& \left[ F^{abc}_d \right]^{s(\Delta^{3})}_{(e,\alpha, \beta) (f,\mu,\nu)} \frac{1}{\sqrt{d_e d_f}}.
\end{eqnarray}
We can then write the path integral as
\begin{align}
\Z(M^3) = \mathcal{D}^{-2N_v} \sum_{\ell} \prod_{\Delta^3 \in \mathcal{I}_{3}} \widetilde{\Tet}(\Delta^3 ; \ell) \prod_{\Delta^1 \in \mathcal{I}_{1}} d(\Delta^1; \ell).
\end{align}
This is the notation originally used in Ref.~\onlinecite{barrett1996}.

\subsubsection{Topological invariance}

To see that $\Z(M^3)$ is a topological invariant, we need to demonstrate that the above state sum is
independent of the branching structure (the local ordering of vertices of the triangulation)
and is also independent of the choice of triangulation of $M^3$.

\begin{figure}[t!]
\includegraphics[width=3.3in]{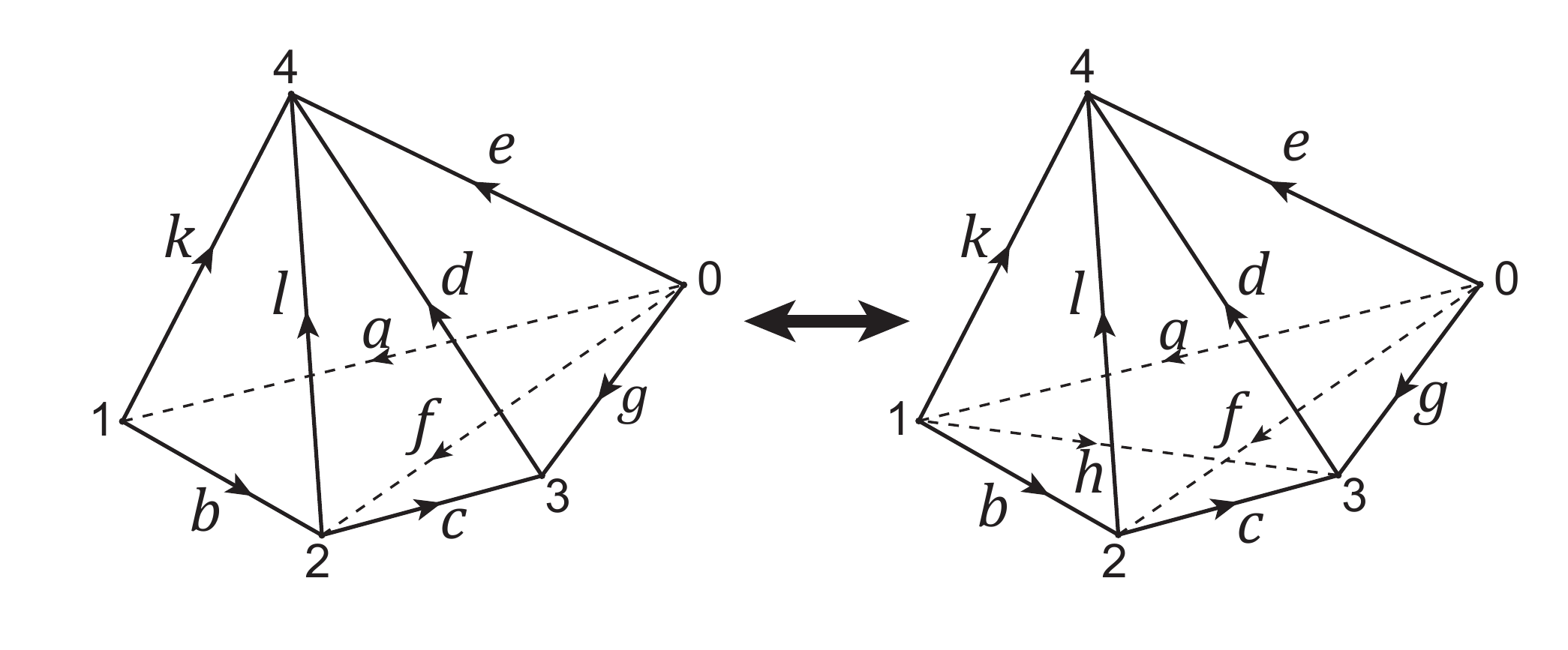}
\caption{The 2-3 Pachner move locally relates a configuration of two tetrahedra to a configuration of three tetrahedra by introducing or removing one edge.}
\label{23PachnerFig}
\end{figure}

\begin{figure}[t!]
\includegraphics[width=3.3in]{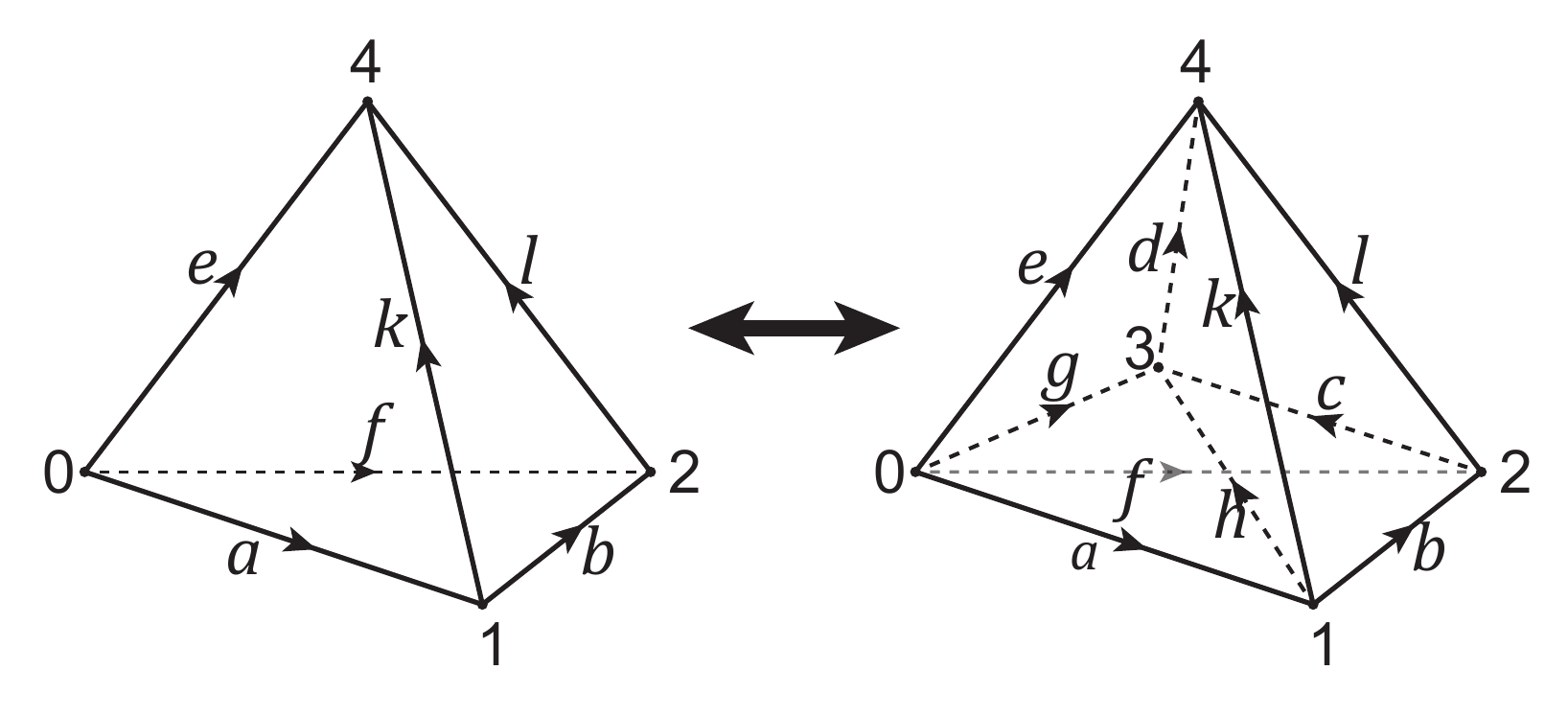}
\caption{The 1-4 Pachner move locally relates a configuration of one tetrahedron to a configuration of four tetrahedra by introducing or removing one vertex and four edges.}
\label{14PachnerFig}
\end{figure}

To demonstrate the independence of the choice of triangulation of $M^3$, it is sufficient to demonstrate the invariance
of $\Z(M^3)$ under the ``2-3'' and ``1-4'' Pachner moves (see Figs.~\ref{23PachnerFig} and \ref{14PachnerFig}).
In what follows, for simplicity we will consider the case where $N_{ab}^c \leq 1$, so that we can write the
$F$-symbols as $F^{abc}_{def} \equiv \left[ F^{abc}_d \right]_{(e,1,1) (f,1,1)} $. The formulae can be easily generalized to the case where $N_{ab}^c$ is not constrained.

We note that a more general discussion that applies to arbitrary cell decompositions (as opposed to triangulations)
is developed in Ref.~\onlinecite{walker2006} using the framework of handle attachments.

The 2-3 Pachner move, shown in Fig.~\ref{23PachnerFig}, is an equality between 2 tetrahedra and 3 tetrahedra, which are related by introducing or removing one edge in the triangulation. Requiring invariance of the state-sum under this move leads to the
pentagon constraint equation for the $F$-symbols
\begin{align}
\label{pentagon}
F^{fcd}_{egl} F^{abl}_{efk} = \sum_{h} F^{abc}_{gfh} F^{ahd}_{egk} F^{bcd}_{khl} .
\end{align}

The 1-4 Pachner move, shown in Fig.~\ref{14PachnerFig}, is an equality between 1 tetrahedron and 4 tetrahedra,
which are related by introducing or removing one vertex and four edges in the triangulation. Requiring invariance under this move leads, after some cancelation of quantum dimensions, to the equation
\begin{align}
\label{14pachner}
F^{abl}_{efk}= \frac{1}{\mathcal{D}^2 }\sum_{c,d,g,h}
F^{abc}_{gfh} F^{ahd}_{egk} F^{bcd}_{khl} [F^{fcd}_{egl}]^*\frac{d_{c} d_{d}}{d_l}
.
\end{align}

Eq.~(\ref{14pachner}) is automatically satified if the $F$-symbols satisfy the pentagon equation [Eq.~(\ref{pentagon})] and the
orthogonality relation (recall we assume here that $N_{ab}^c \leq 1$)
\begin{align}
\sum_f F^{abc}_{def} [F^{abc}_{dgf}]^* = \delta _{eg} N_{ab}^e N_{ec}^d .
\end{align}
To see this, we start with Eq.~(\ref{pentagon}), multiply both sides by
$[F^{abl}_{ezk}]^*$, and sum over $k$. This yields
\begin{align}
N_{ab}^f N_{fl}^e \delta_{fz} F^{fcd}_{egl} = \sum_{k,h} F^{abc}_{gfh} F^{ahd}_{egk} F^{bcd}_{khl} [F^{abl}_{ezk}]^* .
\end{align}
Then, we set $z = f$, multiply both sides by $\frac{d_a d_b}{\mathcal{D}^{2} d_f}$, and sum over $a$ and $b$. Using the identity
$\sum\limits_{a,b} N_{ab}^f \frac{d_a d_b}{d_f} = \mathcal{D}^2$, and the fact that
$F^{fcd}_{egl} = 0$ if $N_{fl}^e = 0$, we obtain Eq.~(\ref{14pachner}).

The independence under changes of branching structure can also be proven, although it is
more involved and requires utilizing the bending factors of fusion diagrams and the spherical condition
on the fusion category $\mathcal{C}$.

\subsubsection{Gauge transformations}

For a given triangle with the associated vector space $V_{ab}^c$, we can consider a unitary transformation on the basis states
\begin{align}
|a,b;c \rangle \rightarrow \Gamma^{ab}_{c} |a,b;c \rangle .
\end{align}
(When $N_{ab}^{c} \leq 1$, $\Gamma^{ab}_{c}$ is simply a phase.)
Such transformations change the $F$-symbols
\begin{align}
[F^{abc}_d]_{ef} \rightarrow \Gamma^{ab}_e \Gamma^{ec}_d [F^{abc}_d]_{ef} [\Gamma^{bc}_f]^{\dagger} [\Gamma^{af}_d]^{\dagger} .
\end{align}
Here, the orientation of the faces of a given tetrahedron determine whether or not the corresponding triangles' states are considered as dual (conjugate) states.
This transformation leaves $\Z(M^3)$ invariant, because a given triangle is the face of exactly two tetrahedra and, as such, it has opposite orientation for these two tetrahedra.

\subsubsection{Wavefunctions and local Hamiltonians}

The state sum can also be used to derive the ground state wavefunctions on a surface $\Sigma^2$ by considering
a manifold $M^3$ whose boundary is $\partial M^3 = \Sigma^2$. Specifically, we have
\begin{widetext}
\begin{align}
\Psi( \ell_{\partial M^3} ) = \mathcal{D}^{-2N_v} \sum_{\ell_{\text{int}(M^3)}} \prod_{\Delta^3 \in \mathcal{I}_{3}} {\Tet}(\Delta^3 ; \ell) \prod_{\Delta^2 \in \mathcal{I}_{2}} \Theta(\Delta^2 ; \ell)^{-1} \prod_{\Delta^1 \in \mathcal{I}_{1}} d(\Delta^1; \ell).
\end{align}
\end{widetext}
Here, the wavefunction $\Psi$ is associated to the configuration $\ell_{\partial M^3}$ of labels on the boundary, which includes the topological charge labels on the edges on the boundary, together with the states associated to each triangle on the boundary, the latter of which can be left implicit when restricting to $N_{ab}^{c}\leq 1$. The summation is over the configurations of labels in the interior of $M^3$.

\begin{figure}[t!]
\includegraphics[width=3.2in]{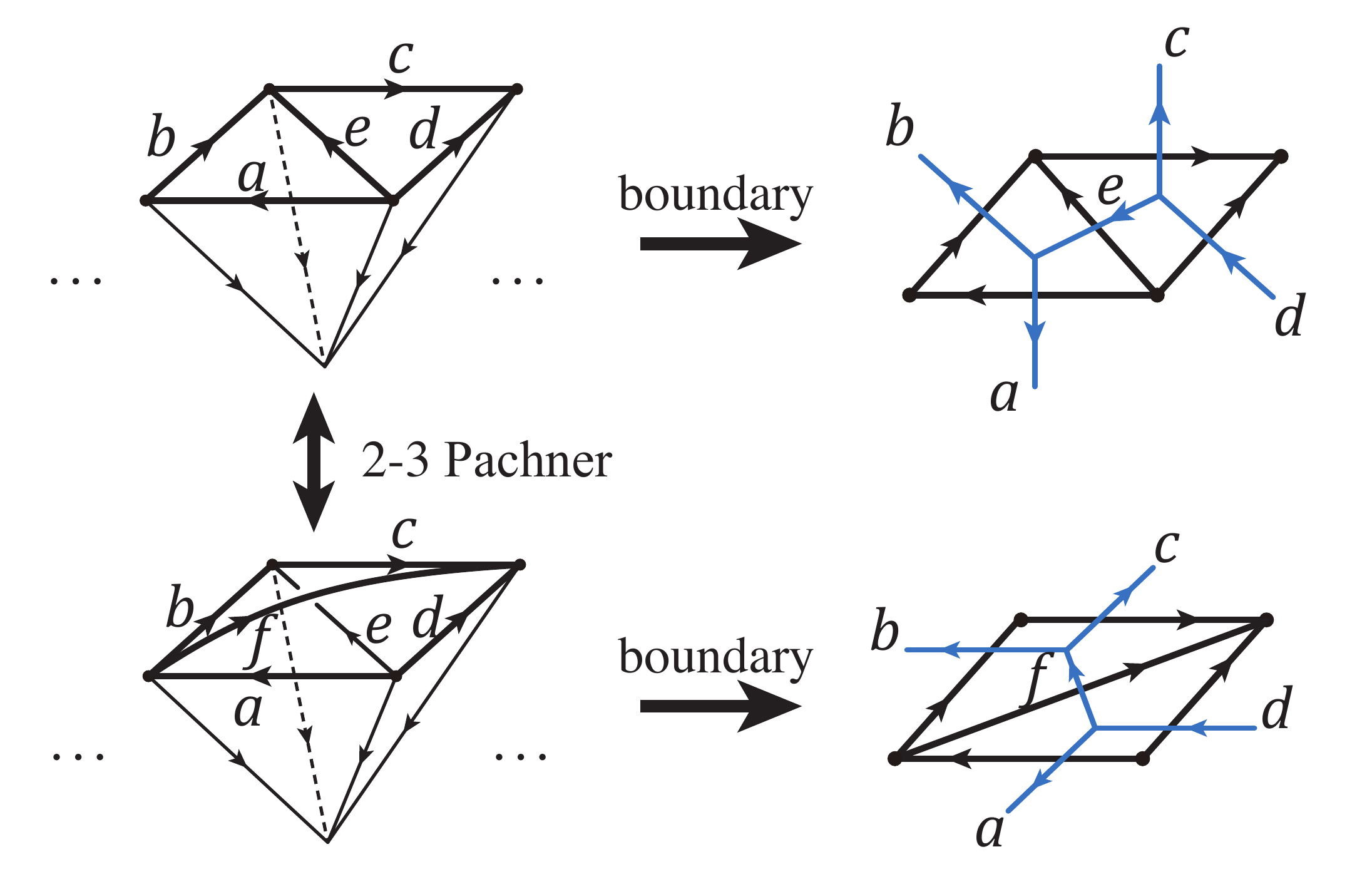}
\caption{Top left: We focus locally on two neighboring tetrahedra in the triangulation. $\cdots$ indicate the rest of the triangulation.
Bold lines indicate the boundary simplexes. Top right: The dual diagram on the boundary triangles. Bottom left: The 2-3 Pachner
move changes the boundary triangulation and, thus, the dual diagram on the boundary (bottom right). This corresponds to applying an $F$-move to the dual diagram on the boundary.}
\label{wfnPachner23Fig}
\end{figure}

\begin{figure}[t!]
\includegraphics[width=3.2in]{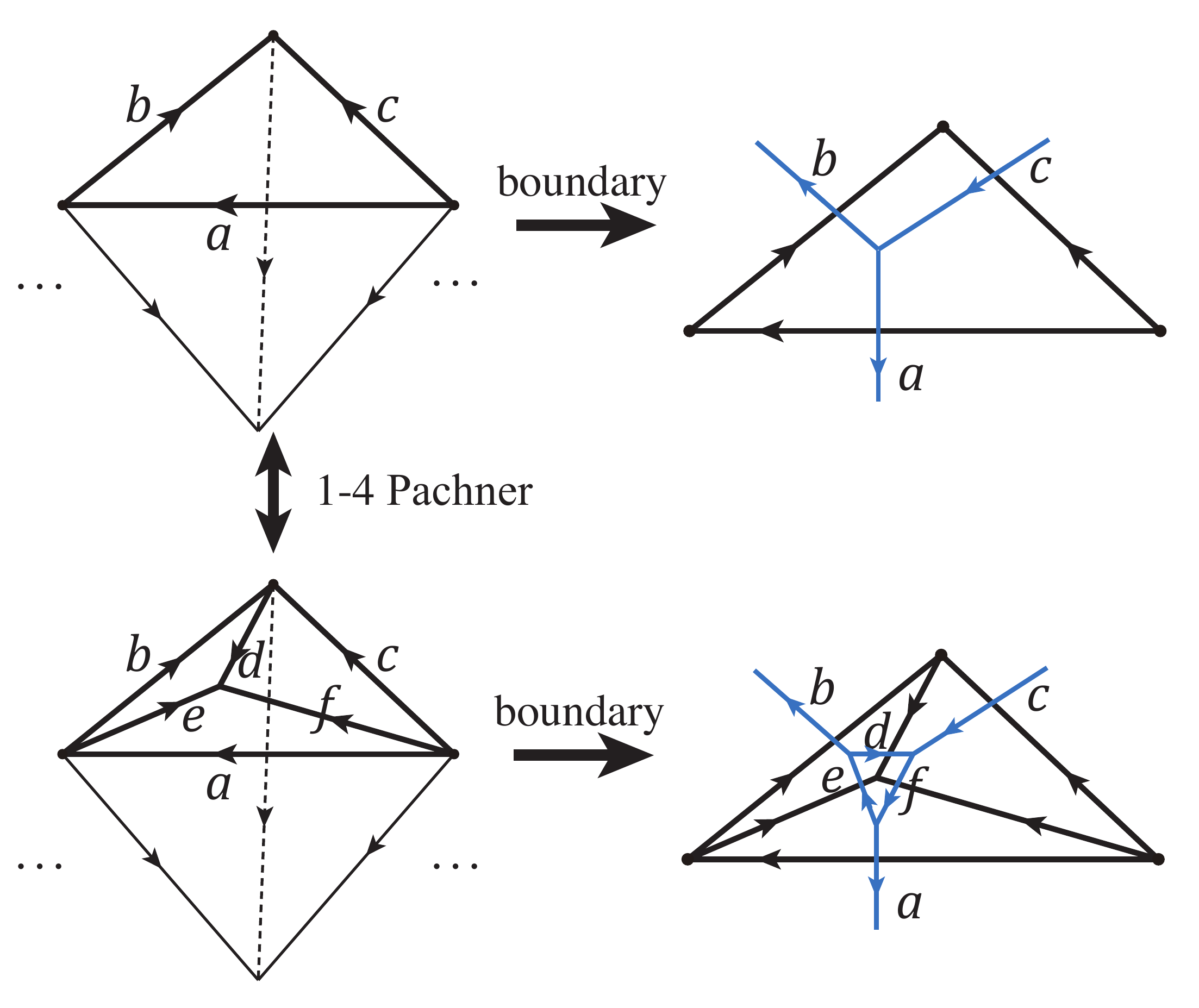}
\caption{Top left: We focus locally on one tetrahedron in the triangulation. $\cdots$ indicate the rest of the triangulation.
Bold lines indicate the boundary simplexes. Top right: The dual diagram on the boundary triangle. Bottom left: The 1-4 Pachner
move changes the boundary triangulation and, thus, the dual diagram on the boundary (bottom right). This corresponds to introducing a bubble and applying $F$-moves to the dual diagram on the boundary.}
\label{wfnPachner14Fig}
\end{figure}

It is convenient to view the wavefunction on the boundary $\partial M^3$ as a wavefunction for degrees of freedom
located on the edges (1-simplexes) of the diagram (cellulation) that is dual to the original triangulation on $\partial M^3$
(see Figs.~\ref{wfnPachner23Fig} and \ref{wfnPachner14Fig}). If we focus on a local patch, it is clear that we have the condition
\begin{align}
\Psi\left(~
\begin{tikzpicture}[baseline={([yshift=-.5ex]current  bounding  box.center)}]
\draw[dualblue,thick,middlearrow={stealth}] (0.8,0) -- (0,0);
\draw[dualblue,thick,middlearrow={stealth reversed}] (-0.3,0.5) -- (0,0);
\draw[dualblue,thick,middlearrow={stealth reversed}] (-0.3,-0.5) -- (0,0);
\draw (0.4,0.15) node {$c$};
\draw (-0.35,0.75) node {$b$};
\draw (-0.35,-0.75) node {$a$};
\end{tikzpicture}
\right)
= 0 \text{\;\;\;if\;\;\;} N_{ab}^c = 0 .
\end{align}
If $N_{ab}^c > 1$, then we have an additional degree of freedom at the vertices of the dual cellulation,
which needs to be specified in the wavefunction.

The invariance of the state sum under the 2-3 and 1-4 Pachner moves implies that the wavefunction satisfies a number of local conditions in the dual diagrams
\begin{align}
\Psi\left(~
\begin{tikzpicture}[baseline={([yshift=-.5ex]current  bounding  box.center)}]
\draw[dualblue,thick,middlearrow={stealth}] (0,0) -- (0,0.8);
\draw[dualblue,thick,middlearrow={stealth reversed}] (0,0) -- (0.5,-0.3) ;
\draw[dualblue,thick,middlearrow={stealth reversed}] (-0.5,-0.3) -- (0,0);
\draw[dualblue,thick,middlearrow={stealth reversed}] (0.5, 1.1) -- (0,0.8) ;
\draw[dualblue,thick,middlearrow={stealth reversed}] (-0.5,1.1) -- (0,0.8);
\draw (0.15,0.4) node {$f$};
\draw (-0.6,1.25) node {$b$};
\draw (-0.6,-0.45) node {$a$};
\draw ( 0.6,1.25) node {$c$};
\draw (0.6,-0.45) node {$d$};
\end{tikzpicture}
\right)
 = \sum_e [F^{abc}_{def}]^*
\Psi\left(~
\begin{tikzpicture}[baseline={([yshift=-.5ex]current  bounding  box.center)}]
\draw[dualblue,thick,middlearrow={stealth reversed}] (0,0) -- (0.8,0);
\draw[dualblue,thick,middlearrow={stealth reversed}] (-0.3,0.5) -- (0,0);
\draw[dualblue,thick,middlearrow={stealth reversed}] (-0.3,-0.5) -- (0,0);
\draw[dualblue,thick,middlearrow={stealth reversed}] (1.1,0.5) -- (0.8,0);
\draw[dualblue,thick,middlearrow={stealth reversed}] (0.8,0) -- (1.1,-0.5);
\draw (0.4,0.15) node {$e$};
\draw (-0.35,0.75) node {$b$};
\draw (-0.35,-0.75) node {$a$};
\draw (1.15,0.75) node {$c$};
\draw (1.15,-0.75) node {$d$};
\end{tikzpicture}
\right)
\end{align}
\begin{align}
\Psi\left(
\begin{tikzpicture}[scale=0.7, baseline={([yshift=-.5ex]current  bounding  box.center)}]
\draw[dualblue,thick,middlearrow={stealth reversed}] (0,-4/3.4) -- (0,-4*0.4/3.4);
\draw[dualblue,thick,middlearrow={stealth reversed}] (-1, 2/3.4 ) -- (-0.4,0.8/3.4);
\draw[dualblue,thick,middlearrow={stealth}] (1,2/3.4) -- (0.4,0.8/3.4);
\draw[dualblue,thick,middlearrow={stealth reversed}] (0.4,0.8/3.4) -- (-0.4,0.8/3.4);
\draw[dualblue,thick,middlearrow={stealth reversed}] (-0.4,0.8/3.4) -- (0,-1.6/3.4);
\draw[dualblue,thick,middlearrow={stealth}] (0.4,0.8/3.4) -- (0,-1.6/3.4) ;
\draw (-1.15,0.75) node {$b$};
\draw (0,-1.4) node {$a$};
\draw (1.15,0.75) node {$c$};
\draw (0, 0.55) node {$d$};
\draw (-0.46,-0.12) node {$e$};
\draw (0.46,-0.12) node {$f$};
\end{tikzpicture}
\right)
= [F^{abd}_{fce}]^* \sqrt{\frac{d_d d_f}{d_c}}
\Psi\left(
\begin{tikzpicture}[scale=0.4, baseline={([yshift=-.5ex]current  bounding  box.center)}]
\draw[dualblue,thick,middlearrow={stealth reversed}] (0,-4/3.4) -- (0,0);
\draw[dualblue,thick,middlearrow={stealth reversed}] (-1, 2/3.4 ) -- (0,0);
\draw[dualblue,thick,middlearrow={stealth}] (1,2/3.4) -- (0,0);
\draw (-1.25,0.8) node {$b$};
\draw (0,-1.7) node {$a$};
\draw (1.25,0.8) node {$c$};
\end{tikzpicture} \right)
\end{align}

The above conditions can be enforced by a local Hamiltonian that is simply a sum of commuting projectors. In Ref.~\onlinecite{levin2005}, Levin and Wen discussed such a Hamiltonian realization of the Turaev-Viro state-sum model for the case where the $F$-symbols satisfy a tetrahedral symmetry (which is equivalent to requiring the diagrams to be isotopy invariant, i.e. bending the diagrams acts trivially on the state space), while some generalizations are discussed in Ref.~\onlinecite{GuPRB2015, lin2014}.

\subsubsection{Resulting TQFT}

The construction presented above defines a topological phase of matter, described by a TQFT. The specific TQFT that is realized is
described by the ``Drinfeld center'' $\D(\mathcal{C})$.~\cite{turaev2010,kirillov2010,balsam2010,balsam2010b}
The Drinfeld center associates a unitary modular tensor category $\D(\mathcal{C})$ to the unitary
fusion category $\mathcal{C}$. Some notable examples are as follows.

When $\mathcal{C}$ describes the fusion properties (including $F$-symbols) of a UMTC $\mathcal{B}$, then
$\D(\mathcal{C}) = \mathcal{B} \times \overline{\mathcal{B}}$, which effectively describes
two decoupled theories, one of which is $\mathcal{B}$, and the other which is its parity-reversed counterpart $\overline{\mathcal{B}}$ (obtained by complex conjugating all of the basic data of $\mathcal{B}$).

As another example, consider the case where the simple objects of $\mathcal{C}$ are simply group elements of a discrete group $\G$ and fusion is defined by
group multiplication. In this case, the $F$-symbols reduce to functions of three group elements $F^{\bf g,h,k}_{\bf ghk, gh, hk}$, and the pentagon
equation reduces to a $3$-cocycle condition for $F^{\bf g,h,k}$. The choice of $F$-symbols, modulo gauge transformations, thus reduces to a choice of $[\omega] \in \mathcal{H}^3(\G, U(1))$. We will denote this fusion category as $\text{Vec}_\G^\omega$.
The TVBW construction yields essentially a discrete gauge theory, known as Dijkgraaf-Witten theory in the full generality. The resulting topological order
is known as the twisted quantum double $\D^{[\omega]}(\G)$ (for a recent exposition, see Ref.~\onlinecite{YTHuTQD}).

Another way to obtain a topological phase described by $\D(\G)$ (without any cohomological twist) is to pick the simple objects in $\mathcal{C}$ to correspond to the irreducible representations of $\G$, where fusion of objects is defined by the tensor product of the irreducible representations, and the $F$-symbols are the usual
$6j$-symbols of the representations. Such a fusion category is denoted by $\mathrm{Rep}(\G)$. The equivalence between the two constructions, $\D(\mathrm{Vec}_\G)$ and $\D(\mathrm{Rep}(\G))$, is demonstrated in Ref.~\onlinecite{BuerschaperPRB2009}.

\subsection{State sum constructions for SET phases}
\label{State_Sum_SETs}

In this section, we generalize the above state sum approaches to capture a general class of SET phases. The topological order
of the SET phase will be described by the Drinfeld center $\D(\mathcal{C}_{\bf 0})$ of a unitary fusion category
$\mathcal{C}_{\bf 0}$. In addition to $\mathcal{C}_{\bf 0}$, we take as input a finite group $G$, which is the symmetry group of the SET phase,
together with a $\mathbb{Z}_2$ grading $\sigma$, as defined in Eq.~(\ref{eq:G_action_sigma}), distinguishing space-time parity even symmetries by $\sigma({\bf g}) = \openone$, and space-time parity odd symmetries by $\sigma({\bf g})=\ast$.

Moreover, as we will see, our state sum will take as input a certain type of $G$-extension of $\mathcal{C}_{\bf 0}$, which we denote $\mathcal{C}_G$:
\begin{align}
\mathcal{C}_G = \bigoplus_{{\bf g}\in G} \mathcal{C}_{\bf g}
\end{align}
that contains $\mathcal{C}_{\bf 0}$ as a subcategory. The fusion rules of simple objects respect group multiplication
\begin{equation}
a_{\bf g} \times b_{\bf h} = \sum_{c_{\bf gh} \in \mathcal{C}_{\bf gh}} N_{ab}^{c} c_{\bf gh}
,
\end{equation}
where we have introduced the shorthand $a_{{\bf g}}$ to indicate that charge $a$ is in $\mathcal{C}_{{\bf g}}$.

When $\sigma({\bf g}) = \openone$ for all ${\bf g} \in G$, we will see that $\mathcal{C}_G$ is a unitary
$G$-graded fusion category. For a discussion of unitary $G$-graded fusion categories written for physicists, see Ref.~\onlinecite{barkeshli2014SDG}.
However, when $G$ has anti-unitary or orientation-reversing elements, the $G$-extension $\mathcal{C}_G$ is not simply a
$G$-graded fusion category. Below we will argue that it can be viewed as a $G$-equivariant 2-category with possibly
anti-unitary $G$-actions, which we will define more precisely.

\begin{figure}[t!]
\includegraphics[width=3.2in]{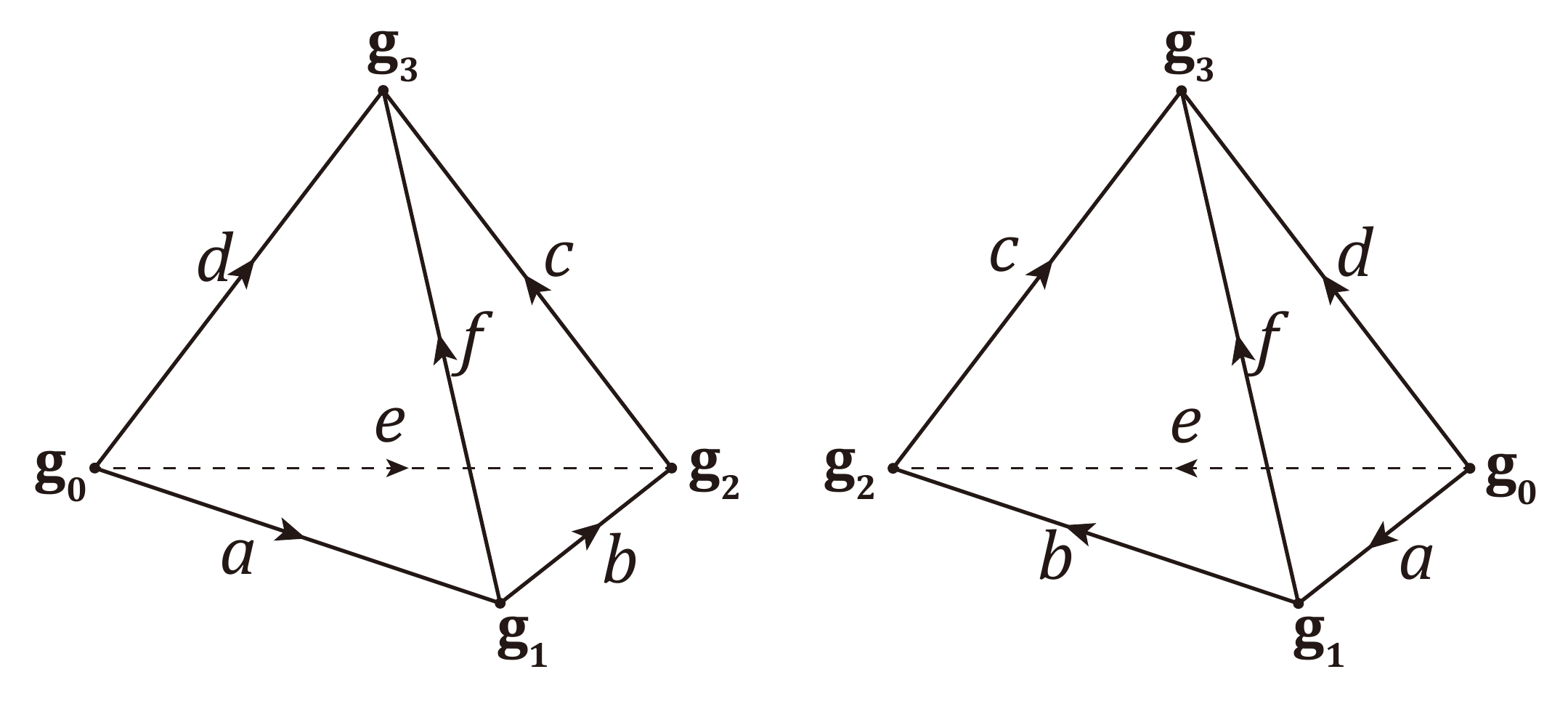}
\caption{Two tetrahedra in the SET state sum, with positive orientation (left) and negative orientation (right). Vertices are labeled by elements of the symmetry group $G$. The edge connecting vertices $j$ and $k$ is labeled by a topological charge in sector $\mathcal{C}_{{\bf g}_{jk}}$ of a 2-category $\mathcal{C}_{G}$, where ${\bf g}_{jk} = {\bf g}_{j}^{-1} {\bf g}_{k}$.}
\label{tetraGFig}
\end{figure}

The construction of our state-sum is essentially a hybrid of the SPT and TVBW state sums described in the previous sections.
When $G$ is trivial, it reduces to the TVBW state sum, and when $\mathcal{C}_{\bf 0}$ is trivial, it reduces to the SPT state sum.

As before, we triangulate the space-time manifold $M^3$, and denote the set of $n$-simplexes of the triangulation by $\mathcal{I}_{n}$ and provide an ordering to the elements of $\mathcal{I}_{0}$, assigning the labels $j=0,1,\ldots,N_{v}-1$ to the vertices (0-simplexes) of the triangulation. Again, this provides a direction to each edge (1-simplex), pointing from the lower ordered endpoint of the edge to the higher one, a branching structure, and an orientation to each 3-simplex $\Delta^{3} \in \mathcal{I}_{3}$.

In this case, we assign a group element ${\bf g}_{j} \in G$ to the $j$th vertex of the triangulation and, defining ${\bf g}_{jk} = {\bf g}_{j}^{-1} {\bf g}_{k}$, we assign
a topological charge label $a \in \mathcal{C}_{{\bf g}_{jk}}$ to the edge $E_{jk} \in \mathcal{I}_{1}$ connecting vertices $j$ and $k$.

For a given triangle $\Delta^{2} \in \mathcal{I}_{2}$, we label its vertices $j_{0}$, $j_{1}$, and $j_{2}$, such that $j_{0}<j_{1}<j_{2}$, and denote the corresponding edges connecting these vertices as $E_{01}$, $E_{12}$, and $E_{02}$. When these edges of the triangle are labeled by the corresponding topological charges $a_{{\bf g}_{01}}$, $b_{{\bf g}_{12}}$, and $c_{{\bf g}_{02}}$, respectively, we assign a vector space $V_{a,b;c}({\bf g}_{j_{0}}, {\bf g}_{j_{1}}, {\bf g}_{j_{2}})$ to the triangle, where $\text{dim } V_{a,b;c} = N_{ab}^c$. Notice that ${\bf g}_{01} {\bf g}_{12}={\bf g}_{02}$. We then assign a state label $\alpha_{\Delta^2}$ of the corresponding vector space to the triangle. The configuration of the group elements labeling the vertices, the topological charge labels on the edges, and corresponding (basis) states on the triangles of the entire triangulation will be collectively denoted as $\ell$.

\begin{figure}[t!]
\includegraphics[width=3.3in]{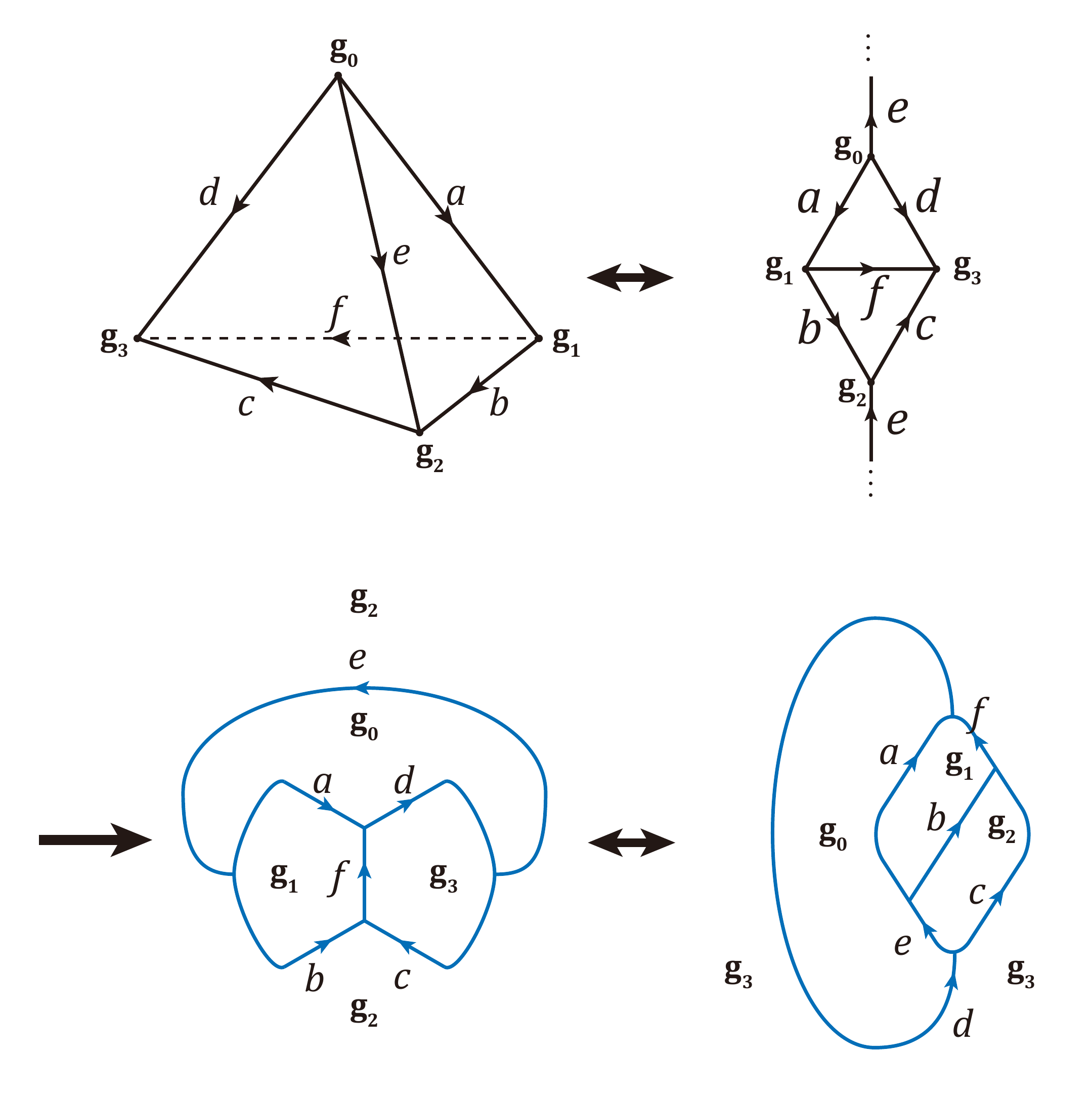}
\caption{Schematic for determining $\Tet(\Delta^3;\ell)$ for a tetrahedron for the SET state sum follows the same procedure as for the TVBW state sum, but with extra structure on the configuration labels.}
\label{graphG}
\end{figure}

Finally, for a configuration $\ell$, we associate an amplitude $\Tet(\Delta^3;\ell)$ to each tetrahedron, a factor $\Theta(\Delta^2;\ell)^{-1}$ to each triangle, and a factor $d(\Delta^1; \ell)$ to each edge. The path integral on the manifold is then
\begin{widetext}
\begin{align}
\Z(M^3) = \mathcal{D}^{-2N_v} \sum_{\ell} \prod_{\Delta^3 \in \mathcal{I}_{3}} \Tet(\Delta^3;\ell) \prod_{\Delta^2 \in \mathcal{I}_{2}} \Theta(\Delta^2; \ell)^{-1} \prod_{\Delta^1 \in \mathcal{I}_{1}} d(\Delta^1; \ell) ,
\end{align}
\end{widetext}
where the sum is over all configurations $\ell$. In this expression
\begin{equation}
\mathcal{D}^{2}_{G} = \sum\limits_{{\bf g} \in G} \sum\limits_{a_{\bf g} \in C_{\bf g}} d_{a_{\bf g}}^2 = |G| \mathcal{D}^{2}_{\bf 0}
\end{equation}
is the total quantum dimension of $\mathcal{C}_G$. The factors $\Tet(\Delta^3; \ell)$ and $\Theta(\Delta^2; \ell)$ appearing
in the state sum can be computed from dual diagrams associated with the 3-simplexes and 2-simplexes, respectively, in a manner identical to that described in Sec.~\ref{sec:TVBW}, as shown in Fig.~\ref{graphG}. These diagrams can be evaluated using a diagrammatic calculus that now contains group elements in the spaces
between the lines. As discussed below, this is naturally interpreted in terms of a 2-category.

As in the TVBW case, we can define
\begin{eqnarray}
\widetilde{\Tet}(\Delta^3;\ell) &=& \frac{ \Tet(\Delta^3; \ell) }{ \sqrt{\Theta(F_{012};\ell) \Theta(F_{013};\ell) \Theta(F_{023};\ell) \Theta(F_{123};\ell)}} \notag \\
&=& [F^{abc}_{def}({\bf g}_0, {\bf g}_1, {\bf g}_2, {\bf g}_3)]^{s(\Delta^3)} \frac{1}{\sqrt{d_e d_f}}
\end{eqnarray}
where the configuration $\ell$ is such that the tetrahedron $\Delta^3$ has vertices $j_{0}, j_{1}, j_{2}, j_{3}$ that are respectively assigned group labels ${\bf g}_0, {\bf g}_1, {\bf g}_2, {\bf g}_3$ and
edges $E_{01},E_{12},E_{23},E_{03},E_{02},E_{13}$ that are respectively labeled by the topological charges $a_{{\bf g}_{01}},b_{{\bf g}_{12}},c_{{\bf g}_{23}},d_{{\bf g}_{03}},e_{{\bf g}_{02}},f_{{\bf g}_{13}}$, as shown in Fig.~\ref{tetraFig}. The faces $F_{012}, F_{023}, F_{123}, F_{013}$ of $\Delta^{3}$ are also respectively labeled by the corresponding basis states $\alpha, \beta, \mu, \nu$, but we leave these implicit while we focus on the cases where there are no fusion multiplicities.
The SET state sum is then given by
\begin{align}
\Z(M^3) = |\mathcal{D}_G|^{-2 N_v} \sum_l \prod_{\Delta^3 \in \mathcal{I}_{3}} \widetilde{\Tet}(\Delta^3; \ell) \prod_{\Delta^1 \in \mathcal{I}_{1}} d (\Delta^1; \ell).
\end{align}

Invariance of the path integral under the $2-3$ Pachner move now gives the equation
\begin{widetext}
\begin{align}
\label{gPentagon}
F^{fcd}_{egl} ({\bf g}_0, {\bf g}_2, {\bf g}_3, {\bf g}_4) F^{abl}_{efk}({\bf g}_0, {\bf g}_1, {\bf g}_2, {\bf g}_4)
= \sum_h F^{abc}_{gfh}({\bf g}_0, {\bf g}_1, {\bf g}_2, {\bf g}_3) F^{ahd}_{egk}({\bf g}_0, {\bf g}_1, {\bf g}_3, {\bf g}_4) F^{bcd}_{khl}({\bf g}_1, {\bf g}_2, {\bf g}_3, {\bf g}_4) .
\end{align}
The $1-4$ Pachner move in this case gives
\begin{align}
F^{abl}_{efk}({\bf g}_0, {\bf g}_1, {\bf g}_2, {\bf g}_4) = \frac{1}{\mathcal{D}_G^2 }\sum_{c,d,g,h}
F^{abc}_{gfh}({\bf g}_0, {\bf g}_1, {\bf g}_2, {\bf g}_3)
F^{ahd}_{egk}({\bf g}_0, {\bf g}_1, {\bf g}_3, {\bf g}_4)
F^{bcd}_{khl}({\bf g}_1, {\bf g}_2, {\bf g}_3, {\bf g}_4)
[F^{fcd}_{egl}({\bf g}_0, {\bf g}_2, {\bf g}_3, {\bf g}_4)]^*\frac{d_{c} d_{d}}{d_l}
.
\end{align}
We note that the topological charges are assigned the following group labels in these expressions: $a_{{\bf g}_{01}},b_{{\bf g}_{12}},c_{{\bf g}_{23}},d_{{\bf g}_{34}}, e_{{\bf g}_{04}},f_{{\bf g}_{02}}, g_{{\bf g}_{03}}, h_{{\bf g}_{13}},k_{{\bf g}_{14}}, l_{{\bf g}_{24}}$.
%
%\begin{align}
%F^{fcd}_{egh}({\bf g}_0, {\bf g}_1, {\bf g}_2, {\bf g}_4) = \frac{1}{\mathcal{D}_G^2 }\sum_{a,b,k,l} F^{ald}_{egk}({\bf g}_0, {\bf g}_1, {\bf g}_3, {\bf g}_4)
%F^{bcd}_{klh}({\bf g}_0, {\bf g}_1, {\bf g}_2, {\bf g}_3)
%F^{abc}_{gfl}({\bf g}_1, {\bf g}_2, {\bf g}_3, {\bf g}_4) [F^{abh}_{efk}({\bf g}_0, {\bf g}_2, {\bf g}_3, {\bf g}_4)]^*  \frac{d_a d_b}{d_f}
%.
%\end{align}

\subsubsection{Ground state wavefunctions and symmetry invariance}

As in the TVBW example, we can obtain ground state wavefunctions by considering the state sum
on a manifold $M^3$ with boundary
\begin{align}
\Psi( \ell_{\partial M^3} ) = \mathcal{D}_G^{-2N_v} \sum_{\ell_{\text{int}(M^3)}} \prod_{\Delta^3 \in \mathcal{I}_{3}}
{\Tet}(\Delta^3; \ell) \prod_{\Delta^2 \in \mathcal{I}_{2}} \Theta(\Delta^2; \ell)^{-1} \prod_{\Delta^1 \in \mathcal{I}_{1}} d (\Delta^1 ; \ell) ,
\end{align}
\end{widetext}
where $\ell_{\partial M^3}$ is the configuration of group elements on the vertices on the boundary and topological charge labels on the edges of the boundary (and basis states on faces when
fusion coefficients are greater than 1). The sum is over all configurations $\ell_{\text{int}(M^3)}$ of the interior of $M^3$. As before,
it is convenient to view the degrees of freedom on the boundary as defined on a dual
cellulation, where $a_{{\bf g}_{jk}}$ is defined on edges (1-cells), ${\bf g}_j$ on plaquettes (2-cells),
and fusion basis states on vertices (0-cells). The wavefunction then satisfies
local relations associated with $F$-moves coming from the 2-3 Pachner move,
and bubbles coming from the 1-4 Pachner move, e.g.
\begin{align}
\Psi\left(
\begin{tikzpicture}[baseline={([yshift=-.5ex]current  bounding  box.center)}]
\draw[dualblue,thick,middlearrow={stealth}] (0,0) -- (0,0.8);
\draw[dualblue,thick,middlearrow={stealth}] (0.5,-0.3) -- (0,0) ;
\draw[dualblue,thick,middlearrow={stealth reversed}] (-0.5,-0.3) -- (0,0);
\draw[dualblue,thick,middlearrow={stealth reversed}] (0.5, 1.1) -- (0,0.8) ;
\draw[dualblue,thick,middlearrow={stealth reversed}] (-0.5,1.1) -- (0,0.8);
\draw (0.15,0.4) node {$f$};
\draw (-0.6,1.25) node {$b$};
\draw (-0.6,-0.45) node {$a$};
\draw ( 0.6,1.25) node {$c$};
\draw (0.6,-0.45) node {$d$};
\draw (0, -0.5) node {${\bf g}_0$};
\draw (-0.5,0.4) node {${\bf g}_1$};
\draw (0, 1.15) node {${\bf g}_2$};
\draw(0.5,0.4) node {${\bf g}_3$};
\end{tikzpicture}
\right)
= \sum_e [F^{abc}_{def}]^*
\Psi\left(
\begin{tikzpicture}[baseline={([yshift=-.5ex]current  bounding  box.center)}]
\draw[dualblue,thick,middlearrow={stealth}] (0.8,0) -- (0,0);
\draw[dualblue,thick,middlearrow={stealth}] (0,0) -- (-0.3,0.5);
\draw[dualblue,thick,middlearrow={stealth}] (0,0) -- (-0.3,-0.5);
\draw[dualblue,thick,middlearrow={stealth}] (0.8,0) -- (1.1,0.5);
\draw[dualblue,thick,middlearrow={stealth}] (1.1,-0.5) -- (0.8,0);
\draw (0.4,0.15) node {$e$};
\draw (-0.5,0) node {${\bf g}_1$};
\draw (0.4, 0.65) node {${\bf g}_2$};
\draw (1.2,0) node {${\bf g}_3$};
\draw (-0.35,0.75) node {$b$};
\draw (0.4, -0.65) node {${\bf g}_0$};
\draw (-0.35,-0.75) node {$a$};
\draw (1.15,0.75) node {$c$};
\draw (1.15,-0.75) node {$d$};
\end{tikzpicture}
\right)
\end{align}

Let us now consider the action of the symmetry group $G$ on the degrees of freedom. In general, for ${\bf h} \in G$ we have
\begin{align}
{\bf h} : \;\; &{\bf g}_j \rightarrow {\bf h} {\bf g}_{{\bf h}(j)}
\nonumber \\
& a_{{\bf g}_{jk}} \rightarrow \,^{\bf h} a_{{\bf g}_{{\bf h}(j){\bf h}(k)} }
.
\end{align}
If ${\bf h}$ is either a unitary or anti-unitary on-site symmetry, we have ${\bf h}(j) = j$. On the other hand, if ${\bf h}$ is a unitary spatial
symmetry, such as reflection or translation, then ${\bf h}(j)$ corresponds to the reflected or translated point
on the lattice. For spatial symmetries, the triangulation on the boundary must therefore respect the lattice
symmetry. We allow the action on topological charge values $\,^{\bf h}a_{{\bf g}_{{\bf h}(j){\bf h}(k)} }$ to be general.

When the $\partial M^3 = S^2$, the resulting Hilbert space is one-dimensional. We require this ground
state wavefunction to be symmetric under the action of ${\bf h}$. Symmetry of the wavefunction implies
\begin{align}
\label{wfnsymm}
\Psi\left( \{ {\bf h} {\bf g}_{j} \}; \{ \,^{\bf h}a_{{\bf g}_{jk}} \} \right)^{\sigma({\bf h})} = \Psi\left(\{ {\bf g}_{j} \}; \{ a_{{\bf g}_{jk}}\} \right) ,
\end{align}
where $\sigma({\bf h})$ is complex conjugation when ${\bf h}$ is a space-time parity odd, and trivial otherwise, as defined in Eq.~(\ref{eq:G_action_sigma}).
For the case when ${\bf h}$ is on-site (unitary or anti-unitary), this should be clear. When ${\bf h}$ corresponds to a unitary spatial reflection symmetry, we obtain this result in two steps. Firstly, the symmetry of the wavefunction requires
\begin{align}
\Psi\left( \{ {\bf h} {\bf g}_{{\bf h}(j)} \}; \{ \,^{\bf h}a_{{\bf g}_{{\bf h}(j){\bf h}(k)}} \} \right) = \Psi\left(\{ {\bf g}_{j} \}; \{ a_{{\bf g}_{jk}}\} \right)
.
\end{align}
Secondly, the path integral definition of the wavefunction has the property that the configuration on the spatially reflected vertex coordinates is related to the configuration on the original coordinates by
\begin{align}
\Psi\left( \{ {\bf h} {\bf g}_{{\bf h}(j)} \}; \{ \,^{\bf h}a_{{\bf g}_{{\bf h}(j){\bf h}(k)}} \} \right) = \Psi\left( \{ {\bf h} {\bf g}_{j} \}; \{ \,^{\bf h}a_{{\bf g}_{jk}} \} \right)^\ast
,
\end{align}
where the complex conjugation arises because the reflection reverses the orientation of every tetrahedron.~\footnote{We note that one can also understand the on-site anti-unitary symmetries in the same way by recognizing them as reflections in the time direction and the boundary $\partial M^3$ upon which the wavefunction is defined as a time-slice of the space-time manifold. The time reflection then gives complex conjugation as a result of reversing the orientation of all the tetrahedra in the state sum.}

In order for the $F$-moves to preserve this symmetry of the wavefunction, the $F$-symbols must satisfy
\begin{align}
\label{Fsym}
[F^{\,^{\bf h}a\,^{\bf h}b\,^{\bf h}c}_{\,^{\bf h}d\,^{\bf h}e\,^{\bf h}f} ({\bf h} {\bf g}_0, {\bf h} {\bf g}_1, {\bf h} {\bf g}_2, {\bf h} {\bf g}_3) ]^{\sigma({\bf h})}
= F^{abc}_{def} ({\bf g}_0, {\bf g}_1, {\bf g}_2, {\bf g}_3)
.
\end{align}
Using this relation, we can drop the arguments in parentheses by taking ${\bf h} = {\bf g}_0^{-1}$,
and replacing $F^{abc}_{def}({\bf g}_0, {\bf g}_1, {\bf g}_2, {\bf g}_3)$ with $[F^{\,^{\bf h}a\,^{\bf h}b\,^{\bf h}c}_{\,^{\bf h}d\,^{\bf h}e\,^{\bf h}f} ]^{\sigma({\bf h})}$, where we define
\begin{equation}
F^{a_{{\bf g}_{01}} b_{{\bf g}_{12}} c_{{\bf g}_{23}} }_{d_{{\bf g}_{03}} e_{{\bf g}_{02}} f_{{\bf g}_{13}}} \equiv F^{abc}_{def} ({\bf 1}, {\bf g}_0^{-1}{\bf g}_1, {\bf g}_0^{-1}{\bf g}_2, {\bf g}_0^{-1}{\bf g}_3)
.
\end{equation}

Using the symmetry condition, the 2-3 Pachner equation can be rewritten as
\begin{align}
\label{gPentagon2}
F^{fcd}_{egl} F^{abl}_{efk} = \sum_h F^{abc}_{gfh} F^{ahd}_{egk} [F^{\,^{{\bf g}_{01}}b \,^{{\bf g}_{01}}c \,^{{\bf g}_{01}}d}_{\,^{{\bf g}_{01}}k \,^{{\bf g}_{01}}h \,^{{\bf g}_{01}}l}]^{\sigma({\bf g}_{01})} .
\end{align}
We note that the topological charges are assigned the following group labels in this expression: $a_{{\bf g}_{01}},b_{{\bf g}_{12}},c_{{\bf g}_{23}},d_{{\bf g}_{34}}, e_{{\bf g}_{04}},f_{{\bf g}_{02}}, g_{{\bf g}_{03}}, h_{{\bf g}_{13}},k_{{\bf g}_{14}}, l_{{\bf g}_{24}}$.

Let us first consider the case where $G$ acts trivially on the objects, i.e. $\,^{\bf h} a_{{\bf g}} = a_{{\bf g}}$ for
all ${\bf h} \in G$ and $a \in \mathcal{C}_G$, and the symmetry group corresponds to unitary, orientation preserving transformations,
i.e. $\sigma({\bf h}) = \openone$ for all ${\bf h}\in G$. In this case, Eq.~(\ref{gPentagon2}) is just the usual pentagon consistency equation for a fusion category.
The additional $G$-graded structure on the objects defined on the $1$-simplexes implies that we have the structure of a unitary $G$-graded
fusion category (see Ref.~\onlinecite{barkeshli2014SDG} for a discussion of $G$-graded fusion categories written for physicists).

In Ref.~\onlinecite{barkeshli2014SDG}, it was argued that (2+1)D SET phases with a unitary on-site symmetry group
$G$ and whose topological order is described by a UMTC $\mathcal{B}$ are classified by
unitary $G$-crossed braided extensions of $\mathcal{B}$, denoted by $\mathcal{B}_G^\times$.
Ref.~\onlinecite{ENO2009} has proven that unitary $G$-graded fusion categories
$\mathcal{C}_G$ are in one-to-one correspondence with $\D(\mathcal{C}_0)_G^\times$, which is a
$G$-crossed braided extension of $\D(\mathcal{C}_0)$. Thus, for purposes of classifying SET phases for a topological order $\D(\mathcal{C}_0)$ and symmetry group $G$ whose action is unitary and orientation-preserving, it is sufficient to consider SET state sum models where the input is a unitary $G$-graded fusion category. As mentioned in the
previous paragraph, $G$-graded fusion categories are obtained if in the above construction we have $G$ act trivially on all the objects,
i.e. $\,^{\bf h} a_{{\bf g}} = a_{{\bf g}}$ for all ${\bf h} \in G$ and $a \in \mathcal{C}_G$. The extra generality of allowing ${\bf h}$ to
permute the objects is therefore not necessary for classifying such (2+1)D SET phases.

However, when $G$ contains anti-unitary or orientation reversing symmetry transformations, we will have $\sigma({\bf h}_{01}) = \ast$
in some cases. The presence of this possible complex conjugation on the RHS of Eq.~(\ref{gPentagon2})
distinguishes it from the usual pentagon consistency equation for a fusion category. Therefore, when $G$ contains anti-unitary
or spatial reflection symmetry transformations, the structure we obtain is not even necessarily that of a fusion category.
Moreover, we no longer have any reason to rule out cases where $\,^{\bf h} a_{{\bf g}} \neq a_{{\bf g}}$, which also distinguishes
Eq.~(\ref{gPentagon2}) from the usual pentagon equation for a fusion category.

This more general mathematical object forms a type of anti-unitary $G$-extension of the category $\mathcal{C}_{\bf 0}$.
We will subsequently argue that this mathematical structure can be viewed as a more general type of $G$-equivariant
2-category with $G$ actions.

\subsubsection{Gauge transformations}
\label{SETgauge}

It is important to also discuss the gauge transformations which keep $\Z(M^3)$ invariant on closed manifolds, and which
do not change the topological phase determined by the ground state wavefunction.

Each triangle is associated with an element in the vector space $V_{a,b;c}({\bf g}_0, {\bf g}_1, {\bf g}_2)$, which maps to either a fusion space $V_{ab}^c({\bf g}_0, {\bf g}_1, {\bf g}_2)$ or its dual (splitting) space $V_c^{ab}({\bf g}_0, {\bf g}_1, {\bf g}_2)$ for a given tetrahedron.
Basis transformations $\Gamma^{ab}_c({\bf g}_0, {\bf g}_1, {\bf g}_2)$ in this space keep $\Z(M^3)$ invariant for closed 3-manifolds $M^3$,
and transform the $F$-symbols as
\begin{widetext}
\begin{align}
F^{abc}_{def}({\bf g}_0, {\bf g}_1, {\bf g}_2, {\bf g}_3) \rightarrow
\Gamma^{ab}_e({\bf g}_0, {\bf g}_1, {\bf g}_2) \Gamma^{ec}_d({\bf g}_0, {\bf g}_2, {\bf g}_3)
F^{abc}_{def}({\bf g}_0, {\bf g}_1, {\bf g}_2, {\bf g}_3)
[\Gamma^{bc}_f({\bf g}_1, {\bf g}_2, {\bf g}_3)]^{-1} [\Gamma^{af}_d({\bf g}_0, {\bf g}_2, {\bf g}_3)]^{-1}
.
\end{align}
\end{widetext}
In order to preserve the symmetry of the wavefunction, the gauge transformations must satisfy
\begin{align}
\Gamma^{\,^{\bf h}a \,^{\bf h}b}_{\,^{\bf h}c}({\bf h} {\bf g}_0, {\bf h} {\bf g}_1, {\bf h} {\bf g}_2) = [\Gamma^{ab}_c({\bf g}_0, {\bf g}_1, {\bf g}_2)]^{s({\bf h})}
.
\end{align}
In this way, we can also replace $\Gamma^{ab}_c({\bf g}_0, {\bf g}_1, {\bf g}_2)$ with $\Gamma^{\,^{\bf h}a \,^{\bf h}b}_{\,^{\bf h}c}$, where ${\bf h} = {\bf g}_{0}^{-1}$, and the gauge transformation may be rewritten in terms of the quantities with group elements removed as
%\begin{widetext}
\begin{align}
F^{abc}_{def} \rightarrow
\Gamma^{ab}_e \Gamma^{ec}_d F^{abc}_{def} \left[ \left(\Gamma^{\,^{{\bf g}_{01}}b \,^{{\bf g}_{01}}c}_{\,^{{\bf g}_{01}}f} \right)^{\sigma({\bf g}_{01})} \right]^{-1} [\Gamma^{af}_d]^{-1}
.
\end{align}
%\end{widetext}

\subsubsection{Diagrammatic calculus and $G$-equivariant 2-categories with $G$-action}
\label{2catSec}

The above construction motivates a new type of diagrammatic calculus, which is naturally associated with the notion of a 2-category with a
$G$-action. Depending on the $\mathbb{Z}_2$ grading of the group elements, this group action can have a unitary or anti-unitary character,
as we will see.

We consider a 2-category $\mathcal{C}$ where the objects, which are elements of a set $\mathcal{C}^0$, correspond to
group elements of $G$. The collection of 1-morphisms $\mathcal{C}^1$ consists of elements
$a_{{\bf g}_i, {\bf g}_j} \in \mathcal{C}^1$, which is a 1-morphism from ${\bf g}_i$ to ${\bf g}_j$. These
1-morphisms are represented diagramatically as lines in a planar graph, where the empty spaces between the lines are labeled
by group elements (the objects in $\mathcal{C}^0$), as shown in Fig.~\ref{graphG}. The 1-morphisms can be composed, so that
$a_{{\bf g}_i, {\bf g}_j} \otimes b_{{\bf g}_j, {\bf g}_k}$ is also a 1-morphism. The collection of  2-morphisms $\mathcal{C}^2$, consists
of morphisms between the 1-morphisms. For example, $f_{ab}^c({\bf g}_i, {\bf g}_j, {\bf g}_k )\in \mathcal{C}^2$ is a map
\begin{align}
f_{ab}^c({\bf g}_i, {\bf g}_j, {\bf g}_k ): a_{{\bf g}_i, {\bf g}_j} \otimes b_{{\bf g}_j, {\bf g}_k} \rightarrow c_{{\bf g}_i, {\bf g}_k}.
\end{align}
The 2-morphisms are represented in the graphical calculus as vertices of the graph (see Fig.~\ref{graphG}). The case where $G$ is trivial reduces
to the case of a fusion category, viewed as a 2-category.

The 2-morphisms $f_{ab}^c({\bf g}_i, {\bf g}_j, {\bf g}_k )$ belong to a vector space
$V_{ab}^c({\bf g}_i, {\bf g}_j, {\bf g}_k)$. The 2-morphisms
\begin{align}
g: a_{{\bf g}_i, {\bf g}_j} \otimes b_{{\bf g}_j, {\bf g}_k} \otimes c_{{\bf g}_k, {\bf g}_l} \rightarrow d_{{\bf g}_i, {\bf g}_l}
\end{align}
thus belong to the vector space
\begin{widetext}
\begin{align}
\bigoplus_e V_{ab}^e({\bf g}_i, {\bf g}_j, {\bf g}_k) \otimes V_{ec}^d({\bf g}_i, {\bf g}_k, {\bf g}_l)
\simeq \bigoplus_f V_{bc}^f ({\bf g}_j, {\bf g}_k, {\bf g}_l)\otimes V_{af}^d({\bf g}_i, {\bf g}_j, {\bf g}_l).
\end{align}
The map between these two spaces is the generalization of the notion of $F$-symbol:
\begin{align}
F^{abc}_{d}({\bf g}_i, {\bf g}_j, {\bf g}_k, {\bf g}_l): \bigoplus_e V_{ab}^e({\bf g}_i, {\bf g}_j, {\bf g}_k) \otimes V_{ec}^d({\bf g}_i, {\bf g}_k, {\bf g}_l)
\rightarrow \bigoplus_f V_{bc}^f ({\bf g}_j, {\bf g}_k, {\bf g}_l)\otimes V_{af}^d({\bf g}_i, {\bf g}_j, {\bf g}_l).
\end{align}
\end{widetext}

We wish to additionally consider the notion of an $H$ action on the 2-category described above, where $H$ is a discrete group.
An $H$ action is defined by a 2-functor $\mathcal{F}_{\bf h}: \mathcal{C} \rightarrow \mathcal{C}$ for every ${\bf h} \in H$. In what follows,
we take $H = G$, so that the group  associated with the group action is equal to the group associated with the objects $\mathcal{C}^0$
of $\mathcal{C}$.

We note that most general definition of a 2-category with $G$ action contains natural transformations
$N_{{\bf g}, {\bf h}}: \mathcal{F}_{\bf g} \circ \mathcal{F}_{\bf h} \rightarrow F_{{\bf gh}}$ and
$M_{{\bf g} , {\bf h}, {\bf k}}:
N_{{\bf g}, {\bf hk}} \circ (\mathcal{F}_{\bf g}\bullet N_{{\bf h}, {\bf k}})
\rightarrow
N_{{\bf gh}, {\bf k}} \circ  (N_{{\bf g}, {\bf h}} \bullet \mathcal{F}_{\bf k})$,
together with 3-cocycle type relations for $M_{{\bf g} , {\bf h}, {\bf k}}$.~\cite{walker2016} (The open dots denote composition of natural transformations and the solid dots denote the ``horizontal'' action of a functor on a natural transformation.) In order to describe the
SET state sum construction presented above, we consider the case where these are trivial and consider a specific type of $G$ action.
We will consider models based on this more general sort of $G$ action elsewhere.~\cite{walker2016}

We consider a $G$ action as follows:
\begin{align}
\mathcal{F}_{\bf h} [{\bf g}] &= {\bf h} {\bf g}, \;\; {\bf g} \in \mathcal{C}^0 = G,
\nonumber \\
\mathcal{F}_{\bf h}[a_{{\bf g}_j, {\bf g}_k}] &= \,^{\bf h}a_{{\bf g}_j, {\bf g}_k}
\nonumber \\
 \mathcal{F}_{\bf h}[f_{ab}^c({\bf g}_i, {\bf g}_j, {\bf g}_k)] &= f_{\,^{\bf h}a\,^{\bf h}b}^{\,^{\bf h} c}({\bf h} {\bf g}_i, {\bf h} {\bf g}_k, {\bf h} {\bf g}_l)
\end{align}
In particular, we have
\begin{widetext}
\begin{align}
\mathcal{F}_{\bf h}[ F^{abc}_{def}({\bf g}_i, {\bf g}_j, {\bf g}_k, {\bf g}_l)]
= [F^{\,^{\bf h} a \,^{\bf h} b \,^{\bf h} c}_{\,^{\bf h} d \,^{\bf h} e \,^{\bf h} f}({\bf h} {\bf g}_i, {\bf h} {\bf g}_j, {\bf h} {\bf g}_k, {\bf h} {\bf g}_l)]^{\sigma({\bf h})}
.
\end{align}
\end{widetext}
Above, $\mathcal{F}_{\bf h}$ acting on the 2-morphism spaces $V_{ab}^c({\bf g}_i, {\bf g}_j, {\bf g}_k)$ can be either a linear or anti-linear map
$\mathcal{F}_{\bf h} : V_{ab}^c({\bf g}_i, {\bf g}_j, {\bf g}_k) \rightarrow V_{\,^{\bf h}a\,^{\bf h}b}^{\,^{\bf h} c}({\bf h} {\bf g}_i, {\bf h} {\bf g}_k, {\bf h} {\bf g}_l)$.
The choice of linear or anti-linear depends on whether ${\bf h}$ acts on the wavefunctions of the SET state in a unitary or anti-unitary fashion.
Time-reversal symmetry, for example, is an anti-unitary symmetry which preserves the orientation of space. Spatial reflection symmetries reverse
the orientation of space and are typically unitary. Symmetries that both change the orientation of space and are anti-unitary are typically considered to
correspond to the product of a time-reversal and spatial reflection symmetry.

The above defines a class of 2-categories with $G$-actions. However, in order for the wavefunction to be invariant
under the symmetry, we actually wish to require that the 2-category with $G$ action also be $G$-equivariant:
\begin{align}
\mathcal{F}_{\bf h}[ F^{abc}_{def}({\bf g}_i, {\bf g}_j, {\bf g}_k, {\bf g}_l)] = F^{abc}_{def}({\bf g}_i, {\bf g}_j, {\bf g}_k, {\bf g}_l).
\end{align}

\subsection{Branch sheets and non-orientable manifolds}
\label{nonOrStateSum}

Given the SET state sum defined above, we can now generalize the definition to the case where the space-time manifold $M^3$
is non-orientable. To do this, we define a two-dimensional surface $Y^2 \subset M^3$, along which we cut $M^3$, obtaining
$\hat{M}^3$. Next, we proceed to define the state sum on $\hat{M}^3$, where the configuration labels associated with the $0$, $1$, and $2$ simplexes
of $\partial \hat{M}^3$ are fixed. These labels break up into two sets of labels, $\ell_L$ and $\ell_R$, associated to the two ``left''
and ``right'' boundary components $\partial \hat{M}^3 = Y^2 \coprod \overline{Y}^2 \coprod \partial M^3$.
This then gives us the state sum $\Z_{\ell_L, \ell_R} (\hat{M}^3)$.

\begin{figure}[t!]
\includegraphics[width=3.0in]{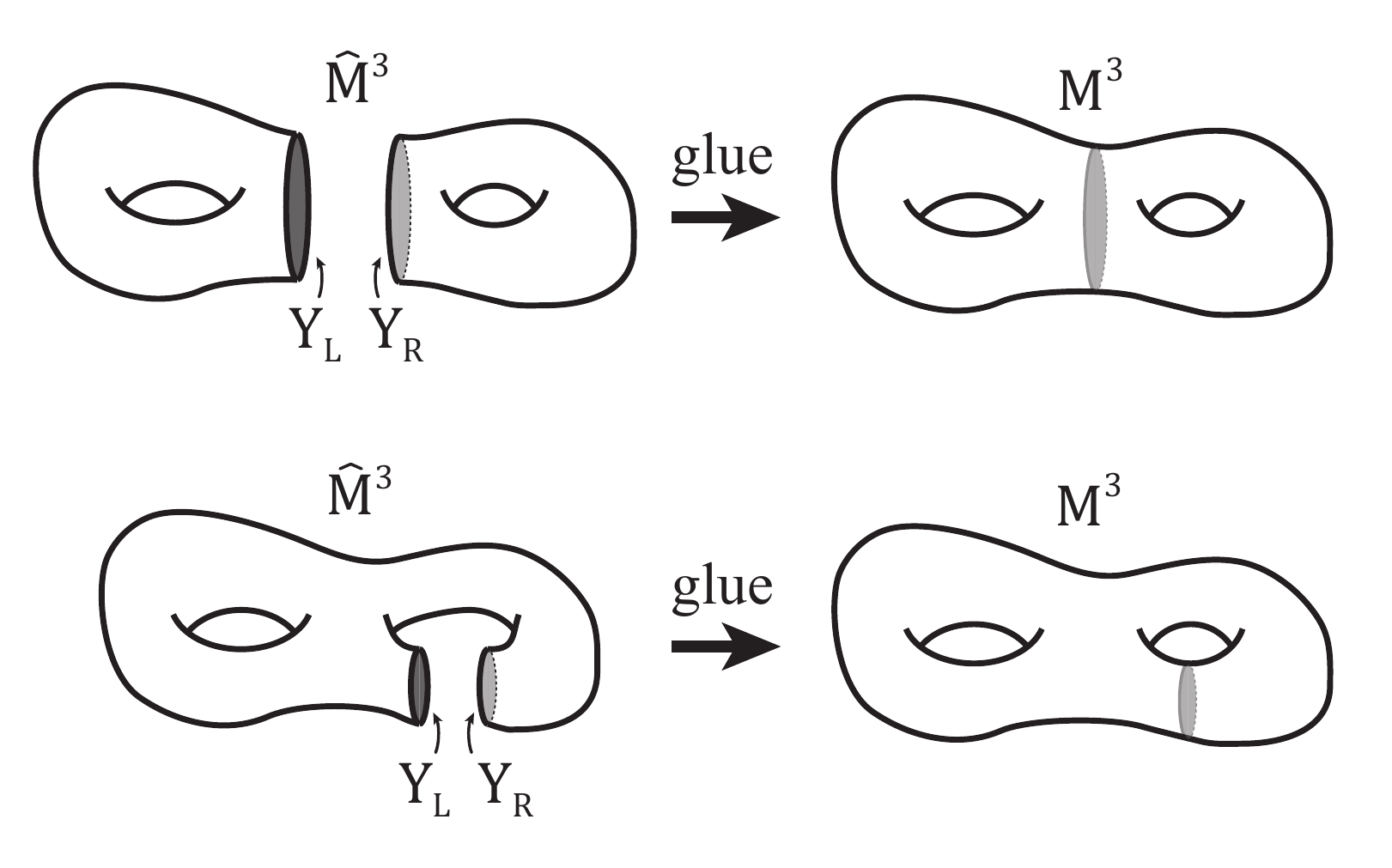}
\caption{The 3-manifold $M^3$ can be obtained from $\hat{M}^3$ by gluing along a surface $Y$. Two examples, where $\hat{M}^3$ is connected and disconnected are shown. $Y_L$ and $Y_R$ are the two surfaces in $\hat{M}^3$ which are glued to each other, as shown. }
\label{cutGlueFig}
\end{figure}

Each group element $\mb{g} \in G$ defines an action on the labels $\ell_L$ and $\ell_R$, described by the functor $\mathcal{F}_\mb{g}$ defined
in the previous section. In the present case, where $M^3$ is non-orientable, we use $\mb{g}$ which corresponds to
an orientation-reversing symmetry, such as a spatial reflection or time-reversal symmetry. We ``re-glue'' the manifold
back together using this $\mb{g}$-action, as follows:
\begin{widetext}
\begin{align}
\Z(M^3) = \mathcal{D}^{-2 N_{v;Y}} \sum_{\ell_Y} \Z_{\ell_{Y}, \mathcal{F}_\mb{g}(\ell_Y)}(\hat{M}^3) \prod_{\Delta^1 \in Y} d(\Delta^1; \ell_Y) \prod_{\Delta^2 \in Y} \Theta(\Delta^2; \ell_Y)^{-1}
.
\end{align}
\end{widetext}

\subsubsection{Invariance under local deformation of branch sheet}

\begin{figure}[t!]
\includegraphics[width=\columnwidth]{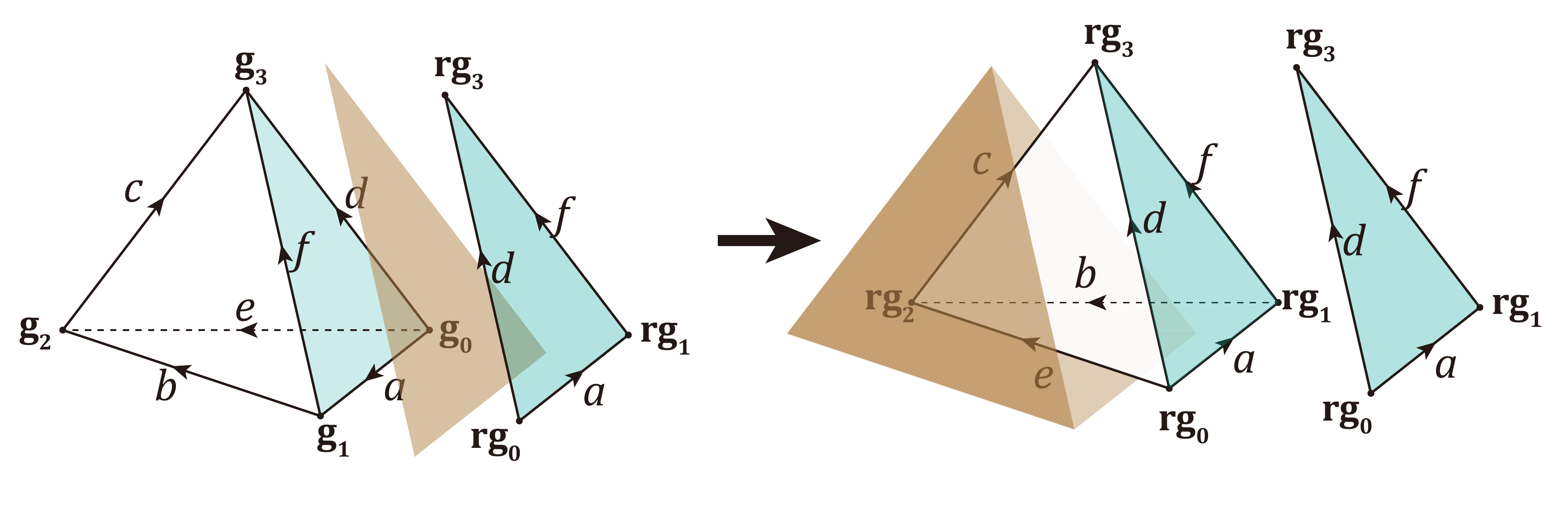}
\caption{Moving the branch sheet (brown) across a 3-simplex}
\label{Fig:bcMove}
\end{figure}

We show that the $G$-equivariant condition allows local deformation of the branch sheet without changing the value of the state sum. Let us consider the local deformation shown in Fig.~\ref{Fig:bcMove}. We start with the branch sheet (brown 2-simplex) going through the 2-simplex $\Delta^2_{\bfg_0 \bfg_1 \bfg_3}$. The amplitude of the 3-simplex is given by
\begin{align}
\widetilde{\Tet}(\Delta^3;\ell) = \frac{1}{\sqrt{d_e d_f}} F^{abc}_{def} (\bfg_0,\bfg_1,\bfg_2,\bfg_3)^*
\end{align}
Now, we deform the branch sheet and push it through a 3-simplex (as is shown in Fig.~\ref{Fig:bcMove}). The $G$-grading of the 3-simplexes has to be changed accordingly in order to be consistent with other 3-simplexes in the triangulation that are not drawn here. After the deformation, the amplitude of the 3-simplex is given by
\begin{align}
\widetilde{\Tet}(\Delta^3;\ell) = 	\frac{1}{\sqrt{d_e d_f}} F^{abc}_{def} (\r \bfg_0,\r \bfg_1,\r \bfg_2,\r \bfg_3) .
\end{align}
The change of the group elements results from the need to keep the elements on each pair of the 0-simplexes that are identified across the branch sheet to differ by $\r$. The complex conjugation $*$ is also needed because the orientation of the 3-simplex is reversed after the branch sheet sweeps through (see Fig.~\ref{Fig:bcMove}).

Therefore, the condition for the invariance of the state sum under local deformation of the branch sheet is given by
\begin{align}
F^{abc}_{def} (\bfg_0,\bfg_1,\bfg_2,\bfg_3)^* =  F^{abc}_{def} (\r \bfg_0,\r \bfg_1,\r \bfg_2,\r \bfg_3) .
\label{Eq:G-equivariant_Condition}
\end{align}
Of course, this is the same as the $G$-equivariant condition for the $G$-graded $F$-symbols.

\subsubsection{Example cellulation: $M^3 = \mathbb{RP}^2 \times S^1$}

\begin{figure}[t!]
\includegraphics[width=3.5in]{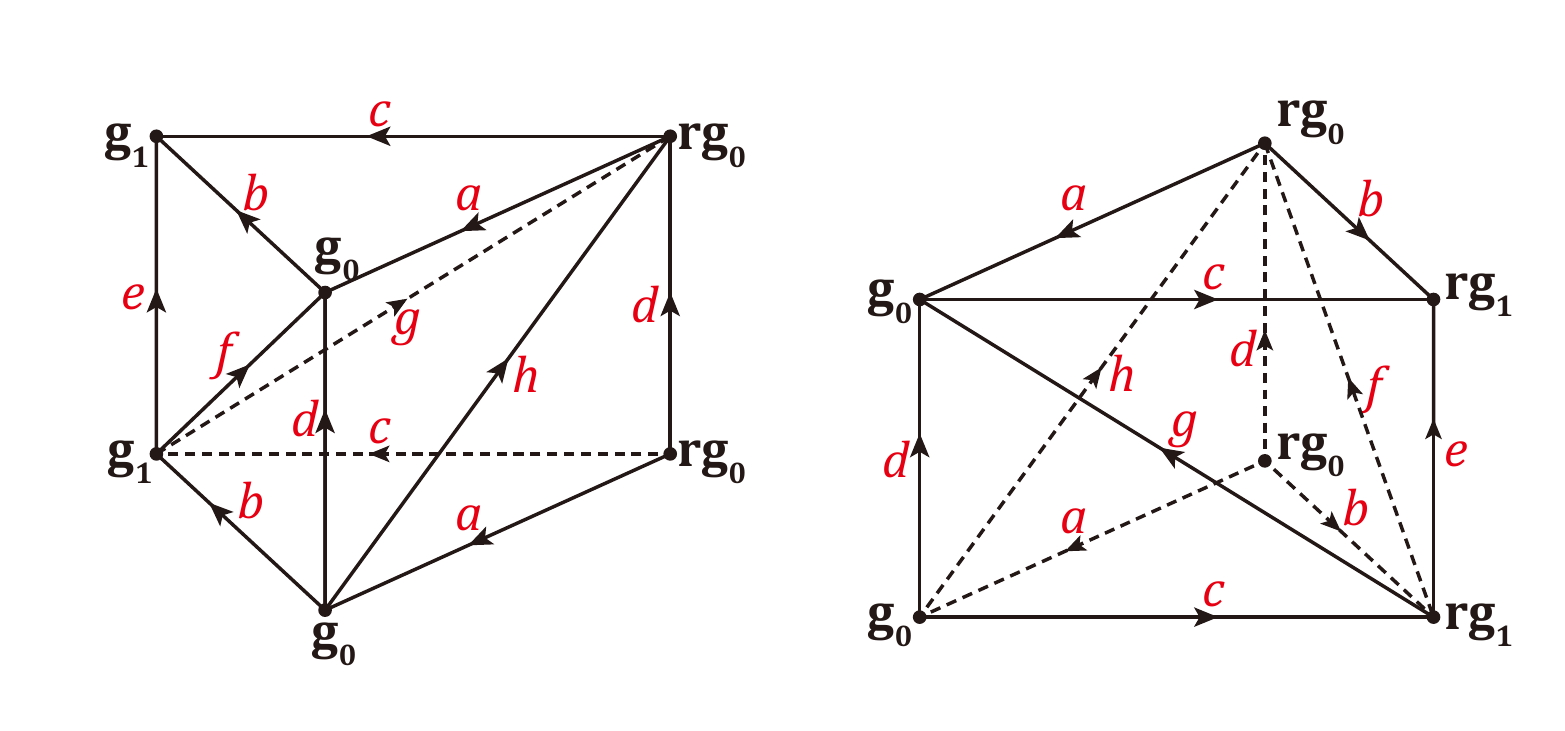}
\caption{A cellulation of $\mathbb{RP}^2 \times S^1$ with $G$-grading.}
\label{Fig:RP2xS1_Complex_Graded}
\end{figure}

As an example, let us consider a simple cellulation of $\mathbb{RP}^2 \times S^1$, containing two distinct vertices and eight edges, as shown in Fig.~\ref{Fig:RP2xS1_Complex_Graded}.
The group $G$ contains an element ${\bf r}$ associated with the orientation reversal, which we use to glue vertices and edges across the branch sheet.
In this example, we take the action on the objects in $\mathcal{C}_G$ to be trivial.
%The subscripts of all the group elements in Fig. \ref{Fig:RP2xS1_Complex_Graded} also reflects the ordering of the 0-simplexes, namely the branching structure. The identification
%of 0-simplexes is given by
%\begin{align}
%& \bfg_1 = \bfg_4 = \bfg_{1'} = \bfg_{4'}, \\
%& \bfg_0 = \bfg_3 = \bfg_{0'} = \bfg_{3'}, \\
%& \bfg_2 = \bfg_5, ~~~ \bfg_{2'} = \bfg_{5'}, \\
%& \bfg_0 = \mathcal{T} \bfg_1, \\
%& \bfg_{2'} = \mathcal{T} \bfg_2,
%\end{align}
%where the last equations are associated to the identification between
%0-simplexes via an orientation reversing map.
The path integral on $\mathbb{RP}^2\times S^1$ is, therefore, given by
\begin{widetext}
\begin{align}
\Z(\mathbb{RP}^2 \times S^1)
%= \frac{1}{\mathcal{D}_G^4} \sum_{\text{config.~} l} Z(\Delta^{3}_{0123}) Z(\Delta^{3}_{1234})Z(\Delta^{3}_{2345}) Z(\Delta^{3}_{0'1'2'3'}) Z(\Delta^{3}_{1'2'3'4'})Z(\Delta^{3}_{2'3'4'5'}) d_a d_b d_c d_d d_e d_f d_g d_h
%\nonumber \\
& = \frac{1}{\mathcal{D}_G^4} \sum_{\ell}
F^{acf}_{dbh} F^{cfa}_{dhg} F^{fac}_{egb}
\left[ F^{abg}_{dch}\right]^*  \left[ F^{bga}_{dhf}\right]^*  \left[ F^{gab}_{efc}\right]^*
\frac{d_a d_d d_e}{d_h}
\label{Eq:RP2_State_Sum_Graded}
\end{align}
\end{widetext}
We emphasize that the $F$-symbols in this expression also depend on the group elements at the vertices of the tetrahedra, although
this is not explicitly labeled in the above equation. Using this cellulation, we can compute $\Z(\mathbb{RP}^2 \times S^1)$ for various examples of SET phases. We present details for specific examples in Sec.~\ref{exampleSec}.

%% file: state-sum3d.tex
\section{(3+1)D TQFTs from braided fusion categories}
\label{3dTQFTSec}

A braided fusion category $\mathcal{B}$ defines a $(3+1)$D TQFT via a path integral
state sum construction due originally to Crane and Yetter\cite{crane1993} and
recently extended to a Hamiltonian construction known as the Walker-Wang model.\cite{walker2012}
In this section, we review the approach of Ref. \onlinecite{walker2006} 
in order to demonstrate how to compute path integrals of this TQFT. 
In the subsequent sections we will apply the tools of this section to the study of 
topological path integrals on non-orientable manifolds. 

\subsection{General construction}

A $(3+1)$D TQFT assigns a complex number $\Z(W^4)$ to a closed $4$-manifold $W^4$. For a $3$-manifold $M^3$,
we have a set of boundary conditions, $\mathcal{C}(M^3)$, which we will define below. 
For manifolds $W^4$ with boundary, 
\begin{align}
\Z(W^4): \mathcal{C}(\partial W^4) \rightarrow \mathbb{C}. 
\end{align}
That is, $\Z(W^4)$ is a map from boundary conditions on $\partial W^4$ into the complex numbers. 
We will write this as $\Z(W^4)[c]$, for $c \in \mathcal{C}(\partial W^4)$. 

\begin{figure}[tb]
\includegraphics[width=1.5in]{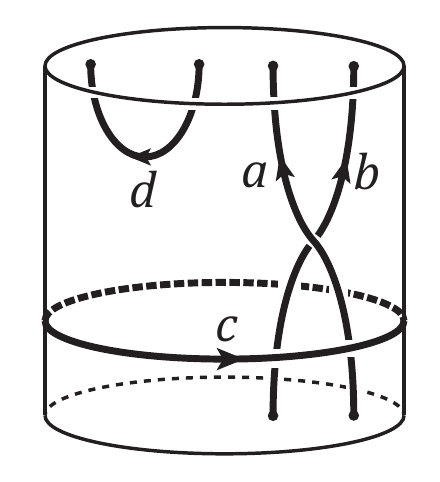}
\caption{A solid cylinder $D^2 \times I = D^3$. The anyon diagram, which is an element of $\mathcal{C}(D^3)$, is shown. 
The boundary condition $c^{(2)}$ in this diagram consists of a total of 6 marked points on the boundary, on which the anyon 
lines in the bulk end, and one loop labeled by $c$. }
\label{diagrams3plus1}
\end{figure}

The set of boundary conditions $\mathcal{C}(M^3)$ for a closed 3-manifold $M^3$ is defined as the set
of all possible anyon diagrams on $M^3$ associated with the braided fusion
category $\mathcal{B}$. For a 3-manifold $M^3$ with boundary, we define $\mathcal{C}(M^3; c^{(2)})$ as the set of all possible anyon diagrams
on $M^3$, with a fixed boundary condition $c^{(2)}$ on the boundary $\partial M^3$. The fixed boundary
condition $c^{(2)}$ consists of an anyon diagram on $\partial M^3$ which can also include anyon lines
from the bulk of the $M^3$ which terminate on the boundary. See Fig. \ref{diagrams3plus1} for an illustration. 
%\meng{What does ``a fixed boundary condition on the anyon diagrams on the two-dimensional 
%boundary mean? I assume it means that anyons line allowed to terminate on the boundary.}. 
When $\partial M^3$ is trivial, we will simply omit writing $c^{(2)}$ in $\mathcal{C}(M^3; c^{(2)})$. 

We define a vector space $\V(M^3;c^{(2)})$ as follows. We consider the space of formal linear superpositions (with complex coefficients) 
of all boundary conditions $c \in \mathcal{C}(M^3;c^{(2)})$ (in other words, formal linear combinations of braided anyon diagrams in $M$ with fixed
boundary $c^{(2)}$), which is denoted $\mathbb{C}[\mathcal{C}(M^3;c^{(2)})]$. Next, we define an
equivalence among these formal superpositions, which correspond to the local relations of the braided fusion category
$\mathcal{B}$. These are the familar fusion of anyon lines, $F$-moves, and $R$-moves. The space $\V(M^3;c^{(2)})$
corresponds to these formal superpositions modulo the equivalence from the local relations:
\begin{align}
\V(M^3;c^{(2)}) = \mathbb{C}[\mathcal{C}(M^3;c^{(2)})]/\sim .
\end{align}
Again, if $M^3$ is closed, we denote the associated vector space as $\V(M^3)$. 

In the Hamiltonian formulation of this TQFT (the Walker-Wang model),\cite{walker2012} the full set of boundary conditions 
$\mathcal{C}(M^3)$ form the basis states of the Hilbert space on $M^3$. The Hamiltonian then determines the ground state 
wavefunctions on $M^3$ (possibly with a fixed boundary condition) as the linear superposition of all possible states with
amplitudes that are related to each other by the diagrammatic rules of the anyon theory. 
This is exactly the space $\V(M^3;c^{(2)})$. 

In what follows, we will use $\emptyset$ to denote the empty diagram.  

\begin{figure}
\includegraphics[width=3.0in]{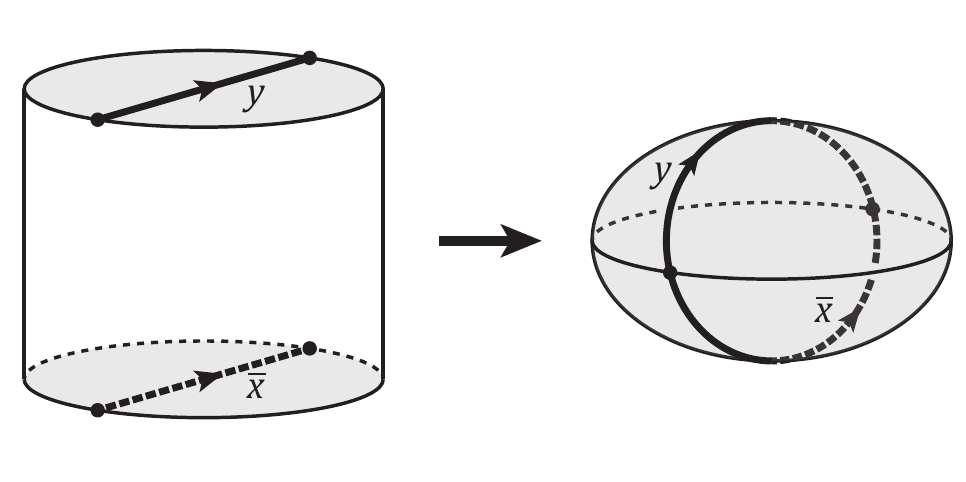}
\caption{We illustrate $D^3 \times I$ by going one dimension lower and displaying
$D^2 \times I$. Left: The boundary conditions $\overline{x}$, $y$ are shown on $D^2 \times I$. 
Right: Identification on the boundary, as explained in the main text. If the boundary conditions
associated with $\overline{x}$ and $y$ match, one gets a closed anyon diagram on
$D^3 \times I = D^4$ ($D^2 \times I = D^3$ is displayed).}
\label{crushM3}
\end{figure}

The inner product in $\V(M^3;c^{(2)})$ is defined in terms of $\Z$ as follows:
\begin{align}
\langle x | y \rangle_{\V(M^3;c^{(2)})} = \Z(M^3 \times I)[\overline{x} \cup y] .
\end{align}
In this context, we define $M^3 \times I$ by identifying points 
$(b, t) \sim (b, s)$ for $b \in \partial M^3$ and $s,t \in I$. 
Thus, by this definition, $\partial (M^3 \times I) = M^3 \cup \overline{M}^3$. 
Furthermore, $x$ is taken to be the boundary condition on one boundary of $M^3 \times I$, and 
$y$ is the boundary condition on the other boundary. The overline implies that the picture $x$ 
has the opposite orientation relative to $y$. Since $x$ and $y$ are defined to have the same boundary condition, 
$c^{(2)}$, the result $\overline{x} \cup y$ in $M^3 \times I$ is a closed anyon diagram. See Fig. \ref{crushM3} for an illustration.

$\Z$ can be evaluated via the following gluing formula. Let $W^4$ be a $4$-manifold obtained after 
gluing $W^4_{\text{cut}}$ along a three-manifold $M^3$. The gluing formula is:
\begin{align}
\Z(W^4)[c] = \sum_{e_{\alpha}} 
\frac{Z(W^4_{\text{cut}})(c_{\text{cut}} \cup e_{\alpha} \cup \overline{e}_{\alpha})}{\langle e_\alpha | e_\alpha \rangle_{\V(M^3;c^{(2)}_{\text{cut}})}} . 
\end{align}
Here, $c_{\text{cut}}$ is the boundary condition after the cut, $\{e_\alpha\}$ is a set of orthonormal basis states for 
$\V(M^3;c^{(2)}_{\text{cut}})$, and $c^{(2)}_{\text{cut}}$ is the restriction of $c_{\text{cut}}$ to $\partial M^3$. 
See Fig. \ref{S1D2inner} for an example. 

\subsection{(2+1)D path integrals from (3+1)D TQFTs}
\label{2dfrom3dSec}

In the case where $\mathcal{B}$ is a UMTC, which is the case of interest to us, the (3+1)D TQFT obtained 
from this approach is trivial in the bulk of the (3+1)D system. This means that there is no intrinsic topological
order in the bulk, although in the presence of symmetries the (3+1)D bulk could correspond to an SPT state. 
Nevertheless, the (3+1)D theory on a manifold with boundary hosts a (2+1)D topological state at its boundary,
whose anyon excitations are described by the UMTC $\mathcal{B}$.

This implies that the path integral $\Z_{2+1}(M^3)$ of a (2+1)D TQFT on a three-manifold $M^3$ can be 
computed in terms of the path integral of the associated (3+1)D TQFT on a $4$-manifold $W^4$, 
such that $\partial W^4 = M^3$:
\begin{align}
\Z_{2+1}( \partial W^4) = \Z_{3+1}(W^4) [\emptyset] . 
\label{eqn:3dfrom4d}
\end{align}
In general, $\Z_{2+1}$ as defined above depends on the choice of the extension $W^4$. 
Two different choices $W^4$ and $\tilde{W}^4$ with $\partial W^4 = \partial \tilde{W}^4$
will differ in their respective path integrals by $\Z_{3+1}(W^4 \cup \tilde{W}^4)$, 
where $W^4 \cup \tilde{W}^4$ is a closed manifold obtained by gluing $W^4$ and $\tilde{W}^4$ along
their boundaries. $\Z_{3+1}(W^4 \cup \tilde{W}^4)$ is in general non-trivial. As we discuss further in Sec. \ref{sec:4dspt}, 
a (2+1)D time-reversal / reflection symmetric topological state is non-anomalous 
(i.e. can exist as a purely (2+1)D phase of matter), if indeed $\Z_{3+1} = 1$ on all closed 4-manifolds. For such
non-anomalous states, (\ref{eqn:3dfrom4d}) is thus independent of the extension $W^4$. 

Anomalous (2+1)D SET states must exist at the surface of a (3+1)D system with $\Z_{3+1} \neq 1$
for some closed 4-manifolds. For such anomalous (2+1)D SET states, $\Z_{2+1}(\partial W^4)$
is only well-defined once one specifies the extension $W^4$. 

We note that in the absence of time-reversal and reflection symmetry, the ambiguity
associated with the extension to $W^4$ is the well-known framing anomaly of
chiral (2+1)D TQFTs.\cite{witten1989,walker1991} In this case the associated topological
phase of matter can still exist as a purely (2+1)D system, although the TQFT requires a framing
for 3-manifolds to give well-defined path integrals. 

%As we discuss later in Sec. \ref{sec:4dspt}, 
%for oriented manifolds (i.e. chiral (2+1)D TQFTs) the ambiguity is associated with a phase factor $e^{\frac{2\pi i }{8}c_-}$ 
%where $c_-$ is the chiral central charge.
%corresponding to the path integral of the (3+1)D theory on $\mathbb{CP}^2$. 
%This is the well-known framing anomaly of (2+1)D TQFTs~\cite{witten1989,walker1991}. 
%We can resolve the ambiguity by considering framed manifolds. However, since we are 
%interested in (2+1)D time-reversal/reflection symmetric theories, the chiral central charge 
%necessarily vanishes and we do not need to worry about the framing anomaly. For 
%unoriented manifolds, we also need to consider another ambiguity coming from $\mathbb{RP}^4$, 
%which we will study in greater details in Sec. \ref{sec:rp4}. It turns out that for all non-anomalous (2+1)D theories 
%this is not an issue, so Eq. \eqref{eqn:3dfrom4d} is well-defined.

\subsection{Computations}

We now perform a number of explicit computations,\cite{walker2006} which will be used as
reference in the subsequent discussion. Let us first consider several examples for $\V(M^3)$. 
From the above definition, it is clear that
\begin{align}
\V(S^3) \simeq \mathbb{C} 
\end{align}
is one-dimensional, spanned by the empty picture in $S^3$, as all closed anyon diagrams 
in $S^3$ can be reduced via local moves to a multiple of the empty diagram. 

It is also clear that
\begin{align}
\V(S^1 \times D^2;\emptyset) \simeq \mathbb{C}^{|\mathcal{B}|}
\end{align}
Here, $|\mathcal{B}|$ corresponds to the number of distinct types of anyons in $\mathcal{B}$. 
The basis vector in $\V(S^1 \times D^2)$ associated to an anyon $a \in \mathcal{B}$ is a loop of the form
$S^1\times \text{pt} \subset S^1\times D^2$ with label $a$, where $\text{pt}$ denotes a point in $D^2$. 
%$\V(S^1 \times D^2) $ is spanned by a loop encircling the $S^1$ associated to each anyon $a \in \mathcal{B}$. 

On the other hand, 
\begin{align}
\V(S^1 \times S^2) \simeq \mathbb{C} .
\end{align}
This follows from modularity of $\mathcal{B}$. As in the case of $\V(S^1 \times D^2; \emptyset)$, it is clear that all diagrams
can be reduced to a single anyon loop encircling the $S^1$. However in this case, this implies that for each point on the $S^1$, 
there is a single puncture on the $S^2$ labeled by $a$. We can consider a loop labeled $b$ encircling this puncture on the $S^2$, 
at a given point on the $S^1$. On the one hand, this loop on $S^2$ is contractible to a point. On the other hand, because it is linking
the $a$ puncture, this diagram is related to the diagram without the $b$ loop by a factor $S_{ab}/S_{0a}$. The braiding non-degeneracy
of a modular category $\mathcal{B}$ implies that there exists at least one $b$ for which the factor $S_{ab}/S_{0a} \neq 1$. Therefore 
consistency demands that $a = 0 \in \mathcal{B}$, and thus $\text{dim } \V(S^1 \times S^2) = 1$. 

Consider $\V(S^2 \times D^1;\emptyset)$. Clearly this is also one-dimensional:
\begin{align}
\V(S^2 \times D^1;\emptyset) \simeq \mathbb{C},
\end{align}
as all pictures can be reduced to a multiple of the empty picture. 

We begin computing some path integrals $\Z(W^4)$ by setting
\begin{align}
\Z(D^4)[\emptyset] = \lambda,
\end{align}
where $\lambda$ is a parameter to be fixed later. 

From the gluing formula,
\begin{align}
\Z(S^4) = \frac{\Z(D^4)[\emptyset] \Z(D^4)[\emptyset]}{\langle \emptyset | \emptyset\rangle_{\V(S^3)} },
\label{eqn:S4}
\end{align}
which implies
\begin{align}
	\langle \emptyset | \emptyset \rangle_{\V(S^3)} = \frac{\lambda^2}{ \Z(S^4)}. 
\end{align}

We can also consider 
\begin{align}
\label{ZD4la}
\Z(D^4)[l_a] = d_a \lambda,
\end{align}
where $l_a$ denotes a loop of $a$ on the boundary of $D^4$, which is $S^3$. By the usual rules for evaluating anyon diagrams, 
the state $|l_a \rangle = d_a |\emptyset \rangle \in \V(S^3)$, where $d_a$ is the quantum dimension of $a$. 

\begin{figure}
\includegraphics[width=1.5in]{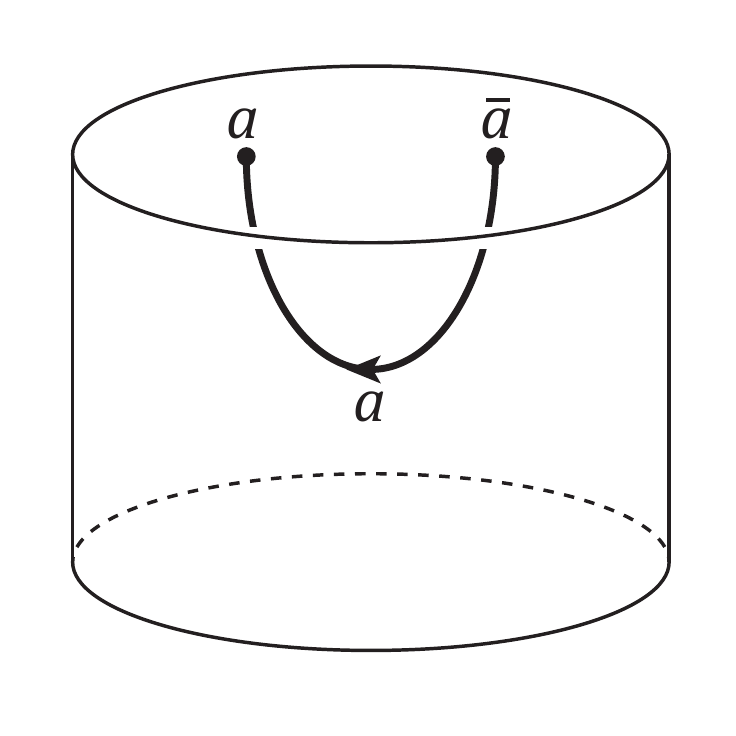}
\caption{$D^3 = D^2 \times I$ is shown, with boundary conditions corresponding to two marked points, $a$ and $\overline{a}$. The
states in $\V(D^3;a,\overline{a})$ are thus spanned by the state $|\text{arc}_a\rangle$, corresponding to an arc connected $a$ and 
$\overline{a}$, as shown.}
\label{arca}
\end{figure}

Now let us consider $\V(D^3; a, b)$, where the boundary conditions on $\partial D^3 = S^2$ consist of
two marked points, labeled by anyons $a$ and $b$. First, observe that
\begin{align}
\label{AD3dim}
\text{dim } \V(D^3;a,b) = \delta_{\overline{a} b} . 
\end{align}
This is because if $b \neq \overline{a}$, there are no such allowed
anyon diagrams in the presence of such boundary conditions. 
When $b = \overline{a}$, $\V(D^3; a, \overline{a})$ is one-dimensional, spanned by an 
arc, which we denote as $\text{arc}_a$, connecting $a$ and $\overline{a}$ (See Fig. \ref{arca}). The inner product is:
\begin{align}
\label{eqn:AD3}
\langle \text{arc}_a | \text{arc}_b \rangle_{\V(D^3;a, \overline{a})} = \Z(D^4)[l_a]= d_a \lambda \delta_{ab} . 
\end{align}
This is illustrated in Fig. \ref{crushM3} . 

\begin{figure}
\includegraphics[width=3.5in]{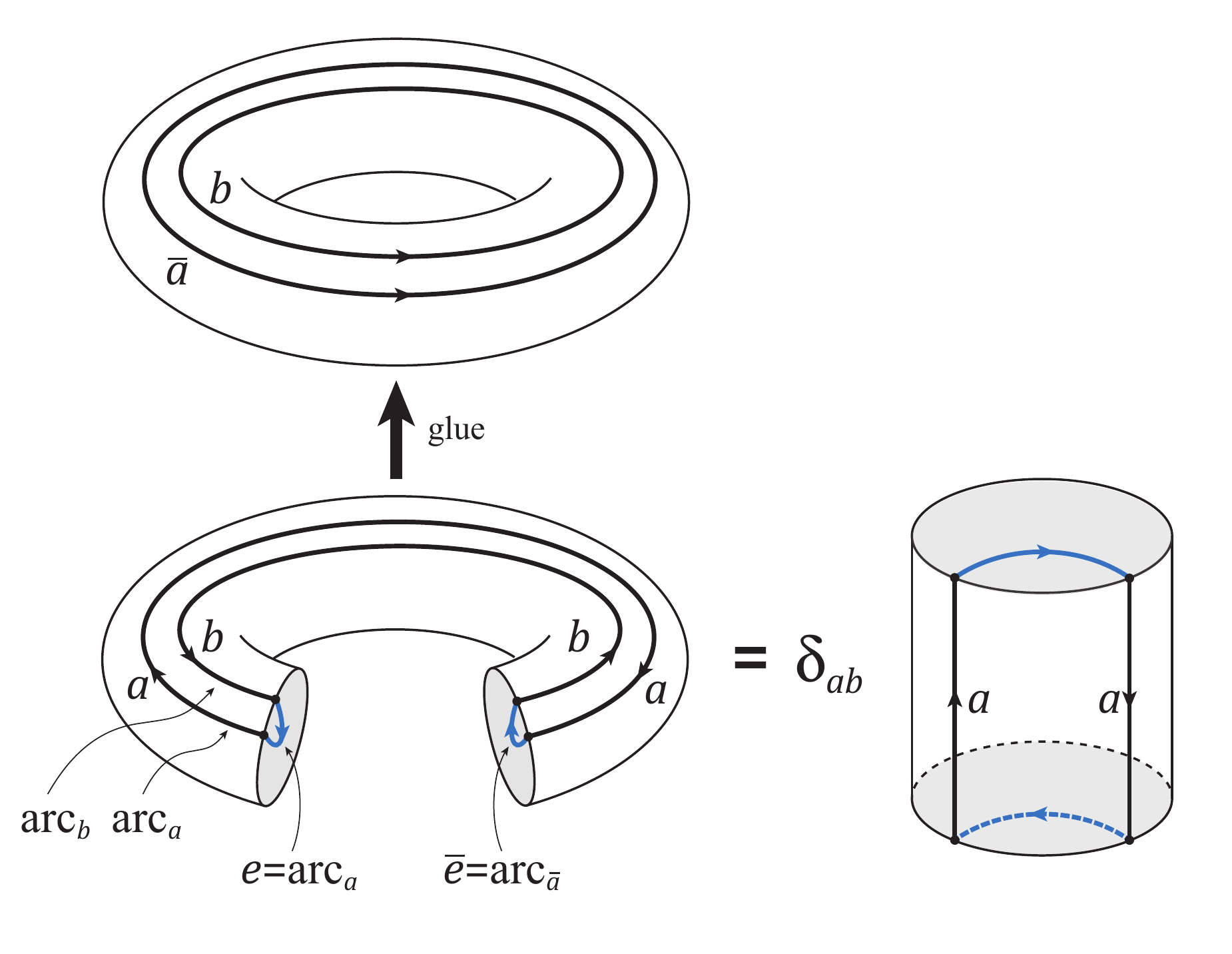}
\caption{Illustration of gluing formula applied to $\mathcal{Z}(S^1 \times D^3)[l_{\overline{a}} \cup l_b]$. 
The illustration is in one dimension down, displaying $S^1 \times D^2$ for convenience. 
The boundary condition contains the loops $l_{\overline{a}} \cup l_b$ as shown in the top figure. 
The top figure is obtained from the bottom figure by gluing along a $D^3$ ($D^2$ shown in the figure). 
The state $e \in \V(D^3; a, \overline{a})$ corresponds to an arc, shown in blue. The cut manifold
with the boundary condition is thus equivalent to a $D^4$ ($D^3$ shown in the bottom right figure), 
with a single loop $l_a$ on the surface of the $D^4$. 
 }
\label{S1D2inner}
\end{figure}

Inner products in $\V(S^1 \times D^2;\emptyset)$ can be computed as follows:
\begin{align}
\label{AS1D2}
 \langle l_a| l_b \rangle_{\V(S^1 \times D^2;\emptyset)} = \Z(S^1 \times D^3)[l_{\overline{a}} \cup l_b] . 
\end{align}
From the gluing formula,
\begin{align}
\label{ZS1D3}
\Z(S^1 &\times D^3)[l_{\overline{a}} \cup l_b] 
\nonumber \\
&=\sum_{e_\alpha \in \V(D^3; a, b )} \frac{\Z(D^4)[\text{arc}_b \cup \overline{e}_{\alpha} \cup \text{arc}_a \cup e_\alpha]}{\langle e_\alpha | e_\alpha\rangle_{\V(D^3; a,b)}}
\nonumber \\
&= \delta_{ab} \frac{\Z(D^4)[\text{arc}_a \cup \text{arc}_{a} \cup \text{arc}_a \cup \text{arc}_a]}{\langle \text{arc}_a | \text{arc}_a \rangle_{\V(D^3; a, \overline{a})}}
\nonumber \\
&= \delta_{ab}  \frac{\Z(D^4)[l_a]}{\langle \text{arc}_a | \text{arc}_a \rangle_{\V(D^3; a, \overline{a})}}
\nonumber \\
&= \delta_{ab}  . 
\end{align}
Here we have used the fact that cutting $l_{\overline{a}} \cup l_b$ gives rise to two arcs, $\text{arc}_b$ and $\text{arc}_a$,
while $\V(D^3; a, \overline{a})$ is spanned by an arc connecting $a$ and $\overline{a}$. Thus the combination 
$\text{arc}_a \cup \text{arc}_{a} \cup \text{arc}_a \cup \text{arc}_a = l_a$. This is illustrated in Fig. \ref{S1D2inner}.

Let us compute $\Z(S^2 \times D^2)[\emptyset]$. In order to compute this, we cut the $S^2$ along a circle $S^1$ to obtain two
disks. Thus:
\begin{align}
\langle \emptyset | \emptyset \rangle_{\V(S^2\times D^1;\emptyset)} & = \Z(S^2 \times D^2)[\emptyset]\\
= 
&\sum_{l_a \in \V(S^1 \times D^2)} \frac{\Z(D^4)[l_a] \Z(D^4)[l_a]}{\langle l_a| l_a \rangle_{\V(S^1 \times D^2;\emptyset)} }
\nonumber \\
&= \sum_a d_a^2 \lambda^2
\end{align}
\label{eqn:S2xD2}

Finally, let us compute $\Z(D^1\times S^3)[\emptyset]$. Cut $S^3$ into two $D^3$ glued along $S^2$:
\begin{align}
\Z(D^1 \times S^3)[\emptyset] &= \frac{\Z(D^4)[\emptyset] \Z(D^4)[\emptyset] }{ \langle \emptyset | \emptyset \rangle_{\V(S^2 \times D^1; \emptyset)}}
\nonumber \\
&= \frac{1}{\mathcal{D}^2} .
\label{eqn:D1xS3}
\end{align}

\subsection{Invertibility and cobordism invariance}

When the input category $\mathcal{B}$ is modular, we obtain an invertible (3+1)D TQFT by a suitable choice
of $\lambda$. This means that on every closed 3-manifold $M^3$ the vector space 
$ \V(M^3)$ is one-dimensional, and on every closed $4$-manifold the path integral is a 
pure phase factor. 

Consider $\Z(S^4)$, which is given in Eq. \eqref{eqn:S4}. The inner product 
$\langle \emptyset|\emptyset\rangle_{\V(S^3)}$ is equal to the path integral $\Z(D^1 \times S^3)[\emptyset]$, 
which is given in Eq. \eqref{eqn:D1xS3} to be $\frac{1}{\mathcal{D}^2}$. Therefore
\begin{equation}
	\Z(S^4)=\lambda^2 \mathcal{D}^2.
	\label{}
\end{equation}
In order for $|\Z(S^4)|=1$, we must choose $|\lambda|=\frac{1}{\mathcal{D}}$. 

Furthermore, we have found in Eq. \eqref{eqn:AD3} that the norm of the state $|\text{arc}_0\rangle$ is $\lambda$. 
In a unitary quantum theory, norms are always positive definite, so $\lambda>0$. Therefore we have determined 
\begin{equation}
	\lambda=\frac{1}{\mathcal{D}}.
	\label{}
\end{equation}
As a result we also have $\Z(S^4)=1$.

As a consequence, we can show that the path integral for closed $4$-manifolds is in fact a cobordism 
invariant. Two 4-manifolds $W_1^4$ and $W_2^4$ are said to be cobordant to each other if there 
exists a five-dimensional manifold $X^5$ such that $\partial X^5 = \overline{W}_1^4 \cup W_2^4$. 
$\Z(W^4)$ is a cobordism invariant if $\Z(W_1^4) = \Z(W_2^4)$ when $W_1^4$ and $W_2^4$ are cobordant
to each other. 

Cobordisms of $4$-manifolds can be generated by the following basic
moves: (0) Removing or adding an $S^4$, (1) replacing an $S^1 \times D^3$ by $S^2 \times D^2$ and vice versa, and (2) replacing $S^0 \times D^4$
with $D^1 \times S^3$ and vice versa. This follows from basic facts about handle decompositions and 
handle cancellations of manifolds.\cite{gompf1999} Thus, in order to show that $\Z$ is a cobordism invariant, we require:
\begin{align}
\Z(S^4) &= 1 
\nonumber \\
\Z(S^1 \times D^3) &= \Z(S^2 \times D^2)
\nonumber \\
\Z(S^0 \times D^4) &= \Z(D^1 \times S^3) . 
\end{align}
These are satisfied with $\lambda=\frac{1}{\mathcal{D}}$, together with the fact that both
$\V(S^1 \times S^2) \simeq \mathbb{C}$ and $\V(S^3) \simeq \mathbb{C}$ when $\mathcal{B}$
is modular.

%% file: gsd.tex
\section{(2+1)D path integrals on non-orientable manifolds}
\label{gsd}

In Sec. \ref{nonOrStateSum}, we provided an explicit definition for the path integral of a TQFT
on a non-orientable manifold $M^3$. An important part of the construction was the use of a 
``branch sheet'' along which the manifold was glued with an orientation reversal. 

The theory in the presence of the ``branch sheet'' has a specific action on anyons that cross
the branch sheet. Let $W_a(\alpha)$ be a Wilson loop operator for a topological charge $a \in \mathcal{B}$,
associated to a loop $\alpha$ that crosses the branch sheet once. Not all such Wilson loops can be consistently defined,
because the topological charge $a$ is transformed to a different topological charge $\ra{a}$ upon traversing the branch sheet,
where $\r \in \text{Aut}_1(\mathcal{B})$ since the branch sheet is orientation reversing.
Thus, $W_a(\alpha)$ can be consistently defined only if $a = \ra{a}$.  

\begin{figure}[tb]
\includegraphics[width=3.0in]{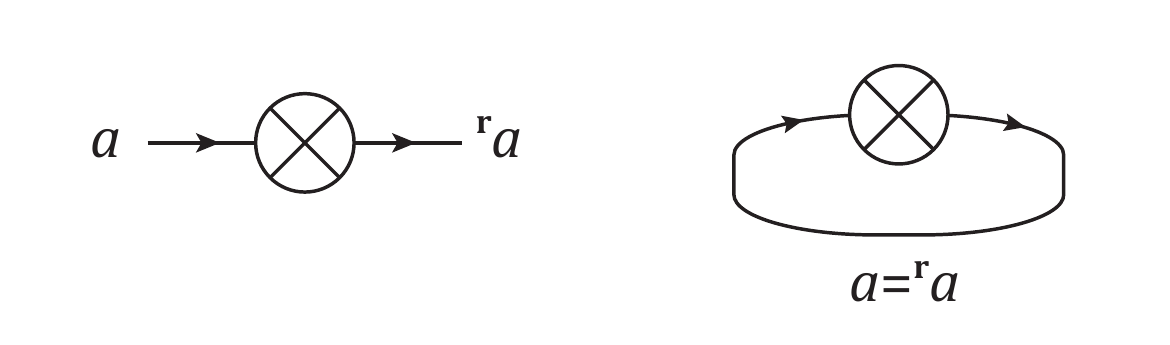}
\caption{Left: An anyon $a$ is permuted to $\ra{a}$ as it traverses a cross-cap, where $\r$ is an orientation-reversing automorphism
of the UMTC $\mathcal{B}$ that describes the anyons. Right: Wilson lines of $a$ can close only if $a = \ra{a}$. }
\label{xcaplinesFig}
\end{figure}

More generally, each non-contractible cycle $\alpha_i$ in $M^3$ is
associated to an $\r_i$-invariant set of anyons whose Wilson loops
around $\alpha_i$ exist. In general, the $\r_i$ can be distinct for different $i$, 
if different actions of reflection are chosen for different branch sheets. 

\subsection{Result}

Here we will provide a formula for the path integral on $\Sigma^2 \times S^1$. 
We specialize to the case where the $\r_i$ defined above are all equal to a fixed 
$\r \in \text{Aut}_1(\mathcal{B})$, for some fixed orientation reversing anti-autoequivalence $\r$.

It is well-known that for a topological phase described by a UMTC $\mathcal{B}$, the path integral on
$\Sigma^2 \times S^1$, where $\Sigma^2$ is an orientable manifold, is equal to the dimension of the Hilbert
space on $\Sigma^2$, and is given by:
\begin{align}
	\Z(\Sigma^2 \times S^1) = \text{dim} \V (\Sigma^2) = \sum_{a \in \mathcal{B}} S_{0a}^{\chi(\Sigma^2)}
	\label{eqn:verlinde-gsd}
\end{align}
Here, $\chi(\Sigma^2) = 2 - 2g$ is the Euler characteristic of $\Sigma^2$ and $g$ is its genus. When instead the surface contains
$n$ punctures, each labelled by a topological charge $a_i \in \mathcal{C}$, then
\begin{align}
\text{dim}\, \V_{\{a_i\}} (\Sigma^2) = \sum_{x \in \mathcal{B}} S_{0x}^{\chi(\Sigma^2)} \prod_{i=1}^n S_{x a_i},
\label{eqn:verlinde-punctured}
\end{align}
where now $\chi(\Sigma^2) = 2 - 2g - n$.

One can generalize the above formulae to the case where symmetry defect lines associated with elements of $\text{Aut}_0(\mathcal{B})$ 
wrap various non-contractible cycles.\cite{barkeshli2014SDG}

In the following, we argue that when $\Sigma^2$ is a closed non-orientable surface, we have
\begin{align}
\label{ZnonOr1}
	\Z(\Sigma^2 \times S^1) = \sum_{a | a= \overline{\ra{a}}} ({\eta}_a^\R S_{0a})^{\chi(\Sigma^2)} ,
\end{align}
where ${\eta}^\R_a$ was defined in Sec. \ref{symfrac}. Note that in principle, one could define branch sheets such that 
that $\r  = \rho_{\bf R} \circ \varphi$, for any order two element of $\varphi \in \text{Aut}_0(\mathcal{B})$ and, as discussed
in Sec. \ref{symfrac}, there is an associated ${\eta}_a^\R$. 

If we interpret $S^1$ as the time-direction, then 
\begin{align}
\Z(\Sigma^2 \times S^1) = \text{dim } \V(\Sigma^2),
\end{align}
where $\V(\Sigma^2)$ is the Hilbert space on $\Sigma^2$. 

If instead we interpret $S^1$ as space and the branch sheet in $\Sigma^2$ is along the time direction, then we should interpret 
$\overline{\ra{a}} = \rho_{\bf T} (a)$ and ${\eta}_a^\mb{T} = \eta_a({\bf T}, {\bf T})$ determines
whether $a$ carries a local Kramers degeneracy. Here, recall $\rho_{\bf T}(a)$ is the action of time-reversal ${\bf T}$
on $a$. The fact that the action of ${\bf T}$ is equivalent to $\r$ together with the charge conjugation map 
can be interpreted to be a consequence of the CPT theorem for Lorentz-invariant field theories.

%Here, the corresponding UMTC $\mathcal{B}$ is a Drinfeld center $\mathcal{B} = \mathcal{Z}(\mathcal{C}_0)$.
%If instead the branch sheet reflects the time coordinate, then
%\begin{align}
%\label{ZnonOr2}
%	\mathcal{Z}(\Sigma^2 \times S^1) = \sum_{a \in \mathcal{B}^\varphi} (\eta_a({\bf T}, {\bf T}) S_{0a})^{\chi(\Sigma^2)} .
%\end{align}
\begin{figure}[tb]
\includegraphics[width=3.0in]{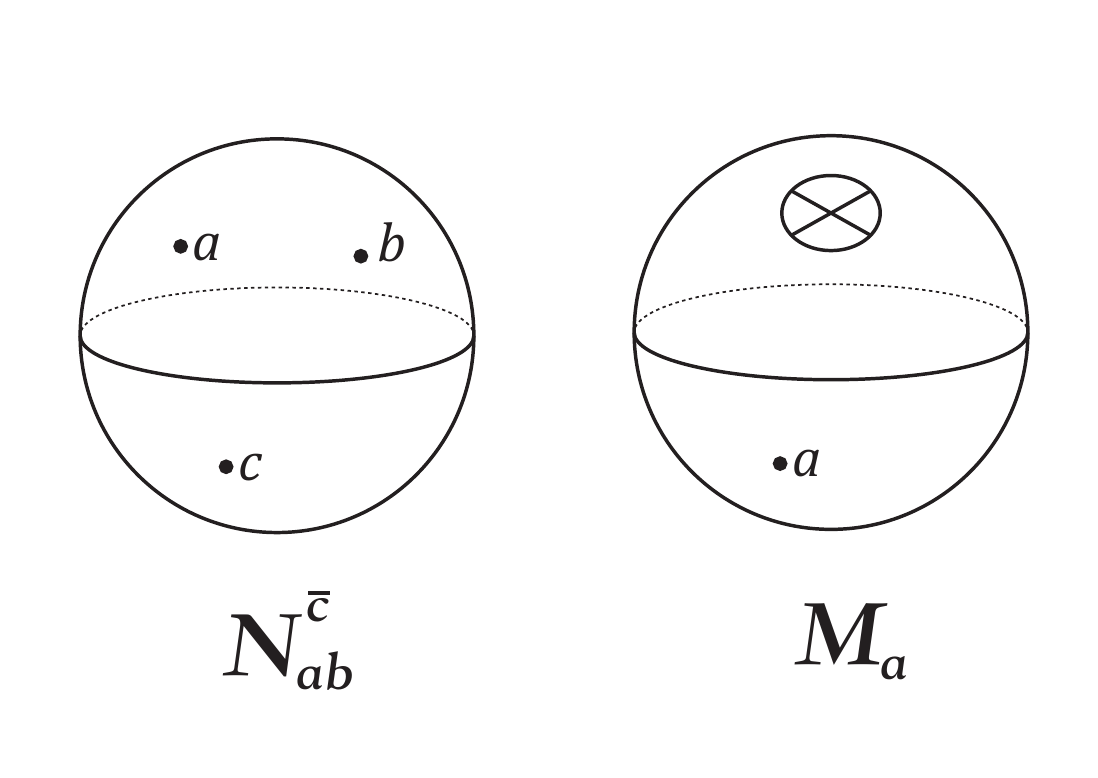}
\caption{Left: 3-punctured sphere, with punctures labelled $a$,$b$,$c$. The dimension of the
Hilbert space if $N_{ab}^{\overline{c}}$. Right: Punctured sphere with a cross-cap, which is equivalent to 
punctured $\mathbb{RP}^2$ (i.e. Mobius band). Puncture labelled by $a$. The dimension of the Hilbert space if $M_a$.}
\label{puncturedMFig}
\end{figure}

Eq. (\ref{eqn:verlinde-gsd}) - (\ref{eqn:verlinde-punctured}) for orientable surfaces can be obtained by gluing together three-punctured spheres,
where we associate a factor $N_{ab}^{\overline{c}}$ when the punctures are labelled by topological charges $a, b, c$. The resulting
sum over intermediate states at the punctures is simplified by using the Verlinde formula
\begin{equation}
	N_{ab}^c=\sum_{x\in \mathcal{B}}\frac{S_{ax}S_{bx}S_{cx}^*}{S_{0x}}.
	\label{verlinde}
\end{equation}

Similarly, we can obtain $\mathcal{Z}(\Sigma^2 \times S^1)$ for non-orientable $\Sigma^2$ by gluing together
three-punctured spheres and cross-caps together. A cross-cap is equivalent to a Mobius band, which is in turn
equivalent to a once-punctured projective plane ($\mathbb{RP}^2$). Thus we associate a factor
$M_a$ to $\mathbb{RP}^2$ with a puncture labelled by $a$. 
In what follows, we demonstrate that
\begin{align}
	M_a \equiv  \mathcal{Z}_a(\mathbb{RP}^2 \times S^1) = \sum_{x | x = \ra{\bar{x}}} S_{ax} {\eta}_x^\R ,
\label{Eq:Ma_RP2}
\end{align}
where $\mathcal{Z}_a(\mathbb{RP}^2 \times S^1) $ is the path integral for the case where
the $\mathbb{RP}^2$ has a single puncture labelled by $a \in
\mathcal{B}$ (see Fig. \ref{puncturedMFig}).

Eq. (\ref{Eq:Ma_RP2}) can be used to derive a general formula for the
path integral on $\Sigma^2 \times S^1$ where now $\Sigma^2$ contains
$n$ punctures, with $n$ Wilson loops of anyons $a_1, \cdots, a_n$
encircling the $S^1$:
\begin{align}
\mathcal{Z}_{a_1,a_2,\cdots,a_n}(\Sigma^2 \times S^1) &= \text{dim }
  \mathcal{V}_{a_1,a_2,\cdots, a_n}(\Sigma^2) 
\nonumber \\
&= \sum_{x | x = \ra{\bar{x}}} (\eta_x^\R)^k S_{0x}^\chi \prod_{i = 1}^n S_{x a_i} .
\end{align}
$\mathcal{V}_{a_1,a_2,\cdots, a_n}(\Sigma^2) $ is the topological
ground state subspace in the presence of $n$ punctures labeled $a_1,
\cdots, a_n$ on the closed non-orientable surface $\Sigma^2$. 
Here $\chi = 2 - 2g - n - k$, where $k$ is the number of cross-caps
that are glued on the genus $g$ surface. 

\subsection{Non-orientable surfaces with even Euler characteristic}
\label{evenEulerSec}

We first observe that we can compute $\mathcal{Z}(\Sigma^2 \times S^1)$ for non-orientable manifolds 
with even Euler characteristic directly by computing fusion diagrams in the presence of parity defects across
non-contractible cycles of genus $g$ surfaces. This generalizing a computation of Ref. \onlinecite{barkeshli2014SDG}
for the ground state degeneracy of a genus $g$ surface in the presence of symmetry defect lines that wrap
non-contractible cycles. 

\begin{figure}[t]
\includegraphics[width=3.7in]{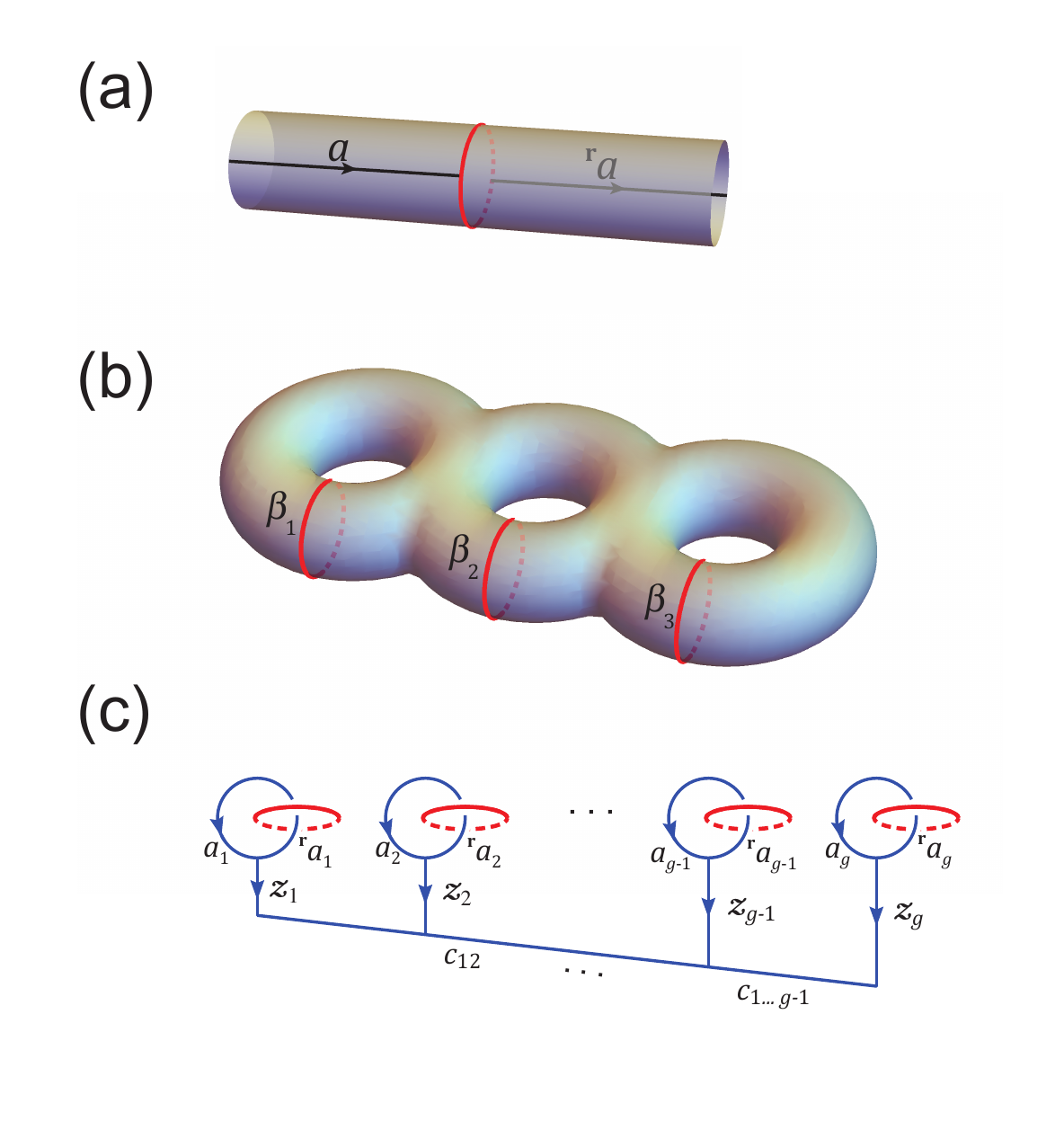}
\caption{(a) A parity defect line (red) wrapping the circumference of the cylinder. An anyon $a$
that traverses it is mapped to $\r(a)$ at the reflected coordinate. (b) A non-orientable
manifold with Euler characteristic $2 - 2g$ can be created by starting with a genus $g$
surface and inserting parity defects along non-contractible cycles. Here we choose
to place the defects along the cycles $\beta_i$, for $i = 1,\cdots, g$, as shown. 
(c) Fusion diagram for computing the ground state degeneracy in the presence of
the parity defects lines. }
\label{paritydefectFig}
\end{figure}

Non-orientable surfaces with even Euler can be understood in terms of oriented genus $g$ surfaces, 
with a ``parity defect line''  wrapping any non-contractible cycle. A parity defect line that wraps a 
cycle $\omega$ is created by cutting the genus $g$ surface along $\omega$ and regluing with a 
reflection twist along $\omega$. As an anyon $a$ traverses the parity defect, it is transformed
to $\ra{a}$ (see Fig. \ref{paritydefectFig}(a)). 

Within this framework, it is straightforward to compute the dimension of the Hilbert space 
by counting fusion diagrams, keeping track of the action of the parity defect, as shown 
in Fig. \ref{paritydefectFig}(c) . 
In particular, let us consider the case where there is a parity defect line wrapping every cycle 
$\beta_i$ of the genus $g$ surface (see Fig. \ref{paritydefectFig}(b)). 
This leads us to the formula:
\begin{align}
\Z(\Sigma^2 \times S^1) = \text{dim } \V(\Sigma^2) 
= \sum N_{z_1 z_2 \cdots z_g}^0 \prod_{j=1}^g N_{a_j \overline{\ra{a_j}}}^{z_j} ,
\end{align} 
where
\begin{align}
N_{z_1 z_2 \cdots z_g}^0 = \sum_{c_{1...j} \in \mathcal{B}} N_{z_1 z_2}^{c_{12}} N_{c_{12} z_3}^{c_{123}} \cdots N^0_{c_{1 ...  g-1}z_g} . 
\end{align}
Using the Verlinde formula (\ref{verlinde}) and the fact that $\overline{\ra{a_j}} = \ra{\bar{a}_j}$, it is straightforward to simplify this:
\begin{align}
\text{dim } \V(\Sigma^2) = \sum_{x, a_j} S_{0x}^{\chi} \prod_{j=1}^g S_{a_j x} S_{\ra{\overline{a}_j} x} .
\end{align}
Next, we use the identity:
\begin{align}
	S_{\ra{\overline{a}_j} x} = S^*_{\overline{a}_j, \ra{x}} = S^*_{a_j, \overline{\ra{x}}} . 
\end{align}
Summing over $a_j$ then gives:
\begin{align}
\label{evenEuler}
\Z(\Sigma^2 \times S^1)= \sum_{x| x = \overline{\ra{x}}} S_{0x}^{\chi}.
\end{align}

\subsection{Dimensional reduction argument}

In order to derive Eq. (\ref{ZnonOr1}) for closed surfaces $\Sigma^2$ of odd Euler characteristic, we need to use
a different approach. One way to demonstrate Eq. (\ref{ZnonOr1}) is via dimensional reduction. 
The idea of dimensional reduction is that the path integral on $\Sigma^2 \times S^1$ can be viewed as
the path integral of a (1+1)D TQFT on $\Sigma^2$:
\begin{align}
	\Z_{2+1}(\Sigma^2 \times S^1) = \Z_{1+1}(\Sigma^2) .
	\label{eqn:dimrec}
\end{align}
Therefore, we simply need to understand which (1+1)D TQFT is obtained upon dimensional reduction, in order to compute
$\Z_{2+1}(\Sigma^2 \times S^1)$. Furthermore, since we are interested in the case where the (1+1) TQFT can be defined on
a non-orientable manifold, we need to consider (1+1)D ``unoriented" TQFTs. These are (1+1)D TQFTs equipped with an action
of reflection symmetry, so that they can be defined on non-orientable manifolds.

\subsubsection{(1+1)D TQFTs}
\label{1dTQFTsec}

Mathematically, a (1+1)D TQFT $\mathcal{E}$ assigns a complex number $\Z_{\mathcal{E}}(\Sigma^2) $ to each closed two-dimensional manifold $\Sigma^2$,
and a vector space $\V_{\mathcal{E}}(S^1)$ to the circle. It is well-known that (1+1)D TQFTs are classified by commutative Frobenius algebras~\cite{MooreSegal, AbramsTQFT},
which can always be decomposed into a direct sum of one-dimensional algebras. This implies that a generic (1+1)D TQFT $\mathcal{E}$ can be considered as a direct
sum $\mathcal{E} = \oplus_i \mathcal{E}_i$, such that
\begin{align}
\V_{\mathcal{E}}(S^1) &= \oplus_i \V_{\mathcal{E}_i}(S^1) ,
\end{align}
with $\text{dim } \V_{\mathcal{E}_i}(S^1) = 1$. Each $\mathcal{E}_i$ is then specified by a complex number $\lambda_i$, 
such that the path integral on a closed oriented surface $\Sigma^2$ is given by
\begin{align}
\Z_{\mathcal{E}} (\Sigma^2) &= \sum_i \Z_{\mathcal{E}_i}  (\Sigma^2)
\nonumber \\
\Z_{\mathcal{E}_i}(\Sigma^2) &=\lambda_i^{\chi(\Sigma^2)}.
	\label{eqn:2dtqft-oriented}
\end{align}

For the theory to be unitary, we also require reflection positivity in the Euclidean space time. For example, for each of the sector $\mathcal{E}_i$, since $\V_{\mathcal{E}_i}(S^1)=1$,
the path integral on a disk $D^2$ defines a state $|\Psi(D^2)\rangle$. The path integral on $S^2$
is equal to the inner product $\langle \Psi(D^2) | \Psi(D^2)\rangle$, which must be positive. From (\ref{eqn:2dtqft-oriented}),
we see that $\Z_{\mathcal{E}_i}(S^2) = \lambda_i^2$, which thus implies that $\lambda_i = \pm |\lambda_i|$. 
From Sec. \ref{1dSPT}, we observe that the sign choice is related to the existence of a (1+1)D invertible TQFT (i.e. a SPT phase), defined by
\begin{equation}
	\Z_\text{SPT}(\Sigma^2)=(-1)^{\chi(\Sigma^2)}.
	\label{}
\end{equation}
In fact, this is nothing but the topological path integral of the non-trivial (1+1)D time-reversal / reflection symmetric SPT.

A (1+1)D unoriented TQFT requires an action of reflection symmetry $\R$. As we discuss in Appendix \ref{1dTQFTSec}, we can 
distinguish two cases. In one case,
\begin{align}
\R: \mathcal{E}_i \rightarrow \mathcal{E}_i, 
\end{align}
so that the action of $\R$ is trivial. In the other case,
\begin{align}
\R: \mathcal{E}_i \oplus \mathcal{E}_{\R(i)} \rightarrow \mathcal{E}_{\R(i)} \oplus \mathcal{E}_i
\end{align}
Each $\mathcal{E}_i$ is completely specified by a complex number $\lambda_i$.

Therefore, a generic (1+1)D unoriented unitary TQFT is described by a set of real numbers $\{\lambda_i\}$, together with the action of
$\R$ on the individual (1+1)D TQFTs $\mathcal{E}_i$. As we show in the Appendix, the path integral $\Z_{\mathcal{E}}(\Sigma^2)$ 
on a non-orientable surface $\Sigma^2$ is
\begin{align}
\label{1dTQFTZ}
\Z_{\mathcal{E}}(\Sigma^2) =\sum_{a | \R(a) = a} \lambda_a^{\chi(\Sigma^2)} . 
\end{align}

\subsubsection{(2+1)D path integrals}

Let us now consider the dimensional reduction of a generic $(2+1)$D topological phase. The similarity of Eq. \eqref{eqn:verlinde-gsd} 
and \eqref{eqn:2dtqft-oriented} suggests that the $(1+1)$D TQFT breaks into sectors labeled by anyon types, with $\lambda_a=S_{0a}$. 
To illustrate this, let us consider the path integral of a (2+1)D TQFT on $M = \Sigma^2 \times S^1$ where $M^2$ is an oriented surface of genus $g$. 
We denote the path integral with a fixed topological charge $a$ as measured through the $S^1$ by $\mathcal{Z}(M, a)$. 
To evaluate $\mathcal{Z}(M, a)$, we can first apply an $S$ transformation, switching to a different basis in which we have a puncture labelled by topological charge $b$ 
on $\Sigma^2$ (this can be created by inserting a Wilson loop of $b$ along the $S^1$):
\begin{equation}
	\begin{split}
		\mathcal{Z}(M, a)&=\sum_{b\in \mathcal{B}}S_{0a}S_{ab}^*\mathcal{Z}_b(M)\\
		&=\sum_{b\in \mathcal{B}} S_{0a}S_{ab}^*\text{dim}\, \V_b(\Sigma^2)\\
		&=\sum_{b\in\mathcal{B}} S_{0a}S_{ab}^*\sum_{x\in\mathcal{B}}S_{0x}^{\chi-1}S_{xb}\\
	&=S_{0a}^{\chi(\Sigma^2)}.
	\end{split}
	\label{}
\end{equation}
Here $Z_b(M) = \text{dim}\, \V_b(\Sigma^2)$ is the path integral in the presence of a puncture on $\Sigma^2$ labelled by $b$. 
The above calculation demonstrates that the physically different sectors in the dimensionally reduced (1+1)D TQFT are labeled by the 
topological charge types as measured through the circle $S^1$.

Combining Eq. (\ref{1dTQFTZ}) and (\ref{evenEuler}), we find that 
\begin{align}
\mathcal{Z}_{2+1}(\Sigma^2 \times S^1) = \sum_{a | a = \overline{\ra{a}}} \lambda_a^{\chi(\Sigma^2)} , 
\end{align}
where 
\begin{align}
S_{0a} = |\lambda_a| .
\end{align}
%Therefore, we can write
%\begin{align}
%\lambda_a=\sigma_a S_{0a},
%\end{align}
%where $\sigma_a = \pm 1$ for the invariant anyons $\overline{\ra{a}} = a$. For the non-invariant anyons, 
%where $\overline{\ra{a}} \neq a$, $\sigma_a$ is irrelevant and can be set to zero 
%because it never enters the computation of $\mathcal{Z}$.

As we discussed in the previous section, the signs
$\lambda_a/|\lambda_a|$ can be interpreted as tensoring a 
(1+1)D reflection SPT phase. We therefore conclude that 
\begin{align}
	\frac{\lambda_a}{|\lambda_a|} = {\eta}_a^\R
\end{align}
Namely, if $\frac{\lambda_a}{|\lambda_a|} =-1$ the Wilson loop of $a$ is ``decorated'' with a (1+1)D reflection SPT state. 
This establishes Eq. (\ref{ZnonOr1}). %- (\ref{ZnonOr2}). 

Let us now consider non-orientable surfaces with punctures. The basic example is a crosscap, or 
RP$^2$ with a puncture. Once we know the GSD on such a manifold, we can build up more 
complicated ones by sewing crosscaps and punctured spheres together. 

Consider a non-orientable manifold $\Sigma^2$ obtained by attaching $k$ crosscaps to a 
sphere with $n$ punctures with topological charges $a_1,a_2, \dots, a_n$. 
The degeneracy can be obtained from fusion:
\begin{equation}
	\sum_{\{x_i,y_i\}\in \mathcal{B}} \prod_{i=1}^k M_{x_i} N_{x_1x_2}^{y_2}N_{y_2x_3}^{y_3}\cdots N_{y_{k-1} x_k}^{b_0} N_{b_0 a_1}^{b_1}\cdots N_{b_{n-1}a_n}^0.
	\label{}
\end{equation}
We can then use the Verlinde formula, Eq. (\ref{verlinde}), to simplify the expression, obtaining
\begin{equation}
	\text{dim}\, \V_{\{a_j\}} (\Sigma^2)=\sum_{x}n_x^k S_{0x}^{\chi}\prod_{i=1}^n S_{a_ix}.
	\label{verlindeNonOrPunctured1}
\end{equation}
Here 
\begin{align}
\chi(\Sigma^2)=2-k-n
\end{align}
is the Euler characteristic, and 
\begin{align}
n_x = \sum_y M_y S_{xy}.
\end{align}

When $k$ is even, we can again directly compute $\text{dim}\, \V_{\{a_i\}} (\Sigma^2)$ 
by counting the fusion diagrams as in Sec. \ref{evenEulerSec} and the result is a straightforward 
generalization of Eq. \eqref{eqn:verlinde-punctured}:
\begin{equation}
	\text{dim}\, \V_{\{a_i\}} (\Sigma^2) = \sum_{x | x = \overline{\ra{x}}} S_{0x}^{\chi(\Sigma^2)} \prod_{i=1}^n S_{x a_i},
	\label{verlindeNonOrPunctured2}
\end{equation}
Considering the limit $k \rightarrow \infty$, we see that (\ref{verlindeNonOrPunctured1}) and (\ref{verlindeNonOrPunctured2}) imply:
\begin{equation}
	n_x = 
	\begin{cases}
		\pm 1 & x = \overline{\ra{x}}\\
		0 & \text{otherwise}
	\end{cases}
	\label{}
\end{equation}
Again the sign can not be resolved since $k$ is restricted to even integers.

Now, setting $a_1=a_2=\dots = a_n=0$, letting $k$ be odd, and comparing 
(\ref{verlindeNonOrPunctured1}) with (\ref{ZnonOr1}), we have
\begin{equation}
	\sum_{x}n_x S_{0x}^{\chi}=\sum_{x\in \mathcal{B}^\varphi} {\eta}_x^\R S_{0x}^\chi.
%	\sum_{x\in \mathcal{B}^{\varphi}}n_x S_{ax}=\sum_{x\in \mathcal{B}^\varphi} \eta_x S_{ax}
	\label{verlindeNonOrPunctured3}
\end{equation}
%Multiplying both sides by $S_{ba}^*$ and summing over $a$ gives
%\begin{align}
%\sum_{x \in \mathcal{B}^{\varphi}} n_x \delta_{bx} = \sum_{x \in \mathcal{B}^{\varphi}} \eta_x \delta_{bx}, 
%\end{align}
%which implies $n_x = \eta_x$ when $x \in \mathcal{B}^{\varphi}$. 
(\ref{verlindeNonOrPunctured3}) strongly suggests that $n_x = {\eta}_x^\R$ when $x = \overline{\ra{x}}$. 
As a result, we have
\begin{equation}
	M_a = \sum_{x | x = \overline{\ra{x}}} {\eta}_x^\R S_{ax}.
	\label{}
\end{equation}
While the above consideration was strongly suggestive of the result for $M_a$, it was not a direct proof
that $n_x = {\eta}_x^\R$. In what follows, we present a direct computation of $M_a$.

\subsection{Computation of $M_a$ using (3+1)D TQFTs}
\label{MaSec}

As discussed in Sec. \ref{2dfrom3dSec}, computations of path integrals on (2+1)D manifolds can be performed
in terms of an associated (3+1)D TQFT. Thus, the formula for $M_a$ presented above can be computed 
explicitly by utilizing the machinery of (3+1)D TQFTs and its relation to (2+1)D TQFTs. 
We wish to compute
\begin{align}
M_a \equiv \Z_a(\mathbb{RP}^2 \times S^1) ,
\end{align}
where $\Z_a(\mathbb{RP}^2 \times S^1)$ is the path integral on $\mathbb{RP}^2 \times S^1$ with a Wilson loop of $a$ inserted along $S^1$, 
implying that there is a puncture on $\mathbb{RP}^2$ labeled by $a$.
We can compute $\Z_a(\mathbb{RP}^2 \times S^1) $ directly by noting that $\mathbb{RP}^2 \times S^1 = \partial ( \mathbb{RP}^2 \times D^2)$, and therefore:
\begin{align}
\Z_a(\mathbb{RP}^2 \times S^1) = \Z(\mathbb{RP}^2 \times D^2)[l_a].
\end{align}
$\Z(\mathbb{RP}^2 \times D^2)[l_a]$ is the path integral of the associated (3+1)D theory on $\mathbb{RP}^2 \times D^2$, where the boundary
condition $l_a$ refers to a loop of anyon $a$ encircling the $S^1$ that lies at the boundary of the $D^2$ in 
$\partial (\mathbb{RP}^2 \times D^2)$. 

\begin{figure}[t!]
\includegraphics[width=\columnwidth]{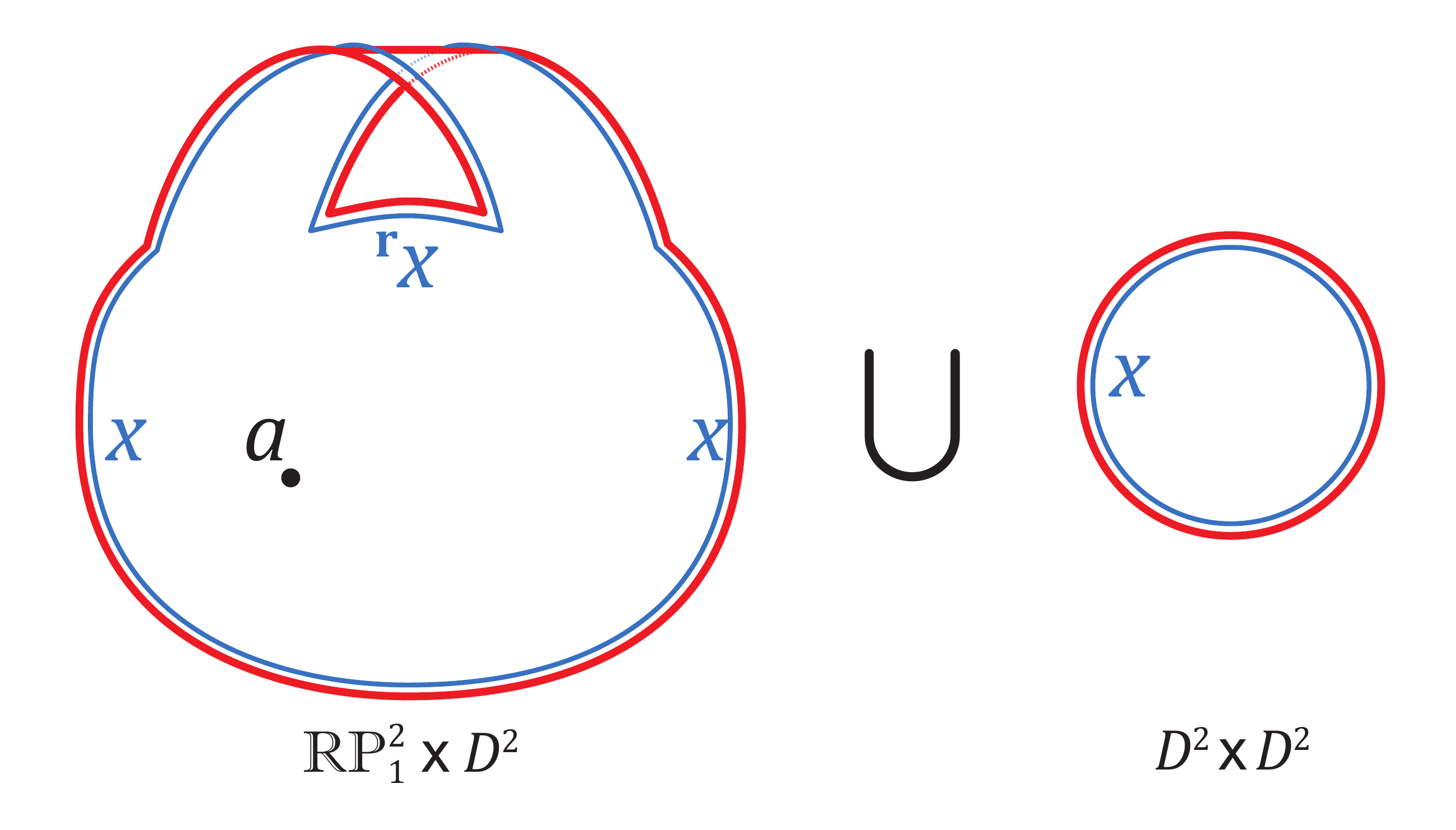}
\caption{Left: $\mathbb{RP}^2_1 \times D^2$, with the $D^2$ not shown. At the boundary of the $D^2$, the $\mathbb{RP}^2_1$ has a
puncture labelled by $a$, with the loop going around the $S^1$ of the $D^2$. There is an anyon loop
$x$ along the boundary of the $\mathbb{RP}^2_1$, which transforms into $\ra{x}$ as it passes through the twist. 
This is glued to $D^2 \times D^2$, shown on the right (with one of the $D^2$'s left implicit). The
two manifolds are glued along $S^1 \times D^2$, shown in red. 
 }
\label{rp212handle}
\end{figure}

To compute $\Z(\mathbb{RP}^2 \times D^2)[l_a]$, we observe that $\mathbb{RP}^2 \times D^2$ has a handle decomposition consisting of a
$0$-handle, a $1$-handle, and a $2$-handle. Recall that the $0$-handle and the $1$-handle consist of 
two $D^4$'s which are glued to each other along $S^0 \times D^3$. Thus the $0$-handle and the $1$-handle 
glued together give $D^2$ times a Mobius band, which we think of as the $1$-skeleton $\mathbb{RP}^2_1$. 
The resulting manifold $\mathbb{RP}^2_1 \times D^2$ is glued to a $2$-handle by gluing $D^4 = D^2 \times D^2$ 
along an $S^1 \times D^2$, where the $S^1$ is the boundary of the Mobius band $\mathbb{RP}^2_1$. See Fig. \ref{rp212handle} for an illustration. 

Therefore, using the gluing formula, we obtain
\begin{align}
\Z(\mathbb{RP}^2 \times D^2)[l_a] &= \sum_{x \in \mathcal{B}} \frac{\Z(D^4)[l_x] \Z( \mathbb{RP}^2_1 \times D^2) [ l_x, l_a] }{\langle l_x | l_x \rangle_{\V(S^1\times D^2)} }
\nonumber \\
&= \sum_{x\in \mathcal{B}} \frac{d_x}{\mathcal{D}} \Z( \mathbb{RP}^2_1 \times D^2) [ l_x, l_a] .
\end{align}
Here, the $l_x$ denotes a loop of anyon $x$ encircling the $S^1$ of the attaching region of the $2$-handle, which corresponds to
the boundary of the Mobius band $\mathbb{RP}^2_1$. The second line was obtained by using Eqs. (\ref{ZD4la}), (\ref{AS1D2}), (\ref{ZS1D3}). 

Observe that $\partial (\mathbb{RP}^2_1 \times D^2) = (S^1 \times D^2) \cup (\mathbb{RP}^2_1 \times S^1)$. 
The $l_x$ and $l_a$ loops in $\Z( \mathbb{RP}^2_1 \times D^2) [ l_x, l_a]$
correspond to loops of anyons $x$ and $a$ encircling the two $S^1$'s on the two parts of the boundary. 
Since these two closed loops link with each other, the anyon diagram can be evaluating using
the usual rules of UMTCs:
\begin{equation}
\includegraphics{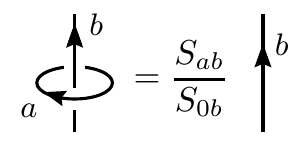}
\end{equation}
Therefore, 
\begin{align}
\Z( \mathbb{RP}^2_1 \times D^2) [ l_x, l_a] = \frac{S_{ax}}{S_{0x}} \Z(\mathbb{RP}^2_1 \times D^2)[l_x] .  
\end{align}

\begin{figure}[t!]
\includegraphics[width=3.5in]{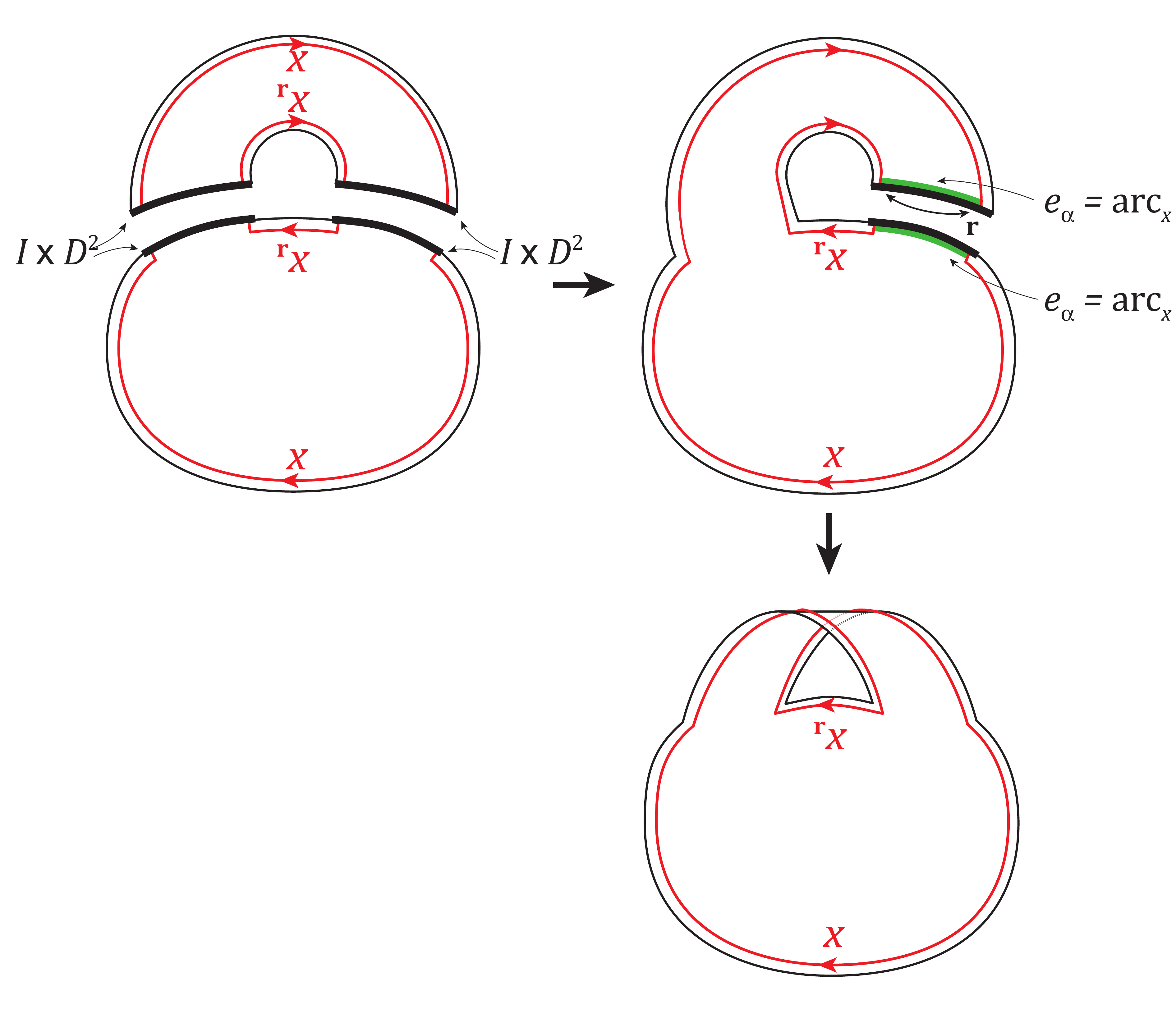}
\caption{
We obtain $\mathbb{RP}^2_1 \times D^2$ by gluing together two $D^4$'s. 
The illustration cuts the dimension in half by not displaying the $D^2$. 
The loop $l_x$ that goes along the boundary of the $\mathbb{RP}^2_1$ is drawn in red, 
which is cut into $\text{arc}_x \cup \text{arc}_{\ra{x}}$ on each $D^4$. 
Top left: the attaching region of the $1$-handle is in bold; the two attaching regions are $I \times D^2 = D^3$. After attaching
along one of the $D^3$'s, we obtain a single $D^4$ as shown in the middle diagram,
with the two arcs $\text{arc}_x \cup \text{arc}_{\ra{x}}$ shown in red. 
The $D^4$ is glued to itself along a $D^3$ with an action of reflection $\R$, to give 
$\mathbb{RP}^2_1 \times D^2$. The intermediate state $e_\alpha = \text{arc}_x \in \V(D^3; a, \ra{a})$ is shown in green. 
Taking into account the fact that $x = \overline{\ra{x}}$, we see directly
that $\text{arc}_x \cup e_\alpha \cup \text{arc}_x \cup \R(e_{\alpha})$ is a single loop $l_x$. 
\label{01handle}
 }
\end{figure}

Next, let us compute $\Z(\mathbb{RP}^2_1 \times D^2)[l_x]$. This consists of the
$0$-handle glued to the $1$-handle (see Fig. \ref{01handle}) , where there is a loop of anyon $x$ running parallel to the boundary of 
the Mobius band $\mathbb{RP}^2_1$. The attaching region is $S^0 \times D^3$, where one of the $D^3$'s is glued
with an action of reflection. Thus we can obtain $\mathbb{RP}^2_1 \times D^2$ by starting
with $D^4 = I \times D^3$ and gluing the two ends of the first interval together, with a reflection twist. In fact, if we temporarily suppress the $D^2$ this is how we obtain a Mobius band from a rectangular strip $D^1\times D^1$. The gluing region is $D^1\times D^2=D^3$, with two points marked by $x$ and $\ra{x}$ in order to correctly produce the anyon line $l_x$ along the boundary.
Now applying the gluing formula:
\begin{widetext}
\begin{align}
	\Z(\mathbb{RP}^2_1 \times D^2)[l_x] = \sum_{e_\alpha \in \V(D^3; x, \ra{x})} \frac{\Z(D^4)[\text{arc}_x \cup e_\alpha \cup \text{arc}_{\ra{x}} \cup \R(e_\alpha) ]}{ \langle e_\alpha | e_\alpha \rangle_{\V(D^3; x, \ra{x})} }
\end{align}
\end{widetext}
From Eq. (\ref{AD3dim}) we see that $\text{dim } \V(D^3; x, \ra{x}) = \delta_{x, \overline{\ra{x}}}$. 
Moreover, as discussed below Eq. (\ref{AD3dim}), we can pick $e_\alpha \in \V(D^3; x, \overline{x})$ to correspond to an arc 
connecting $x$ and $\overline{x}$, and from Eq. \eqref{eqn:AD3}, we have
$\langle \text{arc}_x | \text{arc}_x \rangle_{\V(D^3; x, \ra{x})} = \frac{d_x}{\mathcal{D}}$. 

The boundary condition on the $D^4$ thus reduces to (see Fig. \ref{01handle} ).
\begin{align}
\text{arc}_x \cup e_\alpha \cup \text{arc}_{\overline{x}} \cup \R(e_\alpha) = l_x,
\end{align}
which thus consists of a single loop $l_x$. However, crucially, the $\R(e_{\alpha})$ corresponds to an action of 
reflection on the Hilbert space associated with $\V(D^3; x, \overline{x})$. This is precisely the reflection eigenvalue discussed
in Sec. \ref{symfrac} for the state of the (2+1)D system with two anyons $x$ and $\overline{x}$ on a disk (or equivalently, on opposite ends 
of a cylinder). 
We thus obtain:
\begin{align}
	\Z(D^4)[\text{arc}_x \cup e_\alpha \cup \text{arc}_{\ra{x}} \cup \R(e_\alpha)] = {\eta}_x^\R \Z(D^4)[l_x]. 
\end{align}
Using $\Z(D^4)[l_x] = \frac{d_x}{\mathcal{D}}$, we finally obtain:
\begin{align}
\label{ZRP21D2}
\Z(\mathbb{RP}^2_1 \times D^2)[l_x] = \delta_{x, \overline{\ra{x}}} {\eta}_x^\R . 
\end{align}

Combining these results, we obtain
\begin{align}
	M_a \equiv \Z_a(\mathbb{RP}^2 \times S^1) = \sum_{x | x = \overline{\ra{x}}} S_{ax} {\eta}_x^\R . 
\end{align}

\subsection{Abelian topological phases and loop gas argument}
\label{sec:loop-gas}
Here we provide an alternative argument for the value of $M_a$ for the case of Abelian topological phases.
It is well-known that the ground states of certain Abelian topological states can be obtained as a superposition
of closed loop configurations, for appropriately defined loops. In the following we use this definition of the ground state to
obtain the the ground state degeneracy formula for $M_a$. A related loop gas argument was used recently to study the ground states of the
doubled semion state (without symmetry fractionalization) on non-orientable manifolds.\cite{freedman2016}

\subsubsection{$\mathbb{Z}_2$ toric code}

\begin{center}
  \begin{tabular}{ c c c c c  }
     $(\eta_e^\R, \eta_m^\R)$ & $M_0$ & $M_e$ & $M_m$ & $M_\psi$ \\ \hline
    $ (+1,+1)$ & 2 & 0 & 0 & 0 \\ \hline
    $ (+1,-1)$ & 0 & 2 & 0 & 0 \\ \hline
    $ (-1,+1)$ & 0 & 0 & 2 & 0 \\\hline
$ (-1,-1)$ & 0 & 0 & 0 & 2 \\ \hline
    \hline
  \end{tabular}
\end{center}

Let us begin with the $\mathbb{Z}_2$ toric code state, which is the ground state of the toric code Hamiltonian\cite{kitaev2003} :
\begin{align}
{\bf H} = - \sum_v A_v - \sum_p B_p.
\end{align}
Here, there are spin-1/2 degrees of freedom defined on the links of a square lattice, $A_v = \prod_{l \in s(v)} \sigma^x_l$, $B_p =
\prod_{l \in \partial p} \sigma^z_p$, where $\sigma^a$ for $a = x,y,z$ are the Pauli matrices. Here $l \in s(v)$ are the links
which contain the vertex $v$, while $l \in \partial p$ are the links belonging to plaquette $p$. Working in the $\sigma^x$ basis,
we can define $e$ loops where $\sigma^x = -1$ along the loop, and $\sigma^x = +1$ away from the loop. The ground state on any
closed surface is then a superposition over all possible closed loops configurations:
\begin{align}
|\Psi_0 \rangle = \sum_{C} |C \rangle ,
\end{align}
where $\sum_C$ is a sum over all possible configurations of closed loops. Equivalently, we could go to the dual lattice and work in the $\sigma^z$ basis,
and define $m$ loops where $\sigma^z = -1$ along the loop and $\sigma^z = +1$ away from the loop. In this basis, the
ground state is also a sum over closed $m$ loops. On a torus, the familiar four-fold degeneracy corresponds to whether there
are an odd or an even number of $e$-loops (or, alternatively, $m$ loops in the dual basis) encircling the longitude and meridian of the torus. 

Let us now consider the toric code state on $\mathbb{RP}^2$. Since the first homology $H_1(\mathbb{RP}^2,\mathbb{Z}) = \mathbb{Z}_2$, 
there are two possible states, depending on whether there are an odd or even number of loops enclosing the 
non-contractible cycle. This implies $M_0 = 2$ for the toric code.

\begin{figure}
\includegraphics[width=3.5in]{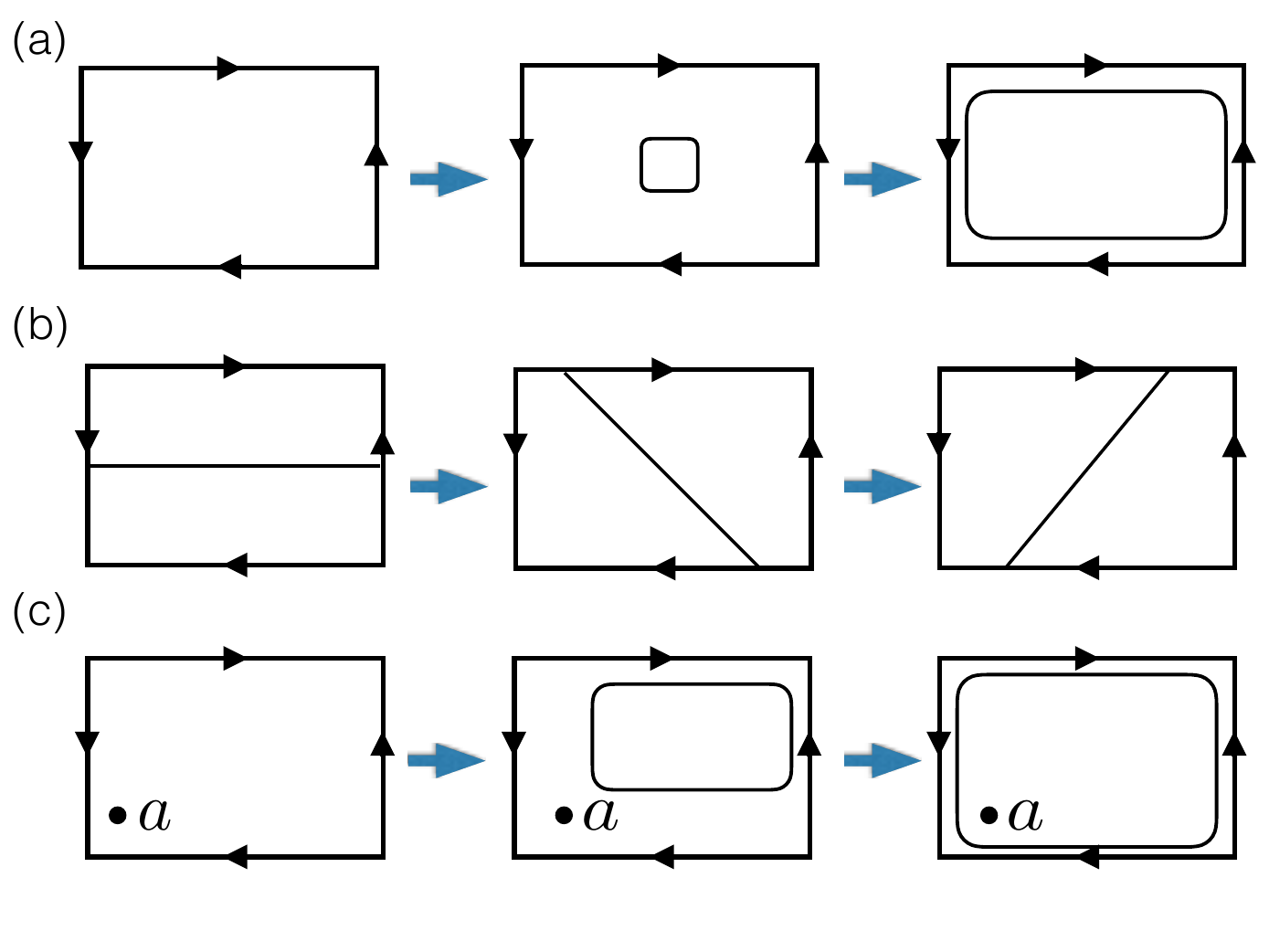}
\caption{$\mathbb{RP}^2$ is equivalent to a square with opposite edges identified as shown.  (a)  A closed loop is created out of the empty picture
and sweeps through $\mathbb{RP}^2$ once. (b) The non-contractible cycle of $\mathbb{RP}^2$ sweeps through all of $\mathbb{RP}^2$ once. (c) Same as
top panel, but with a puncture labelled by $a$. 
 }
\label{loopGas}
\end{figure}

The previous example corresponds to the case where ${\eta}_e^\R = {\eta}_m^\R = 1$. Let us now consider a modification of the
$\mathbb{Z}_2$ toric code, where $e$ has ${\eta}_e^\R = -1$, and $m$ has ${\eta}_m^\R = +1$. We will show that there does not exist any ground
state in this case. Let us begin by considering the empty picture, which contains no $e$ loops. We can consider a sequence of
moves, where an $e$ loop is created out of the vacuum, sweeps through the whole space, and disappears again (see Fig. \ref{loopGas}(a)). The amplitude
for each picture must be equal in order to be a ground state of the Hamiltonian. However, because ${\eta}_e^\R =-1$ we can interpret the $e$ loop as being
decorated with a nontrivial (1+1)D SPT state. Sweeping the $e$ loop around all of $\mathbb{RP}^2$ should then give a factor of $-1$, corresponding
to the path integral of a (1+1)D SPT state on $\mathbb{RP}^2$. Therefore, the empty picture, and consequently every configuration
with an even number of loops across the non-contractible cycle, must have zero weight. For the case of the configuration with
only a single loop around the non-contractible cycle, we can similarly consider the loop to rotate around, sweeping through all
of $\mathbb{RP}^2$ (Fig. \ref{loopGas}(b)). The same argument as above implies that this configuration must also have zero amplitude. We thus conclude
that in this case, $M_0 = 0$. The above argument can be made explicit by considering an exactly solvable model where the
$e$ loops are decorated with (1+1)D SPT states.\cite{zion2016}

Now, we can consider the case where the state on $\mathbb{RP}^2$ has a single $a$ particle at some location, for some $a = e, m, \psi$.
Repeating the above argument, we find that the amplitude for the empty picture need not be zero, because the $e$ loop,
in addition to sweeping through $\mathbb{RP}^2$, must pass through the $a$ particle (see Fig. \ref{loopGas}(c)), 
providing another phase factor associated to the mutual statistics between $a$ and $e$ to the process. 
If $a = m$, it is now consistent to have a sum over all loop configurations, and we
obtain $M_m = 2$. If $a = e$ or $\psi$, repeating the argument for $m$ loops implies that $M_e = M_\psi = 0$.

The argument given in this section is heuristic in nature. In
Sec. \ref{SubSec:decorated_TC} we make this argument precise using a solvable model of the $Z_2$ toric code state
with $\eta_e^{\bf r} = -1$ and $\eta_m^{\bf r} = 1$ defined on $\mathbb{RP}^2$. 

%Let us now generalize the above to the $\mathbb{Z}_N$ toric code. We label the particles as $e^a m^b$, for $a,b = 0 ,\cdots, N-1$, and we consider the case where the $e$ particle is invariant as it encloses the non-contractible cycle of $\mathbb{RP}^2$, but $m$ changes to $-m$.  Since $H_1(\mathbb{RP}^2) = \mathbb{Z}_2$, it follows that if $N$ is odd, the state with zero loops encircling the non-contractible cycle is equivalent to the state with any number of non-contractible loops encircling it. Therefore $M_0 = 1$ if $N$ is odd. When $N$ is even, then only the parity of $e$ loops is well-defined, so we get $M_0 = 2$. This agrees with our formula for $M_0$.

%If $N$ is even, it is possible to have $\tilde{\eta}_e = -1$ and $\tilde{\eta}_{m^{N/2}} = 1$. In this case, then $M_{e^a m^b} = 0$ unless $e^{2\pi i b/ N} = -1$ and $e^{\pi i a} = 1$, i.e. $b=N/2$ and $a = 2$. Thus, $M_x = 2 \sum_{a=0}^{N/2-1} \delta_{x, e^{2a} m^{N/2}}$, which again agrees with our general formula.  

%% file: anomaly.tex
\section{Anomalies of time-reversal and reflection symmetry}
\label{anomalySec}

Not all (2+1)D SETs can exist in purely (2+1)D. It is possible that a given time-reversal or reflection symmetry fractionalization class
is anomalous, and therefore the corresponding (2+1)D SET can only exist at the surface of a (3+1)D SPT state. In this section we develop a theory of
such time-reversal and reflection anomalies. We note that the explicit SET (2+1)D state sum constructions presented in Sec. \ref{statesumSec} therefore
realize anomaly-free (2+1)D SETs. 

\subsection{Dehn twist anomaly on M\"obius band}
\label{dehnAnomaly}

\begin{figure}
\includegraphics[width=3.5in]{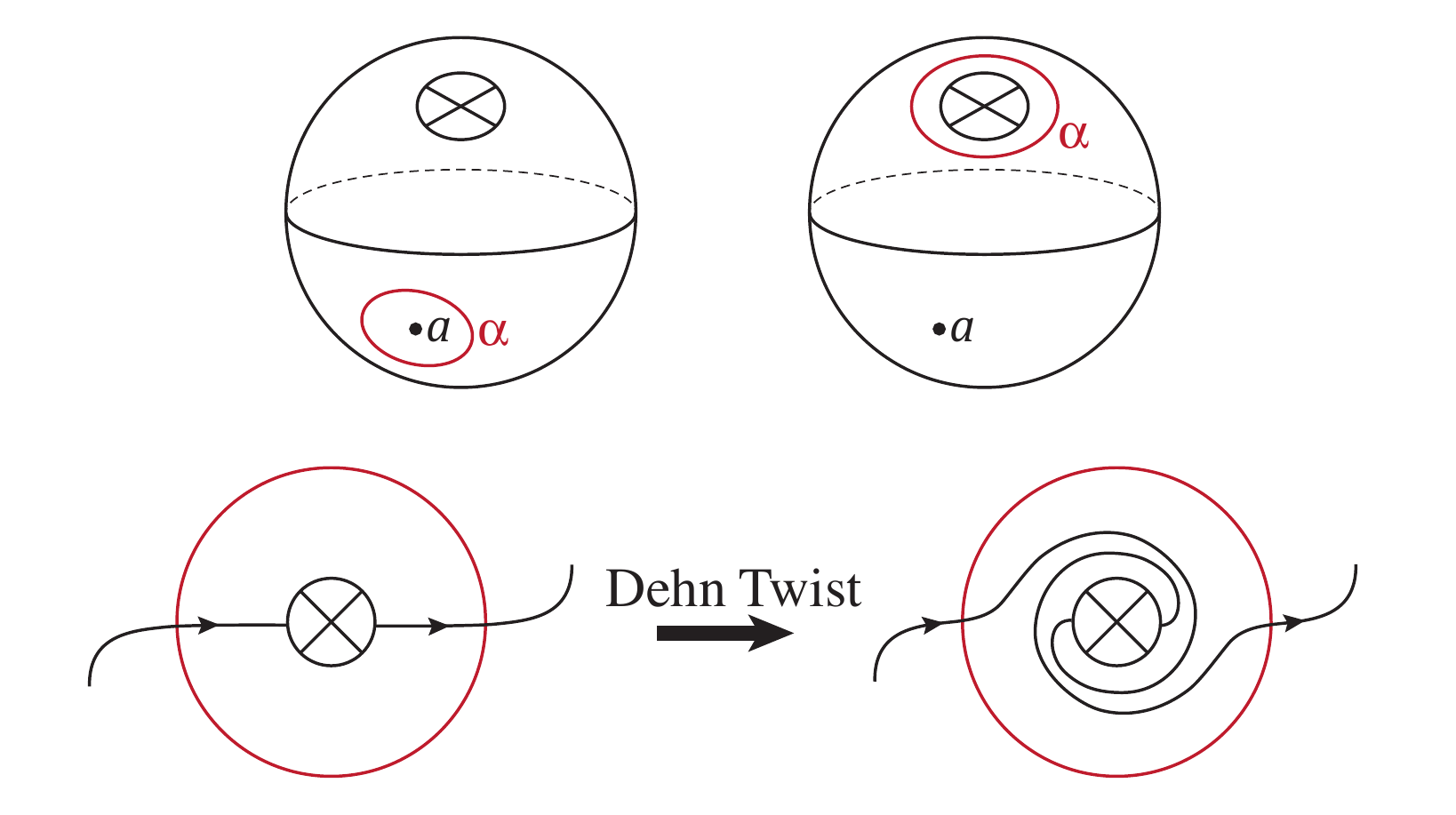}
\caption{Top: A loop $\alpha$ that encircles a puncture labelled
by $a$ on $\mathbb{RP}^2$. The loop is topologically equivalent to a loop
encircling the cross-cap. Bottom: Dehn twist around $\alpha$ and its
effect on a line traversing the cross-cap. It can be seen that the
resulting line can be continuously deformed to the original line, and
therefore the Dehn twist around $\alpha$ is isotopic to the identity
element of the mapping class group.}
\label{ccDehn}
\end{figure}

Let us consider the Hilbert space, $\V_a(\mathbb{RP}^2)$ of the topological state on the projective plane, $\mathbb{RP}^2$, with a single puncture
labelled by a topological charge $a$. We can think of this as a sphere with a cross-cap, together with a puncture
labelled by $a$ (see Fig. \ref{ccDehn}). As mentioned in the previous section, $M_a \equiv \text{ dim } \V_a(\mathbb{RP}^2)$.
Let $|\Psi\rangle \in \V_a(\mathbb{RP}^2)$.

Now we can consider the action of a Dehn twist $\mathcal{T}_\alpha$ along a loop $\alpha$ surrounding $a$, which gives:
\begin{align}
	\mathcal{T}_\alpha |\Psi \rangle = \theta_a |\Psi \rangle ,
\end{align}
where $\theta_a$ is the topological spin of $a$. Importantly, in this case the Dehn twist around $\alpha$ is actually isotopic to the identity.
This can be seen, for example, by considering the effect of $\mathcal{T}_\alpha$ on the non-contractible cycles
and observing that the effect of $\mathcal{T}_\alpha$ can be continuously undone (see Fig. \ref{ccDehn}). 
This implies the constraint:
\begin{align}
\label{consistencyEq}
M_a = \sum_{x | x = \overline{\,\ra{x}}} S_{ax} \eta_x^{\bf r} > 0 \Rightarrow \theta_a = 1 .
\end{align}
The failure to satisfy the above constraint indicates that the theory is inconsistent on non-orientable manifolds, 
and thus that the orientation-reversing symmetry is anomalous. In Sec. \ref{anomalyExamples} we check
this anomaly in a number of examples. 

\subsection{Anomalous SETs and surface of (3+1)D SPT states}
\label{sec:4dspt}

We have found so far that some types of time-reversal or reflection symmetry action are not consistent in 2+1 dimensions by considering
the action of Dehn twists on the punctured projective plane. It is well-known that anomalous symmetry fractionalization classes
can appear at the (2+1)D topologically ordered surface of (3+1)D SPT states. This raises the question of whether we can determine
when a given SET with an action of time reversal or reflection can exist at the surface of a (3+1)D SPT and, if so, which SPT.

Time-reversal and reflection invariant bosonic SPTs have been argued to have a $\mathbb{Z}_2 \times \mathbb{Z}_2$ classification in (3+1)D.\cite{wang2013, kapustin2014}
These SPTs are distinguished by the value of the path integral of the effective TQFT on $\mathbb{RP}^4$ and $\mathbb{CP}^2$:
$\Z(\mathbb{RP}^4) = \pm 1$ and $\Z(\mathbb{CP}^2) = \pm 1$. The bosonic SPT with $\Z(\mathbb{RP}^4) = -1$ and $\Z(\mathbb{CP}^2) = 1$ is the one 
obtained within the group cohomology classification of Ref. \onlinecite{chen2013}. 
The bosonic SPT with $\Z(\mathbb{RP}^4) = 1$ and $\Z(\mathbb{CP}^2) = -1$ is the one that corresponds to the ``beyond group cohomology'' SPT
discussed in Ref. \onlinecite{vishwanath2013}. 

As discussed in Sec. \ref{3dTQFTSec}, a UMTC $\mathcal{B}$ defines a (3+1)D TQFT through the Crane-Yetter-Walker-Wang construction.  In this construction, the bulk intrinsic topological order is trivial, so the bulk is a (3+1)D SPT state, while the (2+1)D surface is topologically
ordered, with an anyon content described by $\mathcal{B}$. Such constructions have been considered in a number of examples of anomalous surface SETs~\cite{Chen2014, BurnellPRB2014, Fidkowski13, chen2014b}.

Given $\mathcal{B}$, together with the actions of time-reversal or reflection symmetry, $\{\tilde{\eta}_a\}$ 
defined in Sec. \ref{symfrac}, we will compute $\Z(\mathbb{RP}^4)$ and $\Z(\mathbb{CP}^2)$ for the (3+1)D SPT defined by $\mathcal{B}$. 
The result is:
\begin{equation}
	\begin{gathered}
	\Z(\mathbb{CP}^2) = \frac{1}{\mathcal{D}} \sum_a d_a^2 \theta_a = e^{\frac{2\pi i}{8}c_-},\\ 
\nonumber \\
\Z(\mathbb{RP}^4) = \frac{1}{\mathcal{D}}\sum_{a | a = \ra{a}} \eta_a^{\bf r} d_a \theta_a .
\end{gathered}
	\label{}
\end{equation}
The result of $\Z(\mathbb{CP}^2)$ is well-known. The value of $\Z(\mathbb{RP}^4)$ and $\Z(\mathbb{CP}^2)$ for a UMTC $\mathcal{B}$ with a particular action of time-reversal or reflection symmetry
determines whether the associated 2+1D topological phase can exist purely in two dimensions (if $\Z(\mathbb{RP}^4) = \Z(\mathbb{CP}^2) = 1$),
or whether it must exist at the surface of the (3+1)D SPT determined by the invariants $\Z(\mathbb{RP}^4)$ and $\Z(\mathbb{CP}^2)$. 

\subsubsection{$\Z(\mathbb{CP}^2)$ }

As reviewed in Appendix \ref{handleSec}, $\mathbb{CP}^2$ can be understood as a $0$-handle glued to a $2$-handle, 
which is in turn glued to a $4$-handle. Importantly, the $S^1$ of the attaching region of the 
$2$-handle is $+1$ framed, to recover the fact that $\mathbb{CP}^2$ has signature $1$. Therefore, 
using the gluing formula:
\begin{align}
\Z(\mathbb{CP}^2) &= \frac{\Z(\mathbb{CP}^2_{2})[\emptyset] \Z(D^4)[\emptyset]}{\langle \emptyset | \emptyset \rangle_{\V(S^3)}}
\nonumber \\
&= \frac{1}{\lambda} \Z(\mathbb{CP}^2_{2})[\emptyset], 
\end{align}
where $\mathbb{CP}^2_2$ is the 2-skeleton of $\mathbb{CP}^2$ (that is, it contains the $0$ and $2$ handles in 
the handle decomposition of $\mathbb{CP}^2)$.  Moreover,
\begin{align}
\Z(\mathbb{CP}^2_{2})[\emptyset] &= \sum_a \frac{\Z(D^4)[l_a^{(+1)}] \Z(D^4)[l_a]}{\langle l_a | l_a \rangle_{\V(S^1 \times D^2;\emptyset)}}
\nonumber \\
&= \sum_a \lambda d_a \theta_a \lambda d_a .
\end{align}
Here, $l_a^{(+1)}$ refers to a loop of anyon $a$ in $D^4$ with a $+1$ twist in its framing. This results from the fact discussed above
that the $2$-handle is glued to the $0$-handle along a circle with a $+1$ framing. 

Therefore, we have
\begin{align}
\Z(\mathbb{CP}^2) = \frac{1}{\mathcal{D}} \sum_a d_a^2 \theta_a.
\end{align}
This is the well-known formula which gives the chiral central charge $c_-$ for UMTCs:
\begin{align}
\frac{1}{\mathcal{D}} \sum_a d_a^2 \theta_a = e^{2\pi i c_- /8} . 
\end{align}

\subsubsection{$\Z(\mathbb{RP}^4)$}
\label{sec:rp4}

The handle decomposition for $\mathbb{RP}^4$ has a single $p$-handle for each $p  =0,1,2,3,4$.
The attachment of the $1$-handle to the $0$-handle has an orientation reversal on one of the $D^3$
of the attaching region, thus giving a M\"obius band inside of $\mathbb{RP}^4$ as required. The $2$-handle
is attached along an $S^1$ that goes around the cycle of the M\"obius band twice, in order to recover the fact that
$H_1(\mathbb{RP}^4;\mathbb{Z}) = \mathbb{Z}_2$. Moreover, the $S^1$ also has a $+1$ framing.

This implies:
\begin{align}
\Z(\mathbb{RP}^4) &= \frac{\Z(\mathbb{RP}^4_{3})[\emptyset] \Z(D^4)[\emptyset]}{\langle\emptyset | \emptyset\rangle_{\V(S^3)}}
\nonumber \\
&= \frac{1}{\lambda} \Z(\mathbb{RP}^4_{3})[\emptyset]
\end{align}
\begin{align}
\Z(\mathbb{RP}^4_{3})[\emptyset] &= \frac{\Z(\mathbb{RP}^4_{2})[\emptyset] \Z(D^4)[\emptyset]}{\langle \emptyset | \emptyset\rangle_{\V(S^2 \times D^1;\emptyset)}}
\nonumber \\
&= \lambda \Z(\mathbb{RP}^4_{2})[\emptyset] .
\end{align}
Thus,
\begin{align}
\Z(\mathbb{RP}^4) = \Z(\mathbb{RP}^4_{2}) [\emptyset]
\end{align}

\begin{figure}
\includegraphics[width=2.5in]{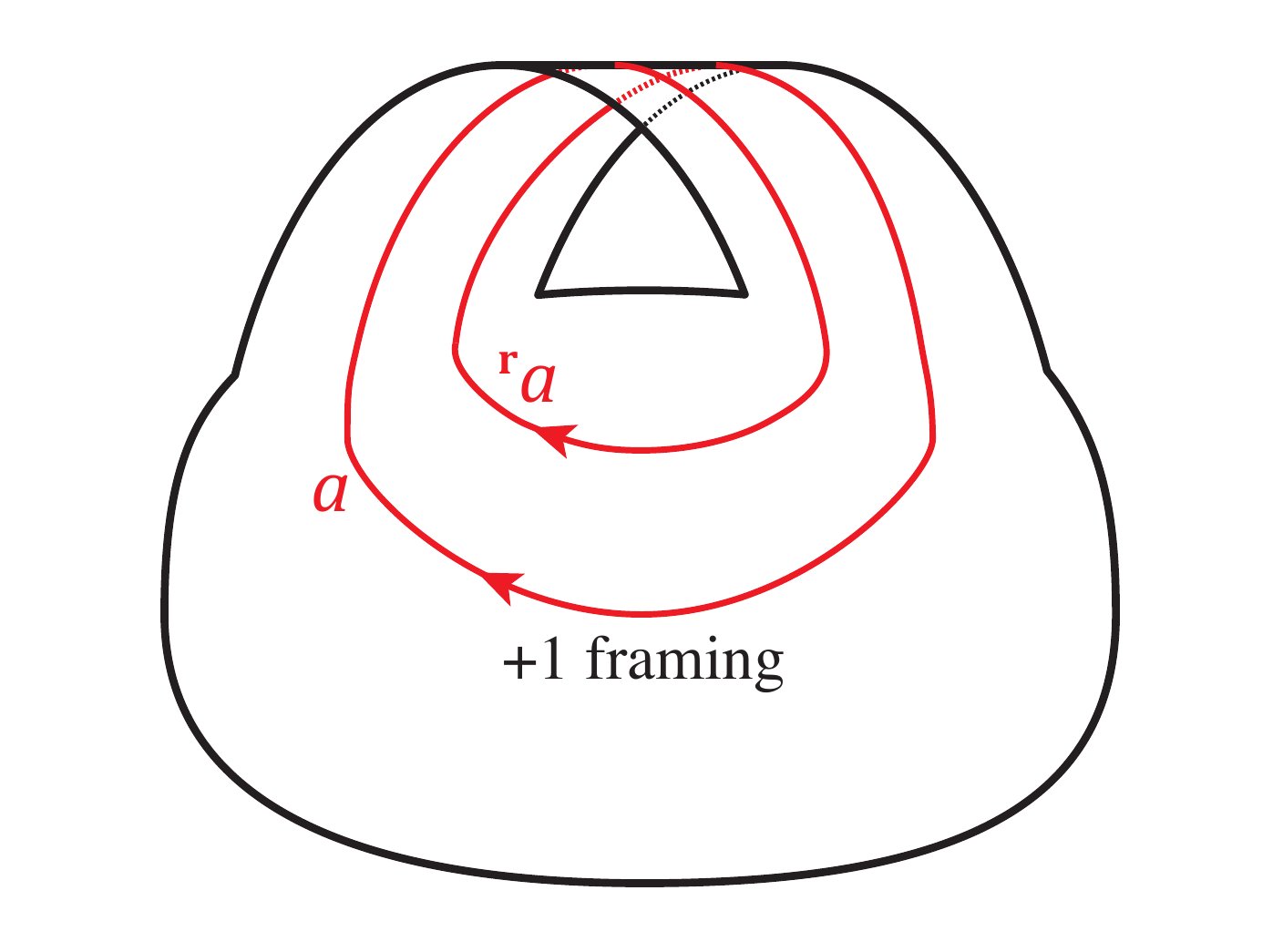}
\caption{The $1$-skeleton of $\mathbb{RP}^4$, denoted $\mathbb{RP}^4_1$, which consists of the $0$ handle glued to the $1$-handle for constructing
$\mathbb{RP}^4$. $\mathbb{RP}^4_1 = S^1 \tilde{\times} D^3$, where the $\tilde{\times}$ indicates a twisted product, because the $D^3$ forms a twisted
bundle over $S^1$. The figure illustrates $S^1 \tilde{\times} D^1$, corresponding to $\mathbb{RP}^4_1$ with only half the dimensions shown,
which in turn corresponds to a M\"obius band. The double loop labelled $a$ with $+1$ framing shows the loop along which the 
$2$-handle is to be glued. 
 }
\label{dlMobius}
\end{figure}

Now,
\begin{align}
\Z(\mathbb{RP}^4_{2})[\emptyset] &= \sum_a \frac{\Z(\mathbb{RP}^4_{1})[l_a^{(+1)}] \Z(D^4)[l_a]}{\langle l_a | l_a \rangle_{\V(S^1 \times D^2;\emptyset)}}
\nonumber \\
&= \sum_a \Z(\mathbb{RP}^4_{1})[l_a^{(+1)}] \lambda d_a 
\nonumber \\
&= \sum_a \Z(\mathbb{RP}^4_{1})[l_a] \theta_a \lambda d_a 
\end{align}
Here $l_a^{(+1)}$ corresponds to the $a$ loop shown in Fig. \ref{dlMobius}. This loop has a $+1$ framing. Moreover, 
half of the loop has its orientation reversed according to the action of reflection. Therefore, as the loop encircles 
the M\"obius band once, the anyon label changes from $a$ to $\ra{a}$, and it comes back to $a$ upon the second traversal. 
The $+1$ framing, recall, comes from the way the $2$-handle must be glued to the $1$-skeleton. This gives an extra 
factor of the topological spin $\theta_a$ relative to the case where the loop has no framing, denoted without any superscript 
for $l_a$. 

%\begin{figure}
%\includegraphics[width=3.5in]{1skel.pdf}
%\caption{
 %}
%\label{1skel}
%\end{figure}

Next, we consider $\Z(\mathbb{RP}^4_1)[l_a]$. The $1$-skeleton $\mathbb{RP}^4_1$ is equal to $\mathbb{RP}^2_1 \times D^2$. That is, 
the $1$-skeleton of $\mathbb{RP}^2_1$, which is a M\"obius band, times $D^2$ is exactly the $1$-skeleton $\mathbb{RP}^4_1$.
The computation of $\Z(\mathbb{RP}^2_1 \times D^2)[l_a]$ was already performed in Sec. \ref{MaSec}. Thus, using Eq. (\ref{ZRP21D2}), we obtain
\begin{align}
\Z(\mathbb{RP}^4_1)[l_a] = \Z(\mathbb{RP}^2_1 \times D^2)[l_a] = \eta^{\bf r}_a \delta_{a \ra{a}}.
\end{align}

Combining the above results, we obtain
\begin{align}
\Z(\mathbb{RP}^4) = \frac{1}{\mathcal{D}}\sum_{a | a = \overline{\ra{a}}} \eta_a^{\bf r} d_a \theta_a
\end{align}

\subsection{Relation between Dehn twist anomaly and $\Z(\mathbb{RP}^4)$, $\Z(\mathbb{CP}^2)$}

In Sec. \ref{dehnAnomaly}, we discussed an important condition, based on consideration of the Dehn twist on a M\"obius band,
that must be satisfied by a non-anomalous SET. In Sec. \ref{sec:4dspt}, we derived two formulae, $\Z(\mathbb{RP}^4)$, $\Z(\mathbb{CP}^2)$,
which determine which (3+1)D SPT hosts a given (2+1)D SET with certain actions of reflection / time-reversal symmetry. 
Here, we consider the Dehn twist anomaly in more detail, and provide a relation to $\Z(\mathbb{RP}^4)$, $\Z(\mathbb{CP}^2)$. 
First, we consider a particular state $|\text{Mb}\rangle$ of the (2+1)D theory defined on a torus $T^2$. 
$|\text{Mb}\rangle$ is defined as 
\begin{align}
	\langle a |\text{Mb}\rangle = \Z_a(\mathbb{RP}^2 \times S^1) = M_a ,
\end{align}
where $|a \rangle$ is the state on $T^2$ with a well-defined topological charge as measured along the meridianal cycle
$\beta$ of the torus. Thus:
\begin{align}
|\text{Mb}\rangle = \sum_{x| x = \overline{\,^{\bf r}x}} \sum_{a} S_{ax} \eta^{\bf r}_x | a \rangle .
\end{align}
$|\text{Mb}\rangle$ can be thought of as the state obtained by the path integral evaluated on $\mathbb{RP}^2_1 \times S^1$. 

Now, let us consider applying a Dehn twist $\mathcal{T}_\beta$ along the meridian $\beta$ of the torus, and computing:
\begin{align}
\frac{\langle \text{Mb}| \mathcal{T}_\beta |\text{Mb}\rangle }{ \langle \text{Mb} |\text{Mb}\rangle } 
= \frac{1}{ |\mathcal{B}^{\bf rc}|} \sum_{a \in \mathcal{B}} \theta_a M_a^2 .
\end{align}
Here, $\langle \text{Mb} | \text{Mb} \rangle =
\mathcal{Z}(\text{Kb}\times S^1) = |\mathcal{B}^{\bf rc}|$ is the number of anyons which satisfy 
$a =\overline{\ra{a}}$ and $\text{Kb}$ refers to the Klein bottle. Remarkably, the following identity holds:
\begin{align}
\label{dehnIdentity}
 \frac{1}{ |\mathcal{B}^{\bf rc}|} \sum_{a \in \mathcal{B}} \theta_a M_a^2 = \Z(\mathbb{RP}^4) \Z(\mathbb{CP}^2) . 
\end{align}

Here we will sketch the proof of Eq. (\ref{dehnIdentity}). First, note that $\Z(X^4)$ for (3+1)D time-reversal or 
reflection symmetric SPT states can be 
considered to correspond to\cite{kapustin2014}
\begin{align}
\label{ZX4}
\Z(X^4) = e^{i \pi  \int_{X^4} (n_2 w_2^2 + n_1 w_4)} ,
\end{align} 
where $w_2$ and $w_4$ are the second and fourth Stieffel-Whitney classes, respectively, and
the integral is over the $4$-manifold $X^4$. $n_1$ and $n_2$ are integers whose value,
modulo $2$, sets the SPT class. The top Stieffel-Whitney class gives the Euler characteristic of the manifold modulo two: 
$e^{i \pi \int_{X^4} w_4} = e^{i \pi \chi(X^4)}$. Since $\mathbb{RP}^4$ and $\mathbb{CP}^2$ both have
odd Euler characteristics, it follows that $\mathcal{Z}(\mathbb{RP}^4) \mathcal{Z}(\mathbb{CP}^2)$ is only 
sensitive to $w_2$. In fact, we have:
\begin{align}
\label{Zn1n2}
\mathcal{Z}(\mathbb{RP}^4) &= e^{i \pi n_1 \int w_4} = (-1)^{n_1}
\nonumber \\
\mathcal{Z}(\mathbb{CP}^2) &= e^{i \pi \int (n_2 w_2^2 + n_1 w_4)} = (-1)^{n_1 + n_2} ,
\end{align}
so that
\begin{align}
\mathcal{Z}(\mathbb{RP}^4) \mathcal{Z}(\mathbb{CP}^2) = (-1)^{n_2}.
\end{align}

Next, we observe that the denominator of Eq. (\ref{dehnIdentity}) is given by 
$|\mathcal{B}^{\bf rc}| = \mathcal{Z}_{3+1}(\text{Kb} \times D^2)[\emptyset ]$,
where Kb denotes the Klein bottle. We can think of the $\text{Kb} \times D^2$ as being 
obtained from gluing together two copies of $\text{Mb} \times D^2$ along $S^1 \times D^2$. 
The numerator of (\ref{dehnIdentity}) is also obtained by gluing together 
two copies of $\text{Mb} \times D^2$ along $S^1 \times D^2$. However in this
latter case, the gluing is done with a $+1$ framing twist, which gives the twist $\theta_a$. 
This $+1$ framing can be shown to precisely shift $\int w_2^2$ by one (relative to gluing 
without the framing twist), without changing the Euler characteristic (The proof
of this fact is beyond the scope of this paper. ) This then implies Eq. (\ref{dehnIdentity}).

%% file: example.tex
\usetikzlibrary{decorations.pathreplacing,decorations.markings}

\def\S{\textrm{S}}
\def\Tr{\text{Tr}}

\makeatletter
\newcommand\xleftrightarrow[2][]{%
  \ext@arrow 9999{\longleftrightarrowfill@}{#1}{#2}}
\newcommand\longleftrightarrowfill@{%
  \arrowfill@\leftarrow\relbar\rightarrow}
\makeatother

%\twocolumngrid
\section{Examples}
\label{exampleSec}

\subsection{Kitaev's quantum double models $\D(\G)$}

In this example section, we study reflection and time-reversal symmetry in Kitaev quantum double models $\D(\G)$ of a 
finite group $\G$.\cite{kitaev2003} These models realize the topological order of a lattice $\G$ gauge theory.  As we will 
see in the following, all these models can be made reflection/time-reversal invariant, and we determine completely the 
symmetry fractionalization class from the microscopic models.

First we briefly review Kitaev's quantum double model. The degrees of freedom are spins living on the edges of a lattice. 
Each edge spin has $|\G|$ different states (i.e. there is a $|\G|$-dimensional Hilbert space on an edge), and we label the basis states 
by group elements of $\G$, i.e. $\ket{g}, g\in \G$. We also need to orient all edges, but the particular choice of the orientations 
is not essential. Flipping the orientation on an edge is equivalent to inverting the group element. The Hamiltonian of the 
quantum double model reads:
\begin{equation}
	\mathbf{H}=-\sum_v A_v -\sum_p B_p.
	\label{}
\end{equation}
Here the vertex operator $A_v= \frac{1}{|\G|}\sum_{h\in \G} A_v^h$, where $A_v^h$ acts on the group elements on 
edges associated to a vertex $v$ by left multiplication. Pictorially, this is defined as:
\begin{equation}
	A_v^h ~
\Bigg|
\begin{tikzpicture}[scale=0.50, baseline={([yshift=-.5ex]current  bounding  box.center)}]
\draw [middlearrow={stealth}] (0,0) -- (0,1);
\draw [middlearrow={stealth}] (0,0) -- (0,-1);
\draw [middlearrow={stealth}] (0,0) -- (1,0);
\draw [middlearrow={stealth}] (0,0) -- (-1,0);
\draw (-1,0) -- (1,0);
\draw (0,1.4) node {$b$};
\draw (0,-1.4) node {$d$};
\draw (-1.4,0) node {$c$};
\draw (1.4,0) node {$a$};
\end{tikzpicture}
\Bigg\rangle
=
\Bigg|
\begin{tikzpicture}[scale=0.50, baseline={([yshift=-.5ex]current  bounding  box.center)}]
\draw [middlearrow={stealth}] (0,0) -- (0,1);
\draw [middlearrow={stealth}] (0,0) -- (0,-1);
\draw [middlearrow={stealth}] (0,0) -- (1,0);
\draw [middlearrow={stealth}] (0,0) -- (-1,0);
\draw (-1,0) -- (1,0);
\draw (0,1.4) node {$hb$};
\draw (0,-1.4) node {$hd$};
\draw (-1.4,0) node {$hc$};
\draw (1.4,0) node {$ha$};
\end{tikzpicture}
\Bigg\rangle
	\label{}
\end{equation}

The plaquette operator $B_p$ is a projector onto the product of group elements around a plaquette being equal to
the identity. Pictorially, this is defined as
\begin{align}
B_p ~
\Bigg|
\begin{tikzpicture}[scale=0.50, baseline={([yshift=-.5ex]current  bounding  box.center)}]
\draw[middlearrow={stealth}] (1,1) -- (-1,1);
\draw[middlearrow={stealth}] (-1,1) -- (-1,-1);
\draw[middlearrow={stealth}] (-1,-1) -- (1,-1);
\draw[middlearrow={stealth}] (1,-1) -- (1,1);
\draw (0,1.4) node {$a$};
\draw (0,-1.4) node {$c$};
\draw (-1.4,0) node {$b$};
\draw (1.4,0) node {$d$};
\end{tikzpicture}
\Bigg\rangle
=\delta_{dcba,1}\Bigg|
\begin{tikzpicture}[scale=0.50, baseline={([yshift=-.5ex]current  bounding  box.center)}]
\draw[middlearrow={stealth}] (1,1) -- (-1,1);
\draw[middlearrow={stealth}] (-1,1) -- (-1,-1);
\draw[middlearrow={stealth}] (-1,-1) -- (1,-1);
\draw[middlearrow={stealth}] (1,-1) -- (1,1);
\draw (0,1.4) node {$a$};
\draw (0,-1.4) node {$c$};
\draw (-1.4,0) node {$b$};
\draw (1.4,0) node {$d$};
\end{tikzpicture}
\Bigg\rangle
\end{align}

As shown in Ref. \onlinecite{kitaev2003}, the quasiparticle types in $\D(\G)$ are labeled by the pair $(C, \pi)$ where $C$ is a 
conjugacy class in $\G$ and $\pi$ is an irreducible representation of the centralizer $Z_{r_C}$ of a representative element 
$r_C \in C$. The trivial quasiparticle is given by the conjugacy class of the identity element $1 \in \G$ and the 
trivial representation (which is labeled as ``1" in the following) of the centralizer $Z_1 = \G$. Therefore, the trivial quasiparticle is labeled by $(1 , 1)$.

\subsubsection{Natural definitions of reflection and time-reversal symmetries}
\label{Sec:Nat_Def_Sym_DG}

We first show that the action of the reflection symmetry ${\bf r}$ should be defined as
\begin{align}
	R_{\bf r} ~
\Bigg|
\begin{tikzpicture}[scale=0.50, baseline={([yshift=-.5ex]current  bounding  box.center)}]
\draw[middlearrow={stealth}] (1,1) -- (-1,1);
\draw[middlearrow={stealth}] (-1,1) -- (-1,-1);
\draw[middlearrow={stealth}] (-1,-1) -- (1,-1);
\draw[middlearrow={stealth}] (1,-1) -- (1,1);
\draw[dotted,thick] (0,-2) -- (0,2);
\draw (0.25,1.4) node {$a$};
\draw (0.25,-1.4) node {$c$};
\draw (-1.4,0) node {$b$};
\draw (1.4,0) node {$d$};
\end{tikzpicture}
\Bigg\rangle
%\equiv
%\Bigg|
%\begin{tikzpicture}[scale=0.50, baseline={([yshift=-.5ex]current  bounding  %box.center)}]
%\draw[middlearrow={stealth reversed}] (1,1) -- (-1,1);
%\draw[middlearrow={stealth reversed}] (-1,1) -- (-1,-1);
%\draw[middlearrow={stealth reversed}] (-1,-1) -- (1,-1);
%\draw[middlearrow={stealth reversed}] (1,-1) -- (1,1);
%\draw[dotted,thick] (0,-2) -- (0,2);
%\draw (0,1.4) node {$a$};
%\draw (0,-1.4) node {$c$};
%\draw (-1.4,0) node {$d$};
%\draw (1.4,0) node {$b$};
%\end{tikzpicture}
%\Bigg\rangle
=
\Bigg|
\begin{tikzpicture}[scale=0.50, baseline={([yshift=-.5ex]current  bounding  box.center)}]
\draw[middlearrow={stealth}] (1,1) -- (-1,1);
\draw[middlearrow={stealth}] (-1,1) -- (-1,-1);
\draw[middlearrow={stealth}] (-1,-1) -- (1,-1);
\draw[middlearrow={stealth}] (1,-1) -- (1,1);
\draw[dotted,thick] (0,-2) -- (0,2);
\draw (0.25,1.4) node {$\bar{a}$};
\draw (0.25,-1.4) node {$\bar{c}$};
\draw (-1.4,0) node {$\bar{d}$};
\draw (1.4,0) node {$\bar{b}$};
\end{tikzpicture}
\Bigg\rangle,
\end{align}
where $a,b,c,d\in \G$, $\bar{g} \equiv g^{-1}$ denotes the inverse of $g \in \G$. The dotted line is the mirror axis of 
the reflection. Under the reflection, the edges together with their labels are permuted. If a pair of edges related by 
reflection have opposite orientations, then after the permutation between them we need to further take inverse group elements 
as their labels. For instance, in the equation above, the group elements $b$ and $d$ are inverted under the reflection, 
since opposite arrows are assigned to the corresponding two edges, which get exchanged by the reflection ${\bf r}$. 
If they had the same arrows, there would be no need to invert the group elements. Such a definition of the reflection 
${\bf r}$ is a canonical choice which preserves both the vertex and the plaquette terms in $\D(\G)$ for a 
generic non-Abelian group $\G$. To see this, we can consider how the plaquette term $B_p$ transforms under the reflection:
\begin{align}
R_{\bf r}^{-1} B_p R_{\bf r} ~
\Bigg|
\begin{tikzpicture}[scale=0.50, baseline={([yshift=-.5ex]current  bounding  box.center)}]
\draw[middlearrow={stealth}] (1,1) -- (-1,1);
\draw[middlearrow={stealth}] (-1,1) -- (-1,-1);
\draw[middlearrow={stealth}] (-1,-1) -- (1,-1);
\draw[middlearrow={stealth}] (1,-1) -- (1,1);
\draw[dotted,thick] (0,-2) -- (0,2);
\draw (0.25,1.4) node {$a$};
\draw (0.25,-1.4) node {$c$};
\draw (-1.4,0) node {$b$};
\draw (1.4,0) node {$d$};
\end{tikzpicture}
\Bigg\rangle
=
\delta_{ \bar{b}\bar{c}\bar{d}\bar{a}, 1}\Bigg|
\begin{tikzpicture}[scale=0.50, baseline={([yshift=-.5ex]current  bounding  box.center)}]
\draw[middlearrow={stealth}] (1,1) -- (-1,1);
\draw[middlearrow={stealth}] (-1,1) -- (-1,-1);
\draw[middlearrow={stealth}] (-1,-1) -- (1,-1);
\draw[middlearrow={stealth}] (1,-1) -- (1,1);
\draw[dotted,thick] (0,-2) -- (0,2);
\draw (0.25,1.4) node {$a$};
\draw (0.25,-1.4) node {$c$};
\draw (-1.4,0) node {$b$};
\draw (1.4,0) node {$d$};
\end{tikzpicture}
\Bigg\rangle,
\end{align}
where $1$ is the identity element in $\G$.
It is not difficult to see that $R_{\bf r}^{-1} B_p R_{\bf r} = B_p$.
We can also easily verify the invariance of the vertex term under $R_{\bf r}$. On the other hand, 
if we try to define a site-centered reflection, the vertex terms are generally not invariant, unless $\G$ is an Abelian group.

The natural assignment of the time-reversal symmetry is $\mb{T}=\mathcal{K}$, where $\mathcal{K}$ denotes complex conjugation.
That is, this assignment of time-reversal symmetry has no action on the group labels, which is possible because every term in the 
Hamiltonian is real.

For certain $\G$ there may be $\mathbb{Z}_2$ electric-magnetic duality symmetries, often implemented as lattice translation symmetries. 
When this is the case, we may compose the reflection symmetry ${\bf r}$ with any of the $\mathbb{Z}_2$ electric-magnetic symmetries 
to obtain a distinct action of reflection symmetry. Here, to keep the discussion general, we will focus on the canonical choices of ${\bf r}$ and $\mb{T}$ in this section.

\subsubsection{Symmetry action on quasiparticles}
\label{Sec:Sym_Action_QP}

We can also derive the symmetry action on quasiparticles, using the explicit forms of string operators given in
Refs. \onlinecite{kitaev2003, BombinPRB2008, beigi2011}. Since the derivation is quite technical we refer the 
interested readers to Appendix \ref{sec:string} for details. Here we summarize the results.

The actions of reflection and time-reversal symmetry on the anyon types in $\D(\G)$ are given by
\begin{align}
\label{rTaction}
	& {\bf r}:~ (C,\pi) \rightarrow (C^{-1},\pi), \\
& \mb{T}:~ (C,\pi) \rightarrow (C,\pi^*).
\end{align}
A remark is in order for the precise meaning of $(C^{-1}, \pi)$. Recall that $\pi$ is an irreducible representation of the centralizer $Z_{r_C}$ where $r_C$ is a representative element of $C$. For $C^{-1}$, we need to pick  $r_C^{-1}$ as the representative element for $C^{-1}$. The centralizer group is canonically the same $Z_{r_C} = Z_{r_C^{-1}}$. Hence, the anyon label $(C^{-1},\pi)$ indeed makes sense.

An anyon $\{(C,\pi)\}$ can have well-defined $\eta^{\bf r}_{(C,\pi)}$ or $\eta^{\bf T}_{(C, \pi)}$ if 
it satisfies respectively the conditions:
\begin{align}
	\ra{(C,\pi)}  = \overline{(C,\pi)},   \label{Eq:QP_Reflection_Inv_Condition} \\
{}^\mb{T}  (C,\pi)  = (C,\pi). \label{Eq:QP_TR_Inv_Condition}
\end{align}
Here $\overline{(C,\pi)} = (C^{-1}, \pi^*)$ is the anti-particle (topological charge conjugate) of $(C, \pi)$. Thus, using Eq. (\ref{rTaction}), we see that
both conditions in Eqs. (\ref{Eq:QP_Reflection_Inv_Condition})-(\ref{Eq:QP_TR_Inv_Condition}) reduce to $\pi\simeq\pi^*$. Therefore we actually have $\mathcal{B}^{\mb{rc}}=\mathcal{B}^\mb{T}$, 
and we will use $\mathcal{B}^{\bf rc}$ and $\mathcal{B}^\mb{T}$ interchangeably in the following discussion of $\D(\G)$. (Note that here
$\mathcal{B} = \D(\G)$, and recall $\mathcal{B}^{\bf rc}$ is the set of anyons which satisfy $\,^{\mb r}\overline{a} = a$, while $\mathcal{B}^{\mb T}$
is the set of anyons which satisfy $\,^{\bf T}a = a$.) 

For an anyon $(C,\pi)$ which satisfies $\pi = \pi^*$, we find 
\begin{align}
	{\eta}_{(C,\pi)}^{\bf r} = \eta_{(C, \pi)}^\mb{T}=\nu_\pi.
\end{align}
Here $\nu_\pi$ is the Frobenius-Schur indicator of the representation $\pi$. An explicit expression of $\nu_\pi$ is given by
\begin{equation}
	\nu_\pi=\frac{1}{|Z_{r_C}|} \sum_{h\in Z_{r_C}} \Tr\left( \pi(h^2)\right).
	\label{Eq:FS_indicator}
\end{equation}
$\nu_\pi$ is $1$ for a real representation, and $-1$ for pseudo-real. For complex representations, $\nu_\pi$ is automatically $0$. 

Our results also illustrate the ``CPT'' theorem in the TQFT, because we see that the action of ${\bf rc}$ on the anyons is identical to that of $\mb{T}$.
This has an implication on the Euclidean path integral of $\D(\G)$ given by the state sum method described in Sec. \ref{statesumSec}. The state sum 
allows us to consider non-orientable space-time manifolds by inserting branch sheets along which a direction of space-time is reversed. 
We can interpret the reversed direction as space or time, and either one gives us the same result provided we replace ${\bf rc}$ with $\mb{T}$ in our 
expressions. 

\subsubsection{State sum model of $\D(\G)$}

The Euclidean path integral of $\D(\G)$ is given by the state sum model. Before we consider time-reversal or reflection symmetry, 
the input data to the state sum is basically the group $\G$, but viewed as a fusion category. We will denote this fusion category 
by $\text{Vec}_\G$. The simple objects in $\text{Vec}_\G$ are given by the group elements. The fusion coefficients are directly 
given by the group multiplication, namely
\begin{align}
N^c_{ab} = \delta_{ab,c},
\end{align}
where $ab$ on the right hand side of the equation is understood as the product of the two group elements $a$ and $b$. Notice that for a 
non-Abelian group $\G$, the fusion is generally non-commutative, i.e. $N^c_{ab} \neq N^c_{ba}$. The quantum dimensions $d_a = 1$ for 
all $a \in \G$. The $F$-symbols are all trivial, namely $1$ when their labels are compatible with the fusion and $0$ otherwise:
\begin{align}
F^{abc}_{def} = \delta_{ab,e} \delta_{ec,d} \delta_{bc,f} \delta_{af,d}.
\label{Eq:F-Symbol_DG}
\end{align}
When we include time-reversal or reflection symmetry of $\D(\G)$ , as is shown in
Sec. \ref{State_Sum_SETs}, we need to consider a $\mathbb{Z}_2^{\bf r}$-extension $C_{\mathbb{Z}_2^{\bf r}}$ of the fusion category $\text{Vec}_\G$.  To be compatible with the symmetry action given in Sec. \ref{Sec:Nat_Def_Sym_DG}, we choose the trivial $\mathbb{Z}_2^{\bf r}$ extension which is given directly by two copies of $\text{Vec}_\G$:
\begin{align}
	\mathcal{C}_{\mathbb{Z}_2^{\bf r}} = \text{Vec}_\G \oplus \text{Vec}_\G,
\end{align}
where the two $\text{Vec}_\G$'s carry different $\mathbb{Z}_2^{\bf r}$ gradings. The fusion coefficients in $\mathcal{C}_{\mathbb{Z}_2^{\bf r}}$ are still given by the group multiplication of $\G$ (on top of the grading). The $F$-symbols in $\mathcal{C}_{\mathbb{Z}_2^{\bf r}}$ are independent of their $\mathbb{Z}_2^{\bf r}$ gradings:
\begin{align}
F^{abc}_{def}(\mb{g}_0, \mb{g}_1, \mb{g}_2, \mb{g}_3)= F^{abc}_{def},
\end{align}
where $\mb{g}_0, \mb{g}_1, \mb{g}_2, \mb{g}_3 \in \mathbb{Z}_2^{\bf r}$ and $ F^{abc}_{def}$ is the 
$F$-symbol given in Eq. (\ref{Eq:F-Symbol_DG}). More complicated choices of the $F$-symbols 
$F^{abc}_{def}(\mb{g}_0, \mb{g}_1, \mb{g}_2, \mb{g}_3)$ could be considered, although these will yield
reflection SETs that are distinct from that realized in the Kitaev quantum double Hamiltonian that we are studying here. 

As an example, we can calculate the Euclidean path integral of $\D(\G)$ on $\mathbb{RP}^2 \times S^1$. With the cellulation of $\mathbb{RP}^2 \times S^1$ in Fig. \ref{Fig:RP2xS1_Complex_Graded}, the path integral of $\D(\G)$ is given by Eq. (\ref{Eq:RP2_State_Sum_Graded}) which can be easily simplified to
\begin{align}
\Z(\mathbb{RP}^2 \times S^1) = \frac{1}{|\G|}\sum_{a\in \G}\sum_{b\in \G} \delta_{a^2, 1} \delta_{ab,ba}.
\label{Eq:DG_Partition_RP2_a}
\end{align}
In this expression, when we view $\D(\G)$ as a discrete gauge field with gauge group $\G$, the group elements $a$ and $b$ that are being summed over represent the gauge fluxes through the two non-trivial cycles of $\mathbb{RP}^2 \times S^1$, which are the non-trivial cycle in $\mathbb{RP}^2$ and the $S^1$ cycle respectively.

Physically the result in Eq. \eqref{Eq:DG_Partition_RP2_a} is quite suggestive. When we view $\D(\G)$ as a discrete gauge theory, 
each state on $\mathbb{RP}^2$ can be labeled by a conjugacy class $C$ of $\G$, which is the gauge flux through the non-trivial 
cycle on $\mathbb{RP}^2$. Since twice this non-trivial cycle is homologically trivial (i.e. contractible to a point
due to the fact that the first homology $H_1(\mathbb{RP}^2,\mathbb{Z})=\mathbb{Z}_2$), the gauge flux threading the non-trivial cycle should square to the 
identity element $1$ in $\G$. Note that, for a given conjugacy class $C$, if one element in $C$ squares to $1$, all elements in $C$ square to $1$, 
in which case we will just say the conjugacy class $C$ squares to $1$. Therefore, we expect that $\Z(\mathbb{RP}^2 \times S^1)$, the topological ground state 
degeneracy of $\D(\G)$ on $\mathbb{RP}^2$, is equal to the number of conjugacy classes $C$ that square to $1$. In Appendix \ref{sec:dgRP2}, 
we show that Eq. \eqref{Eq:DG_Partition_RP2_a} indeed gives this result.

On the other hand, since we know $\nu_\pi={\eta}_{(C,\pi)}^{\bf r}= \eta_{(C, \pi)}^\mb{T}$, Eq. \eqref{Eq:Ma_RP2} gives:
\begin{align}
\Z(\mathbb{RP}^2 \times S^1) = \sum_{(C,\pi) \in \mathcal{B}^\mb{rc}} S_{(1,1) (C,\pi)} \nu_\pi.
\label{Eq:DG_Partition_RP2_b}
\end{align}
In Appendix \ref{sec:dgRP2}, we also prove that the two seemingly different expressions of $\Z(\mathbb{RP}^2 \times S^1) $, 
Eqs. (\ref{Eq:DG_Partition_RP2_a}) and (\ref{Eq:DG_Partition_RP2_b}), are in fact equal. Since 
$\nu_\pi={\eta}_{(C,\pi)}^{\bf r}= \eta_{(C, \pi)}^\mb{T}$ is derived from a Hamiltonian approach, the consistency between 
the two expression of $\Z(\mathbb{RP}^2 \times S^1) $ shows that the $\mathbb{Z}_2^{\bf r}$-graded state sum construction 
of $\D(\G)$ is compatible with the Hamiltonian formalism with its natural definition of reflection/time-reversal symmetry.  

\subsubsection{Example: $\D(Q_8)$}

The simplest finite group with a pseudo-real representation is the quaternion group $Q_8$. 
Physically, it is the subgroup of $\mathrm{SU}(2)$ generated by $\pi$ spin rotations. Formally it has the presentation
\begin{equation}
Q_8 = \langle -1, i, j, k | (-1)^2 =1, i^2, j^2, k^2 = ijk =-1\rangle.
	\label{}
\end{equation}
$Q_8$ has a two-dimensional representation descending from the spin-$1/2$ representation of $\mathrm{SU}(2)$:
\begin{equation}
	i\mapsto i\sigma_x, j\mapsto i\sigma_y, k\mapsto i\sigma_z,
	\label{}
\end{equation}
where here the $\sigma_a$ for $a = x,y,z$ are the Pauli matrices. It is straightforward to check that this 
two-dimensional representation is pseudo-real. According to our analysis, in the Kitaev quantum double model 
of $\D(Q_8)$, the gauge charge corresponding to this two-dimensional irrep has $\eta^\R=-1$ and $\eta^\mb{T}=-1$.

There is another way to realize the topological order of a discrete $\G$ gauge theory in lattice models: namely a 
Levin-Wen string-net model\cite{levin2005} with the input unitary fusion category being $\text{Rep}(\G)$. For 
$\G=Q_8$, the Levin-Wen model is apprently time-reversal invariant: because all $F$ symbols are real 
(see Appendix \ref{app:data:Q8} for details), we can just define $\mb{T}=\mathcal{K}$. Correspondingly, 
the state-sum construction can be done using the data of $\text{Rep}(\G)$.

However, using the state-sum model to calculate the ground-state degeneracy on $\mathbb{RP}^2$, we find that for the 
$\text{Vec}_{Q_8}$ category, $\mathcal{Z}(\mathbb{RP}^2\times S^1)=2$ and for $\text{Rep}(Q_8)$ we get 
$\mathcal{Z}(\mathbb{RP}^2\times S^1)=3$. Therefore we conclude that the Levin-Wen model and the Kitaev quantum double model 
actually realize distinct reflection and time-reversal SET phases. In fact, the result is consistent with all invariant 
anyons having $\eta^\R=1$ in the Levin-Wen realization of $\D(Q_8)$.

\subsection{$\mathbb{Z}_N$ toric code}
\label{ZNexample}

In this section we examine the $\mathbb{Z}_N$ toric code topological order with reflection and/or time-reversal symmetry. 
First we shall consider the toric code model in its original form, and then describe how to place the lattice model on 
$\mathbb{RP}^2$. We then systematically study various symmetry fractionalization classes for the $\mathbb{Z}_N$ toric code. 
We also systematically study the associated $\mathbb{Z}_2^{\bf r}$ extensions of $\mathbb{Z}_N$ which correspond to the 
$\mathbb{Z}_2^{\bf r}$-equivariant 2-categories that are used as input into our SET state sum model to yield topological path integrals. 

\subsubsection{$\mathbb{Z}_N$ toric code on a crosscap}

The Hamiltonian for the $\mathbb{Z}_N$ toric code state on a square lattice is given by
\begin{equation}
	\mathbf{H}=-\sum_p B_p - \sum_v A_v,
	\label{}
\end{equation}
where the plaquette terms $B_p$ and vertex terms $A_v$ are defined as
\begin{align}
B_p=\begin{tikzpicture}[scale=0.50, baseline={([yshift=-.5ex]current  bounding  box.center)}]
\draw (1,1) -- (-1,1);
\draw (-1,1) -- (-1,-1);
\draw (-1,-1) -- (1,-1);
\draw (1,-1) -- (1,1);
\draw (0,1.4) node {$Z$};
\draw (0,-1.4) node {$Z$};
\draw (-1.45,0) node {$Z^\dag$};
\draw (1.45,0) node {$Z^\dag$};
\end{tikzpicture} + \text{h.c.}\\
A_v = \begin{tikzpicture}[scale=0.50, baseline={([yshift=-.5ex]current  bounding  box.center)}]
\draw (0,-1) -- (0,1);
\draw (-1,0) -- (1,0);
\draw (0,1.4) node {$X$};
\draw (0,-1.4) node {$X$};
\draw (-1.4,0) node {$X$};
\draw (1.4,0) node {$X$};
\end{tikzpicture} + \text{h.c.}
\end{align}
Here $Z$ and $X$ are $\mathbb{Z}_N$ clock and shift operators acting on $\mathbb{Z}_N$ spins on the edges, 
satisfying $ZX=e^{\frac{2\pi i}{N}}XZ$. One can identify two types of excitations of the Hamiltonian: the magnetic 
(electric) excitations corresponding to violations of plaquette (vertex) terms. The elementary magnetic (electric) 
particle is denoted by $m$ ( $e$). Note that in this definition, a vertex violation where $A_v$ has eigenvalue $e^{2\pi i /N}$
is an $e$ particle on even sites of the square lattice while it is an $\bar{e}$ particle on odd sites. 

The Hamiltonian has both site-centered and plaquette-centered reflection symmetries. We will focus on the latter. 
We can define $\mb{r}$ by just reflecting the coordinates, without any action on the spins. Let us first determine 
how the electric and magnetic excitations transform under the reflection. To create a pair of quasiparticles, $e$ and $\bar{e}$, we can apply a single $Z$:
\begin{equation}
	\begin{tikzpicture}[scale=0.50, baseline={([yshift=-.5ex]current  bounding  box.center)}]
\draw (-0.7,-1) -- (-0.7,1);
\draw (0.7,-1) -- (0.7,1);
\draw (-1.5,0) -- (1.5,0);
\draw[dotted,thick] (0,-1.5) -- (0,1.5);
\draw (0.25,0.28) node {$Z$};
\end{tikzpicture}.
	\label{}
\end{equation}
This generates a two-anyon state consisting of $e$ and $\bar{e}$ separated horizontally. The mirror plane of the reflection is taken to be along the dashed 
line. Since the operator $Z$ in this case acts on a spin that lies on the mirror plane, it commutes with the reflection operator $R_{\bf r}$,
implying that the resulting two-anyon state is reflection symmetric. However this requires that under reflection, we have: 
\begin{align}
\ra{e} = \bar{e}. 
\end{align}
In order to satisfy the fact that for all anyons we must have $\theta_{\ra{a}} = \theta_a^*$, we see that we must therefore have
\begin{align}
\ra{m} = m . 
\end{align}
Form the above considerations we can further deduce that $\eta_e^\R=1$ because the state created with $Z$ manifestly has a reflection eigenvalue of $1$. 

\begin{figure}[t]
	\centering
	\includegraphics[width=1\columnwidth]{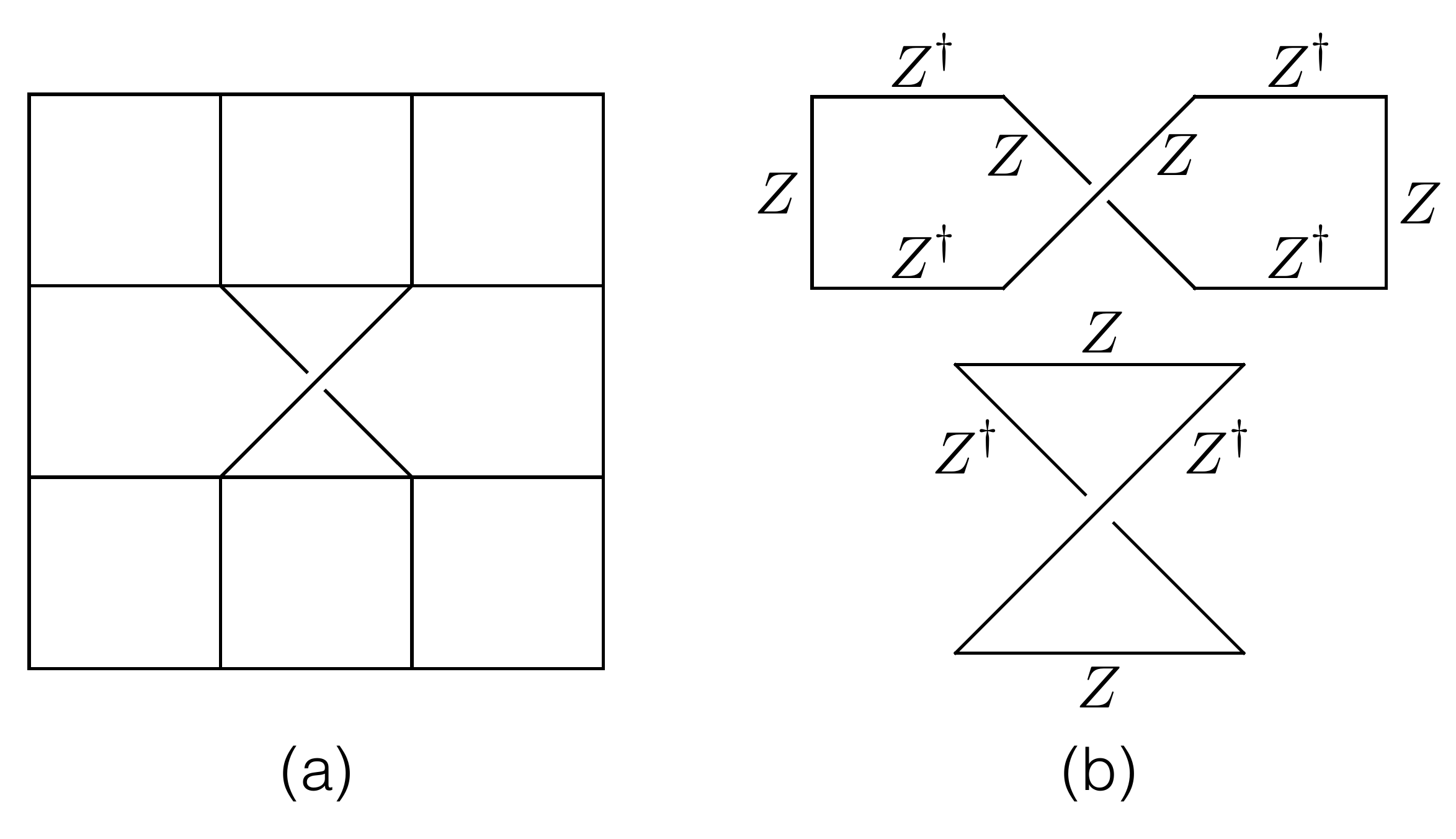}
	\caption{$\mathbb{Z}_N$ toric code on a crosscap (punctured $\mathbb{RP}^2$). 
(a) The lattice used to represent a crosscap. (b) Two ``twisted'' plaquette terms.}
	\label{fig:TC}
\end{figure}

To create a lattice discretization of $\mathbb{RP}^2$ with a single puncture, we insert a ``crosscap'' into a square lattice with boundary. 
Pictorially, imagine we first make a hole in the lattice and then identify ``antipodal'' points to create a crosscap. The 
identification is done by reconnecting sites related to each other by the reflection, see Fig. \ref{fig:TC}. Notice that the 
middle row now has two plaquettes, one of which has $8$ edges.  To see that this lattice is indeed topologically 
equivalent to $\mathbb{RP}^2$ with a single puncture, we can calculate the Euler characteristic $\chi$:
The number of vertices, edges, and plaquettes are $N_0=16, N_1=8, N_2=24$, respectively, so $\chi=N_0-N_1+N_2=0$, as expected. 

We can similarly write a Hamiltonian for the toric code on this new lattice. As we mentioned, the reflection does not act on the 
spins. The vertex terms stay the same (i.e. $XXXX$ for each vertex); for the corner vertex terms we multiply $X$ from the two adjacent
edges, and for the other boundary vertex terms we multiply all $X$ on the adjacent three edges. The plaquette terms are 
also straightforward to write down, and are illustrated in Fig. \ref{fig:TC}

Since the vertex and plaquette terms all commute with each other, we can determine the ground state 
degeneracy (GSD) of this system by counting stabilizers; this provides a computation of 
$M_a$, where $a$ is the topological charge measured along the boundary of the system. Since each spin 
is defined on an edge, there are $N_1$ spins, while there are $N_0$ stabilizers from the vertices and $N_2$ 
stabilizers from the plaquettes. However, not all stabilizers are independent. It is not hard to see that there is 
no constraint on the plaquette stabilizers (multiplying all of them gives a closed $e$ string operator 
along the boundary of the lattice). For vertex stabilizers, the situation is a little more complicated. 

On the usual square lattice (without the ``twisted'' plaquettes), with e.g. $N_0=16, N_1=24, N_2=9$, we can take a product 
of vertex stabilizers over all vertices, and this gives the identity operator:
$\prod_{v \text{ even}} A_v \prod_{v \text{ odd}} A_v^\dagger = {\bf 1}$, where $\prod_{v \text{ even}} $ is a product over even sites
and $\prod_{v \text{ odd}}$ is a product over odd sites. Thus in the usual square lattice case there would be 
only $N_0 - 1$ independent vertex stabilizers. Therefore, the GSD would be given by 
$N^{N_1- (N_0-1+N_2)}=1$ which is the expected answer on a disk. 

On the other hand, in the present case, with the twisted plaquettes, one can see that for odd $N$ there is 
no constraint on the product of all vertex stabilizers anymore (the previous $\prod_{v \text{ even}} A_v \prod_{v \text{ odd}} A_v^\dagger$,
for example, is no longer well-defined because the lattice is not bipartite). Therefore, 
the GSD$=N^{N_1-N_0-N_2}=1$. 

For even $N$ we still have a constraint $\prod_v A_v^{N/2}= {\bf 1}$. If we first apply the 
stabilizer constraints associated with $N_0 -1$ vertices, we get $N$ total states left over. From the relation
$\prod_v A_v^{N/2}= {\bf 1}$, only $N/2$ of these remaining states are associated with the eigenvalues 
of the remaining vertex term. After imposing that this remaining vertex term have $+1$ eigenvalue in order to
be in the ground state subspace, we find a residual $2$-fold degeneracy. As discussed below, this residual $2$-fold
degeneracy can be associated with the eigenvalues of Wilson loop operators that cross the crosscap. 

We can further see in this computation that the topological charge on the boundary is trivial, so the above computations 
imply $M_0 = 2$ or $1$ for $N$ even or odd. To see this, observe that the product over all plaquette
operators is equal to an $e$ string around the boundary. Combined with the fact that this product has eigenvalue one in the ground state, we deduce that
the boundary cannot have any magnetic charge associated to it. On the other hand, given the boundary conditions of this model (``smooth boundaries''), 
any $e$ topological charge will cost finite energy as it will be a violation of the vertex terms. So we conclude that the boundary 
topological charge in the computation of the ground state degeneracy is indeed $0$.

\begin{figure}[t]
	\centering
	\includegraphics[width=\columnwidth]{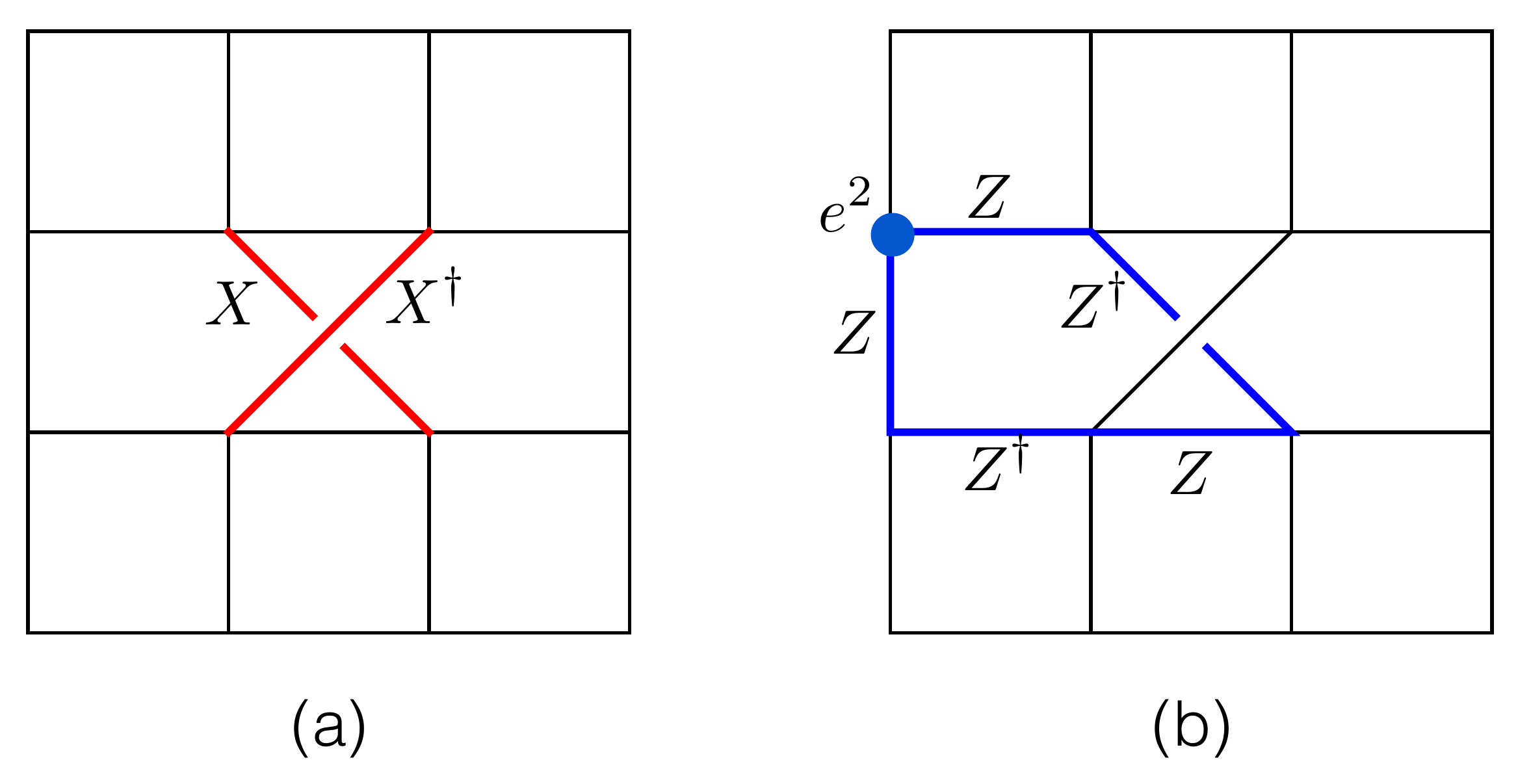}
	\caption{Illustration of Wilson operators. (a) A closed $m$ loop passing through the crosscap. Applying $X^\dagger$ creates a pair of
$m$ and $\bar{m}$ on the two twisted plaquettes. Applying $X$ on the other edge then annihilates them in such a way that one can interpret
the resulting process as a single $m$ excitation traversing a loop that passes through the crosscap. 
(b) An $e$ string passing through the crosscap, turning into $\bar{e}$.}
	\label{fig:wilsonTC}
\end{figure}

We can alternatively derive the GSD using the algebra of Wilson loop (string) operators, which also illustrates what happens to a 
quasiparticle when it moves across the crosscap. In Fig. \ref{fig:wilsonTC}(a), we construct a Wilson operator $W_m$ of 
$m$ which passes through the crosscap. This is of course consistent with the reflection action $\ra{m}=m$, because otherwise one
would not be able to create a closed string operator of $m$ that passes through the crosscap. If we try to do the same for $e$ along 
a loop passing through the crosscap, e.g. the thick path shown in Fig. \ref{fig:wilsonTC}(b), 
we find that we cannot close the loop: recall that an $e$ (or $\bar{e}$) string is an alternating product of $Z$ and $Z^\dagger$ along 
a path on the lattice. However, the path in Fig. \ref{fig:wilsonTC}(b) has an odd length, so it is not possible to close up a $e$ Wilson 
line without creating any excitations (i.e in Fig. \ref{fig:wilsonTC}(b), an $e^2$ excitation is created). The only way out is that for even 
$N$, we can form the product of $Z^{N/2}$ along the path to create a closed Wilson operator $W_{e^{N/2}}$ for $e^{N/2}$. Since the paths 
of $W_{e^{N/2}}$ and $W_m$ overlap exactly on one edge, we have the following algebra:
\begin{equation}
	W_{e^{N/2}}W_m=-W_m W_{e^{N/2}}.
	\label{}
\end{equation}
This algebra then implies a topologically protected two-fold GSD for even $N$. For odd $N$, since there is no closed $e$ loop, 
there is no non-trivial algebra, and thus the GSD is one. 

\subsubsection{Other fractionalization classes for $\mathbb{Z}_N$ toric code}

Let us now discuss more general time-reversal and reflection symmetry fractionalization classes for the 
$\mathbb{Z}_N$ toric code. We use the notation $(a,b)\equiv e^am^b$ to label anyons in 
$\D(\mathbb{Z}_N)$. We will still assume that the reflection acts as $\ra{e}=\bar{e}, \ra{m}=m$. 
The set $\mathcal{B}^{\mb{rc}}$ is given by
\begin{equation}
	\mathcal{B}^{\mb{rc}}=
	\begin{cases}
		\{e^i\}_{i=0}^{N-1} & \text{$N$ is odd}\\
		\{e^i, e^im^{N/2}\}_{i=0}^{N-1} & \text{$N$ is even}
	\end{cases}
	\label{}
\end{equation}
Therefore the reflection symmetry fractionalization quantum numbers $\{\eta_a^{\R}\}$ are completely 
specified by $\eta_e^{\R}$ for $N$ odd, and $\eta_e^{\R}, \eta_{m^{N/2}}^{\R}$ for $N$ even.

 We can compute the GSD on a crosscap from (\ref{Eq:Ma_RP2}). Let us consider odd $N$ first. 
From the fusion rules we obtain the constraint $(\eta_e^\R)^N=1$, which implies that $\eta_e^\R=1$. So we have
 \begin{equation}
	 \begin{split}
		 M_{(a,b)}&=\sum_{n = 0}^{N-1} S_{(a,b), (n,0)} = \frac{1}{N}\sum_{n=0}^{N-1}e^{\frac{2\pi i na}{N}}=\delta_{a0}.
	 \end{split}
	 \label{}
 \end{equation}
This is of course what we found using the explicit Hamiltonian construction in the previous section.

 For even $N$, we have
\begin{align}
	M_{(a,b)} &= \sum_{n = 0}^{N-1} [S_{(a,b), (n,0)} (\eta_e^\R)^n + S_{(a,b),(n,N/2)} \eta_{m^{N/2}}(\eta_e^\R)^n ]
\nonumber \\
& =\frac{1 + (-1)^a \eta_{m^{N/2}}^\R}{N}\sum_{n=0}^{N-1} e^{\frac{2\pi i bn}{N}}(\eta_e^\R)^n \\
& =
\begin{cases}
	1+(-1)^a\eta_{m^{N/2}}^\R & \text{ if }e^{\frac{2\pi i b}{N}}\eta_e^\R=1\\
	0 & \text{otherwise}
\end{cases}
\end{align}
So $M_{(a,b)}$ does not vanish in the following cases:
\begin{align}
\label{ZNMa}
	(\eta_{e}^\R,  \eta_{m^{N/2}}^\R)  &= (+1, +1), \;\; M_{(a, 0)}=2  \text{  for even } a.
\nonumber \\
	(\eta_{e}^\R,  \eta_{m^{N/2}}^\R)  &= (-1, +1), \;\;  M_{(a, N/2)}=2 \text{  for even } a.
\nonumber \\
	(\eta_{e}^\R,  \eta_{m^{N/2}}^\R)  &= (+1, -1), \;\;  M_{(a,0)}=2 \text{  for odd } a.
\nonumber \\
	(\eta_{e}^\R,  \eta_{m^{N/2}}^\R)  &= (-1, -1), \;\; M_{(a, N/2)}=2 \text{  for odd } a.
\end{align}
The last case, with $\eta_e^\R=\eta_{m^{N/2}}^\R=-1$, is anomalous, as we will discuss in Sec. \ref{anomalyExamples}.

Let us now apply the theory developed in Sec. \ref{State_Sum_SETs}: we consider the SET state sum construction for the 
topological path integral of the $\mathbb{Z}_{N}$ toric code SETs. We focus on the case where $N$ is even.  
Recall that the input to this construction, in this case, is a certain type of $\mathbb{Z}_2$ extension of the fusion category
associated with $\mathbb{Z}_N$; more mathematically, it is a $\mathbb{Z}_2^{\R}$ equivariant 2-category. Here we will 
systematically solve for these $\mathbb{Z}_2^{\R}$ extensions, discuss the resulting fractionalization quantum numbers $\{\eta_a^{\R}\}$ 
and explicitly compute the topological path integral $\mathcal{Z}(\mathbb{RP}^2 \times S^1)$. We note that the results 
are essentially the same as if we replace ${\bf rc}$ with time-reversal ${\bf T}$ below. 

Starting from the fusion category $\mathcal{C}_\mb{0}=\text{Vec}_{\mathbb{Z}_{N}}$, we consider the possible extensions
\begin{align}
\mathcal{C}_{\mathbb{Z}_2^{\R}} = \mathcal{C}_{\bf 0} \oplus \mathcal{C}_{\R} . 
\end{align}
These categorical extensions are closely related to group extensions of $\mathbb{Z}_N$ by $\mathbb{Z}_2$. 
We use the labeling $a_\mb{g} \in \mathcal{C}_{\mathbb{Z}_2^{\R}}$, where ${\bf g} = {\bf 0}, {\R}$, 
indicates the grading for the simple objects in the extension and $a=0,1,\dots, N-1$.
We will also use the label $a_{\mb 0} = [a]$. 

For these $\mathbb{Z}_2^{\R}$ extensions, we consider the case where the $\mathbb{Z}_2^{\R}$-action on the objects $a_{\bf g}$
is trivial (see Sec. \ref{State_Sum_SETs}): $\,^{\bf h}a_{\bf g} = a_{\bf g}$, for ${\bf g},{\bf h} \in \mathbb{Z}_2^{\R}$. 

The fusion rules of the extension in general take the following form:
\begin{equation}
	a_\mb{g}\times b_\mb{h} = [a\times b\times\coho{w}(\mb{g,h})]_{\mb{gh}}.
	\label{}
\end{equation}
One can show that $\coho{w}(\mb{g,h})$ are classified by $\mathcal{H}^2[\mathbb{Z}_2, \mathbb{Z}_{N}]=\mathbb{Z}_2$, 
with the $2$-cocycle $\coho{w}(\R, \R)=[0]$ or $[1]$. In the former case, the fusion rules of $\mathcal{C}_{\mathbb{Z}_2^{\R}}$
are associated with group multiplication in $\mathbb{Z}_{N}\times \mathbb{Z}_2$, while for the latter the case the fusion rules are associated with
$\mathbb{Z}_{2N}$.

We further need to solve for the $F$-symbols $F^{abc}_{def}({\bf g}_0, {\bf g}_1, {\bf g}_2, {\bf g}_3)$
of the extension. To this end, by using the symmetry condition, Eq. (\ref{Fsym}), solving the ``twisted'' pentagon equation, (\ref{gPentagon2}), 
and fixing a particular choice of gauge (see Sec. \ref{SETgauge}), we find 
\begin{equation}
	F^{a_\mb{g}b_\mb{h}c_\mb{k}}=\chi_c(\mb{g,h}),
	\label{}
\end{equation}
where $\chi_c(\mb{g,h})$ is a phase factor. Note that the lower subscripts on the $F$-symbols can be omitted because 
they are determined uniquely by the superscripts. Using the gauge freedom of the $F$-symbols, we can further 
choose the gauge where $\chi_c(\mb{g,0}) = \chi_c(\mb{0,h}) = 1$ for all $c$. The only remaining freedom is therefore in the 
choice of $\chi_a(\R, \R)$. The twisted pentagon equation, (\ref{gPentagon2}), further leads to  the following conditions: 
(1) $\chi_a$ is a character on $\mathbb{Z}_{N}$, so we can write $\lambda\equiv \chi_{[1]}(\R, \R)$ 
and $\chi_a(\R, \R)=\lambda^a$. (2) $\chi_a(\R, \R)=\chi_a^*(\R, \R)$, thus $\lambda=\pm 1$. (3) The following 
obstruction-vanishing condition must hold:
\begin{equation}
	\chi_{\cohosub{w}(\R, \R)}(\R, \R)=1.
	\label{}
\end{equation}
Therefore we find that the there are in total three distinct $\mathbb{Z}_2^{\R}$ extensions $\mathcal{C}_{\mathbb{Z}_2^{\R}}$. 
These correspond to $\coho{w}(\R, \R)=[0], \lambda=\pm 1$ and $\coho{w}(\R, \R)=[1], \lambda=1$. These three 
solutions correspond exactly to the three different (non-anomalous) assignments of $\eta^\R$ (or $\eta^{\bf T}$) values 
to the $\mathbb{Z}_{N}$ toric code: $(\eta^{\R}_{e}, \eta^{\R}_{m^{N/2}}) = (1,1),\, (1, -1),\, (-1, 1)$.

Using this data, we can now explicitly compute $M_0 =\text{dim } \mathcal{V}(\mathbb{RP}^2) = \mathcal{Z}(\mathbb{RP}^2 \times S^1)$
by explicitly evaluating the topological path integral using the cellulation given in Fig. \ref{Fig:RP2xS1_Complex_Graded} and the corresponding 
expression, Eq. (\ref{Eq:RP2_State_Sum_Graded}). We find that if $\coho{w}(\R, \R) = [1]$,  it is not possible to 
satisfy the $\mathbb{Z}_2$-graded fusion rules on the cellulation of $\mathbb{RP}^2$ and thus $M_0 = 0$. When 
$\coho{w}(\R, \R)=[0]$, the state sum evaluates to $1+\lambda$. Therefore:
\begin{equation}
	\mathcal{Z}(\mathbb{RP}^2\times S^1)=\delta_{\cohosub{w}(\R, \R), [0]}(1+\lambda).
	\label{}
\end{equation}
This is consistent with the results of Eq. (\ref{ZNMa}). The three choices for the $\mathbb{Z}_2^{\bf r}$ extensions
give a value for $\mathcal{Z}(\mathbb{RP}^2\times S^1)$ that matches the result obtained in Eq. (\ref{ZNMa}) for the three non-anomalous
symmetry fractionalization classes. 

In fact, we can identify exactly which of the three solutions corresponds to the three different choices of $(\eta^{\R}_{e}, \eta^{\R}_{m^{N/2}})$
as follows. In the cellulation of $\mathbb{RP}^2 \times S^1$ in Fig. \ref{Fig:RP2xS1_Complex_Graded} that we use to evaluate the state sum, there 
are two distinct edges along the $S^1$ direction, which in the current example must both be equal to $d$ in order for the amplitude of the state
sum to be non-zero. 

We can view the state sum as a lattice gauge theory, where the link variables 
are gauge connections. The holonomy along $S^1$ (i.e. the value of the Wilson loop 
along $S^1$) is a gauge-invariant quantity. To make a correspondence with the usual convention of labeling in toric code, we notice that the state sum 
is actually ``dual'' to the usual Hamiltonian construction: the vertex constraint in the Hamiltonian construction corresponds 
to the constraint on a plaquette in the state sum. So a Wilson loop operator in the state sum construction should correspond 
to a magnetic Wilson loop in the Hamiltonian construction. Therefore, $d$ actually labels the electric charge threading the $S^1$. 
Suppose that $d$ corresponds to $n_e$ units of electric charge (i.e. $d = e^{n_e}$). 
Computing the state sum with a fixed $d$ (i.e. not summing over $d$ in Eq.  (\ref{Eq:RP2_State_Sum_Graded})), we find
\begin{equation}
	\begin{split}
	\mathcal{Z}(\mathbb{RP}^2\times S^1)|_{n_e}&=\frac{2}{N}\chi_{n_e}(\R, \R)\delta_{\cohosub{w}(\R,\R),[0]}\\
	&=\frac{2}{N}\lambda^{n_e} \delta_{\cohosub{w}(\R,\R),[0]}.
	\end{split}
	\label{}
\end{equation}

We have only fixed the electric charge measured through the $S^1$, while the magnetic charge is left unspecified. 
Therefore $\mathcal{Z}(\mathbb{RP}^2\times S^1)|_{n_e}$ is equal to
\begin{multline}
\mathcal{Z}(\mathbb{RP}^2\times S^1, e^{n_e}) + \mathcal{Z}(\mathbb{RP}^2\times S^1, e^{n_e} m^{N/2}),
%	=\frac{1}{N}(\eta_e^\R)^{n_e} (1+\eta_{m^{N/2}}^\R).
	\label{}
\end{multline}
where recall from Sec. \ref{gsd} that $\mathcal{Z}(\mathbb{RP}^2 \times S^1, x) = S_{0x} \eta_x^{\bf r}$ is the value of the topological 
path integral with a fixed topological charge $x$ as measured along the $S^1$. 
Thus, we find:
\begin{align}
\frac{2}{N}\lambda^{n_e} \delta_{\cohosub{w}(\R,\R),[0]} = \frac{1}{N} (\eta_{e}^{\bf r} )^{n_e} (1 + (\eta_{m}^{\bf r})^{N/2}) ,
\end{align}
where we have used the fact that $\eta_{e^a m^b}^{\bf r} = (\eta_{e}^{\bf r})^a (\eta_m^{\bf r})^b$. We thus conclude:
\begin{equation}
	(\eta_e^\R, \eta_{m^{N/2}}^{\R}) = (\lambda, (-1)^{\cohosub{w}(\R,\R )}).
	\label{}
\end{equation}
Here we have slightly abused notation and denoted $(-1)^{\cohosub{w}(\R,\R )} = 1$ if ${\coho{w}(\R,\R )}=[0]$
and $(-1)^{\cohosub{w}(\R,\R )} = -1$ if ${\coho{w}(\R,\R )}=[1]$.

\subsection{Decorated toric code model}
\label{SubSec:decorated_TC}

In this section we consider a specific Hamiltonian realization of a $\mathbb{Z}_2$ toric code with the following reflection symmetry action:
\begin{equation}
	\ra{e}=e, \ra{m}=m, \eta_{e}^\R=-1.
	\label{}
\end{equation}
The construction generalizes that of Ref. \onlinecite{zion2016} to reflection symmetry. In particular, we will demonstrate how to put the model on a crosscap, which provides a formal derivation of the heuristic loop gas argument given in Sec. \ref{sec:loop-gas}.

\subsubsection{ (1+1)D cluster state as a reflection SPT}
	The (1+1)D cluster state is defined on a chain with a spin-$\frac{1}{2}$ degree of freedom per site. Let $X_i$, $Y_i$, and $Z_i$ denote the Pauli matrices on the $i$th site. The Hamiltonian for the cluster state is given by
\begin{align}
\mathbf{H}_\text{cluster} = -\sum_i Z_{i-1} X_i Z_{i+1}.
\label{eqn:cluster}
\end{align}
$\mathbf{H}_\text{cluster}$ can be related to the Hamiltonian of the a trivial paramagnet $\mathbf{H}_\text{para} \equiv  -\sum_i X_i$ via local unitary gates:
\begin{align}
\mathbf{H}_\text{cluster} = \mathbf{U}_\text{cp} H_\text{para} \mathbf{U}_\text{cp}^\dag,
\end{align}
where
\begin{align}
\mathbf{U}_\text{cp} = \mathbf{U}_\text{cp} ^\dag=
\prod_{i} \CZ_{i,i+1}\end{align}
is a product of local control-Z gates:
\begin{align}
	{}\CZ_{i,i+1} \equiv e^{i\pi \left(\frac{1-Z_i}{2}\right)\left(\frac{1-Z_{i+1}}{2}\right)}.
\end{align}
The ground state of $\mathbf{H}_\text{cluster}$, namely the cluster state, can be easily constructed from that of $\mathbf{H}_\text{para}$, namely the paramagnet state. The latter can be written as $|\text{PM}+ \rangle \equiv \prod_i |+_i\rangle$, where we define $|+_i\rangle$ to be the state on the $i$th site such that $|+_i\rangle = X_i |+_i\rangle$. The cluster state $|\text{CS}\rangle$ can then be written as
\begin{align}
|\text{CS}\rangle  = \left( \prod_{i} \CZ_{i,i+1} \right) \left( \prod_i |+_i\rangle \right).
\end{align}

Now we consider the {\it site-centered} reflection symmetry in this one-dimensional chain. We assume the sites are labeled by integers and the reflection $\r$ permutes the sites via $\r: i\rightarrow -i$. We define the reflection symmetry action to be generated by the following operator:
\begin{align}
R_\r = \left( \prod_{i} X_i \right)  R_{0\r},
\label{Eq:Reflection_Op}
\end{align}
where $R_{0\r}$ is the ``site-permutation" operator such that
\begin{align}
R_{0\r} X_i R_{0\r}^\dag = X_{\r(i)},~~~~R_{0\r} Z_i R_{0\r}^\dag = Z_{\r(i)}.
\end{align}
It is easily to check that the cluster state Hamiltonian $H_\text{cluster} $ is
invariant under the action of $R_\r $.

We consider the cluster states on a finite 1D chain with the site index $i\in[-N,N]$. Notice that the summation of $i$ in the Hamiltonian Eq. \eqref{eqn:cluster} does not contain the two outer most sites, which give rise to 4-fold degenerate ground state Hilbert space $\mathcal{H}_\text{cs}$ of $\mathbf{H}_\text{cluster}$. The four states can be written as
\begin{align}
|\text{CS}, x_{-N}, x_N \rangle = \mathbf{U}_\text{cp}  \Big( |x_{-N} \rangle \otimes |\text{PM}+\rangle \otimes  |x_{N} \rangle \Big),
\end{align}
where $x_{-N}, x_N$ can independently take values $+$ and $-$, and it is understood that $|\text{PM}+\rangle \equiv \prod_{i \in [-N+1,N-1]} |+_i\rangle$. We can see that there are effectively 2-fold degrees of freedom localized on each boundary. Now, we calculate the topological invariant $\Tr_{\mathcal{H}_\text{cs}} R_{\r}$ that signifies the non-trivial reflection-symmetric SPT character of the cluster states:
\begin{widetext}
\begin{align}
\Tr_{\mathcal{H}_\text{cs}} R_{\r}
& = \sum_{ x_{-N}, x_N = +, -}   \langle \text{CS}, x_{-N}, x_N | R_{\r} |\text{CS}, x_{-N}, x_N \rangle
\nonumber \\
& = \sum_{ x_{-N}, x_N = +, -}   \Big( \langle x_{-N} | \otimes \langle \text{PM}\!+\! | \otimes  \langle x_{N} | \Big)
\mathbf{U}_\text{cp}^\dag R_{\r}    \mathbf{U}_\text{cp}
\Big( |x_{-N} \rangle \otimes |\text{PM}+\rangle \otimes  |x_{N} \rangle \Big)
\nonumber \\
& = \sum_{ x_{-N}, x_N = +, -} x_{-N} x_N  \Big( \langle x_{-N} | \otimes \langle \text{PM}\!+\! | \otimes  \langle x_{N} | \Big)
 Z_{-N} Z_N R_{0\r}
\Big( |x_{-N} \rangle \otimes |\text{PM}+\rangle \otimes  |x_{N} \rangle \Big)
\nonumber \\
& = - 2,
\label{Eq:Cluster_Trace_U_r}
\end{align}
\end{widetext}
where all the non-zero contribution comes from the case with $x_{-N} = - x_N$. In deriving Eq. \ref{Eq:Cluster_Trace_U_r}, we've used following relations
\begin{align}
R_{\r}  \mathbf{U}_\text{cp} R_{\r}^\dag
& = \left( \prod_{i} X_i \right)   \left( \prod_{i} \CZ_{i,i+1} \right) \left( \prod_{i} X_i \right)
\nonumber \\
& =  \left( \prod_{i} -Z_i Z_{i+1}\CZ_{i,i+1} \right)
\nonumber \\
& = (-1)^\text{\# of bonds} Z_{-N} Z_{N} \prod_{i}\CZ_{i,i+1}
\nonumber \\
& = Z_{-N} Z_{N} \mathbf{U}_\text{cp} ,
\label{Eq:Commutator_U_r_U_cp}
\end{align}
where we used the fact that ``$\#$ of bonds" in the site-centered reflection symmetric geometry is even. This is enough to show that the cluster state is indeed a nontrivial reflection SPT.
%Physically, $| \Tr_{\mathcal{H}_\text{cs}} \mathbf{U}_{\r} |$ calculates the dimension of the boundary degrees of freedom (on a single boundary), which is {\it not} universal. Instead, $ \Tr_{\mathcal{H}_\text{cs}} \mathbf{U}_{\r} /| \Tr_{\mathcal{H}_\text{cs}} \mathbf{U}_{\r} | = -1$ is the {\it universal} character of the cluster state as a non-trivial 1+1D reflection symmetric SPT.

We would like to note that the conclusion crucially depends on the fact that the reflection operation is site-centered. Instead, if we consider bond-centered reflection, the minus sign in Eq. \ref{Eq:Commutator_U_r_U_cp} disappears due to the fact that ``$\#$ of bonds" in the bond-centered reflection symmetric geometry is odd.

\subsubsection{Toric code with electric strings decorated by cluster states}
We again consider the toric code models on the square lattice. We will use the labels $v$, $l$ and $p$ are vertices, links and plaquettes of the square lattice.

 Let $|\vac\rangle$ denote the configuration with no strings, namely $\sigma_l^z|\vac\rangle= |\vac\rangle $ for all links $l$. The ground state of $H_\text{tc}$ is given by the equal-weight superposition of all string configuration, which can be written as
\begin{align}
\left( \prod_p \frac{1+B_p}{\sqrt{2}} \right)|\vac \rangle.
\end{align}

Now, we generalize the idea of Ref. \onlinecite{zion2016} and construct the decorated toric code (dTC) model. We introduce spin-$\frac{1}{2}$ degrees of freedom on each vertex.  The Hamiltonian is given by
\begin{align}
\mathbf{H}_\text{dtc} =
 - \sum_{\text{vertex}~ v} A_v
 - \sum_{\text{vertex}~ v} \left( \frac{1+A_v}{2} \right) C_v
 - \sum_{\text{plaquette}~ p} \tilde{B}_p ,
\label{Eq:Ham_dTC}
\end{align}
The definition of $A_v$ remains the same as the one in the standard toric code Hamiltonian.  To define $C_v$, we first introduce
\begin{align}
s_l = \frac{1-\sigma^z_l}{2}, ~~~~ U_{\text{cp},l} = \CZ_{\partial_1 l,\partial_2 l},
\end{align}
where the two vertices connected to the link $l$ are denoted by $\partial_1 l$ and $\partial_2 l$, respectively. $s_l$ can be thought of as the occupation number of strings. The term $C_v$ is given by
\begin{align}
C_v \equiv \left(\prod_{l\in v} (U_{\text{cp},l})^{s_l} \right) X_v \left(\prod_{l\in v} (U_{\text{cp},l})^{s_l} \right)^\dag.
\end{align}
Apparently $C_v$ commutes with $A_{v'}$ for all vertices $v'$. The term $\left( \frac{1+A_v}{2} \right) C_v$ ensures the following properties in the ground state: (1) if no string passes through the vertex $v$, the spin on $v$ is polarized to $|+_v\rangle$ and (2) if the vertex is passed through by one or more strings, the spin on $v$ is a part of the cluster state along the strings. To summarize, the $C_v$ terms enforce that the cluster states are aligned with electric strings.

Finally, the definition of the operator $\tilde{B}_p$ is given by
\begin{align}
\tilde{B}_p = \prod_{l \in p} \sigma^x_l U_{\text{cp},l}.
\end{align}
Notice that $\tilde{B}_p$ itself is Hermitian and $\tilde{B}_p^2=\mathds{1}$. Physically, $\tilde{B}_p$ creates an electric string decorated by the 1D cluster state, fuses the strings to the boundary of the plaquette $p$. It is crucial that $U_{\text{cp},l} ^2 = 1$ consistent with the $\mathbb{Z}_2$ nature of strings.  It is straightforward to show that the operators $\tilde{B}_p$ commute with all the $A_v$ and $C_v$ terms. The ground state of $H_\text{dtc}$ is given by
\begin{align}
\left( \prod_p \frac{1+\tilde{B}_p}{\sqrt{2}} \right)\Big( |\vac \rangle \otimes |\text{PM}+\rangle \Big),
\end{align}
where $|\text{PM}+\rangle \equiv \prod_{\text{vertex}~ v} |+_v\rangle$. This ground state wave function is the equal-weight superposition of all (decorated) string configurations. Since the electric strings are now decorated by the 1D cluster states, the electric charge $e$ has $\eta^\r_e = -1$ under the vertex-centered reflection $\r$. The magnetic charge $m$ still has $\eta_m^\R=1$.

\subsubsection{Decorated toric code model on non-orientable manifolds}

\begin{figure}[tb]
\subfloat[][]{
\begin{tikzpicture}[scale=1.5][baseline={([yshift=-.5ex]current  bounding  box.center)}]
\draw [thick] (0,0) -- (0,1) -- (1,1) -- (1,0)--(0,0);
\draw (0.5,-0.2) node {$l$};
\draw (0.5, 1.2 ) node {$l'$};
\end{tikzpicture}
}
~~~~~~~~~~~~~
\subfloat[][]{
\begin{tikzpicture}[scale=1.5][baseline={([yshift=-.5ex]current  bounding  box.center)}]
\draw [thick] (0,0) -- (1,0);
\draw [thick] (0,1) -- (1,1);
\draw [thick] (1,0) -- (0,1);
\draw [line width = 7, white] (0.1,0.1) -- (0.9,0.9);
\draw [thick] (0,0) -- (1,1);
\draw (0.5,-0.2) node {$l$};
\draw (0.5, 1.2 ) node {$l'$};
\end{tikzpicture} }
\caption{(a) regular plaquette connecting $l$ and $l'$ with the same orientation. (b) twistd plaquette connecting $l$ and $l'$ with opposite orientation reversed.}.
\label{Fig:regular_twisted_plaquettes}
\end{figure}
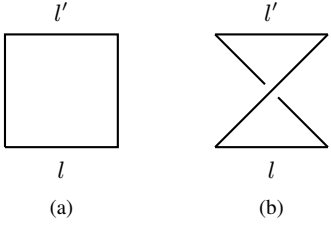

\begin{figure}[tb]
\includegraphics[width= 1\columnwidth]{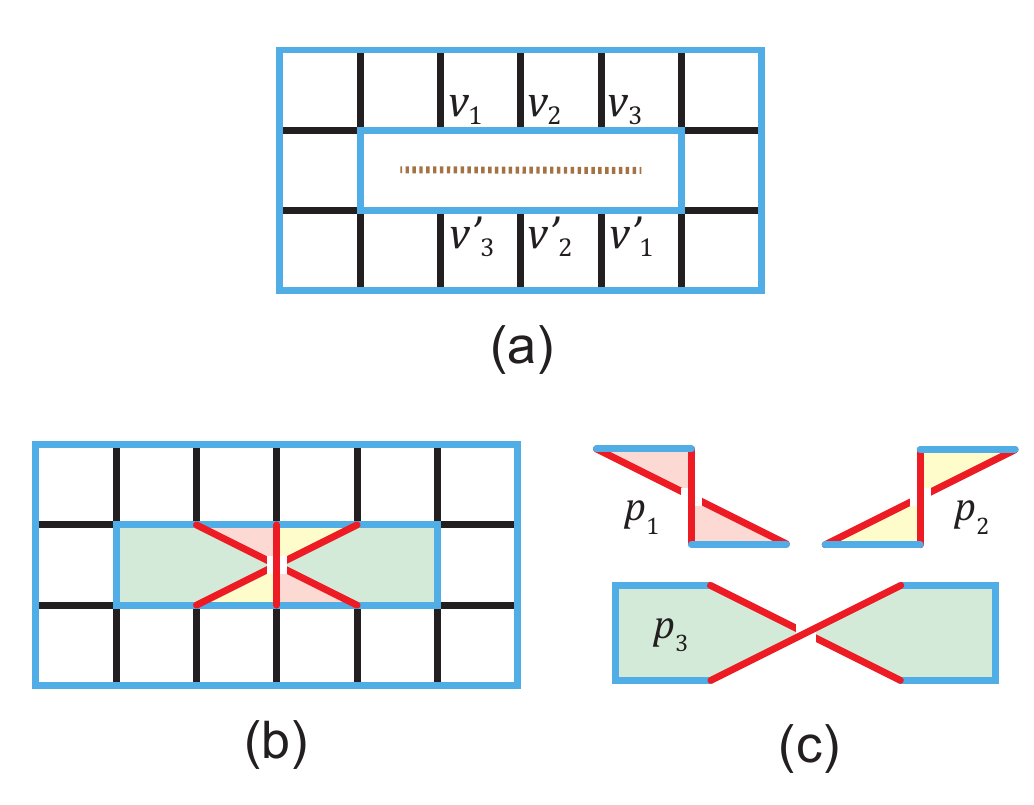}
\caption{
(a) An annulus with two pieces of its boundary colored blue. To form a crosscap, the interior boundary will be glued to itself via the identification of antipodal points. Equivalently, we can just connect the pairs of vertices $(v_i, v'_i)$ with $i=1,2,3$ across the branch cut (brown dash line). (b) Crosscap geometry with three links (red) crossing the branch cut and three twisted plaquettes $p_1$ (red), $p_2$ (yellow) and $p_3$ (green). (c) The three twisted plaquettes in the crosscap geometry.
}
\label{Fig:dtcCrosscap}
\end{figure}

First, we consider putting the dTC model on a crosscap. We follow a similar prescription as to create a lattice realization of a crosscap, but now with site-centered reflection. Again we start from an oriented manifold (i.e. a disk), and remove some edges so that there is a boundary. We then re-glue the boundary but with an ``orientation-reversal branch cut'' (e.g. reconnecting sites to the ``antipodal'' counterparts in Fig. \ref{Fig:dtcCrosscap} (a) and (b)). Certain plaquettes across the branch cut need to be twisted. In fact, the twisted plaquette should be viewed as a regular plaquette with a part of its boundary being reflected. For example, the twisted plaquette in Fig. \ref{Fig:regular_twisted_plaquettes} (b) can be obtained from the regular plaquette in Fig. \ref{Fig:regular_twisted_plaquettes} (a) by reflecting the link $l$ about its center. With this perspective we can construct the plaquette operators acting on the twisted plaquettes.

For the convenience of discussion, we first introduce a graphical representation of the operator $\tilde{B}_p$ on a regular plaquette $p$:
\begin{align}
\tilde{B}_p  =
\begin{tikzpicture}[scale=1.2,baseline={([yshift=-0.5ex]current  bounding  box.center)}]
\draw [line width = 2, blue] (0,0) -- (0,1) -- (1,1) -- (1,0)--(0,0)--(0,1);
\draw (0.5,0.5) node{$p$};
\draw (0.5,-0.2) node {$l_3$};
\draw (-0.2,0.5) node {$l_2$};
\draw (0.5, 1.2 ) node {$l_1$};
\draw (1.2,0.5) node {$l_4$};
\end{tikzpicture},
~~\text{where}~~
\begin{tikzpicture}[baseline={([yshift=-.5ex]current  bounding  box.center)}][scale=2]
\draw [line width = 2, blue] (0,0) -- (0,1);
\draw (-0.2,0.5) node {$l$};
\end{tikzpicture}
~~ =\sigma^x_l U_{\text{cp},l}
\label{Eq:Reg_Plaquette}
\end{align}
Here, we write $\tilde{B}_p $ as the product of four ``link operators" represented by the blue lines. As an example, we demonstrate how to construct the plaquette operator on the twist plaquette shown in Fig. \ref{Fig:regular_twisted_plaquettes} (b), which can be obtained from Eq. \ref{Eq:Reg_Plaquette} by having the reflection only acting on the link $l_3$. According to the defintion of the reflection symmetry in Eq. \ref{Eq:Reflection_Op}, the reflection not only permutes the sites but also acts non-trivially on the spins. Therefore, the twisted plaquette operator is given by
\begin{align}
\tilde{B}_p & =
X_{\partial_1 l_3} X_{\partial_2 l_3}
\left(
\begin{tikzpicture}[scale=1.3,baseline={([yshift=-0.5ex]current  bounding  box.center)}]
\draw [line width = 2, blue] (0,0) -- (1,0) -- (0,1) -- (1,1);
\draw [line width = 7, white] (0.3,0.3) -- (0.7,0.7);
\draw [line width = 2, blue] (1,0) -- (0,0) -- (1,1) -- (0,1);
\draw (0.5,0.75) node{$p$};
\draw (0.5,-0.2) node {$l_3$};
\draw (0.0,0.7) node {$l_2$};
\draw (0.5, 1.2 ) node {$l_1$};
\draw (1,0.7) node {$l_4$};
\end{tikzpicture}\right)
X_{\partial_1 l_3} X_{\partial_2 l_3}
 \nonumber \\
& =
- Z_{\partial_1 l_2} Z_{\partial_2 l_2}  Z_{\partial_1 l_4} Z_{\partial_2 l_4}
\left(
\begin{tikzpicture}[scale=1.3,baseline={([yshift=-0.5ex]current  bounding  box.center)}]
\draw [line width = 2, blue] (0,0) -- (1,0) -- (0,1) -- (1,1);
\draw [line width = 7, white] (0.3,0.3) -- (0.7,0.7);
\draw [line width = 2, blue] (1,0) -- (0,0) -- (1,1) -- (0,1);
\draw (0.5,0.75) node{$p$};
\draw (0.5,-0.2) node {$l_3$};
\draw (0.0,0.7) node {$l_2$};
\draw (0.5, 1.2 ) node {$l_1$};
\draw (1,0.7) node {$l_4$};
\end{tikzpicture}\right)
 \nonumber \\
&
= -
\begin{tikzpicture}[scale=1.3,baseline={([yshift=-0.5ex]current  bounding  box.center)}]
\draw [line width = 2, blue] (0,0) -- (1,0) -- (0,1) -- (1,1);
\draw [line width = 2, white] (1,0.0) -- (0,1);
\draw [line width = 2, dotted, blue] (1.0,0.0) -- (0,1);
\draw [line width = 12, white] (0.3,0.3) -- (0.7,0.7);
\draw [line width = 2, blue] (1,0) -- (0,0) -- (1,1) -- (0,1);
\draw [line width = 2, white] (0.0,0.0) -- (1,1);
\draw [line width = 2, dotted, blue] (0.0,0.0) -- (1,1);
\draw (0.5,0.75) node{$p$};
\draw (0.5,-0.2) node {$l_3$};
\draw (0.0,0.7) node {$l_2$};
\draw (0.5, 1.2 ) node {$l_1$};
\draw (1,0.7) node {$l_4$};
\end{tikzpicture}.
\label{Eq:Twisted_Plaquette}
\end{align}
Here we have absorbed the factors of $\Big( Z_{\partial_1 l_2} Z_{\partial_2 l_2} \Big)$ and $ \Big( Z_{\partial_1 l_4} Z_{\partial_2 l_4} \Big) $ into the ``twisted" link operators:
\begin{align}
\begin{tikzpicture}[scale=1,baseline={([yshift=-0.5ex]current  bounding  box.center)}]
\draw [line width = 2, dotted, blue] (1.0,0.0) -- (0,1);
\draw (0.2,0.5) node{$l$};
\end{tikzpicture}
 ~~ &= ~~
\begin{tikzpicture}[scale=1,baseline={([yshift=-0.5ex]current  bounding  box.center)}]
\draw [line width = 2, dotted, blue] (1.0,1) -- (0,0);
\draw (0.2,0.5) node{$l$};
\end{tikzpicture}
\\
&= \Big( Z_{\partial_1 l}  Z_{\partial_2 l} \Big)  U_{\text{cp},l}  ~ \sigma^x_l \\
&= \Big( Z_{\partial_1 l}  Z_{\partial_2 l} \Big)  \CZ_{\partial_1 l,\partial_2 l}  ~ \sigma^x_l .
\end{align}
 Notice that the overall minus sign in twisted plaquette operator $\tilde{B}_p$ results from an odd number of links (i.e. $l_3$) being reflected in the construction. This example in fact illustrates the general prescription to write down plaquette operators on twisted plaquettes.

In order to have a well-defined lattice model on non-orientable manifolds, we should have the following requirements:
\begin{enumerate}
\item There are no vertices located on branch cut, while links and plaquettes are allowed to cross the branch cut.

\item When a plaquette crosses the branch cut, the branch cut only intersects the boundary of the plaquette on two non-adjacent links.

\item The lattice remains bipartite.
\end{enumerate}

Let us write down the Hamiltonian for the dTC model on a lattice version of a non-orientable manifold. It is not difficult to see that the $A_v$ term remains unchanged.  The plaquette operators $\tilde{B}_p$ are given by
\begin{align}
\tilde{B}_p \equiv \sgn(p)  \prod_{l \in p} \sigma^x_l \tilde{U}_{\text{cp},l},
\label{Eq:All_Plaquette_Op}
\end{align}
where
\begin{align}
\tilde{U}_{\text{cp},l} \equiv
\begin{cases}
Z_{\partial_1 l}  Z_{\partial_2 l}   \CZ_{\partial_1 l,\partial_2 l}, &  \text{if link $l$ crosses the branch cut,}\\
      \CZ_{\partial_1 l,\partial_2 l}, & \text{otherwise}. \\
      \end{cases}
\label{Eq:dTC_twiste_link}
\end{align}
Here $\sgn(p)$ is defined as follows: if plaquette $p$ does not cross the branch cut, $\sgn(p)=1$; if plaquette $p$ crosses the branch cut, then $\sgn(p)$ is $+1/-\!1$ if the number of links between the two links on $\partial p$ crossing the branch cut is even/odd.

Notice that requirement of bi-partition of the lattice ensures that any plaquette $p$ must contain even number of links. Therefore, ``\# of links between the two links" modulo 2 and, consequently, $\sgn(p)$ are always well-defined. The way that the operator $\tilde{U}_{\text{cp},l}$ is defined in Eq. \ref{Eq:dTC_twiste_link} as well as the signs $\text{sign}(p)$ is a natural generalization of the example given in Eq. \ref{Eq:Twisted_Plaquette}. It is easy to convince ourselves that definition of
$\tilde{U}_{\text{cp},l}$ is indeed suitable for the most general situations. Using these operators $\tilde{U}_{\text{cp},l}$, we can define the operator $C_v$ as
\begin{align}
C_v \equiv \left(\prod_{l\in v} (\tilde{U}_{\text{cp},l})^{s_l} \right) X_v \left(\prod_{l\in v} (\tilde{U}_{\text{cp},l})^{s_l} \right)^\dag.
\end{align}

We can finally define the dTC Hamiltonian on a crosscap.  To be precise, the crosscap is defined as follows: we start with an ``annulus'' as shown in Fig. \ref{Fig:dtcCrosscap} (a). The two pieces of its boundary are colored blue. To form a crosscap, the inner boundary must be glued to itself via the identification of antipodal points. As shown in Fig. \ref{Fig:dtcCrosscap} (b), the gluing is equivalent to connecting the pairs of vertices $(v_i, v'_i)$ with $i=1,2,3$ via links (red lines in Fig. \ref{Fig:dtcCrosscap} (b)) across the branch cut (brown dash line in Fig. \ref{Fig:dtcCrosscap} (a)). The resulting crosscap geometry is shown in Fig. \ref{Fig:dtcCrosscap} (b). Notice that there are an {\it odd} number of links across the branch cut, as one would expect for site-centered reflection (i.e. the sites on one side of the branch cut, say the $v_i$'s, when viewed as a 1d chain, are ``glued" to the 1D chain formed by the sites on the other side of the branch cut via a {\it site-centered} reflection map). As a result, the lattice remains bipartite.

The crosscap in Fig. \ref{Fig:dtcCrosscap} (b) has three links (red) crossing the branch cut and three twisted plaquettes $p_1$, $p_2$ and $p_3$ (Fig. \ref{Fig:dtcCrosscap} (c)). It is easy to see that
\begin{align}
\sgn(p_1) = \sgn(p_2) =\sgn(p_3) = -1,
\end{align}
which implies that
\begin{align}
\tilde{B}_{p_1}  \tilde{B}_{p_2}  \tilde{B}_{p_3} = - \tilde{B}_{p_1 \cup p_2 \cup p_3},
\label{Eq:Twisted_Plaquette_Union}
\end{align}
where $p_1 \cup p_2 \cup p_3$ is the regular (or un-twisted) plaquette formed by joining $p_1$, $p_2$ and $p_3$, and $\tilde{B}_{p_1 \cup p_2 \cup p_3}$ can be naturally defined via Eq. \ref{Eq:All_Plaquette_Op}. Notice there is a $-1$ factor on the r.h.s of Eq. \ref{Eq:Twisted_Plaquette_Union}. This $-1$ factor is universal, and is tied to the fact that there is always an {\it odd} number of links crossing the branch cut connecting {\it odd} number of pairs of vertices. This $-1$ factor can be reproduced in any other realization of the crosscap, as long as the conditions on the lattice and that on the number of links crossing the branch cut are satisfied.

Physically, this $-1$ factor means that when we create a decorated electric string in the center of the crosscap and pass it through all the twisted plaquettes, the wavefuntion acquires an $-1$ phase factor. In fact, this $-1$ factor is nothing but the $\r^2$ eigenvalue $\eta^\r_e = -1$ for the electric charge $e$. Now, if we view the crosscap as a punctured $\mathbb{RP}^2$ (with the outer boundary of the crosscap the boundary of the puncture), the $-1 $ phase factor implies that there must be an $m$ particle localized on the puncture. As a consequence, the ground state degeneracy on $\mathbb{RP}^2$ (without punctures) is 0.

Before we end the discussion of the dTC model on non-orientable manifolds, a couple of remarks are in order. First, if we {\it violate} the bi-partition condition on the lattice and consider a crosscap geometry with an even number of links crossing the branch cut, the $-1$ phase factor does {\it not} come up. In this case, when a decorated electric string passes though the twisted plaquettes, it effectively only undergoes a link-centered reflection, under which the 1D cluster state is a trivial SPT. It is also straightforward to generalize the construction to $\mathbb{Z}_N$ toric code with even $N$.

\subsection{Anomalous SETs: eTmT, efmf, $\D(S_3)$ and gauged T-Pfaffian}
\label{anomalyExamples}

In this subsection we study more examples of anomalous SETs with reflection/time-reversal symmetry. One of them, the gauged T-Pfaffian,
is well-known in the literature of surface topological order for (3+1)D bosonic SPT phases, and ``microscopic'' constructions of these SETs 
have been proposed. We also find a new anomalous SET associated with $\D(S_3)$ in which the time-reversal/reflection symmetry 
permutes anyons in an unconventional way.

\subsubsection{eTmT and its $\mathbb{Z}_N$ generalizations}

The eTmT state is the $\mathbb{Z}_2$ toric code state where $e$ and $m$ both carry a local Kramers degeneracy, so that
$\eta_e^{\bf T} = \eta_m^{\bf T} = -1$. For reflection symmetry, we can consider a similar state, where $\eta_e^{\bf r } = \eta_m^{\bf r} = -1$. 

Here we can further consider a generalization to the $\mathbb{Z}_N$ toric code state. Recall that we use the notation
$(a,b) = (e^a, m^b)$ for the composite of $a$ e particles and $b$ m-particles. As discussed in Sec. \ref{ZNexample}, here we take the reflection
symmetry action to be $\ra{e} = \overline{e}$ and $\ra{m} = m$, which implies that for $N$ even we have the reflection
symmetry fractionalization quantum numbers $\eta_e^{\bf r}$, $\eta_{m^{N/2}}^{\bf r}$. The generalization of the eTmT state is the case where
$\eta_e^{\bf r} = \eta_{m^{N/2}}^{\bf r} = -1$ (and analogously if reflection is replaced by time-reversal). In fact, condensing $(2,0)$ 
induces a topological phase transition to the eTmT state (the deconfined anyons are $(0,0), (1,0), (0,N/2)$ and $(N/2,1)$).

First we can see that this state possesses the Dehn twist anomaly: From Eq. (\ref{ZNMa}), and from the fact that 
$\theta_{(a,N/2)}=-1$ for odd $a$, we see that the constraint (\ref{consistencyEq}) is violated. Thus $\eta_{e}^\R=\eta_{m^{N/2}}^\R=-1$ is anomalous. 

We further find that for all these states,
\begin{align}
\mathcal{Z}(\mathbb{RP}^4) &= -1,
\nonumber \\
\mathcal{Z}(\mathbb{CP}^2) &= 1,
\end{align}
which tells us that the eTmT state, together with its generalizations, are all surface topological orders associated with the 
surface of the ``within group cohomology'' (3+1)D SPT state. 

\subsubsection{efmf}

efmf refers to a surface topological order of a ``beyond group cohomology'' time-reversal invariant (3+1)D SPT phase, 
first conceived in Ref. \onlinecite{vishwanath2013}, and further confirmed by an exactly solvable model~\cite{BurnellPRB2014} 
and a layer construction~\cite{wang2013}. There are three nontrivial quasiparticles in this topological phase, all of which are fermions, and
which are mutual semions (that is, a full braid of one around another gives a phase of $-1$). The quasiparticles form a $\mathbb{Z}_2\times\mathbb{Z}_2$ 
fusion algebra, similar to a $\mathbb{Z}_2$ toric code. In its simplest incarnation, time-reversal/reflection symmetry acts trivially on the 
quasiparticles, i.e. all $\eta_a^\mb{T}$ or ${\eta}_a^\R$ are equal to $1$.\cite{BurnellPRB2014, wang2013}

For such a trivial action of time-reversal and reflection symmetry, we find that for the efmf state,
\begin{equation}
	\Z(\mathbb{CP}^2)=\Z(\mathbb{RP}^4)=-1.
	\label{}
\end{equation}
$\Z(\mathbb{CP}^2)=-1$ implies that a (2+1)D realization of this topological order has to have $c_-=\pm 4$, 
which necessarily breaks time-reversal/reflection symmetry. This was the original argument in 
Ref. \onlinecite{vishwanath2013} for why the SET state is anomalous. It has not been entirely clear, however, 
whether this is the only anomaly possessed by the SET. Following suggestions of Ref. \onlinecite{kapustin2014}, 
Ref. \onlinecite{Thorngren2015} argued that the SPT phase with efmf surface should have 
$\Z(\mathbb{CP}^2)=-1$ but $\Z(\mathbb{RP}^4)=1$.\footnote{Formally, Ref. \onlinecite{kapustin2014} and 
\onlinecite{Thorngren2015} argued that the path integral of the bulk SPT phase is given by 
$\Z(M^4)=\exp\left( \pi i\int_{M^4}w_2^2 \right)$, where $w_2$ is the second Stieffel-Whitney class. 
From Eqs. (\ref{ZX4})-(\ref{Zn1n2}), this implies $\mathcal{Z}(\mathbb{RP}^4) = 1$ and $\mathcal{Z}(\mathbb{CP}^2) = -1$. }
Our result does not agree with Ref. \onlinecite{Thorngren2015}. We also note that Ref. \onlinecite{metlitski2015} arrived 
at the same conclusion as ours although with a very different argument.

\subsubsection{$\D(S_3)$}

$S_3$ is the symmetric group of order $6$ (the permutation group on 3 elements). It can be represented as
\begin{equation}
	S_3=\{r,s| r^3=s^2=1, srs=r^{-1}\}.
	\label{}
\end{equation}
The quantum double of $S_3$, or $S_3$ gauge theory, is a topological phase with $8$ quasiparticle types.  
Let us now enumerate the anyon types. For $S_3$, there are three conjugacy class $[1], [r], [s]$, whose centralizers 
are $S_3, \mathbb{Z}_3$ and $\mathbb{Z}_2$ respectively. $S_3$ has three irreducible representations of dimensions 
$1, 1$ and $2$. We will denote the nontrivial 1D representation by $B$ and the 2D representation by $C$. For the other 
conjugacy classes, we have $([r], \omega^n)$ where $\omega=e^{\frac{2\pi i}{3}}$ and $n=0,1,2$, and $([s], \pm)$. 
The fusion rules, $S$ and $T$ matrices can be found in Ref. \onlinecite{Cui2014}. Notice that all anyons are 
self-dual (each is its own anti-particle). 

Under an orientation-reversing symmetry, the complex representations have to be conjugated, so we must have 
$([r], \omega)\leftrightarrow ([r], \omega^2)$ under either ${\bf T}$ or ${\bf r}$. Morevoer, $\D(S_3)$ has an order two element in
$\text{Aut}_{0,0}(\D(S_3))$ (i.e. a unitary $\mathbb{Z}_2$ topological symmetry): \cite{Beigi2010, barkeshli2014SDG}
\begin{equation}
	([1], C)\leftrightarrow ([r], 1).
	\label{S3emdual}
\end{equation}
Notice that the symmetry exchanges a pure gauge charge with a pure gauge flux, so Eq. \eqref{S3emdual}
can be regarded as an ``electromagnetic duality'' in the discrete gauge theory.

Therefore, we can consider the following action of time-reversal symmetry:
\begin{align}
	 \,^{\bf T}([1], C) &= ([r],1)
\nonumber \\
\,^{\bf T} ( [r],1) &= ( [1],C)
\nonumber \\
\,^{\bf T} ( [r],\omega) &= ( [r], \omega^2)
\nonumber \\
\,^{\bf T} ([r], \omega^2) &= ([r], \omega)
%\leftrightarrow ([r], 1), ([r], \omega)\leftrightarrow ([r], \omega^2).
	\label{eqn:s3permutation}
\end{align}
Namely we compose the unitary $\mathbb{Z}_2$ topological symmetry with the conjugation of the complex representations. 

Below we will frame the discussion in terms of time-reversal symmetry, but since the theory has a CPT invariance, we expect
that time-reversal can be replaced by reflection symmetry everywhere in the discussion below. 

For the rest of the section we will follow the notation of Ref. \onlinecite{Cui2014} 
for the anyon types in $\D(S_3)$:
\begin{equation}
	\begin{gathered}
	A \equiv ([1], 1), \;\; B \equiv ([1],B),  \;\; C \equiv ([1],C)\\
	D \equiv ([s], +), \;\; E \equiv  ([s], -)\\
	F \equiv ([r], 1),\;\; G \equiv ([r], \omega), \;\; H \equiv ([r], \omega^2).
	\end{gathered}
	\label{}
\end{equation}

The modular ${S}$ and ${T}$ matrices of $\D(S_3)$ are given by:
\begin{equation}
\label{eq:ds3-S}
{S} = \frac{1}{6}
\begin{bmatrix}
1 & 1 & 2 & 3 & 3 & 2 & 2 & 2 \\
1 & 1 & 2 & -3 & -3 & 2 & 2 & 2 \\
2 & 2 & 4 & 0 & 0 & -2 & -2 & -2 \\
3 & -3 & 0 & 3 & -3 & 0 & 0 & 0 \\
3 & -3 & 0 & -3 & 3 & 0 & 0 & 0 \\
2 & 2 & -2 & 0 & 0 & 4 & -2 & -2 \\
2 & 2 & -2 & 0 & 0 & -2 & -2 & 4 \\
2 & 2 & -2 & 0 & 0 & -2 & 4 & -2 \\
\end{bmatrix}
\end{equation}

\begin{equation}
\label{eq:ds3-T}
{T} = \text{diag}(1,1,1,1,-1,1,\omega,\omega^2),
\end{equation}
where all rows and columns are ordered alphabetically, $A-H$.

There are three nontrivial particles invariant under $\mb{T}$: $B, D, E$. Remarkably, we find that
with the action of time-reversal symmetry given in \eqref{eqn:s3permutation}, we are forced
into the constraint that
\begin{align}
\label{etaTB}
\eta_B^\mb{T} = -1.
\end{align}
This result follows from a careful analysis of the time-reversal symmetry action on the topological data following the 
framework presented in Ref. \onlinecite{barkeshli2014SDG};  we leave the details of the derivation to 
Appendix \ref{sec:appDS3}.  From the fusion rule $B\times D=E$, together with Eq. (\ref{etaTB}), it follows that 
\begin{align}
\eta_{E}^\mb{T}=-\eta_{D}^\mb{T}. 
\end{align}
The remaining choice in the definition of the $\eta_a^\mb{T}$ values is 
\begin{align}
\eta_D^\mb T = \pm 1
\end{align}

This binary choice for $\eta_E^{\mb T} =-\eta^\mb{T}_D = \mp 1$, for this particular choice of time-reversal symmetry action
$\rho_{\mb T}$, is consistent with the cohomology classification of symmetry fractionalization
 $\mathcal{H}^2_{[\rho]}(\mathbb{Z}_2^{\mb T}, \mathcal{A})$.\cite{barkeshli2014SDG}
The Abelian group $\mathcal{A} = \mathbb{Z}_2$ in this case, because $B$ is the only Abelian anyon and it fuses with itself to 
the trivial particle. In this case, $\mathcal{H}^2_{[\rho]}(\mathbb{Z}_2^{\mb T}, \mathbb{Z}_2) = \mathbb{Z}_2$. The two cohomology
elements can be distinguished by two possible choices for the two-cocycle $\coho{w}(\mb{T,T}) = A$ or $B$. 
It can be shown\cite{barkeshli2014SDG} that $\eta^{\mb T}_E$ is determined by the mutual braiding
phase between $E$ and $\coho{w}(\mb{T,T})$, which is either one if $\coho{w}(\mb{T,T}) = A$, or $-1$ if $\coho{w}(\mb{T,T}) = B$. 

We now check the Dehn twist anomaly, according to the condition presented in Sec. \ref{dehnAnomaly}. 
There are three non-bosonic particles $E, G, H$. We compute the values of $M_a = \mathcal{Z}_a(\mathbb{RP}^2 \times S^1)$ 
using Eq. (\ref{Eq:Ma_RP2}). We find that $M_a = 0$ for all $a$ except the following:
\begin{align}
	M_{D} &=1+\eta_{D}^\mb{T}, 
\nonumber \\
M_E &=1-\eta_D^\mb{T}.
	\label{}
\end{align}
If $\eta_{D}^\mb{T}=-1$ then we find $M_{E} > 0$ and $\theta_{E}=-1$, indicating an anomaly.

Let us further compute the (3+1)D path integral $\Z(\mathbb{RP}^4)$ ($\mathcal{Z}(\mathbb{CP}^2) = 1$ because $c_- = 0$ for $\D(S_3)$ ). 
With the given data, we find
\begin{equation}
	\Z(\mathbb{RP}^4)=\eta_{D}^\mb{T}.
	\label{}
\end{equation}
Therefore, the anomalous SET with $\eta_{D}^\mb{T}=-1$ exists on the surface of a (3+1)D 
bosonic time-reversal SPT phase with $\mathcal{Z}(\mathbb{RP}^4) = -1$ and $\mathcal{Z}(\mathbb{CP}^2) = 1$. This is the same
(3+1)D SPT whose (2+1)D surface can also host the eTmT state. 

We also directly compute $\mathcal{Z}(\mathbb{RP}^2 \times S^1)$ for this $\D(S_3)$ SET 
using the path integral state sum presented in Sec. \ref{State_Sum_SETs}. Recall that without any
symmetry, the input into the associated TVBW path integral can be taken to be 
$\text{Rep}(S_3)$. Here $\text{Rep}(S_3)$ is the fusion category associated with the 3 
irreducible representations of $S_3$; these are the $A$, $B$, $C$ objects presented above. 
$\D(S_3)$ then coincides with the Drinfeld center of $\text{Rep}(S_3)$. 

For the SET state sum construction, the input into the path integral is a $G$-equivariant 2-category
with $G$ action. In the present context, where $G = \mathbb{Z}_2^{\mb T}$, we therefore need a 
type of $\mathbb{Z}_2^{\mb T}$ extension of $\text{Rep}(S_3)$. In other words, we need a category
\begin{align}
\mathcal{C}_{\mathbb{Z}_2} = \mathcal{C}_{\bf 0} \oplus \mathcal{C}_{\bf T}, 
\end{align}
where $\mathcal{C}_{\bf 0} = \text{Rep}(S_3)$. 

We find that the appropriate extension $\mathcal{C}_{\mathbb{Z}_2}$ which, when input into the SET state sum construction,
gives $\D(S_3)$ with the time-reversal symmetry fractionalization discussed above is the following. We take
$\mathcal{C}_{\mathbb{Z}_2}$ to correspond to the fusion category associated with SU(2)$_4$. This fusion category has 
5 simple objects, labeled by SU(2) spin $j = 0, \frac{1}{2}, 1, \frac{3}{2}, 2$. The integer spin objects
$j = 0,1,2$ correspond exactly to the three irreducible representations of $S_3$,
where $\{j = 0, j= 1, j=2\}$ is mapped to $\{A, B, C\}$. The two half-integer spin objects then form the $\mathcal{C}_\mb{T}$ part of the category. 

The $F$-symbols $F^{abc}_{def}(\mb{g}_0, \mb{g}_1, \mb{g}_2, \mb{g}_2)$, for $\mb{g}_0, \mb{g}_1, \mb{g}_2, \mb{g}_3 \in
\mathbb{Z}_2^\mb{T}$ of $\mathcal{C}_{\mathbb{Z}_2}$ are given in terms of the $F$-symbols of SU(2)$_4$. 
As discussed below Eq. (\ref{Fsym}), the symmetry of the $F$-symbols can be used to replace 
$F^{abc}_{def}(\mb{g}_0, \mb{g}_1, \mb{g}_2, \mb{g}_2)$ with $[F^{abc}_{def}]^{\sigma(\mb{g}_{01})}$, 
where we take the group action on the labels $a,b,c,d,e,f$ to be trivial. Recall that $\sigma(\mb{T}) = *$ and $\sigma(\mb{0}) = 1$. 
Since the $F$-symbols of SU(2)$_4$ can be chosen to all be real,\cite{Bonderson07b} we can ignore the complex conjugation $\sigma(\mb{g}_{01})$. 
The resulting $F$-symbols $F^{abc}_{def}$ for $\mathcal{C}_{\mathbb{Z}_2}$ can be taken to be the $F$-symbols of SU(2)$_4$. 

Carrying out the state sum explicitly using Eq. (\ref{Eq:RP2_State_Sum_Graded}), we indeed find $\mathcal{Z}(\mathbb{RP}^2\times S^1)=0$, 
in agreement with the above results using Eq. (\ref{ZnonOr1}).

\subsubsection{Gauged T-Pfaffian}

\begin{center}
\begin{table*}
	\begin{tabular}{|c|c|c|c|c|c|c|c|c|c|c|c|c|c|c|c|c|c|c|}
		\hline
		$a$ & $I_0$ & $\psi_0$ & $I_2$ & $\psi_2$ & $I_4$ & $\psi_4$ & $I_6$ & $\psi_6$ & $\sigma_1$ & $\sigma_3$ & $\sigma_5$ & $\sigma_7$ & $s_1$ & $s_3$ & $s_5$ & $s_7$ & $s\sigma_0$ & $s\sigma_2$\\
		\hline
		$\theta_a$ & $1$ & $-1$ & $-i$ & $i$ & $1$ & $-1$ & $-i$ & $i$ & $1$ & $-1$ & $-1$ & $1$ & $1$ & $-1$ & $-1$ & $1$ & $e^{\frac{i\pi}{4}}$ & $e^{-\frac{i\pi}{4}}$\\
		\hline
		$\overline{a}$ & $I_0$ & $\psi_0$ & $I_6$ & $\psi_6$ & $I_4$ & $\psi_4$ & $I_2$ & $\psi_2$ & $\sigma_7$ & $\sigma_5$ & $\sigma_3$ & $\sigma_1$ & $s_7$ & $s_5$ & $s_3$ & $s_1$ & $s\sigma_0$ & $s\sigma_2$\\
		\hline
		${}^\mb{T}a$ & $I_0$ & $\psi_0$ & $\psi_2$ & $I_2$ & $I_4$ & $\psi_4$ & $\psi_6$ & $I_6$ & $\sigma_1$ & $\sigma_3$ & $\sigma_5$ & $\sigma_7$ & $s_7$ & $s_5$ & $s_3$ & $s_1$ & $s\sigma_2$ & $s\sigma_0$\\
		\hline
		$\eta_a^\mb{T}$ & $1$ & $1$ &  &  & $-1$ & $-1$ &  &  & $\eta$ & $-\eta$ & $-\eta$ & $\eta$ & & & & & &\\
		\hline
	\end{tabular}
	\caption{Anyon types and time-reversal action in gauged T-Pfaffian$_{\eta}$, where $\eta = \pm 1$. We only list $\eta_a^\mb{T}$ for $\mb{T}$-invariant anyons.}
	\label{tab:TPfaffian}
\end{table*}
\end{center}

T-Pfaffian is a fermionic topological phase that can exist on the surface of $(3+1)$D electronic 
topological insulators~\cite{Bonderson13d, chen2014b}, preserving the $\mathrm{U}(1)\rtimes \mathbb{Z}_2^{\bf T}$ 
symmetry. There are actually two versions of T-Pfaffian states, which will be denoted by T-Pfaffian$_\pm$, differing 
in the assignment of $\{\eta^{\bf T}_a\}$ values. It has been shown in Ref. \onlinecite{metlitski2015}
%, exploiting the recently discovered duality between free Dirac fermions and QED$_3$~\cite{maxashvin, wangsenthil1}, 
that T-Pfaffian$_+$ can be realized on the surface of the free-fermion topological insulator (TI), while T-Pfaffian$_-$ has to exist on the 
surface of a more complicated state which can be thought of as stacking the free-fermion TI together with the eTmT bosonic SPT phase. 
One can further argue that if we ignore the $\mathrm{U}(1)$ symmetry, then the bulk of the free-fermion TI can be adiabatically connected 
to a trivial fermionic band insulator. This implies that it is possible to gauge the fermion parity in T-Pfaffian$_\pm$ without breaking 
time-reversal symmetry. We will focus on the resulting bosonic topological phases, which are called the gauged T-Pfaffian phases.

First let us briefly review quasiparticle content of the T-Pfaffian phases. We start from the 
$\text{Ising}\times \mathrm{U}(1)_{-8}$ state. Topological charges in $\mathrm{U}(1)_{-8}$ 
are labeled by an integer mod $8$, denoted by $[j]_8$, while the topological charges in Ising
have the standard labeling $\{I, \sigma, \psi\}$, where $I$ is the identity particle, $\psi$ is the fermion,
and $\sigma$ is the non-Abelian quasiparticle; these have the usual fusion rules
$\sigma \times \sigma = I + \psi$, $\psi \times \psi = 1$. 
We use the notation $a_j = (a, [j]_8)$, where $a = I, \sigma, \psi$. 
The quasiparticles of the T-Pfaffian phase are described  by a subset of $\text{Ising}\times \mathrm{U}(1)_{-8}$, 
restricted by the following rule: $\sigma$ can only be combined with odd $j$, while $I, \psi$ can be combined with even $j$. 
This way we get 12 quasiparticle types, and $\psi_4$ is identified as the electron. The $\mb{T}$ symmetry action can be 
almost determined completely from general considerations~\cite{Bonderson13d,chen2014b}, except the local 
$\mb{T}^2$ value of $\sigma_1$. We thus label the two choices as T-Pfaffian$_\eta$, where $\eta\equiv \eta_{\sigma_1}^\mb{T}$. 
See Table \ref{tab:TPfaffian} for a summary of the relevant datum.

In the gauged theory, we need to introduce 6 more topological charges with non-trivial mutual braiding 
statistics with the electron $\psi_4$. These are the fermion parity fluxes. We follow the notations of 
Ref.~\onlinecite{Bonderson13d} and denote them by $s_1, s_3, s_5, s_7, s\sigma_0, s\sigma_2$. 
 None of these topological charges are invariant under $\mb{T}$, as shown in Table \ref{tab:TPfaffian}.
%\begin{align}
%\rho_{\bf T}: &s_1 \leftrightarrow s_7, 
%\nonumber \\
%&s_3 \leftrightarrow s_5, 
%\nonumber \\
%& s\sigma_0 \leftrightarrow s\sigma_2.
%\label{}
%\end{align}

We can compute $M_a$, and we find all are zero except for
\begin{align}
M_{s_1}&=M_{s_7}=1+\eta 
\nonumber \\
M_{s_3}&=M_{s_5}=1-\eta.
\label{}
\end{align}
As stated above, we have $\theta_{s_3}=\theta_{s_5}=-1$. Therefore $\eta=-1$ results in Dehn twist anomaly on a 
M\"obius band. It is also interesting to notice that to reveal the anomaly from $M_a$ one needs to consider 
the case where $a$ is one of the fermion parity fluxes (i.e. $s_3$ or $s_5$); the original particles of the T-Pfaffian
do not by themselves reveal the anomaly. 

We also find
\begin{align}
\Z(\mathbb{RP}^4) &=\eta, 
\nonumber \\
\Z(\mathbb{CP}^2) &=1,
\end{align}
which implies that the gauged T-Pfaffian$_-$ state has the same anomaly as the eTmT state. 

Since gauged T-Pfaffian$_+$ is free of any anomaly, we can ask whether it can be realized using a state-sum construction. 
Unfortunately this appears to not be possible, and thus the gauged T-Pfaffian$_+$ presents an example of a time-reversal (or reflection)
symmetric SET which lies outside of the state-sum construction discussed in Sec. \ref{State_Sum_SETs}.\footnote{We thank Max Metlitski for point this out to us.} 
In fact, even in the absence time-reversal symmetry, this topological order itself can not be realized in any kind of quantum double/string-net model. 

The reason is that if a topological order can be realized through such models, it must admit a gappable boundary.\cite{kong2014, kitaev2012} Although 
gauged T-Pfaffian$_+$ has zero chiral central charge, $c_- = 0$, the (non-chiral) edge modes are still not gappable.
To see this, let us condense the Abelian boson $I_4$, using the rules for topological Bose condensation discussed in Ref. \onlinecite{bais2009}. 
After $I_4$ condenses, $\sigma_1,\sigma_3,\sigma_5,\sigma_7,s_1,s_3,s_5,s_7$ are confined, and we have the following identification
\begin{equation}
I_k \sim I_{[k+4]_8}, \;\;\; \psi_k \sim\psi_{[k+4]_8}, \quad k=0,2,4,6
\label{}
\end{equation}
Therefore the original Abelian sector reduces to $I_0, I_2, \psi_0, \psi_2$. The $s\sigma_0, s\sigma_2$ are the 
fixed-points of $I_4$, and thus each of them has to split into two Abelian anyons $s\sigma_{0}^\pm$ and 
$s\sigma_{2}^\pm$. It is not hard to see that the resulting theory is $\mathrm{U}(1)_{-2}\times\mathrm{U}(1)_4$.
This theory does not admit a gapped boundary according to the criterion proven in Ref.~\onlinecite{levin2013}. 
As a result, the gauged T-Pfaffian$_+$ topological order does not admit gapped boundary either.

%% file: parity-eigenvalue.tex
\section{Derivation of reflection eigenvalue in (1+1)D SPT states}
\label{ReigenvalueAppendix}

Ground states of (1+1)D translationally invariant gapped systems can
be efficiently approximated by a matrix product state (MPS) representation:
\begin{equation}
	\ket{\Psi}=\sum_{i_1, \cdots, i_L} B^\mathsf{T}A^{i_1}A^{i_2}\cdots A^{i_L}B\ket{i_1,i_2,\cdots, i_L}.
	\label{}
\end{equation}
Here $i_j$ labels basis states of the local Hilbert space at site $j$, and $A^i$ is a $d\times d$ matrix. The $d$-dimensional vector $B$
determines the boundary condition. We shall follow the convention of Ref. \onlinecite{PollmannTurner} and represent the matrix $A$ as a product of a 
$d\times d$ matrix $\Gamma$ and a positive real diagonal matrix $\Lambda$: $A^i=\Gamma_i \Lambda$, such that
it is already in the canonical form (so that $\Lambda$ contains the Schmidt eigenvalues). Since we consider a 
short-ranged correlated state, we assume that the MPS is injective. This representation is most convenient in 
infinite systems, but we will use it to frame our discussion of a finite chain. Our derivation closely follows the 
calculation of partial reflection in Ref. \onlinecite{PollmannTurner}; it is also closely related to the discussion in 
Ref. \onlinecite{zaletel2015}. 

On a finite chain with $N$ sites, a reflection-symmetric MPS requires
\begin{equation}
	\Gamma_j^\mathsf{T} = e^{i\theta} U_j^\dag \Gamma_{N-j+1} U_j,
	\label{}
\end{equation}
for a unitary matrix $U_j$. In Ref. \onlinecite{pollmann2010, ChenPRB2011a} it was shown that
\begin{equation}
	U U^*= \pm \mathds{1},
	\label{}
\end{equation}
where the minus sign indicates a nontrivial SPT phase.

\begin{figure}[t!]
	\centering
	\includegraphics[width=\columnwidth]{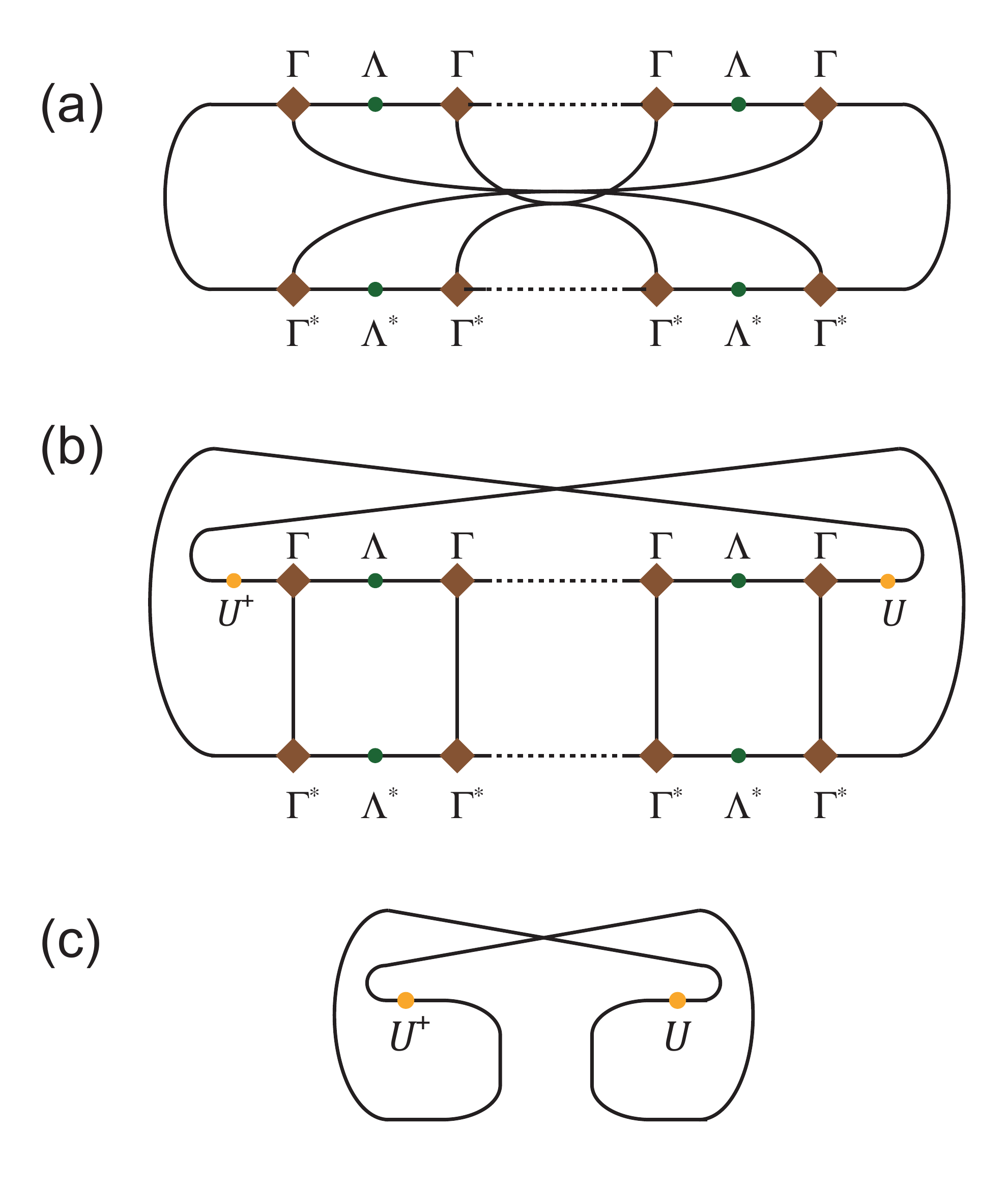}
	\caption{Diagrammatic evaluation of the trace of the reflection operator.}
	\label{fig:diagramR}
\end{figure}

The evaluation of the trace of the reflection operator, which consists of summing over all boundary conditions $B$, 
can be represented in the diagram shown in Fig \ref{fig:diagramR}(a). We can then reverse the top chain to untwist 
the diagram, which introduces factors of $U$ and $U^\dag$ accumulating at the ends, see Fig. \ref{fig:diagramR}(b). 
The ladder in the bulk of the diagram is a product of many copies of the $d^2 \times d^2$ transfer matrix 
$T = \sum_i A^i \otimes (A^i)^*$, such that $T_{\alpha \beta, \gamma \delta} = \sum_i A^i_{\alpha\gamma} (A^i_{\beta \delta})^*$. 
For a long chain, due to injectivity the effect is to project onto the unique eigenvector of the transfer matrix 
with the largest eigenvalue (which we can take to be unity). In the canonical form, we can take the
eigenvector to be proportional to $\delta_{\alpha\beta} \delta_{\gamma \delta}$, where $\alpha,\beta,\delta,\gamma = 1, \cdots, d$.\cite{MPS} 
This allows a futher simplification of the diagram to Fig. \ref{fig:diagramR}(c). At this point we can directly read off the desired result:
\begin{equation}
	\Tr \,R_\mb{r} = \Tr\, U^\mathsf{T} U^\dag = \pm \Tr\, \mathds{1}=\pm d.
	\label{}
\end{equation}

We also observe that the derivation applies to the case of a fixed boundary condition as well. Denote the state by 
$|\Psi_B\rangle$. It is amount to insert a projector to $|B\rangle\langle B|$ into the two outermost lines in the 
evaluation of the trace, so in the end we have
\begin{equation}
	\langle\Psi_B|R_\mb{r}|\Psi_B\rangle = \langle B| U^\mathsf{T} U^\dag|B\rangle = \pm \langle B| \mathds{1}|B\rangle=\pm 1.
	\label{}
\end{equation}

%% file: 1dTQFT.tex
\section{(1+1)D TQFTs}
\label{1dTQFTSec}

Some basic properties of (1+1)D TQFTs were reviewed in Sec. \ref{1dTQFTsec}. Here,
we briefly review some additional aspects of (1+1)D TQFTs, from the point of view of extended TQFTs 
and category theory. While this perspective utilizes several mathematical results,
the subsequent computations do not require familiarity with these theorems. 
This perspective provides us a simple way to generalize to unoriented (1+1)D TQFTs 
and to compute the path integral on non-orientable surfaces. 

A (1+1)D extended TQFT assigns a vector space $\V(I; a, b)$ to the interval $I$, with boundary conditions $a$ and $b$
associated to the left and right boundary. The boundary conditions $a$ and $b$ are chosen from a given set of 
allowed boundary conditions $\{a,b,\cdots\}$. The (1+1)D TQFT is further equipped with a ``gluing map''
\begin{align}
g: \V(I; a,b) \otimes \V(I; b, c)\rightarrow \V(I; a,c) .
\end{align}
We use the notation 
\begin{align}
e_\alpha \cup e_\beta \equiv g(e_\alpha, e_\beta)
\end{align}
for $e_\alpha \in \V(I; a, b)$ and $e_\beta \in \V(I; b,c)$. 

This provides us the structure of a 1-category $\mathcal{C}$, where the objects of $\mathcal{C}$ are the set of boundary conditions, 
the one-morphisms are the vector spaces $\V(I;a,b)$, and the composition of morphisms is the gluing map. 
Note that the morphisms form a vector space over complex numbers,
which implies that $\mathcal{C}$ is a $\mathbb{C}$-linear $1$-category.

The (1+1)D TQFT also assigns a vector space $\V(S^1)$ to the circle. \footnote{$\V(S^1)$ is the degree zero Hochschild homology of the 1-category, although this is not important for the present discussion.}
%There is a ``cutting'' map:
%\begin{align}
%k: \V(S^1) \rightarrow \V(I; a,a) \otimes \V(I; a, a),
%\end{align}
%associated with ``cutting'' the the circle into two intervals, and also 
We have the gluing map to $\V(S^1)$:
\begin{align}
\tilde{g}: \V(I; a,a) \rightarrow \V(S^1) . 
\end{align}
We will also extend the notation
\begin{align}
e_\alpha \cup e_\beta \in \V(S^1 \sqcup S^1),
\end{align}
for $e_\alpha, e_\beta \in \V(S^1)$. 

Thus, we see that the classification of (1+1)D TQFTs is closely related to the classification of 1-categories. 
It is a theorem that the $1$-categories of interest to us can all be related to a direct sum of trivial 
categories.\footnote{More specifically, since we are interested in cases where $\V(I; a,b)$ and $\V(S^1)$ 
are finite vector spaces, the 1-category $\mathcal{C}$ is a $\mathbb{C}$-linear semi-simple 1-category. 
Moreover, one can show that two 1-categories that are Morita equivalent give rise to the same 
(1+1)D TQFT. It is a well-known theorem that $\mathbb{C}$-linear semi-simple 1-categories 
are Morita equivalent to a direct sum of trivial 1-categories, which contain a single object and a 
single morphism.}
This means that we can take $\mathcal{C}$ to consist of a single object $a$, and $n$ morphisms 
from that object into itself. In this case, since there is only one object, we can drop the boundary condition 
label in $\V(I; a,a)$ and write simply $\V(I)$. We have:
\begin{align}
\text{dim } \V(I) = \text{dim } \V(S^1) = n, 
\end{align}
with each morphism corresponding to a state in $\V(I)$. We let $e_\alpha$ form a basis of states in $\V(I)$ . 
$\tilde{g}(e_\alpha)$ therefore forms a basis in $\V(S^1)$. In what follows, we will sometimes drop the
$\tilde{g}$ in the notation and keep it implicit. We further have the
important requirement that, for $e_\alpha , e_\beta \in \mathcal{V}(I)$,
\begin{align}
\label{orthoMorphisms}
e_\alpha \cup e_\beta = \delta_{\alpha \beta} e_\alpha.
\end{align}
Eq. (\ref{orthoMorphisms}) implies that $e_\alpha \in \V(I)$ are idempotents, because 
they square to themselves under the gluing map.

A (1+1)D TQFT assigns a complex number, the path integral $\Z(\Sigma^2)$, to 
closed surfaces $\Sigma^2$. For surfaces with boundary, $\Z(\Sigma^2)$ is a map:
\begin{align}
\Z(\Sigma^2): \V( \partial \Sigma^2) \rightarrow \mathbb{C}. 
\end{align}
As we explain, it suffices to specify the path integral on a disk:
\begin{align}
\Z(D^2): \V(S^1) \rightarrow \mathbb{C} . 
\end{align}
We define:
\begin{align}
\Z(D^2)[e_\alpha] = \lambda_\alpha ,
\end{align}
where $\lambda_\alpha$, for $\alpha = 1,\cdots, \text{dim } \V(S^1)$, are the parameters of the (1+1)D TQFT. 
$\Z(D^2)$ defines an inner product on $\V(I)$ as follows:
\begin{align}
\langle e_\alpha | e_\beta \rangle_{\V(I)} \equiv \Z( D^2) [e_\alpha \cup e_\beta ] = \lambda_\alpha \delta_{\alpha \beta} . 
\end{align}

\begin{figure}
\includegraphics[width=3.0in]{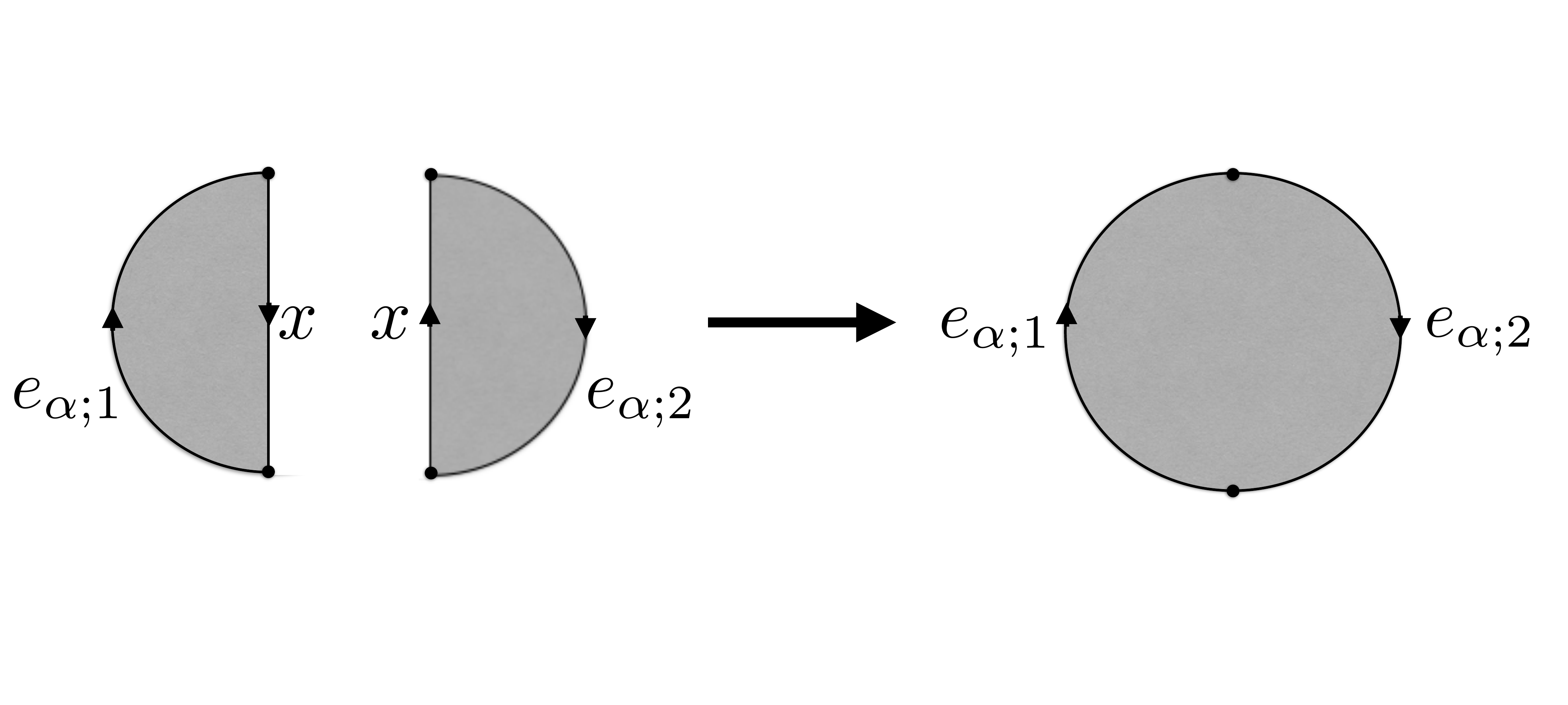}
\caption{Gluing two disks along an interval to obtain a single disk. }
\label{glueDiskFig}
\end{figure}

Path integrals for TQFTs satisfy an important gluing formula. Let us consider a surface $\Sigma^2_{\text{cut}}$ which,
when glued along a one-manifold $M^1$, gives rise to a surface $\Sigma^2$. Fig. \ref{glueDiskFig} shows an example where 
$\Sigma^2 = D^2$, $\Sigma^2_{\text{cut}} = D^2 \bigsqcup D^2$, and $M^1 = I$. The gluing formula is
\begin{align}
\Z(\Sigma^2)[ e_\alpha ] = \sum_{x \in \V(M^1)} \frac{\Z(\Sigma^2_{\text{cut}})[e_{\alpha;1} \cup x \cup e_{\alpha;2}  \cup x ]}{\langle x |x \rangle_{\V(M^1)}} . 
\end{align}
Here, $e_\alpha \in \V(\partial \Sigma^2)$, $\sum_{x\in \V(M^1)}$ is a sum over orthogonal states $x \in \V(M^1)$, and 
$e_{\alpha;1} \cup x \cup e_{\alpha;2}  \cup x$ is a state in $\V(\partial \Sigma^2_{\text{cut}})$. 
Importantly, the gluing maps imply 
\begin{align}
e_\alpha = e_{\alpha;1} \cup e_{\alpha;2}  .
\end{align}

\begin{figure}
\includegraphics[width=3.0in]{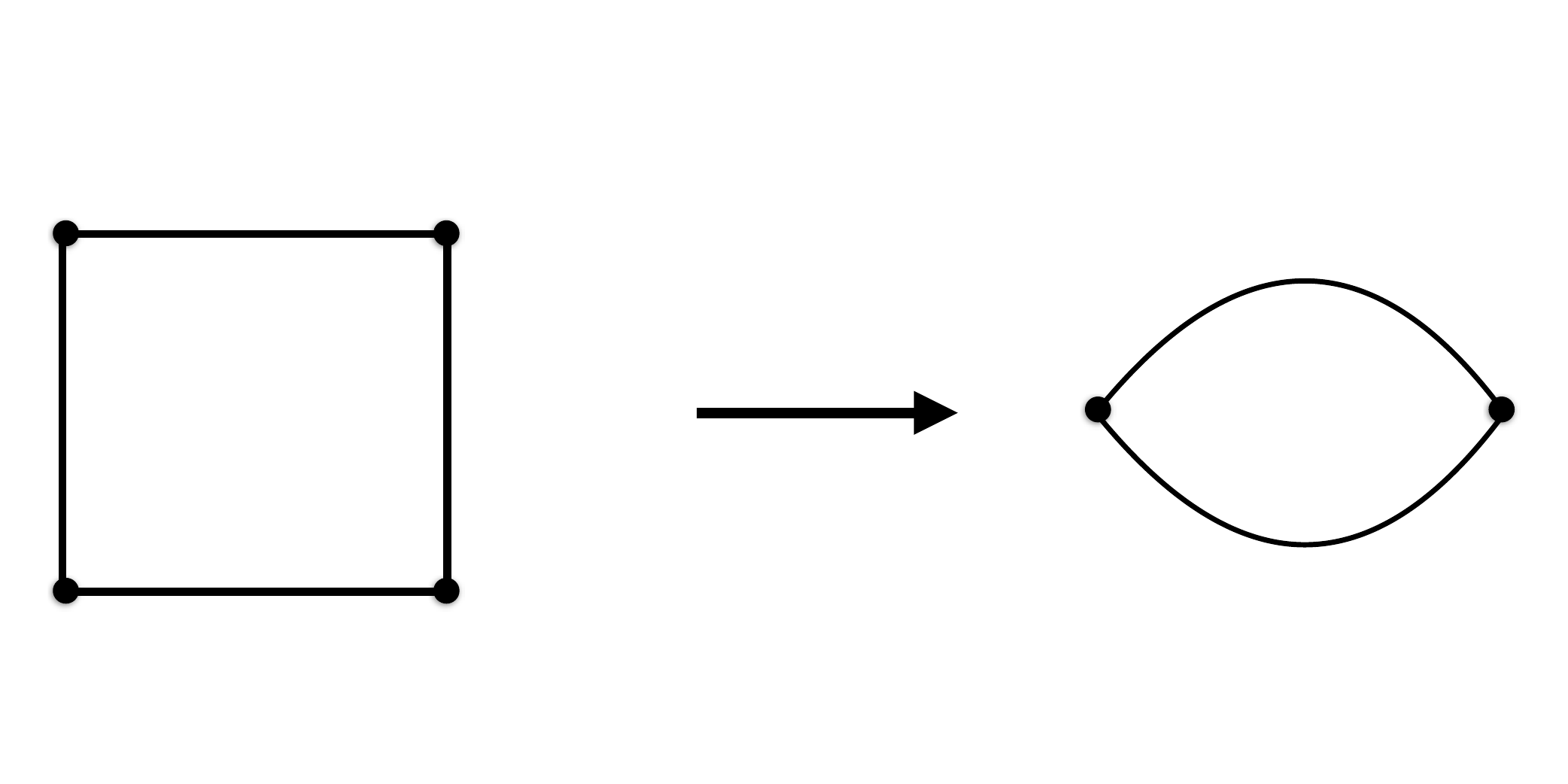}
\caption{Left: $I \times I$. Right: Boundary is ``crushed'' as explained in the text. }
\label{crushFig}
\end{figure}

The inner product on $\V(M^1)$ can be defined in terms of the path integral as
\begin{align}
\langle x | y \rangle_{\V(M^1)} = \Z(M^1 \times I)[x \cup y].
\end{align}
In this context, we define $M^1 \times I$ by ``crushing'' the boundary, that is, by identifying points 
$(b, t) \sim (b, s)$ for $b \in \partial M^1$ and $s,t \in I$ (see Fig. \ref{crushFig} for an illustration for the case $M^1 = I$). 
Thus, by this definition, $\partial (M^1 \times I) = M^1 \cup \overline{M}^1$. 

\subsection{Path integral on oriented closed surfaces}

From $\Z(D^2)$ and the gluing formula, we can determine the path integral on any surface $\Sigma^2$. The
result is:
\begin{align}
\label{z2dOriented}
\Z(\Sigma^2) = \sum_\alpha \lambda_\alpha ^{\chi(\Sigma^2)},
\end{align}
where $\chi(\Sigma^2)$ is the Euler characteristic of $\Sigma^2$. 

Below we will derive this from the gluing formula. 
As a warmup, let us first consider the following applications of the gluing formula. 
First, let us consider gluing a disk to itself along an interval, to obtain an annulus:
\begin{align}
\label{annulusGlue}
\Z(S^1 \times I)[e_\alpha \cup e_\beta] &= \sum_{x \in \V(I)} \frac{\Z(D^2)[e_\alpha \cup x \cup e_\beta \cup x]}{\langle x | x \rangle_{\V(I)}}
\nonumber \\
&= \sum_{x \in \V(I)} \delta_{\alpha x} \delta_{x \beta} \frac{\lambda_\alpha}{\lambda_x } 
\nonumber \\
&= \delta_{\alpha \beta}. 
\end{align}
Here, we used the fact that $\langle x | x \rangle_{\V(I)} = \Z(D^2)(x \cup x) = \lambda_x$, and
the fact that $e_\alpha \cup x \cup e_\beta \cup x = \delta_{\alpha x} \delta_{x \beta} e_\alpha$. 

Next, let us consider gluing an annulus to a disk to obtain a disk:
\begin{align}
\Z(D^2)[e_\alpha] = \sum_{e_\beta} \frac{\Z(S^1 \times I)[e_\alpha \cup e_\beta] \Z(D^2)[e_\beta]}{\langle e_\beta | e_\beta \rangle_{\V(S^1)} }
\end{align}
Observe that 
\begin{align}
\label{S1inner}
\langle e_\beta | e_\beta \rangle_{\V(S^1)} = \Z(S^1 \times I)[e_\beta \cup e_\beta] . 
\end{align}
Using Eq. (\ref{annulusGlue}) and (\ref{S1inner}) then gives $\Z(D^2)[e_\alpha] = \Z(D^2)[e_\alpha]$, as expected. 

\begin{figure}
\includegraphics[width=3.0in]{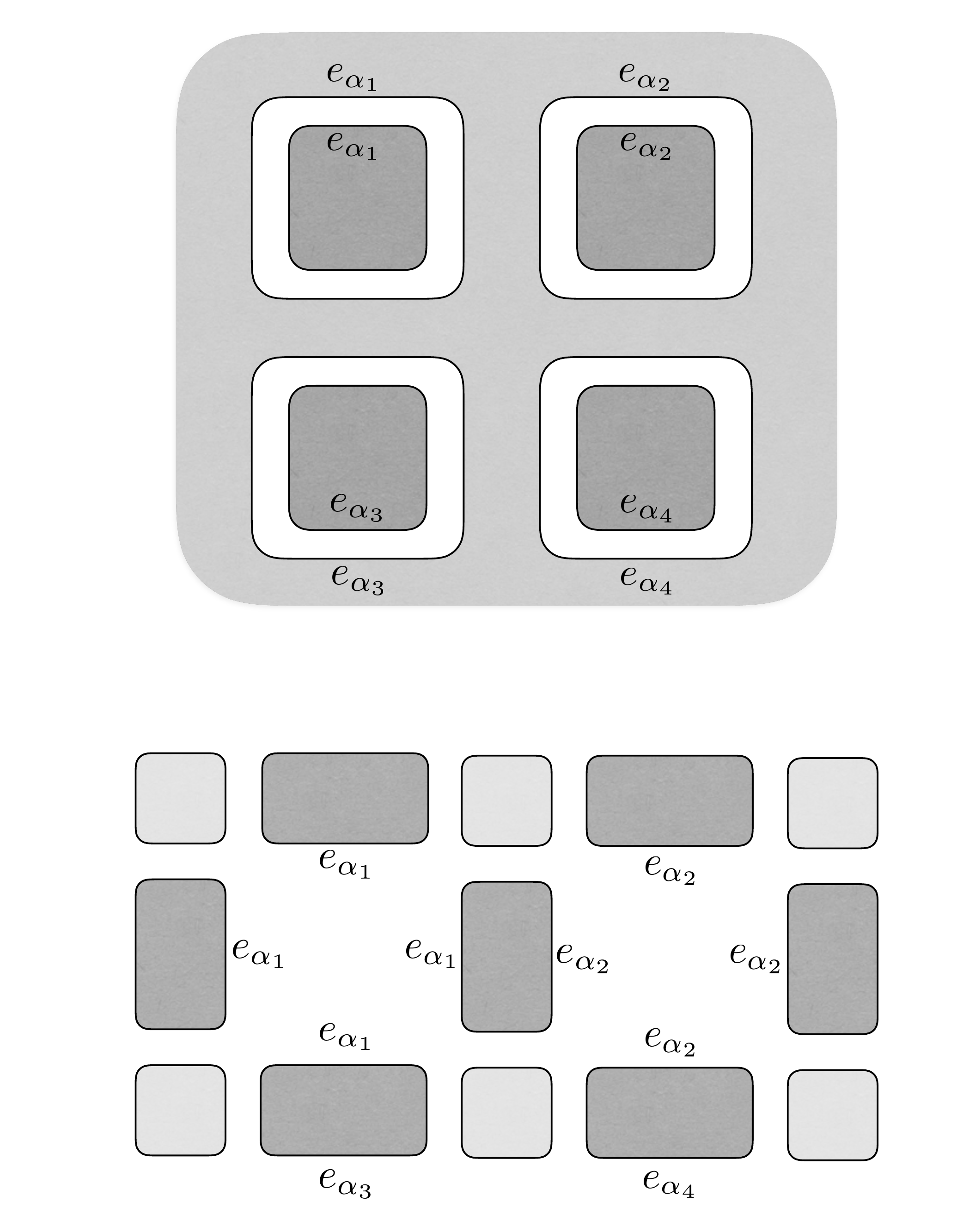}
\caption{Top: Gluing $2$-cells (shaded dark disks) to the $1$-skeleton. The boundary conditions $e_{\alpha_i}$ are labelled. Bottom:
Gluing the $1$-cells to the $0$-cells, with some of the boundary conditions labelled. }
\label{cellFig}
\end{figure}

To derive (\ref{z2dOriented}), we pick a handle decomposition for $\Sigma^2$. This corresponds to 
taking a triangulation of $\Sigma^2$ and thickening the vertices, edges, and triangles all to disks. 
The thickened vertices form the $0$-cells and the thickened edges form the $1$-cells, which are glued 
to the $0$-cells along two disjoint intervals, forming the $1$-skeleton
$\Sigma^2_1$. $\Sigma^2$ is then formed by gluing the faces of the triangulation, which are the $2$-cells, to
$\Sigma^2_1$. For an illustration, see Fig. \ref{cellFig}. 

We can compute $\Z(\Sigma^2)$ by using this handle decomposition together with the gluing formula. 
First we consider gluing the $2$-cells to $\Sigma^2_1$ (see Fig. \ref{cellFig}):
\begin{align}
\Z(\Sigma^2) = \sum_{e_\alpha^{(N_2)} \in \V(S^1)^{N_2}} \frac{\Z(\Sigma^2_1)[e_\alpha^{(N_2)}] \Z(\text{2-cells})[e_\alpha^{(N_2)}]}{\langle e_\alpha^{(N_2)} | e_\alpha^{(N_2)} \rangle_{\V(S^1)^{N_2}}} .
\end{align}
Here, $N_2$ is the number of 2-cells (i.e. the number of faces of the triangulation), and 
$e_\alpha^{(N_2)} = \bigcup_i e_{\alpha_i}$ is a state in $\V(S^1)^{N_2}$, with $e_{\alpha_i} \in \V(S^1)$ and $i = 1,\cdots, N_2$. 
We have 
\begin{align}
\Z(\text{2-cells})[e_\alpha^{(N_2)}] = \prod_i \lambda_{\alpha_i}. 
\end{align}
Moreover, 
\begin{align}
\langle e_{\alpha}| e_{\beta} \rangle_{\V(S^1)} &= \Z(S^1 \times I)[e_{\alpha} \cup e_{\beta}]  
\nonumber \\
&= \delta_{\alpha \beta} 
%\frac{ \Z(D^2)[e_\alpha]}{\langle e_\alpha| e_\alpha \rangle_{\V(I)} }
%\nonumber \\
%&= 1 ,
\end{align}
where in the second line we used Eq. (\ref{annulusGlue}). %the gluing formula. 
Therefore,
\begin{align}
\Z(\Sigma^2) = \sum_{e_\alpha^{(N_2)} \in \mathcal{V}(S^1)^{N_2}} \Z(\Sigma^2_1)[e_\alpha^{(N_2)}] \prod_{i=1}^{N_2} \lambda_{\alpha_i}  .
\end{align}

Applying the gluing formula again, it is straightforward to see that $\Z(\Sigma^2_1)[e_\alpha^{(N_2)}] = 0$ unless
all $e_{\alpha_i}$ are equal to each other: $e_{\alpha_i} = e_\alpha$ (see Fig. \ref{cellFig}). This follows from 
Eq. (\ref{orthoMorphisms}). Thus:
\begin{align}
\label{1cellGluing}
\Z(\Sigma^2_1)[e_\alpha^{(N_2)}] = 
\sum_{e_\alpha \in \mathcal{V}(S^1)} \frac{\Z(\Sigma^2_0)[e_\alpha^{(N_0)}] \Z(\text{1-cells})[e_\alpha^{(N_1)}]}{\langle e_\alpha | e_\alpha \rangle_{\V(I)}^{2N_1}} .
\end{align}
Here, $\Sigma^2_0$ is the collection of $0$-cells and $e_\alpha^{(N_0)} = \bigcup_{i=1}^{N_0} e_\alpha$ is the union of the boundary conditions
for $\Sigma^2_0$. Similarly, $N_1$ is the number of $1$-cells and $e_{\alpha}^{(N_1)} = \bigcup_{i=1}^{N_1} e_{\alpha}$ is the union
of boundary conditions for the collection of $1$-cells. 

Thus $\Z(\Sigma^2_0)[e_\alpha^{(N_0)}] = \lambda_\alpha^{N_0}$ and $\Z(\text{1-cells})[e_\alpha] = \lambda_\alpha^{N_1}$. 
The denominator of Eq. (\ref{1cellGluing}) is raised to the $2 N_1$ power because each $1$-cell is glued to the 
$0$-skeleton along two disjoint intervals (see Fig. \ref{cellFig}). Putting
everything together we get Eq. (\ref{z2dOriented}), where the Euler characteristic is determined by
\begin{align}
\chi(\Sigma^2) = N_0 - N_1 + N_2 .
\end{align}

\subsection{Unoriented (1+1)D TQFTs}

Now let us proceed to generalize the above discussion to the case of unoriented (1+1)D TQFTs, 
which we can define on non-orientable manifolds. In the unoriented case, the (1+1)D TQFT possesses
a reflection map:
\begin{align}
{\bf r}: \V(I)\rightarrow \V(I) ,
\end{align}
such that ${\bf r}^2$ is the identity on $\V(I)$. The reflection map also acts on elements of $\V(S^1)$. 

One can prove\cite{walker2016} that, up to the appropriate flavor of
Morita equivalence, we may assume that ${\bf r}$ merely permutes the idempotents 
\begin{align}
{\bf r} (e_\alpha) = e_{{\bf r}(\alpha)} , 
\end{align}
for $e_\alpha \in \V(I)$. Therefore, either $e_\alpha$ is fixed under ${\bf r}$: ${\bf r} (e_\alpha) = e_\alpha$, 
or two idempotents $e_\alpha$ and $e_\beta$ are mapped to each other under ${\bf r}$. 

With the action of reflection, we can now define the path integral on a non-orientable manifold via
the gluing formula. Importantly, to obtain the path integral on a non-orientable manifold, we must pick a 
one-dimensional manifold along which we apply an orientation reversal, using the action of ${\bf r}$. 

Proceeding through the gluing constructions, it is immediately clear that, for non-orientable $\Sigma^2$,
only those boundary conditions $e_\alpha$ such that ${\bf r} (\alpha) = \alpha$ can contribute to the sum.
(For unoriented but orientable $\Sigma^2$, all $e_\alpha$ contribute to the sum.)

Therefore, for a generic (1+1)D TQFT described by a set of complex numbers $\{\lambda_i\}$ and the action
of reflection, we have
\begin{align}
\Z(\Sigma^2) = \sum_{\alpha | \alpha = {\bf r}(\alpha)} \lambda_\alpha^{\chi(\Sigma^2)} . 
\end{align}

%% file: 4dPF.tex
\section{Handle decompositions of manifolds}
\label{handleSec}

\begin{figure}
\includegraphics[width=3.5in]{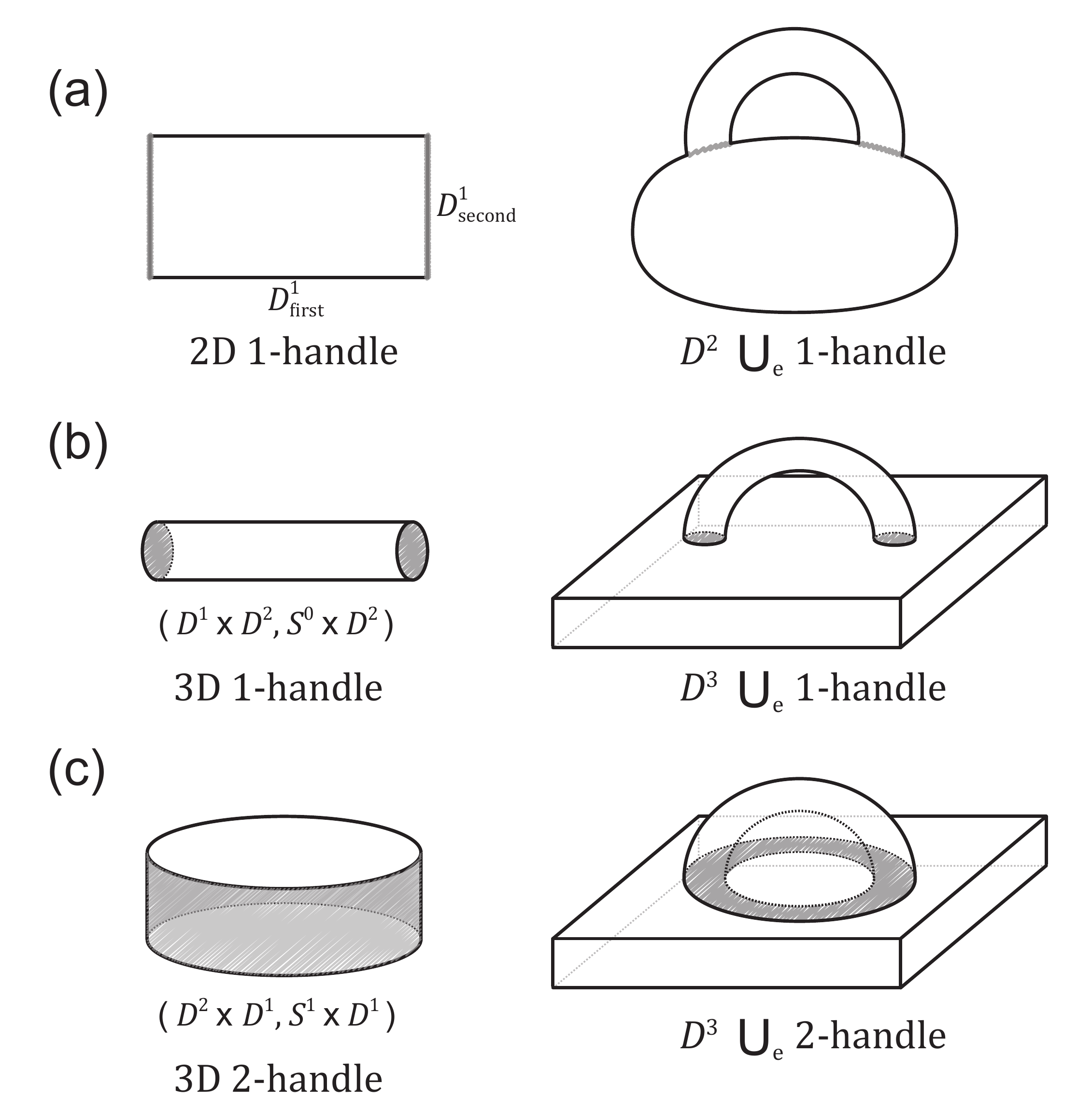}
\caption{ Examples of handles in various dimensions, with the
  attaching regions shown in bold. }
\label{handlesFig}
\end{figure}

A $d$-dimensional $p$-handle is a pair $(D^p \times D^q, (\partial D^p) \times D^q)$, with $p+q = d$.
The submanifold $(\partial D^p) \times D^q \subset \partial(D^p\times D^q)$ is called the ``attaching region" of the $p$-handle.
We can think of a $d$-dimensional $p$-handle as a $p$-cell which has been thickened up to be $d$-dimensional.
%$d$-dimensional ball, which is convenient to write as
%$D^p \times D^{q}$, with $p + q = d$. It has a distinguished subset referred to as the ``attaching region'' on its boundary:
%\begin{align}
%\text{d-dimensional p-handle } h: (D^p \times D^{q}, \partial D^p \times D^q )
%\end{align}
The non-attaching region is $D^p \times \partial D^q$. Given a $d$-manifold with boundary $(M, \partial M)$, 
we attach $h$ by embedding $\partial D^p \times D^q \overset{e}{\hookrightarrow}\partial M$. 
The effect on the boundary is $\partial M \rightarrow (\partial M \setminus e(\partial D^p \times D^q)) \cup (D^p \times \partial D^q)$. 
See Figs. \ref{handlesFig} for illustration. 

\begin{figure}
\includegraphics[width=1.5in]{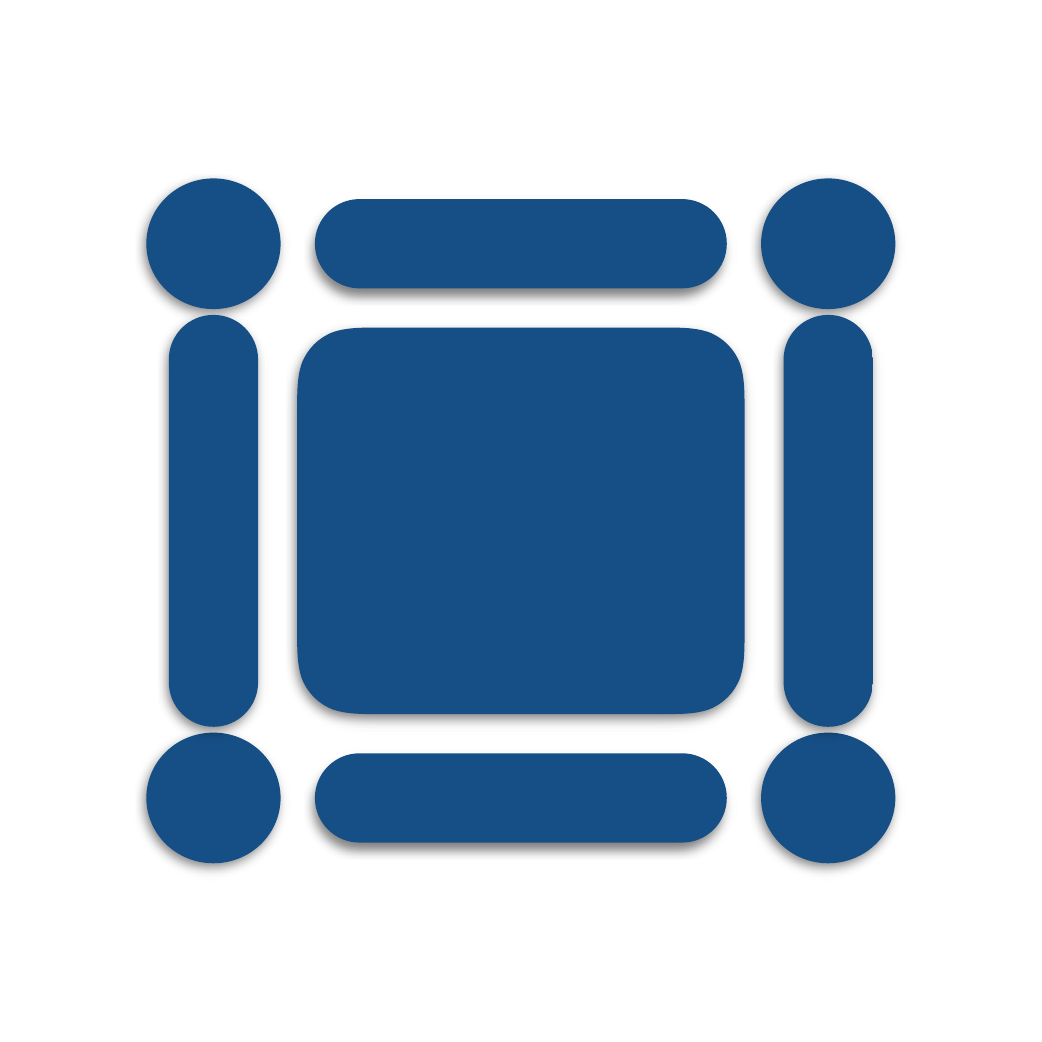}
\caption{A cell decomposition can be viewed built from handle attachments as illustrated. The small circles correspond to 
0-handles, the edges connecting them to 1-handles, and the larger square corresponds to a 2-handle. }
\label{cellManifold}
\end{figure}

Every manifold has a handle decomposition. For example, one can take a cellulation of
a $d$-dimensional manifold, and thicken the $0$-cells into $d$-balls $D^d$. Next,
one can take the $1$-cells and thicken them to $d$-balls as well, and glue them to 
the one cells along two points, $S^0 \times D^{d-1}$. The $2$-cells can be thickened to 
$d$-balls, and glued to the $0$- and $1$- cells along $S^1 \times D^{d-2}$, and so on. 
See Fig. \ref{cellManifold} for an illustration. 

\subsection{Handle decomposition of $\mathbb{CP}^2$}

$\mathbb{CP}^2$ can be obtained by gluing a $0$-handle, a $2$-handle, and a $4$-handle together. Since the $2$-handle is glued 
along $S^1 \times D^2$, one must choose a circle $S^1$ in the boundary of the $0$-handle along which to glue the $2$-handle.
The key property of $\mathbb{CP}^2$ is that this circle has a $+1$ framing. In other words, if we thicken the $S^1$ of the attaching
region to a ribbon, this ribbon twists around itself once along the $S^1$. 

The $+1$ framing ensures that the signature of the resulting manifold
is $+1$ . The $2$-handle introduces a second homology class to the manifold: the 2-handle introduces a 
sphere $S^2$ inside of the resulting manifold which is not the boundary of any $3$-manifold that is also inside of the 
manifold. 

\subsection{Handle decomposition of $\mathbb{RP}^4$}

\begin{figure}
\includegraphics[width=3.0in]{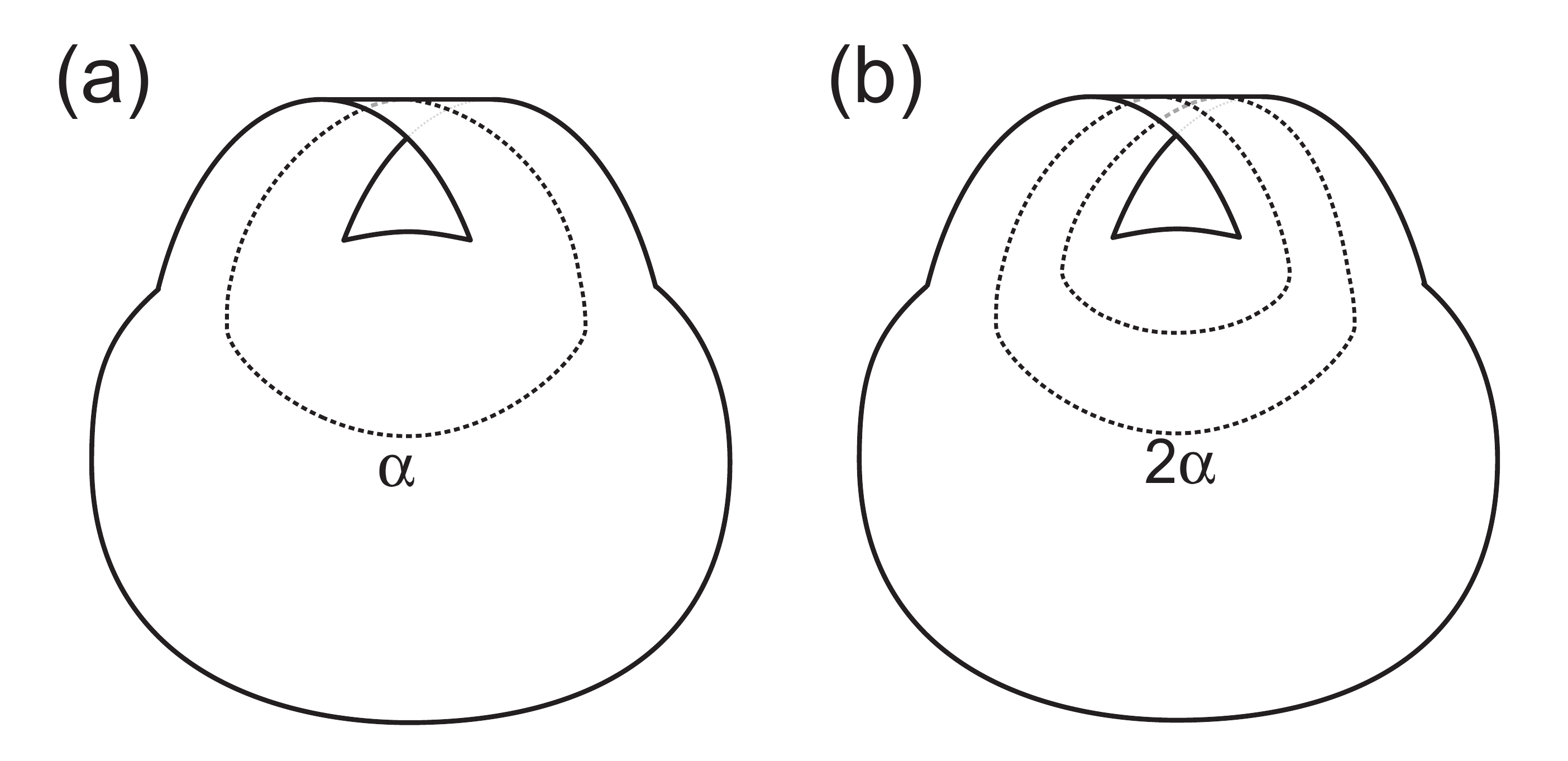}
\caption{The loops $\alpha$ and $2\alpha$ on the Mobius band}
\label{mobiusFig}
\end{figure}

$\mathbb{RP}^4$ can be obtained by gluing together one handle of every index. First, one glues the $1$-handle
to the $0$-handle, along an $S^0 \times D^3$ of the $1$-handle. Importantly, one of the $D^3$'s is
glued with an orientation reversal. The result is $\text{Mb} \times D^2$, where $\text{Mb}$ refers to the Mobius
band. This attachment is illustrated in Fig. \ref{mobiusFig}. 

Next, we attach the $2$-handle along an $S^1 \times D^2$. To see which $S^1$, observe that attaching a 
$2$-handle to a particular circle renders that circle contractible, and thus a trivial element of the
first homology of the resulting manifold. The previous $1$-handle introduced a first homology 
to the manifold, generated by a loop that we denote $\alpha$; in order to ensure that 
$H_1(\mathbb{RP}^4 ; \mathbb{Z}) = \mathbb{Z}_2$, we thus glue the $2$-handle along $2\alpha$, shown in Fig. \ref{mobiusFig}.
Moreover, we must attach the $2$-handle to $2\alpha$ with a $+1$ framing, similar to the case of $\mathbb{CP}^2$.
(We want the self-intersection of the $\mathbb{RP}^2$ inside $\mathbb{RP}^4$ to be 1.)

Finally, the $3$-handle is attached along $S^2 \times D^1$ and the manifold is closed off with a 
$4$-handle attached along the final boundary, $S^3$.

%% file: string.tex
\section{String operators and their transformations under reflection/time-reversal symmetries}
\label{sec:string}
\begin{figure}
\includegraphics[width=3.0in]{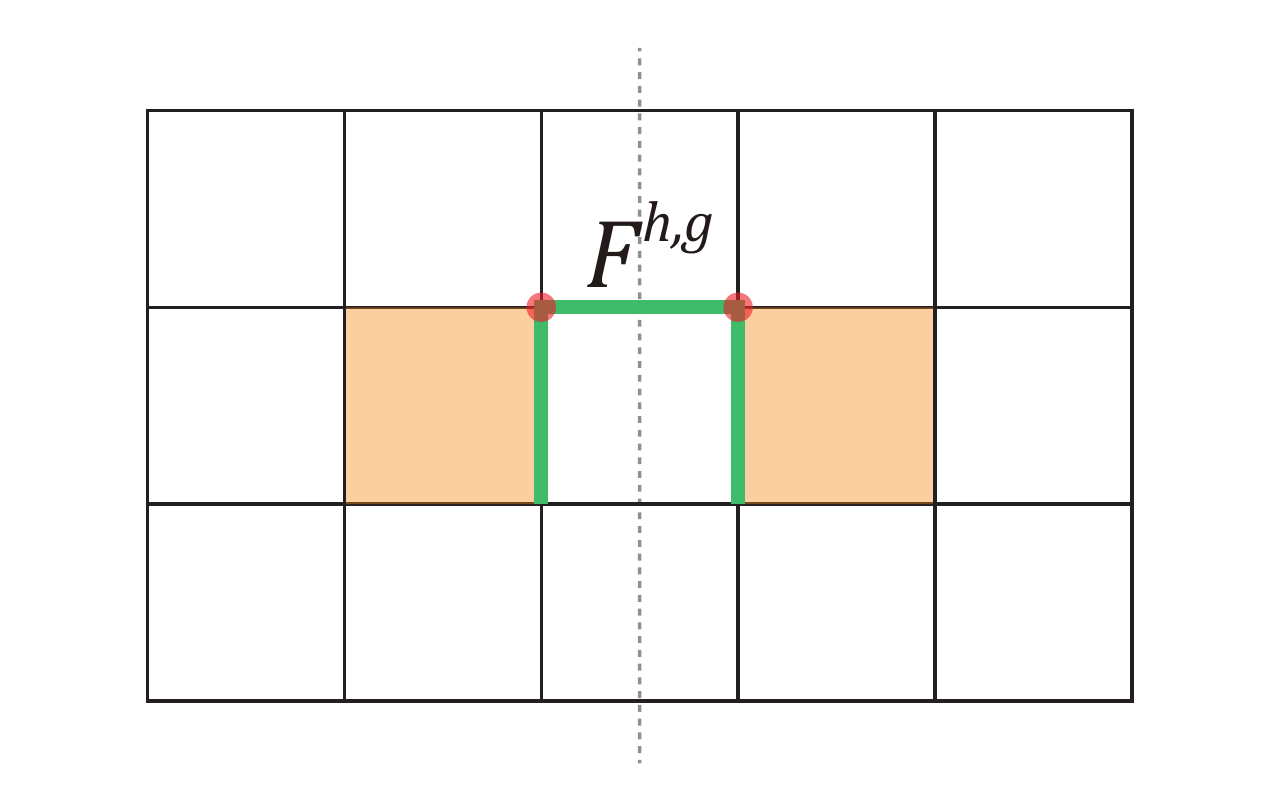}
\caption{The (minimal) string operator $F^{h,g} $ acts only on the green edges. When apply $F^{h,g} $ to the ground state, it creates a vertex (red) and a plaquette (orange) excitation on each end of the string. The dotted line is the mirror plane for the reflection symmetry $\r$}
\label{Fig:minimal_string}
\end{figure}

In the Kitaev quantum double model $\D(\G)$, the quasiparticles can be created in pair by string operator. Following the treatment in Ref. \onlinecite{Beigi2010}, we can construct the (minimal) string operator $F^{h,g}$ which acts on the green edges in Fig. \ref{Fig:minimal_string} such that
\begin{align}
F^{h,g} ~ 
\Bigg| 
\begin{tikzpicture}[scale=0.50, baseline={([yshift=-.5ex]current  bounding  box.center)}]
\draw[middlearrow={stealth reversed}] (1,1) -- (-1,1);
\draw[middlearrow={stealth}] (-1,1) -- (-1,-1);
\draw[middlearrow={stealth reversed}] (1,-1) -- (1,1);
\draw[dotted,thick] (0,-2) -- (0,2);
\draw (0.25,1.4) node {$y$};  
\draw (-1,-1.4) node {$x_1$}; 
\draw (1,-1.4) node {$x_2$};
\end{tikzpicture} 
\Bigg\rangle
=
\delta_{g,y_1}
\Bigg| 
\begin{tikzpicture}[scale=0.50, baseline={([yshift=-.5ex]current  bounding  box.center)}]
\draw[middlearrow={stealth reversed}] (1,1) -- (-1,1);
\draw[middlearrow={stealth}] (-1,1) -- (-1,-1);
\draw[middlearrow={stealth reversed}] (1,-1) -- (1,1);
\draw[dotted,thick] (0,-2) -- (0,2);
\draw (0.25,1.4) node {$y$};  
\draw (-1,-1.4) node {$h x_1$}; 
\draw (1,-1.4) node {$~\bar{g} h g x_2$};
\end{tikzpicture} 
\Bigg\rangle,
\end{align}
where $h,g\in \G$. As is shown in Fig. \ref{Fig:minimal_string}, when acted on the ground state, $F^{h,g} $ creates a vertex (red) and a plaquette (orange) excitation on each end of the string. It is easy to check that under the reflection $\r$ about the dotted line, the string operators transform as
\begin{align}
	{R_\R}^{-1} F^{h,g} R_\R = F^{\bar{g}hg,\bar{g}},
\end{align} 
where $R_\R$ is the operator that generates the reflection $\r$.

However we notice that the string $F^{h,g}$ does not create excitations with a definite topological charge type. But we can use them to build string operators associated to a certain anyon type.  For $C = \{c_1, c_2, ...\}$, we define $\{q_1, q_2,...\}$ such that
\begin{align}
c_i = q_i r_C \bar{q}_i, 
\end{align}
with $i$ labeling the elements in $C$. Also, we will label the basis of the vector space associated to the irrep $\pi$ with $j$ (and $j'$). The string operator that creates a pair of quasi-particles with a fixed anyon type $(C, \pi)$ and its anti-particle is given by
\begin{align}
F_{(C,\pi)}^{(i',j');(i,j)} = \sum_{h,g} \delta_{c_{i'},h} \delta_{c_i,\bar{g}h g} \pi(\bar{q}_{i'} g q_i )_{j'j} F^{h,g}.
\end{align}
Here $(i',j')$ and $(i,j)$ each labels the basis of the ``local" Hilbert spaces associated to the anyons ($(C,\pi)$ and its anti-particle) created on the ends of the string operator. Under the reflection $\R$, the string operator transforms as
\begin{align}
	R_\R^{-1} F_{(C,\pi)}^{(i',j');(i,j)} R_\R & = \sum_{h,g} \delta_{c_{i'},h} \delta_{c_i,\bar{g}h g} \pi(\bar{q}_{i'} g q_i )_{j'j} F^{\bar{g}hg,\bar{g}} 
\nonumber \\
& = \sum_{h,g} \delta_{c_{i'}, \bar{g}hg} \delta_{c_i,h } \left( \pi(\bar{q}_{i} g q_{i'} )^\dag\right)_{j'j} F^{h,g},
\nonumber \\
& = \sum_{h,g} \delta_{c_{i'}, \bar{g}hg} \delta_{c_i,h }  \pi(\bar{q}_{i} g q_{i'} )^*_{jj'} F^{h,g},
\nonumber \\
& = \mathcal{K} F_{(C,\pi)}^{(i,j);(i',j')} \mathcal{K}
\label{String_Reflection}
\end{align}
where we've made a change of variables $\bar{g}hg\rightarrow h$ and $\bar{g}\rightarrow g$ in the second line, and $\mathcal{K}$ is the complex conjugation. Notice that the action of $\mathcal{K}$ does not change the conjugacy class $C$ and only act on the representation $\pi$. We can say that under $\R$ the anyon $(C, \pi)$ becomes $(C^{-1}, \pi)$. Notice we pick $r_C$ and $r_C^{-1}$ as the representative elements for $C$ and $C^{-1}$. They share the same centralizer group $Z_{r_C} = Z_{r_C^{-1}}$. Hence, the anyon label $(C^{-1},\pi)$ indeed makes sense.

%Eq. \ref{String_Reflection} can be understood as follows. The reflection $\mb{R}$ switches the positions of the anyon $(C,\pi)$ and its anti-particle, therefore reversing the flux (i.e. the pair $C$ and $C^{-1}$ becomes $C^{-1}$ and $C$). It also complex-conjugates the irrep $\pi$. 
%In more concise terms, apart from the action on spatial positions, reflection $R$ changes the gauge flux of the anyon to its inverse value and keep the gauge charge intact (up to local deformations). 

A set of string operators $F_{(C,\pi)}^{(i',j');(i,j)}$ are symmetric under $\R$ if and only if there exists local unitary transformations $U$ and $U'$ (acting on the vector space associated to $\pi$) such that
\begin{align}
R_\R^{-1} F_{(C,\pi)}^{(i',j');(i,j)} R_\R = \sum_{k,k'} U'_{j'k'} U_{jk} F_{(C,\pi)}^{(i,k);(i',k')}.
\label{Symmetric_String_Condition}
\end{align}
For reflection symmetric string operators, the fact that $R^2$ is an trivial action leads to the condition that
\begin{align}
U U' = U' U = \pm \mathds{1}.
\label{UV_Local_Transformation}
\end{align}
In fact, $UU'$ (or $U'U$) are exactly the accumulated ``local transformations'' on a single anyon $(C,\pi)$ (or its anti-particle) after we perform
the reflection $\R$ twice. We will label the $\pm$ sign in this equation as $\eta^{\r}_{(C,\pi)}$.  Eq. (\ref{String_Reflection}) and \ref{Symmetric_String_Condition}, in fact, require that the following must hold for symmetric string operators: 
\begin{align}
\pi(\bar{q}_{i} g q_{i'} )^*_{jj'} = \sum_{k,k'} U_{jk} U'_{j'k'} \pi(\bar{q}_{i} g q_{i'} )_{kk'},
\end{align}
for all $g\in \G$ such that $c_{i'} = \bar{g} c_i g$ or equivalently $r_C = \bar{q}_{i'}\bar{g}q_i r_C \bar{q}_i g q_{i'}$. When we pick $\bar{q}_i g q_{i'} = 1$, we obtain that 
\begin{align}
	U {U'}^\mathsf{T} = \mathds{1}.
\end{align}
Therefore, the transformation given by $U$ and $U'$ is in fact conjugation of the irrep $\pi$ by $U$. Such $U$ exists only when the irrep $\pi$ is real or pseudoreal. We can now rewrite Eq. (\ref{UV_Local_Transformation}) as 
\begin{align}
	UU^\mathsf{T} = \eta^{\r}_{(C,\pi)} \cdot \mathds{1}. 
\end{align}

To summarize, a string operator creating $(C, \pi)$ is symmetric if $\pi$ is self-conjugate and we can associate a $\r^2$ eigenvalue to the anyon $(C, \pi)$, given by the value of $UU^\mathsf{T}$.
For a real irrep $\pi$, $\eta^\r_{(C,\pi)}= 1$. For a pseudoreal irrep $\pi$, $\eta^\r_{(C,\pi)}= -1$. In fact, $\eta^\r_{(C,\pi)}$ coincides exactly with the Frobenius-Schur indicator of the irrep $\pi$:
\begin{align}
	\eta^\r_{(C,\pi)} = \frac{1}{|Z_{r_C}|} \sum_{g\in Z_{r_C}} \Tr\left( \pi(g^2)\right),
\label{R2_eigenvalues}
\end{align}
where $|Z_{r_C}|$ is the order of $Z_{r_C}$. This expression automatically evaluates to $0$  when $\pi$ is neither real or pseudoreal. 

In Sec. \ref{globalsym}, we define $\eta^\r_{(C,\pi)}$ a (1+1)D SPT invariant, or the eigenvalue of $\r$ of the two-anyon state. Let $|\text{vac}\rangle$ denotes the vacuum state of $\D(\G)$ on the sphere. We can consider the following states 
\begin{align}
| (i',j');(i,j) \rangle \equiv F_{(C,\pi)}^{(i',j');(i,j)} |\text{vac}\rangle.
\end{align}
We can think of the states $\{| (i',j');(i,j) \rangle  \}$ as the states on two-punctured sphere, which is topologically equivalent to a cylinder. Once we fixed $(C,\pi)$, we can view the cylinder as an effective (1+1)D system and analyze its (1+1)D SPT properties under reflection $\r$. In this case, the labels $(i',j')$ and $(i,j)$ are simply labeling the local degrees of freedom lying on the two ends of the cylinder. The requirement Eq. (\ref{Symmetric_String_Condition}) that the string operator is symmetric under $\r$ becomes 
\begin{align}
	R_\r | (i',j');(i,j) \rangle = \sum_{k,k'} U_{jk} U'_{j'k'} | (i,k);(i',k') \rangle,
\end{align}
which is of course the condition for the effective (1+1)D system (together with the degrees of freedom localized on the boundary) to be reflection invariant. The reflection SPT invariant can be calculated via analyzing how the local degrees of freedom $(i',j')$ and $(i,j)$ transform under $\r$, as we did above, or can be obtained through the trace of $R_\r$ in the Hilber space spanned by $\{| (i',j');(i,j) \rangle  \}$:
\begin{align}
	\sum_{i'j'ij} &\langle (i',j');(i,j)| R_\r | (i',j');(i,j) \rangle \\
& = \left( \sum_{ii'} \delta_{ii'} \right) \left( \sum_{jj'} U_{jj'} U'_{j'j}\right) \\
&= |C| \dim \pi \times \eta^\r_{(C,\pi)}\\
& = d_{(C,\pi)} \eta^\r_{(C,\pi)},
\end{align}
where $\dim \pi$ is the dimension of the irrep. $\pi$ and $d_{(C,\pi)}$ is the quantum dimension of the anyon $(C,\pi)$.

It is then straightforward to derive
\begin{equation}
 \mb{T}^{-1} F^{h,g} \mb{T} = F^{h,g},
	\label{}
\end{equation}
and
\begin{align}
 \mb{T}^{-1} &F_{(C,\pi)}^{(i',j');(i,j)} \mb{T} \\
&= \sum_{h,g} \delta_{c_{i'},h} \delta_{c_i,\bar{g}h g} \pi(\bar{q}_{i'} g q_i )^*_{j'j} F^{h,g}  \\
&= \mathcal{K} F_{(C,\pi)}^{(i',j');(i,j)} \mathcal{K}.
\end{align}
From the last equation, we see that the TR symmetry is quite similar to the reflection symmetry except that it neither exchanges the two ends of the string operator nor does it change the gauge flux to its inverse. The TR symmetry generates the following mapping among anyons:
\begin{align}
\mb{T}:~ (C,\pi) \rightarrow (C,\pi^*).
\end{align}
For TR-invariant anyons, $\pi$ has to be real or pseudoreal. The eigenvalue of the TR-invariant anyon $(C, \pi)$ under $\mb{T}^2$ is again given by the Frobenius-Schur indicator 
\begin{align}
\eta^{\mb{T}}_{(C,\pi)} = \frac{1}{|Z_{r_C}|} \sum_{h\in Z_{r_C}} \Tr\left( \pi(g^2)\right).
\label{T2_eigenvalues}
\end{align}

%% file: dgRP2.tex
\section{Consistency of $\Z(\mathbb{RP}^2 \times S^1)$ in Kitaev's quantum double model $\D(\G)$}
\label{sec:dgRP2}

Eq. (\ref{Eq:DG_Partition_RP2_a}) and Eq. (\ref{Eq:DG_Partition_RP2_b}) provide two different expression of $\Z(\mathbb{RP}^2 \times S^1)$ for the 
Kitaev's quantum double model $\D(\G)$. In the following, we will prove that the two expression are in fact equal to each other. The topological $S$-matrix element $S_{(1,1) (C,\pi)}$ in Eq. \ref{Eq:DG_Partition_RP2_b} is given by
\begin{align}
S_{(1,1) (C,\pi)} = \frac{|C| \dim \pi }{|\G|},
\end{align}
which is the ratio between the quantum dimension $d_{(C,\pi)} = |C| \dim \pi$ of the quasiparticle $(C,\pi)$ and the total quantum dimension of $\D(\G)$ given by $|\G|$. Here, $|C|$ is the cardinality of the conjugacy class $C$ and $\dim \pi$ is the dimension of the representation $\pi$. 

Let us start from Eq. (\ref{Eq:DG_Partition_RP2_b}). 
Combining the expression of $S_{(1,1) (C,\pi)} $ with Eq. (\ref{Eq:FS_indicator}), we have
\begin{align}
& \Z(\mathbb{RP}^2 \times S^1) 
\nonumber \\
= & 
\sum_{(C,\pi) \in \mathcal{B}^\mb{T}} S_{(1,1) (C,\pi)} \nu_\pi,
\nonumber \\
= &
\sum_{\text{all} ~(C,\pi)} S_{(1,1) (C,\pi)} \nu_\pi, 
\nonumber \\
= &
\sum_{\text{all} ~(C,\pi)}  \frac{|C| \dim \pi }{|\G|} \frac{1}{|Z_{r_C}|} \sum_{g\in Z_{r_C}} \Tr\left( \pi(g^2)\right) 
\nonumber \\
= &
\sum_{\text{conj. class} ~C} \sum_{g\in Z_{r_C}}  \frac{|C| }{|\G|} \frac{1}{|Z_{r_C}|} \sum_{\pi \in \text{irrep. of } Z_{r_C}}\Tr\left( \pi(1)\right)  \Tr\left( \pi(g^2)\right) 
\nonumber \\
= &
\sum_{\text{conj. class} ~C} \sum_{g\in Z_{r_C}}  \frac{|C| }{|\G|} \delta_{1,g^2}
\nonumber \\
= &
\sum_{g\in \G} \sum_{\text{conj. class} ~C} \frac{|C| }{|\G|} \delta_{1,g^2} \delta_{r_C g,g r_C}
\nonumber \\
= &
\sum_{g\in \G} \sum_{h \in \G} \frac{1 }{|\G|} \delta_{1,g^2} \delta_{h g,g h}
\end{align}
Several keys steps in this derivation are as follows. From the second line to the third line, we've used the fact that for $(C,\pi) \notin \mathcal{B}^\mb{T}$, $\nu_{\pi}$ is automatically $0$. In the fourth line, we've plugged in the expression of $S_{(1,1) (C,\pi)} $ and Eq. (\ref{Eq:FS_indicator}). From the forth line to the fifth line, we've rewritten $\dim\pi$ as $\Tr\left( \pi(1)\right)$ (where $1$ stand for identity element in $\G$). In deriving the sixth line, we've made use of the identify that 
\begin{align}
\frac{1}{|Z_{r_C}|} \sum_{\pi \in \text{irrep. of } Z_{r_C}}\Tr\left( \pi(1)\right)  \Tr\left( \pi(g^2)\right) = \delta_{1,g^2}.
\end{align}
As we see, in the end of this derivation, we arrive at Eq. (\ref{Eq:DG_Partition_RP2_a}). Therefore, Eq. (\ref{Eq:DG_Partition_RP2_a}) and Eq. (\ref{Eq:DG_Partition_RP2_b}) are equal to each other.

Now, we prove that Eq. (\ref{Eq:DG_Partition_RP2_a}) is equal to the number of conjugacy classes $C$ that square to $1$. We notice that 
\begin{align}
\Z(\mathbb{RP}^2 \times S^1)  
& = \sum_{g\in \G} \sum_{h \in \G} \frac{1 }{|\G|} \delta_{1,g^2} \delta_{h g,g h} 
\nonumber \\
& = \sum_{g\in \G}  \frac{|Z_g| }{|\G|} \delta_{1,g^2} 
\nonumber \\
& = \sum_{g\in \G}  \frac{1}{|[g]|} \delta_{1,g^2},
\end{align}
where $Z_g = \{h| gh = hg, h\in \G\}$ is the centralizer of $g$ and $[g]$ denotes the conjugacy class that contains $g$. The cardinality (or the number of element) of $Z_g $ and $[g]$ are denoted as $|Z_g |$ and $|[g]|$ respectively. In the derivation above, we've used the fact that $|\G| = |Z_g| \cdot |[g]|$ for any group element $g\in \G$. Notice that if $g^2 =1$, any element in $[g]$ also squares to $1$. Therefore, the expression above counts the number of conjugacy classes that square to $1$.

%% file: DS3.tex
\section{Fractionalization constraints for $\D(S_3)$ with unconventional symmetry action}
\label{sec:appDS3}

In this section, we derive the condition 
\begin{align}
\eta_B^\mb{T} \equiv \eta_{B}(\mb{T,T})=-1,
\label{etaBconstraint}
\end{align}
for the time-reversal symmetry action given in Eq. \eqref{eqn:s3permutation}, which we 
repeat here for convenience:
\begin{align}
	T : \;\;  &C \leftrightarrow F, 
\nonumber\\
& G \leftrightarrow H .
	\label{}
\end{align}
Here we use the notation $\eta_B(\mb{T,T})$ to connect with the discussion presented in Ref. \onlinecite{barkeshli2014SDG}.
Eq. (\ref{etaBconstraint}) is a consequence of the following two consistency conditions on $\eta_a(\mb{T,T})$ and 
$U_\mb{T}(a,b,c)$, which were derived in Ref. \onlinecite{barkeshli2014SDG, Bonderson13d}:
\begin{align}
\label{consistency1}
		\frac{U_{\mb{T}}(a, b; e)U_{\mb{T}}(e, c; d)}{U_{\mb{T}}(b,c; f)U_{\mb{T}}(a,f; d)}F^{abc}_{def}= \left( F^{\,^\mb{T}a\,^\mb{T}b\,^\mb{T}c}_{\,^\mb{T}d \,^\mb{T}e \,^\mb{T}f} \right)^* ,
\end{align}
\begin{align}
	\frac{U_\mb{T}(\,^\mb{T}a,\,^\mb{T}b; \,^\mb{T}c)}{U_\mb{T}(a,b; c)}=\frac{\eta_a(\mb{T,T})\eta_b(\mb{T,T})}{\eta_c(\mb{T,T})}.
	\label{consistency2}
\end{align}
To see how Eq. (\ref{etaBconstraint}) follows from the above conditions, we need to use the fusion rules 
and $F$-symbols of $\D(S_3)$, which can be found in Ref. \onlinecite{Cui2014}. Applying Eq. (\ref{consistency2}) to the 
fusion rule $B\times G = G$, we obtain
\begin{equation}
	\eta_B(\mb{T,T})=\frac{U_\mb{T}(B, H; H)}{U_\mb{T}(B, G; G)} .
	\label{etaBformula}
\end{equation}
Using the $F$-symbol $F^{BGG}_{BGA}=1, F^{BHH}_{BHA}=-1$, we have
\begin{equation}
	\frac{U_\mb{T}(B,G;G)U_\mb{T}(G, G; B)}{U_\mb{T}(G, G; A)}=-1.
	%U_\mb{T}(B,H,H)U_\mb{T}(H, H, B)U^{-1}_\mb{T}(H, H, I)=-1, 
	\label{eqn:F1}
\end{equation}
where we have also used the fact that $U_{\bf T}(B, A;B) = 1$ (note that for all $a$, $U_{\bf g}(a, 0;a) = U_{\bf g}(0, a; a) = 1$ in general\cite{barkeshli2014SDG}). 
Furthermore, we need the following $F$ matrices:
\begin{equation}
	F^{GGH}_H=-\frac{1}{\sqrt{2}}
	\begin{pmatrix}
		1 & 1\\
		1 & -1
	\end{pmatrix}, 
	F^{HHG}_G=-\frac{1}{\sqrt{2}}
	\begin{pmatrix}
		1 & 1\\
		-1 & 1
	\end{pmatrix}.
	\label{}
\end{equation}
The rows of the matrices are indexed as $A, B$ and the columns as $C,F$. Applying Eq. (\ref{consistency1}) to each
entry of $F^{GGH}_H$, $F^{HHG}_G$ and using $U_{\bf T}(A, H;H) = 1$, leads to
\begin{equation}
	U_\mb{T}(G, G; A)=U_\mb{T}(G, G; B) U_\mb{T}(B, H ; H).
	\label{eqn:F2}
\end{equation}
Combining Eqs. \eqref{eqn:F2}, \eqref{eqn:F1}, and \eqref{etaBformula}, we obtain $\eta_B(\mb{T}, \mb{T})=\frac{U_\mb{T}(B, H; H)}{U_\mb{T}(B, G; G)}= -1$.

%% file: data.tex
\section{$Q_8$ and $\text{Rep}(Q_8)$} 
\label{app:data:Q8}

Here we collect some facts about $Q_8$ and $\text{Rep}(Q_8)$. 
The group $Q_8$ is the quaternion group of order $8$. It has 8 group
elements and can be given by the group presentation
\begin{align}
Q_8 = \langle -1, i, j, k | (-1)^2 =1, i^2, j^2, k^2 = ijk =-1\rangle,
\end{align}
where 1 is the identity element and $-1$ commutes with all other elements of the group. 

$Q_8$ has 5 distinct irreducible representations, which means there are 5 simple objects in $\text{Rep}(Q_8)$. We label them as $1$, $X$, $Y$, $Z$ and $\sigma$. $1$ correspond to the trivial representation of $Q_8$. $X$, $Y$ and $Z$ label three different 1-dimensional representations. $\sigma$ stands for the 2-dimensional representation. The non-trivial fusion rules in $\text{Rep}(Q_8)$ are given by
\begin{align}
& X \times X = Y \times Y = Z \times Z = 1,   \nonumber \\
& X \times Y = Y \times X = Z \nonumber \\
& Y \times Z = Z \times Y = X \nonumber \\
& Z \times X = X \times Z = Y \nonumber \\
& a \times \sigma = \sigma \times a =  \sigma, \;\;\;a=X,Y,Z\nonumber \\
& \sigma \times \sigma = 1 + X + Y + Z.
\end{align}
This is an example of the Tambara-Yamagami category for Abelian group
$A = \mathbb{Z}_2 \times   \mathbb{Z}_2$.\cite{Tambara1998} Here, the 
Abelian group $A$ refers to the Abelian group $\{1, X, Y, Z\}$ with
group multiplication given by fusion. Following the general classification of Ref. \onlinecite{Tambara1998}, the non-trivial F-symbols of $\text{Rep}(Q_8)$ are:
\begin{align}
& \left[ F^{a\sigma b}_\sigma \right]_{\sigma \sigma} = \left[ F^{\sigma a \sigma}_b \right]_{\sigma \sigma} = \chi(a,b) \nonumber \\
& \left[ F^{\sigma \sigma \sigma}_\sigma \right]_{ab}= \frac{\nu_{\sigma}}{\sqrt{|A| } \chi(a,b)}, 
\label{eqn:Q8F}
\end{align}
where $\nu_{\sigma}=-1$ is the Frobenius-Schur indicator of the representation $\sigma$. 
Here $a, b = 1, X, Y, Z$ and $\chi(a,b)$ is a symmetric bicharacter on $A$, where
$\chi(a,b) = \chi(b,a)$ and
\begin{align}
& \chi(1,1) = \chi(1,X) = \chi(1,Y) = \chi(1,Z) = 1, \nonumber \\
& \chi(X,X) = \chi(Y,Y) = \chi(Z,Z) = 1, \nonumber \\
& \chi(X,Y) = \chi(Y,Z) = \chi(Z,X) = -1. \nonumber \\ 
\label{eqn:bichar}
\end{align}

The Drinfield center $\D(\text{Rep}(Q_8))$, which is equivalent to the quantum double $\D(Q_8)$, has 22 simple objects. 

The group $Q_8$ is closely related to the group $D_8$, the dihedral group of $8$ elements which corresponds to the group of symmetries of a square. 
The two groups have the same character table. As fusion categories, $\text{Rep}(Q_8)$ and $\text{Rep}(D_8)$ are only slightly different. The difference only 
lies in the choice of Frobenius-Schur indicator $\nu_\sigma$, where $\nu_\sigma = 1$ for $\text{Rep}(D_8)$ and $-1$ for $\text{Rep}(Q_8)$.

%\subsection{Data of $D_8$ and $\text{Rep}(D_8)$} 
%The group $D_8$ is the dihedral group of order $8$. It has 8 group elements and can be given by the group presentation
%\begin{align}
%D_8 = \langle r, s | r^4 = s^2 = \text{id}, rsrs = \text{id}  \rangle,
%\end{align}

%$D_8$ has 5 different representations, which means there are 5 simple objects in $\text{Rep}(D_8)$. We label them as $1$, $X$, $Y$, $Z$ and $\sigma$. $1$ correspond to the trivial representation of $D_8$. $X$, $Y$ and $Z$ label three different 1-dimensional representations. $\sigma$ stands for the 2-dimensional representation. The fusion rules is exactly the same as that of $\text{Rep}(Q_8)$, since $Q_8$ and $D_8$ share exactly the same character table.  The non-trivial F-symbols of $\text{Rep}(D_8)$ take exactly the same form as given in Eq. \eqref{eqn:Q8F}, but with a trivial Frobenius-Schur indicator $\nu_\sigma=1$.

%The Drinfield center $\mathcal{Z}(\text{Rep}(D_8))$ is equivalent to the quantum double $\mathfrak{D}(Q_8)$. 

%% file: draftMain.bbl
\begin{thebibliography}{120}
\expandafter\ifx\csname natexlab\endcsname\relax\def\natexlab#1{#1}\fi
\expandafter\ifx\csname bibnamefont\endcsname\relax
  \def\bibnamefont#1{#1}\fi
\expandafter\ifx\csname bibfnamefont\endcsname\relax
  \def\bibfnamefont#1{#1}\fi
\expandafter\ifx\csname citenamefont\endcsname\relax
  \def\citenamefont#1{#1}\fi
\expandafter\ifx\csname url\endcsname\relax
  \def\url#1{\texttt{#1}}\fi
\expandafter\ifx\csname urlprefix\endcsname\relax\def\urlprefix{URL }\fi
\providecommand{\bibinfo}[2]{#2}
\providecommand{\eprint}[2][]{\url{#2}}

\bibitem[{\citenamefont{Pollmann et~al.}(2010)\citenamefont{Pollmann, Turner,
  Berg, and Oshikawa}}]{pollmann2010}
\bibinfo{author}{\bibfnamefont{F.}~\bibnamefont{Pollmann}},
  \bibinfo{author}{\bibfnamefont{A.~M.} \bibnamefont{Turner}},
  \bibinfo{author}{\bibfnamefont{E.}~\bibnamefont{Berg}}, \bibnamefont{and}
  \bibinfo{author}{\bibfnamefont{M.}~\bibnamefont{Oshikawa}},
  \bibinfo{journal}{Phys. Rev. B} \textbf{\bibinfo{volume}{81}},
  \bibinfo{pages}{064439} (\bibinfo{year}{2010}).

\bibitem[{\citenamefont{Chen et~al.}(2011{\natexlab{a}})\citenamefont{Chen, Gu,
  and Wen}}]{chen2011}
\bibinfo{author}{\bibfnamefont{X.}~\bibnamefont{Chen}},
  \bibinfo{author}{\bibfnamefont{Z.-C.} \bibnamefont{Gu}}, \bibnamefont{and}
  \bibinfo{author}{\bibfnamefont{X.-G.} \bibnamefont{Wen}},
  \bibinfo{journal}{Phys. Rev. B} \textbf{\bibinfo{volume}{83}},
  \bibinfo{pages}{035107} (\bibinfo{year}{2011}{\natexlab{a}}).

\bibitem[{\citenamefont{Fidkowski and Kitaev}(2011)}]{fidkowski2011}
\bibinfo{author}{\bibfnamefont{L.}~\bibnamefont{Fidkowski}} \bibnamefont{and}
  \bibinfo{author}{\bibfnamefont{A.}~\bibnamefont{Kitaev}},
  \bibinfo{journal}{Phys. Rev. B} \textbf{\bibinfo{volume}{83}},
  \bibinfo{pages}{075103} (\bibinfo{year}{2011}).

\bibitem[{\citenamefont{Schuch et~al.}(2011)\citenamefont{Schuch,
  P\'erez-Garc\'{\i}a, and Cirac}}]{schuch2011}
\bibinfo{author}{\bibfnamefont{N.}~\bibnamefont{Schuch}},
  \bibinfo{author}{\bibfnamefont{D.}~\bibnamefont{P\'erez-Garc\'{\i}a}},
  \bibnamefont{and} \bibinfo{author}{\bibfnamefont{I.}~\bibnamefont{Cirac}},
  \bibinfo{journal}{Phys. Rev. B} \textbf{\bibinfo{volume}{84}},
  \bibinfo{pages}{165139} (\bibinfo{year}{2011}).

\bibitem[{\citenamefont{Fannes et~al.}(1992)\citenamefont{Fannes, Nachtergaele,
  and Werner}}]{fannes1992}
\bibinfo{author}{\bibfnamefont{M.}~\bibnamefont{Fannes}},
  \bibinfo{author}{\bibfnamefont{B.}~\bibnamefont{Nachtergaele}},
  \bibnamefont{and} \bibinfo{author}{\bibfnamefont{R.~F.}
  \bibnamefont{Werner}}, \bibinfo{journal}{Comm. Math. Phys.}
  \textbf{\bibinfo{volume}{144}}, \bibinfo{pages}{443} (\bibinfo{year}{1992}).

\bibitem[{\citenamefont{Verstraete and Cirac}(2006)}]{Verstraete2006}
\bibinfo{author}{\bibfnamefont{F.}~\bibnamefont{Verstraete}} \bibnamefont{and}
  \bibinfo{author}{\bibfnamefont{J.~I.} \bibnamefont{Cirac}},
  \bibinfo{journal}{Phys. Rev. B} \textbf{\bibinfo{volume}{73}},
  \bibinfo{pages}{094423} (\bibinfo{year}{2006}).

\bibitem[{\citenamefont{Hastings}(2007)}]{Hastings1D}
\bibinfo{author}{\bibfnamefont{M.~B.} \bibnamefont{Hastings}},
  \bibinfo{journal}{J. Stat. Mech.} p. \bibinfo{pages}{P08024}
  (\bibinfo{year}{2007}).

\bibitem[{\citenamefont{Wen}(2004)}]{wen04}
\bibinfo{author}{\bibfnamefont{X.-G.} \bibnamefont{Wen}},
  \emph{\bibinfo{title}{Quantum Field Theory of Many-Body Systems}}
  (\bibinfo{publisher}{Oxford Univ. Press}, \bibinfo{address}{Oxford},
  \bibinfo{year}{2004}).

\bibitem[{\citenamefont{Nayak et~al.}(2008)\citenamefont{Nayak, Simon, Stern,
  Freedman, and Sarma}}]{nayak2008}
\bibinfo{author}{\bibfnamefont{C.}~\bibnamefont{Nayak}},
  \bibinfo{author}{\bibfnamefont{S.~H.} \bibnamefont{Simon}},
  \bibinfo{author}{\bibfnamefont{A.}~\bibnamefont{Stern}},
  \bibinfo{author}{\bibfnamefont{M.}~\bibnamefont{Freedman}}, \bibnamefont{and}
  \bibinfo{author}{\bibfnamefont{S.~D.} \bibnamefont{Sarma}},
  \bibinfo{journal}{Rev. Mod. Phys.} \textbf{\bibinfo{volume}{80}},
  \bibinfo{pages}{1083} (\bibinfo{year}{2008}).

\bibitem[{\citenamefont{Chen et~al.}(2012)\citenamefont{Chen, Gu, Liu, and
  Wen}}]{XieScience2012}
\bibinfo{author}{\bibfnamefont{X.}~\bibnamefont{Chen}},
  \bibinfo{author}{\bibfnamefont{Z.-C.} \bibnamefont{Gu}},
  \bibinfo{author}{\bibfnamefont{Z.-X.} \bibnamefont{Liu}}, \bibnamefont{and}
  \bibinfo{author}{\bibfnamefont{X.-G.} \bibnamefont{Wen}},
  \bibinfo{journal}{Science} \textbf{\bibinfo{volume}{338}},
  \bibinfo{pages}{1604} (\bibinfo{year}{2012}).

\bibitem[{\citenamefont{Chen et~al.}(2013)\citenamefont{Chen, Gu, Liu, and
  Wen}}]{chen2013}
\bibinfo{author}{\bibfnamefont{X.}~\bibnamefont{Chen}},
  \bibinfo{author}{\bibfnamefont{Z.-C.} \bibnamefont{Gu}},
  \bibinfo{author}{\bibfnamefont{Z.-X.} \bibnamefont{Liu}}, \bibnamefont{and}
  \bibinfo{author}{\bibfnamefont{X.-G.} \bibnamefont{Wen}},
  \bibinfo{journal}{Phys. Rev. B} \textbf{\bibinfo{volume}{87}},
  \bibinfo{pages}{155114} (\bibinfo{year}{2013}).

\bibitem[{\citenamefont{Lu and Vishwanath}(2012)}]{lu2012}
\bibinfo{author}{\bibfnamefont{Y.-M.} \bibnamefont{Lu}} \bibnamefont{and}
  \bibinfo{author}{\bibfnamefont{A.}~\bibnamefont{Vishwanath}},
  \bibinfo{journal}{Phys. Rev. B} \textbf{\bibinfo{volume}{86}},
  \bibinfo{pages}{125119} (\bibinfo{year}{2012}).

\bibitem[{\citenamefont{Kapustin}(2014)}]{kapustin2014}
\bibinfo{author}{\bibfnamefont{A.}~\bibnamefont{Kapustin}}
  (\bibinfo{year}{2014}), \eprint{arXiv:1403.1467}.

\bibitem[{\citenamefont{Senthil}(2015)}]{senthil2015}
\bibinfo{author}{\bibfnamefont{T.}~\bibnamefont{Senthil}},
  \bibinfo{journal}{Annual Review of Condensed Matter Physics}
  \textbf{\bibinfo{volume}{6}}, \bibinfo{pages}{299} (\bibinfo{year}{2015}).

\bibitem[{\citenamefont{Freed}(2014)}]{freed2014}
\bibinfo{author}{\bibfnamefont{D.~S.} \bibnamefont{Freed}}
  (\bibinfo{year}{2014}), \eprint{arXiv:1404.7224}.

\bibitem[{\citenamefont{Freed and Hopkins}(2016)}]{freed2016}
\bibinfo{author}{\bibfnamefont{D.~S.} \bibnamefont{Freed}} \bibnamefont{and}
  \bibinfo{author}{\bibfnamefont{M.~J.} \bibnamefont{Hopkins}}
  (\bibinfo{year}{2016}), \eprint{arXiv:1604.06527}.

\bibitem[{\citenamefont{Wen}(2002)}]{wen2002psg}
\bibinfo{author}{\bibfnamefont{X.-G.} \bibnamefont{Wen}},
  \bibinfo{journal}{Phys. Rev. B} \textbf{\bibinfo{volume}{65}},
  \bibinfo{pages}{165113} (\bibinfo{year}{2002}).

\bibitem[{\citenamefont{Levin and Stern}(2012)}]{LevinPRB2012}
\bibinfo{author}{\bibfnamefont{M.}~\bibnamefont{Levin}} \bibnamefont{and}
  \bibinfo{author}{\bibfnamefont{A.}~\bibnamefont{Stern}},
  \bibinfo{journal}{Phys. Rev. B} \textbf{\bibinfo{volume}{86}},
  \bibinfo{pages}{115131} (\bibinfo{year}{2012}).

\bibitem[{\citenamefont{Essin and Hermele}(2013)}]{essin2013}
\bibinfo{author}{\bibfnamefont{A.~M.} \bibnamefont{Essin}} \bibnamefont{and}
  \bibinfo{author}{\bibfnamefont{M.}~\bibnamefont{Hermele}},
  \bibinfo{journal}{Phys. Rev. B} \textbf{\bibinfo{volume}{87}},
  \bibinfo{pages}{104406} (\bibinfo{year}{2013}).

\bibitem[{\citenamefont{Mesaros and Ran}(2013)}]{mesaros2013}
\bibinfo{author}{\bibfnamefont{A.}~\bibnamefont{Mesaros}} \bibnamefont{and}
  \bibinfo{author}{\bibfnamefont{Y.}~\bibnamefont{Ran}},
  \bibinfo{journal}{Phys. Rev. B} \textbf{\bibinfo{volume}{87}},
  \bibinfo{pages}{155115} (\bibinfo{year}{2013}), \eprint{arXiv:1212.0835}.

\bibitem[{\citenamefont{Lu and Vishwanath}(2016)}]{lu2013}
\bibinfo{author}{\bibfnamefont{Y.-M.} \bibnamefont{Lu}} \bibnamefont{and}
  \bibinfo{author}{\bibfnamefont{A.}~\bibnamefont{Vishwanath}},
  \bibinfo{journal}{Phys. Rev. B} \textbf{\bibinfo{volume}{93}},
  \bibinfo{pages}{155121} (\bibinfo{year}{2016}), \eprint{arXiv:1302.2634}.

\bibitem[{\citenamefont{Barkeshli et~al.}(2014)\citenamefont{Barkeshli,
  Bonderson, Cheng, and Wang}}]{barkeshli2014SDG}
\bibinfo{author}{\bibfnamefont{M.}~\bibnamefont{Barkeshli}},
  \bibinfo{author}{\bibfnamefont{P.}~\bibnamefont{Bonderson}},
  \bibinfo{author}{\bibfnamefont{M.}~\bibnamefont{Cheng}}, \bibnamefont{and}
  \bibinfo{author}{\bibfnamefont{Z.}~\bibnamefont{Wang}}
  (\bibinfo{year}{2014}), \eprint{arXiv:1410.4540}.

\bibitem[{\citenamefont{Tarantino et~al.}(2016)\citenamefont{Tarantino,
  Lindner, and Fidkowski}}]{Tarantino_SET}
\bibinfo{author}{\bibfnamefont{N.}~\bibnamefont{Tarantino}},
  \bibinfo{author}{\bibfnamefont{N.}~\bibnamefont{Lindner}}, \bibnamefont{and}
  \bibinfo{author}{\bibfnamefont{L.}~\bibnamefont{Fidkowski}},
  \bibinfo{journal}{New J. Phys.} \textbf{\bibinfo{volume}{18}},
  \bibinfo{pages}{035006} (\bibinfo{year}{2016}), \eprint{arXiv:1506.06754}.

\bibitem[{\citenamefont{Chen et~al.}(2015)\citenamefont{Chen, Burnell,
  Vishwanath, and Fidkowski}}]{Chen2014}
\bibinfo{author}{\bibfnamefont{X.}~\bibnamefont{Chen}},
  \bibinfo{author}{\bibfnamefont{F.~J.} \bibnamefont{Burnell}},
  \bibinfo{author}{\bibfnamefont{A.}~\bibnamefont{Vishwanath}},
  \bibnamefont{and}
  \bibinfo{author}{\bibfnamefont{L.}~\bibnamefont{Fidkowski}},
  \bibinfo{journal}{Phys. Rev. X} \textbf{\bibinfo{volume}{5}},
  \bibinfo{pages}{041013} (\bibinfo{year}{2015}).

\bibitem[{\citenamefont{Lan et~al.}(2015)\citenamefont{Lan, Wang, and
  Wen}}]{Lan2015}
\bibinfo{author}{\bibfnamefont{T.}~\bibnamefont{Lan}},
  \bibinfo{author}{\bibfnamefont{J.~C.} \bibnamefont{Wang}}, \bibnamefont{and}
  \bibinfo{author}{\bibfnamefont{X.-G.} \bibnamefont{Wen}},
  \bibinfo{journal}{Phys. Rev. Lett.} \textbf{\bibinfo{volume}{114}},
  \bibinfo{pages}{076402} (\bibinfo{year}{2015}).

\bibitem[{\citenamefont{Lan et~al.}(2016)\citenamefont{Lan, Kong, and
  Wen}}]{Lan2016}
\bibinfo{author}{\bibfnamefont{T.}~\bibnamefont{Lan}},
  \bibinfo{author}{\bibfnamefont{L.}~\bibnamefont{Kong}}, \bibnamefont{and}
  \bibinfo{author}{\bibfnamefont{X.-G.} \bibnamefont{Wen}}
  (\bibinfo{year}{2016}), \eprint{arXiv:1602.05946}.

\bibitem[{\citenamefont{Turaev}()}]{turaev2000}
\bibinfo{author}{\bibfnamefont{V.}~\bibnamefont{Turaev}},
  \eprint{math/0005291}.

\bibitem[{\citenamefont{Kirillov}()}]{kirillov2004}
\bibinfo{author}{\bibfnamefont{A.~J.} \bibnamefont{Kirillov}},
  \eprint{math/0401119}.

\bibitem[{\citenamefont{Turaev}(2010)}]{turaev2010book}
\bibinfo{author}{\bibfnamefont{V.}~\bibnamefont{Turaev}},
  \emph{\bibinfo{title}{Homotopy Quantum Field Theory}}
  (\bibinfo{publisher}{European Mathematical Society}, \bibinfo{year}{2010}).

\bibitem[{\citenamefont{Etingof et~al.}(2009)\citenamefont{Etingof, Nikshych,
  and Ostrik}}]{ENO2009}
\bibinfo{author}{\bibfnamefont{P.}~\bibnamefont{Etingof}},
  \bibinfo{author}{\bibfnamefont{D.}~\bibnamefont{Nikshych}}, \bibnamefont{and}
  \bibinfo{author}{\bibfnamefont{V.}~\bibnamefont{Ostrik}}
  (\bibinfo{year}{2009}), \eprint{arXiv:0909.3140}.

\bibitem[{\citenamefont{Cheng et~al.}(2015)\citenamefont{Cheng, Zaletel,
  Barkeshli, Vishwanath, and Bonderson}}]{cheng2015}
\bibinfo{author}{\bibfnamefont{M.}~\bibnamefont{Cheng}},
  \bibinfo{author}{\bibfnamefont{M.}~\bibnamefont{Zaletel}},
  \bibinfo{author}{\bibfnamefont{M.}~\bibnamefont{Barkeshli}},
  \bibinfo{author}{\bibfnamefont{A.}~\bibnamefont{Vishwanath}},
  \bibnamefont{and} \bibinfo{author}{\bibfnamefont{P.}~\bibnamefont{Bonderson}}
  (\bibinfo{year}{2015}), \eprint{arXiv:1511.02263}.

\bibitem[{\citenamefont{Hsieh et~al.}(2014{\natexlab{a}})\citenamefont{Hsieh,
  Sule, Cho, Ryu, and Leigh}}]{hsieh2014}
\bibinfo{author}{\bibfnamefont{C.-T.} \bibnamefont{Hsieh}},
  \bibinfo{author}{\bibfnamefont{O.~M.} \bibnamefont{Sule}},
  \bibinfo{author}{\bibfnamefont{G.~Y.} \bibnamefont{Cho}},
  \bibinfo{author}{\bibfnamefont{S.}~\bibnamefont{Ryu}}, \bibnamefont{and}
  \bibinfo{author}{\bibfnamefont{R.~G.} \bibnamefont{Leigh}},
  \bibinfo{journal}{Phys. Rev. B} \textbf{\bibinfo{volume}{90}},
  \bibinfo{pages}{165134} (\bibinfo{year}{2014}{\natexlab{a}}).

\bibitem[{\citenamefont{Hsieh et~al.}(2014{\natexlab{b}})\citenamefont{Hsieh,
  Morimoto, and Ryu}}]{Hsieh2014b}
\bibinfo{author}{\bibfnamefont{C.-T.} \bibnamefont{Hsieh}},
  \bibinfo{author}{\bibfnamefont{T.}~\bibnamefont{Morimoto}}, \bibnamefont{and}
  \bibinfo{author}{\bibfnamefont{S.}~\bibnamefont{Ryu}},
  \bibinfo{journal}{Phys. Rev. B} \textbf{\bibinfo{volume}{90}},
  \bibinfo{pages}{245111} (\bibinfo{year}{2014}{\natexlab{b}}).

\bibitem[{\citenamefont{Zaletel et~al.}(2015)\citenamefont{Zaletel, Lu, and
  Vishwanath}}]{zaletel2015}
\bibinfo{author}{\bibfnamefont{M.}~\bibnamefont{Zaletel}},
  \bibinfo{author}{\bibfnamefont{Y.-M.} \bibnamefont{Lu}}, \bibnamefont{and}
  \bibinfo{author}{\bibfnamefont{A.}~\bibnamefont{Vishwanath}}
  (\bibinfo{year}{2015}), \eprint{arXiv:1501.01395}.

\bibitem[{\citenamefont{Qi and Fu}(2015)}]{YangPRL2015}
\bibinfo{author}{\bibfnamefont{Y.}~\bibnamefont{Qi}} \bibnamefont{and}
  \bibinfo{author}{\bibfnamefont{L.}~\bibnamefont{Fu}}, \bibinfo{journal}{Phys.
  Rev. Lett.} \textbf{\bibinfo{volume}{115}}, \bibinfo{pages}{236801}
  (\bibinfo{year}{2015}).

\bibitem[{\citenamefont{Hermele and Chen}(2016)}]{hermele2016}
\bibinfo{author}{\bibfnamefont{M.}~\bibnamefont{Hermele}} \bibnamefont{and}
  \bibinfo{author}{\bibfnamefont{X.}~\bibnamefont{Chen}},
  \bibinfo{journal}{Phys. Rev. X} \textbf{\bibinfo{volume}{6}},
  \bibinfo{pages}{041006} (\bibinfo{year}{2016}), \eprint{arXiv:1508.00573}.

\bibitem[{\citenamefont{Cheng et~al.}(2016)\citenamefont{Cheng, Gu, Jiang, and
  Qi}}]{cheng2016}
\bibinfo{author}{\bibfnamefont{M.}~\bibnamefont{Cheng}},
  \bibinfo{author}{\bibfnamefont{Z.-C.} \bibnamefont{Gu}},
  \bibinfo{author}{\bibfnamefont{S.}~\bibnamefont{Jiang}}, \bibnamefont{and}
  \bibinfo{author}{\bibfnamefont{Y.}~\bibnamefont{Qi}} (\bibinfo{year}{2016}),
  \eprint{arXiv:1606.08482}.

\bibitem[{\citenamefont{Song et~al.}()\citenamefont{Song, Huang, Fu, and
  Hermele}}]{song2016}
\bibinfo{author}{\bibfnamefont{H.}~\bibnamefont{Song}},
  \bibinfo{author}{\bibfnamefont{S.-J.} \bibnamefont{Huang}},
  \bibinfo{author}{\bibfnamefont{L.}~\bibnamefont{Fu}}, \bibnamefont{and}
  \bibinfo{author}{\bibfnamefont{M.}~\bibnamefont{Hermele}},
  \eprint{arXiv:1604.08151}.

\bibitem[{\citenamefont{Cho et~al.}(2015)\citenamefont{Cho, Hsieh, Morimoto,
  and Ryu}}]{cho2015}
\bibinfo{author}{\bibfnamefont{G.~Y.} \bibnamefont{Cho}},
  \bibinfo{author}{\bibfnamefont{C.-T.} \bibnamefont{Hsieh}},
  \bibinfo{author}{\bibfnamefont{T.}~\bibnamefont{Morimoto}}, \bibnamefont{and}
  \bibinfo{author}{\bibfnamefont{S.}~\bibnamefont{Ryu}},
  \bibinfo{journal}{Phys. Rev. B} \textbf{\bibinfo{volume}{91}},
  \bibinfo{pages}{195142} (\bibinfo{year}{2015}).

\bibitem[{\citenamefont{Metlitski}(2015)}]{metlitski2015}
\bibinfo{author}{\bibfnamefont{M.~A.} \bibnamefont{Metlitski}}
  (\bibinfo{year}{2015}), \eprint{arXiv:1510.05663}.

\bibitem[{\citenamefont{Witten}(2016)}]{witten2016}
\bibinfo{author}{\bibfnamefont{E.}~\bibnamefont{Witten}}
  (\bibinfo{year}{2016}), \eprint{arXiv:1605.02391}.

\bibitem[{\citenamefont{Vishwanath and Senthil}(2013)}]{vishwanath2013}
\bibinfo{author}{\bibfnamefont{A.}~\bibnamefont{Vishwanath}} \bibnamefont{and}
  \bibinfo{author}{\bibfnamefont{T.}~\bibnamefont{Senthil}},
  \bibinfo{journal}{Phys. Rev. X} \textbf{\bibinfo{volume}{3}},
  \bibinfo{pages}{011016} (\bibinfo{year}{2013}).

\bibitem[{\citenamefont{Wang and Senthil}(2013)}]{wang2013}
\bibinfo{author}{\bibfnamefont{C.}~\bibnamefont{Wang}} \bibnamefont{and}
  \bibinfo{author}{\bibfnamefont{T.}~\bibnamefont{Senthil}},
  \bibinfo{journal}{Phys. Rev. B} \textbf{\bibinfo{volume}{87}},
  \bibinfo{pages}{235122} (\bibinfo{year}{2013}).

\bibitem[{\citenamefont{Metlitski et~al.}(2013)\citenamefont{Metlitski, Kane,
  and Fisher}}]{MetlitskiPRB2013}
\bibinfo{author}{\bibfnamefont{M.~A.} \bibnamefont{Metlitski}},
  \bibinfo{author}{\bibfnamefont{C.~L.} \bibnamefont{Kane}}, \bibnamefont{and}
  \bibinfo{author}{\bibfnamefont{M.~P.~A.} \bibnamefont{Fisher}},
  \bibinfo{journal}{Phys. Rev. B} \textbf{\bibinfo{volume}{88}},
  \bibinfo{pages}{035131} (\bibinfo{year}{2013}).

\bibitem[{\citenamefont{{Bonderson} et~al.}(2013)\citenamefont{{Bonderson},
  {Nayak}, and {Qi}}}]{Bonderson13d}
\bibinfo{author}{\bibfnamefont{P.}~\bibnamefont{{Bonderson}}},
  \bibinfo{author}{\bibfnamefont{C.}~\bibnamefont{{Nayak}}}, \bibnamefont{and}
  \bibinfo{author}{\bibfnamefont{X.-L.} \bibnamefont{{Qi}}},
  \bibinfo{journal}{Journal of Statistical Mechanics: Theory and Experiment}
  \textbf{\bibinfo{volume}{9}}, \bibinfo{eid}{09016} (\bibinfo{year}{2013}),
  \eprint{arXiv:1306.3230}.

\bibitem[{\citenamefont{Chen et~al.}(2014)\citenamefont{Chen, Fidkowski, and
  Vishwanath}}]{chen2014b}
\bibinfo{author}{\bibfnamefont{X.}~\bibnamefont{Chen}},
  \bibinfo{author}{\bibfnamefont{L.}~\bibnamefont{Fidkowski}},
  \bibnamefont{and}
  \bibinfo{author}{\bibfnamefont{A.}~\bibnamefont{Vishwanath}},
  \bibinfo{journal}{Phys. Rev. B} \textbf{\bibinfo{volume}{89}},
  \bibinfo{pages}{165132} (\bibinfo{year}{2014}), \eprint{arXiv:1306.3250}.

\bibitem[{\citenamefont{Wang et~al.}(2013)\citenamefont{Wang, Potter, and
  Senthil}}]{wang2013b}
\bibinfo{author}{\bibfnamefont{C.}~\bibnamefont{Wang}},
  \bibinfo{author}{\bibfnamefont{A.~C.} \bibnamefont{Potter}},
  \bibnamefont{and} \bibinfo{author}{\bibfnamefont{T.}~\bibnamefont{Senthil}},
  \bibinfo{journal}{Phys. Rev. B} \textbf{\bibinfo{volume}{88}},
  \bibinfo{pages}{115137} (\bibinfo{year}{2013}), \eprint{arXiv:1306.3223}.

\bibitem[{\citenamefont{Metlitski et~al.}()\citenamefont{Metlitski, Fidkowski,
  Chen, and Vishwanath}}]{metlitski2014}
\bibinfo{author}{\bibfnamefont{M.~A.} \bibnamefont{Metlitski}},
  \bibinfo{author}{\bibfnamefont{L.}~\bibnamefont{Fidkowski}},
  \bibinfo{author}{\bibfnamefont{X.}~\bibnamefont{Chen}}, \bibnamefont{and}
  \bibinfo{author}{\bibfnamefont{A.}~\bibnamefont{Vishwanath}},
  \eprint{arXiv:1406.3032}.

\bibitem[{\citenamefont{Metlitski et~al.}(2015)\citenamefont{Metlitski, Kane,
  and Fisher}}]{MetlitskiPRB2015}
\bibinfo{author}{\bibfnamefont{M.~A.} \bibnamefont{Metlitski}},
  \bibinfo{author}{\bibfnamefont{C.~L.} \bibnamefont{Kane}}, \bibnamefont{and}
  \bibinfo{author}{\bibfnamefont{M.~P.~A.} \bibnamefont{Fisher}},
  \bibinfo{journal}{Phys. Rev. B} \textbf{\bibinfo{volume}{92}},
  \bibinfo{pages}{125111} (\bibinfo{year}{2015}).

\bibitem[{\citenamefont{{Fidkowski} et~al.}(2013)\citenamefont{{Fidkowski},
  {Chen}, and {Vishwanath}}}]{Fidkowski13}
\bibinfo{author}{\bibfnamefont{L.}~\bibnamefont{{Fidkowski}}},
  \bibinfo{author}{\bibfnamefont{X.}~\bibnamefont{{Chen}}}, \bibnamefont{and}
  \bibinfo{author}{\bibfnamefont{A.}~\bibnamefont{{Vishwanath}}},
  \bibinfo{journal}{Phys. Rev. X} \textbf{\bibinfo{volume}{3}},
  \bibinfo{eid}{041016} (\bibinfo{year}{2013}), \eprint{arXiv:1305.5851}.

\bibitem[{\citenamefont{Seiberg and Witten}()}]{SW1}
\bibinfo{author}{\bibfnamefont{N.}~\bibnamefont{Seiberg}} \bibnamefont{and}
  \bibinfo{author}{\bibfnamefont{E.}~\bibnamefont{Witten}},
  \eprint{arXiv:1602.04251}.

\bibitem[{\citenamefont{Cho et~al.}(2014)\citenamefont{Cho, Teo, and
  Ryu}}]{ChoPRB2014}
\bibinfo{author}{\bibfnamefont{G.~Y.} \bibnamefont{Cho}},
  \bibinfo{author}{\bibfnamefont{J.~C.~Y.} \bibnamefont{Teo}},
  \bibnamefont{and} \bibinfo{author}{\bibfnamefont{S.}~\bibnamefont{Ryu}},
  \bibinfo{journal}{Phys. Rev. B} \textbf{\bibinfo{volume}{89}},
  \bibinfo{pages}{235103} (\bibinfo{year}{2014}).

\bibitem[{\citenamefont{Kapustin and Thorngren}(2014)}]{kapustin2014b}
\bibinfo{author}{\bibfnamefont{A.}~\bibnamefont{Kapustin}} \bibnamefont{and}
  \bibinfo{author}{\bibfnamefont{R.}~\bibnamefont{Thorngren}}
  (\bibinfo{year}{2014}), \eprint{arXiv:1404.3230}.

\bibitem[{\citenamefont{Wang et~al.}(2016)\citenamefont{Wang, Lin, and
  Levin}}]{WangPRX2016}
\bibinfo{author}{\bibfnamefont{C.}~\bibnamefont{Wang}},
  \bibinfo{author}{\bibfnamefont{C.-H.} \bibnamefont{Lin}}, \bibnamefont{and}
  \bibinfo{author}{\bibfnamefont{M.}~\bibnamefont{Levin}},
  \bibinfo{journal}{Phys. Rev. X} \textbf{\bibinfo{volume}{6}},
  \bibinfo{pages}{021015} (\bibinfo{year}{2016}).

\bibitem[{\citenamefont{Cui et~al.}(2016)\citenamefont{Cui, Galindo, Plavnik,
  and Wang}}]{cui2016}
\bibinfo{author}{\bibfnamefont{S.~X.} \bibnamefont{Cui}},
  \bibinfo{author}{\bibfnamefont{C.}~\bibnamefont{Galindo}},
  \bibinfo{author}{\bibfnamefont{J.~Y.} \bibnamefont{Plavnik}},
  \bibnamefont{and} \bibinfo{author}{\bibfnamefont{Z.}~\bibnamefont{Wang}},
  \bibinfo{journal}{Comm. Math. Phys.} \textbf{\bibinfo{volume}{348}},
  \bibinfo{pages}{1043} (\bibinfo{year}{2016}).

\bibitem[{\citenamefont{Turaev and Viro}(1992)}]{turaev1992}
\bibinfo{author}{\bibfnamefont{V.}~\bibnamefont{Turaev}} \bibnamefont{and}
  \bibinfo{author}{\bibfnamefont{O.}~\bibnamefont{Viro}},
  \bibinfo{journal}{Topology} \textbf{\bibinfo{volume}{31}},
  \bibinfo{pages}{865} (\bibinfo{year}{1992}).

\bibitem[{\citenamefont{Barrett and Westbury}(1996)}]{barrett1996}
\bibinfo{author}{\bibfnamefont{J.~W.} \bibnamefont{Barrett}} \bibnamefont{and}
  \bibinfo{author}{\bibfnamefont{B.~W.} \bibnamefont{Westbury}},
  \bibinfo{journal}{Trans. Amer. Math. Soc.} \textbf{\bibinfo{volume}{348}},
  \bibinfo{pages}{3997} (\bibinfo{year}{1996}).

\bibitem[{\citenamefont{Turaev}(1994)}]{turaev1994}
\bibinfo{author}{\bibfnamefont{V.}~\bibnamefont{Turaev}},
  \emph{\bibinfo{title}{Quantum invariants of knots and 3-manifolds}}
  (\bibinfo{publisher}{Walter de Gruyter \& Co.}, \bibinfo{address}{Berlin},
  \bibinfo{year}{1994}).

\bibitem[{\citenamefont{Roberts}(1995)}]{Roberts1995}
\bibinfo{author}{\bibfnamefont{J.}~\bibnamefont{Roberts}},
  \bibinfo{journal}{Topology} \textbf{\bibinfo{volume}{34}},
  \bibinfo{pages}{771} (\bibinfo{year}{1995}).

\bibitem[{\citenamefont{Levin and Wen}(2005)}]{levin2005}
\bibinfo{author}{\bibfnamefont{M.~A.} \bibnamefont{Levin}} \bibnamefont{and}
  \bibinfo{author}{\bibfnamefont{X.-G.} \bibnamefont{Wen}},
  \bibinfo{journal}{Phys. Rev. B} \textbf{\bibinfo{volume}{71}},
  \bibinfo{pages}{045110} (\bibinfo{year}{2005}).

\bibitem[{\citenamefont{Turaev and Virelizier}(2010)}]{turaev2010}
\bibinfo{author}{\bibfnamefont{V.}~\bibnamefont{Turaev}} \bibnamefont{and}
  \bibinfo{author}{\bibfnamefont{A.}~\bibnamefont{Virelizier}}
  (\bibinfo{year}{2010}), \eprint{arXiv:1006.3501}.

\bibitem[{\citenamefont{Kirillov and Balsam}(2010)}]{kirillov2010}
\bibinfo{author}{\bibfnamefont{A.}~\bibnamefont{Kirillov}} \bibnamefont{and}
  \bibinfo{author}{\bibfnamefont{B.}~\bibnamefont{Balsam}}
  (\bibinfo{year}{2010}), \eprint{arXiv:1004.1533}.

\bibitem[{\citenamefont{Balsam}(2010{\natexlab{a}})}]{balsam2010}
\bibinfo{author}{\bibfnamefont{B.}~\bibnamefont{Balsam}}
  (\bibinfo{year}{2010}{\natexlab{a}}), \eprint{arXiv:1010.1222}.

\bibitem[{\citenamefont{Balsam}(2010{\natexlab{b}})}]{balsam2010b}
\bibinfo{author}{\bibfnamefont{B.}~\bibnamefont{Balsam}}
  (\bibinfo{year}{2010}{\natexlab{b}}), \eprint{arXiv:1012.0560}.

\bibitem[{\citenamefont{Heinrich et~al.}(2016)\citenamefont{Heinrich, Burnell,
  Fidkowski, and Levin}}]{heinrich2016}
\bibinfo{author}{\bibfnamefont{C.}~\bibnamefont{Heinrich}},
  \bibinfo{author}{\bibfnamefont{F.}~\bibnamefont{Burnell}},
  \bibinfo{author}{\bibfnamefont{L.}~\bibnamefont{Fidkowski}},
  \bibnamefont{and} \bibinfo{author}{\bibfnamefont{M.}~\bibnamefont{Levin}}
  (\bibinfo{year}{2016}), \eprint{arXiv:1606.07816}.

\bibitem[{\citenamefont{Turaev and Virelizier}(2012)}]{turaev2012}
\bibinfo{author}{\bibfnamefont{V.}~\bibnamefont{Turaev}} \bibnamefont{and}
  \bibinfo{author}{\bibfnamefont{A.}~\bibnamefont{Virelizier}}
  (\bibinfo{year}{2012}), \eprint{arXiv:1202.6292}.

\bibitem[{\citenamefont{Witten}(1989)}]{witten1989}
\bibinfo{author}{\bibfnamefont{E.}~\bibnamefont{Witten}},
  \bibinfo{journal}{Comm. Math. Phys.} \textbf{\bibinfo{volume}{121}},
  \bibinfo{pages}{351} (\bibinfo{year}{1989}).

\bibitem[{\citenamefont{Moore and Seiberg}(1989)}]{moore1989b}
\bibinfo{author}{\bibfnamefont{G.}~\bibnamefont{Moore}} \bibnamefont{and}
  \bibinfo{author}{\bibfnamefont{N.}~\bibnamefont{Seiberg}},
  \bibinfo{journal}{Comm. Math. Phys.} \textbf{\bibinfo{volume}{123}},
  \bibinfo{pages}{177} (\bibinfo{year}{1989}).

\bibitem[{\citenamefont{Freedman and Meyer}(2001)}]{freedman2001}
\bibinfo{author}{\bibfnamefont{M.~H.} \bibnamefont{Freedman}} \bibnamefont{and}
  \bibinfo{author}{\bibfnamefont{D.~A.} \bibnamefont{Meyer}},
  \bibinfo{journal}{Foundations of Computational Mathematics}
  \textbf{\bibinfo{volume}{1}}, \bibinfo{pages}{325} (\bibinfo{year}{2001}).

\bibitem[{\citenamefont{Kitaev}(2003)}]{kitaev2003}
\bibinfo{author}{\bibfnamefont{A.}~\bibnamefont{Kitaev}},
  \bibinfo{journal}{Annals Phys.} \textbf{\bibinfo{volume}{303}},
  \bibinfo{pages}{2} (\bibinfo{year}{2003}), \eprint{arXiv:quant-ph/9707021}.

\bibitem[{\citenamefont{Shor}(1995)}]{shor1995}
\bibinfo{author}{\bibfnamefont{P.~W.} \bibnamefont{Shor}},
  \bibinfo{journal}{Phys. Rev. A} \textbf{\bibinfo{volume}{52}},
  \bibinfo{pages}{R2493} (\bibinfo{year}{1995}).

\bibitem[{\citenamefont{Freedman and Hastings}(2016)}]{freedman2016}
\bibinfo{author}{\bibfnamefont{M.~H.} \bibnamefont{Freedman}} \bibnamefont{and}
  \bibinfo{author}{\bibfnamefont{M.~B.} \bibnamefont{Hastings}},
  \bibinfo{journal}{Comm. Math. Phys.} \textbf{\bibinfo{volume}{347}},
  \bibinfo{pages}{389} (\bibinfo{year}{2016}).

\bibitem[{\citenamefont{Freedman et~al.}(2004)\citenamefont{Freedman, Nayak,
  Shtengel, Walker, and Wang}}]{freedman2004}
\bibinfo{author}{\bibfnamefont{M.}~\bibnamefont{Freedman}},
  \bibinfo{author}{\bibfnamefont{C.}~\bibnamefont{Nayak}},
  \bibinfo{author}{\bibfnamefont{K.}~\bibnamefont{Shtengel}},
  \bibinfo{author}{\bibfnamefont{K.}~\bibnamefont{Walker}}, \bibnamefont{and}
  \bibinfo{author}{\bibfnamefont{Z.}~\bibnamefont{Wang}},
  \bibinfo{journal}{Ann. Phys.} \textbf{\bibinfo{volume}{310}},
  \bibinfo{pages}{428 } (\bibinfo{year}{2004}).

\bibitem[{\citenamefont{Chan et~al.}(2016)\citenamefont{Chan, Teo, and
  Ryu}}]{chan2016}
\bibinfo{author}{\bibfnamefont{A.~P.~O.} \bibnamefont{Chan}},
  \bibinfo{author}{\bibfnamefont{J.~C.~Y.} \bibnamefont{Teo}},
  \bibnamefont{and} \bibinfo{author}{\bibfnamefont{S.}~\bibnamefont{Ryu}},
  \bibinfo{journal}{New J. Phys.} \textbf{\bibinfo{volume}{18}},
  \bibinfo{pages}{035005} (\bibinfo{year}{2016}).

\bibitem[{\citenamefont{Ben-Zion et~al.}(2016)\citenamefont{Ben-Zion, Das, and
  McGreevy}}]{zion2016}
\bibinfo{author}{\bibfnamefont{D.}~\bibnamefont{Ben-Zion}},
  \bibinfo{author}{\bibfnamefont{D.}~\bibnamefont{Das}}, \bibnamefont{and}
  \bibinfo{author}{\bibfnamefont{J.}~\bibnamefont{McGreevy}},
  \bibinfo{journal}{Phys. Rev. B} \textbf{\bibinfo{volume}{93}},
  \bibinfo{pages}{155147} (\bibinfo{year}{2016}).

\bibitem[{\citenamefont{Chen et~al.}(2011{\natexlab{b}})\citenamefont{Chen, Gu,
  and Wen}}]{ChenPRB2011a}
\bibinfo{author}{\bibfnamefont{X.}~\bibnamefont{Chen}},
  \bibinfo{author}{\bibfnamefont{Z.-C.} \bibnamefont{Gu}}, \bibnamefont{and}
  \bibinfo{author}{\bibfnamefont{X.-G.} \bibnamefont{Wen}},
  \bibinfo{journal}{Phys. Rev. B} \textbf{\bibinfo{volume}{83}},
  \bibinfo{pages}{035107} (\bibinfo{year}{2011}{\natexlab{b}}).

\bibitem[{\citenamefont{Chen et~al.}(2011{\natexlab{c}})\citenamefont{Chen, Gu,
  and Wen}}]{ChenPRB2011b}
\bibinfo{author}{\bibfnamefont{X.}~\bibnamefont{Chen}},
  \bibinfo{author}{\bibfnamefont{Z.-C.} \bibnamefont{Gu}}, \bibnamefont{and}
  \bibinfo{author}{\bibfnamefont{X.-G.} \bibnamefont{Wen}},
  \bibinfo{journal}{Phys. Rev. B} \textbf{\bibinfo{volume}{84}},
  \bibinfo{pages}{235128} (\bibinfo{year}{2011}{\natexlab{c}}).

\bibitem[{\citenamefont{Kapustin and Turzillo}(2015)}]{kapustin2015}
\bibinfo{author}{\bibfnamefont{A.}~\bibnamefont{Kapustin}} \bibnamefont{and}
  \bibinfo{author}{\bibfnamefont{A.}~\bibnamefont{Turzillo}}
  (\bibinfo{year}{2015}), \eprint{arXiv:1504.01830}.

\bibitem[{\citenamefont{Shiozaki and Ryu}(2016)}]{ShiozakiMPS}
\bibinfo{author}{\bibfnamefont{K.}~\bibnamefont{Shiozaki}} \bibnamefont{and}
  \bibinfo{author}{\bibfnamefont{S.}~\bibnamefont{Ryu}} (\bibinfo{year}{2016}),
  \eprint{arXiv:1607.06504}.

\bibitem[{\citenamefont{Pollmann and
  Turner}(2012{\natexlab{a}})}]{pollmann2012}
\bibinfo{author}{\bibfnamefont{F.}~\bibnamefont{Pollmann}} \bibnamefont{and}
  \bibinfo{author}{\bibfnamefont{A.~M.} \bibnamefont{Turner}},
  \bibinfo{journal}{Phys. Rev. B} \textbf{\bibinfo{volume}{86}},
  \bibinfo{pages}{125441} (\bibinfo{year}{2012}{\natexlab{a}}).

\bibitem[{\citenamefont{Pollmann et~al.}(2012)\citenamefont{Pollmann, Berg,
  Turner, and Oshikawa}}]{PollmannPRB2012}
\bibinfo{author}{\bibfnamefont{F.}~\bibnamefont{Pollmann}},
  \bibinfo{author}{\bibfnamefont{E.}~\bibnamefont{Berg}},
  \bibinfo{author}{\bibfnamefont{A.~M.} \bibnamefont{Turner}},
  \bibnamefont{and} \bibinfo{author}{\bibfnamefont{M.}~\bibnamefont{Oshikawa}},
  \bibinfo{journal}{Phys. Rev. B} \textbf{\bibinfo{volume}{85}},
  \bibinfo{pages}{075125} (\bibinfo{year}{2012}).

\bibitem[{\citenamefont{Kitaev and Kong}(2012)}]{kitaev2012}
\bibinfo{author}{\bibfnamefont{A.}~\bibnamefont{Kitaev}} \bibnamefont{and}
  \bibinfo{author}{\bibfnamefont{L.}~\bibnamefont{Kong}},
  \bibinfo{journal}{Comm. Math. Phys.} \textbf{\bibinfo{volume}{313}},
  \bibinfo{pages}{351} (\bibinfo{year}{2012}).

\bibitem[{\citenamefont{Barkeshli et~al.}(2013)\citenamefont{Barkeshli, Jian,
  and Qi}}]{barkeshli2013defect2}
\bibinfo{author}{\bibfnamefont{M.}~\bibnamefont{Barkeshli}},
  \bibinfo{author}{\bibfnamefont{C.-M.} \bibnamefont{Jian}}, \bibnamefont{and}
  \bibinfo{author}{\bibfnamefont{X.-L.} \bibnamefont{Qi}},
  \bibinfo{journal}{Phys. Rev. B} \textbf{\bibinfo{volume}{88}},
  \bibinfo{pages}{235103} (\bibinfo{year}{2013}).

\bibitem[{\citenamefont{Atiyah}(1988)}]{atiyah}
\bibinfo{author}{\bibfnamefont{M.}~\bibnamefont{Atiyah}},
  \bibinfo{journal}{Publications Mathématiques de l’Institut des Hautes
  Scientifiques} \textbf{\bibinfo{volume}{68}}, \bibinfo{pages}{175}
  (\bibinfo{year}{1988}).

\bibitem[{\citenamefont{Levin and Gu}(2012)}]{levin2012}
\bibinfo{author}{\bibfnamefont{M.}~\bibnamefont{Levin}} \bibnamefont{and}
  \bibinfo{author}{\bibfnamefont{Z.-C.} \bibnamefont{Gu}},
  \bibinfo{journal}{Phys. Rev. B} \textbf{\bibinfo{volume}{86}},
  \bibinfo{pages}{115109} (\bibinfo{year}{2012}).

\bibitem[{\citenamefont{Else and Nayak}(2014)}]{ElsePRB2014}
\bibinfo{author}{\bibfnamefont{D.~V.} \bibnamefont{Else}} \bibnamefont{and}
  \bibinfo{author}{\bibfnamefont{C.}~\bibnamefont{Nayak}},
  \bibinfo{journal}{Phys. Rev. B} \textbf{\bibinfo{volume}{90}},
  \bibinfo{pages}{235137} (\bibinfo{year}{2014}).

\bibitem[{\citenamefont{Dijkgraaf et~al.}(1989)\citenamefont{Dijkgraaf, Vafa,
  Verlinde, and Verlinde}}]{dijkgraaf1989}
\bibinfo{author}{\bibfnamefont{R.}~\bibnamefont{Dijkgraaf}},
  \bibinfo{author}{\bibfnamefont{C.}~\bibnamefont{Vafa}},
  \bibinfo{author}{\bibfnamefont{E.}~\bibnamefont{Verlinde}}, \bibnamefont{and}
  \bibinfo{author}{\bibfnamefont{H.}~\bibnamefont{Verlinde}},
  \bibinfo{journal}{Comm. Math. Phys.} \textbf{\bibinfo{volume}{123}},
  \bibinfo{pages}{485} (\bibinfo{year}{1989}).

\bibitem[{\citenamefont{Ponzano and Regge}(1968)}]{ponzano1968}
\bibinfo{author}{\bibfnamefont{G.}~\bibnamefont{Ponzano}} \bibnamefont{and}
  \bibinfo{author}{\bibfnamefont{T.}~\bibnamefont{Regge}}, in
  \emph{\bibinfo{booktitle}{Spectroscopic and Group Theoretical Methods in
  Physics}}, edited by \bibinfo{editor}{\bibfnamefont{F.}~\bibnamefont{Bloch}}
  (\bibinfo{publisher}{(Amsterdam: North-Holland)}, \bibinfo{year}{1968}), pp.
  \bibinfo{pages}{1--58}.

\bibitem[{\citenamefont{Kogut}(1979)}]{kogut1979}
\bibinfo{author}{\bibfnamefont{J.~B.} \bibnamefont{Kogut}},
  \bibinfo{journal}{Rev. Mod. Phys.} \textbf{\bibinfo{volume}{51}},
  \bibinfo{pages}{659} (\bibinfo{year}{1979}).

\bibitem[{\citenamefont{Walker}(2006)}]{walker2006}
\bibinfo{author}{\bibfnamefont{K.}~\bibnamefont{Walker}}
  (\bibinfo{year}{2006}), \urlprefix\url{http://canyon23.net/math/tc.pdf}.

\bibitem[{\citenamefont{Kitaev}(2006)}]{kitaev2006}
\bibinfo{author}{\bibfnamefont{A.}~\bibnamefont{Kitaev}},
  \bibinfo{journal}{Ann. Phys.} \textbf{\bibinfo{volume}{321}},
  \bibinfo{pages}{2 } (\bibinfo{year}{2006}).

\bibitem[{\citenamefont{Bonderson}(2007)}]{Bonderson07b}
\bibinfo{author}{\bibfnamefont{P.~H.} \bibnamefont{Bonderson}}, Ph.D. thesis,
  \bibinfo{school}{California Institute of Technology} (\bibinfo{year}{2007}).

\bibitem[{\citenamefont{Gu et~al.}(2015)\citenamefont{Gu, Wang, and
  Wen}}]{GuPRB2015}
\bibinfo{author}{\bibfnamefont{Z.-C.} \bibnamefont{Gu}},
  \bibinfo{author}{\bibfnamefont{Z.}~\bibnamefont{Wang}}, \bibnamefont{and}
  \bibinfo{author}{\bibfnamefont{X.-G.} \bibnamefont{Wen}},
  \bibinfo{journal}{Phys. Rev. B} \textbf{\bibinfo{volume}{91}},
  \bibinfo{pages}{125149} (\bibinfo{year}{2015}).

\bibitem[{\citenamefont{Lin and Levin}(2014)}]{lin2014}
\bibinfo{author}{\bibfnamefont{C.-H.} \bibnamefont{Lin}} \bibnamefont{and}
  \bibinfo{author}{\bibfnamefont{M.}~\bibnamefont{Levin}},
  \bibinfo{journal}{Phys. Rev. B} \textbf{\bibinfo{volume}{89}},
  \bibinfo{pages}{195130} (\bibinfo{year}{2014}).

\bibitem[{\citenamefont{Hu et~al.}(2013)\citenamefont{Hu, Wan, and
  Wu}}]{YTHuTQD}
\bibinfo{author}{\bibfnamefont{Y.}~\bibnamefont{Hu}},
  \bibinfo{author}{\bibfnamefont{Y.}~\bibnamefont{Wan}}, \bibnamefont{and}
  \bibinfo{author}{\bibfnamefont{Y.-S.} \bibnamefont{Wu}},
  \bibinfo{journal}{Phys. Rev. B} \textbf{\bibinfo{volume}{87}},
  \bibinfo{pages}{125114} (\bibinfo{year}{2013}).

\bibitem[{\citenamefont{Buerschaper and Aguado}(2009)}]{BuerschaperPRB2009}
\bibinfo{author}{\bibfnamefont{O.}~\bibnamefont{Buerschaper}} \bibnamefont{and}
  \bibinfo{author}{\bibfnamefont{M.}~\bibnamefont{Aguado}},
  \bibinfo{journal}{Phys. Rev. B} \textbf{\bibinfo{volume}{80}},
  \bibinfo{pages}{155136} (\bibinfo{year}{2009}).

\bibitem[{\citenamefont{Walker}(2016)}]{walker2016}
\bibinfo{author}{\bibfnamefont{K.}~\bibnamefont{Walker}}
  (\bibinfo{year}{2016}), \bibinfo{note}{unpublished}.

\bibitem[{\citenamefont{Crane and Yetter}(1993)}]{crane1993}
\bibinfo{author}{\bibfnamefont{L.}~\bibnamefont{Crane}} \bibnamefont{and}
  \bibinfo{author}{\bibfnamefont{D.}~\bibnamefont{Yetter}}, in
  \emph{\bibinfo{booktitle}{Quantum Topology}}, edited by
  \bibinfo{editor}{\bibfnamefont{L.}~\bibnamefont{Kauffman}} \bibnamefont{and}
  \bibinfo{editor}{\bibfnamefont{R.}~\bibnamefont{Baadhio}}
  (\bibinfo{publisher}{World Scientific}, \bibinfo{address}{Singapore},
  \bibinfo{year}{1993}), \eprint{arXiv:hep-th/9301062}.

\bibitem[{\citenamefont{Walker and Wang}(2012)}]{walker2012}
\bibinfo{author}{\bibfnamefont{K.}~\bibnamefont{Walker}} \bibnamefont{and}
  \bibinfo{author}{\bibfnamefont{Z.}~\bibnamefont{Wang}},
  \bibinfo{journal}{Frontier of Physics} \textbf{\bibinfo{volume}{7}},
  \bibinfo{pages}{150} (\bibinfo{year}{2012}).

\bibitem[{\citenamefont{Walker}(1991)}]{walker1991}
\bibinfo{author}{\bibfnamefont{K.}~\bibnamefont{Walker}}
  (\bibinfo{year}{1991}), \eprint{http://canyon23.net/math/1991TQFTNotes.pdf}.

\bibitem[{\citenamefont{Gompf and Stipsicz}(1999)}]{gompf1999}
\bibinfo{author}{\bibfnamefont{R.~E.} \bibnamefont{Gompf}} \bibnamefont{and}
  \bibinfo{author}{\bibfnamefont{A.}~\bibnamefont{Stipsicz}},
  \emph{\bibinfo{title}{4-manifolds and Kirby Calculus}}
  (\bibinfo{publisher}{American Mathematical Society}, \bibinfo{year}{1999}).

\bibitem[{\citenamefont{Moore and Segal}(2006)}]{MooreSegal}
\bibinfo{author}{\bibfnamefont{G.}~\bibnamefont{Moore}} \bibnamefont{and}
  \bibinfo{author}{\bibfnamefont{G.}~\bibnamefont{Segal}}
  (\bibinfo{year}{2006}), \eprint{arXiv:hep-th/0609042}.

\bibitem[{\citenamefont{Abrams}(1996)}]{AbramsTQFT}
\bibinfo{author}{\bibfnamefont{L.}~\bibnamefont{Abrams}}, \bibinfo{journal}{J.
  Knot Theory Ramications} \textbf{\bibinfo{volume}{5}}, \bibinfo{pages}{569}
  (\bibinfo{year}{1996}).

\bibitem[{\citenamefont{Burnell et~al.}(2014)\citenamefont{Burnell, Chen,
  Fidkowski, and Vishwanath}}]{BurnellPRB2014}
\bibinfo{author}{\bibfnamefont{F.~J.} \bibnamefont{Burnell}},
  \bibinfo{author}{\bibfnamefont{X.}~\bibnamefont{Chen}},
  \bibinfo{author}{\bibfnamefont{L.}~\bibnamefont{Fidkowski}},
  \bibnamefont{and}
  \bibinfo{author}{\bibfnamefont{A.}~\bibnamefont{Vishwanath}},
  \bibinfo{journal}{Phys. Rev. B} \textbf{\bibinfo{volume}{90}},
  \bibinfo{pages}{245122} (\bibinfo{year}{2014}).

\bibitem[{\citenamefont{Bombin and Martin-Delgado}(2008)}]{BombinPRB2008}
\bibinfo{author}{\bibfnamefont{H.}~\bibnamefont{Bombin}} \bibnamefont{and}
  \bibinfo{author}{\bibfnamefont{M.~A.} \bibnamefont{Martin-Delgado}},
  \bibinfo{journal}{Phys. Rev. B} \textbf{\bibinfo{volume}{78}},
  \bibinfo{pages}{115421} (\bibinfo{year}{2008}).

\bibitem[{\citenamefont{Beigi et~al.}(2011{\natexlab{a}})\citenamefont{Beigi,
  Shor, and Whalen}}]{beigi2011}
\bibinfo{author}{\bibfnamefont{S.}~\bibnamefont{Beigi}},
  \bibinfo{author}{\bibfnamefont{P.~W.} \bibnamefont{Shor}}, \bibnamefont{and}
  \bibinfo{author}{\bibfnamefont{D.}~\bibnamefont{Whalen}},
  \bibinfo{journal}{Comm. Math. Phys.} \textbf{\bibinfo{volume}{306}},
  \bibinfo{pages}{663} (\bibinfo{year}{2011}{\natexlab{a}}).

\bibitem[{\citenamefont{Thorngren}(2015)}]{Thorngren2015}
\bibinfo{author}{\bibfnamefont{R.}~\bibnamefont{Thorngren}},
  \bibinfo{journal}{J. High Energ. Phys.} \textbf{\bibinfo{volume}{2015}},
  \bibinfo{pages}{152} (\bibinfo{year}{2015}).

\bibitem[{\citenamefont{Cui et~al.}(2015)\citenamefont{Cui, Hong, and
  Wang}}]{Cui2014}
\bibinfo{author}{\bibfnamefont{S.~X.} \bibnamefont{Cui}},
  \bibinfo{author}{\bibfnamefont{S.-M.} \bibnamefont{Hong}}, \bibnamefont{and}
  \bibinfo{author}{\bibfnamefont{Z.}~\bibnamefont{Wang}},
  \bibinfo{journal}{Quantum Information Processing}
  \textbf{\bibinfo{volume}{14}}, \bibinfo{pages}{2687} (\bibinfo{year}{2015}),
  \eprint{arXiv:1401.7096}.

\bibitem[{\citenamefont{Beigi et~al.}(2011{\natexlab{b}})\citenamefont{Beigi,
  Shor, and Whalen}}]{Beigi2010}
\bibinfo{author}{\bibfnamefont{S.}~\bibnamefont{Beigi}},
  \bibinfo{author}{\bibfnamefont{P.~W.} \bibnamefont{Shor}}, \bibnamefont{and}
  \bibinfo{author}{\bibfnamefont{D.}~\bibnamefont{Whalen}},
  \bibinfo{journal}{Comm. Math. Phys.} \textbf{\bibinfo{volume}{306}},
  \bibinfo{pages}{663} (\bibinfo{year}{2011}{\natexlab{b}}).

\bibitem[{\citenamefont{Kong}(2014)}]{kong2014}
\bibinfo{author}{\bibfnamefont{L.}~\bibnamefont{Kong}}, \bibinfo{journal}{Nucl.
  Phys. B} \textbf{\bibinfo{volume}{886}}, \bibinfo{pages}{436 }
  (\bibinfo{year}{2014}).

\bibitem[{\citenamefont{Bais and Slingerland}(2009)}]{bais2009}
\bibinfo{author}{\bibfnamefont{F.~A.} \bibnamefont{Bais}} \bibnamefont{and}
  \bibinfo{author}{\bibfnamefont{J.~K.} \bibnamefont{Slingerland}},
  \bibinfo{journal}{Phys. Rev. B} \textbf{\bibinfo{volume}{79}},
  \bibinfo{pages}{045316} (\bibinfo{year}{2009}).

\bibitem[{\citenamefont{Levin}(2013)}]{levin2013}
\bibinfo{author}{\bibfnamefont{M.}~\bibnamefont{Levin}},
  \bibinfo{journal}{Phys. Rev. X} \textbf{\bibinfo{volume}{3}},
  \bibinfo{pages}{021009} (\bibinfo{year}{2013}).

\bibitem[{\citenamefont{Barkeshli}()}]{barkeshli2016kitp}
\bibinfo{author}{\bibfnamefont{M.}~\bibnamefont{Barkeshli}},
  \bibinfo{note}{10/19/2016},
  \urlprefix\url{http://online.kitp.ucsb.edu/online/topoquant_c16/barkeshli/}.

\bibitem[{\citenamefont{Wang and Levin}(2016)}]{wang2016}
\bibinfo{author}{\bibfnamefont{C.}~\bibnamefont{Wang}} \bibnamefont{and}
  \bibinfo{author}{\bibfnamefont{M.}~\bibnamefont{Levin}}
  (\bibinfo{year}{2016}), \eprint{arXiv:1610.04624}.

\bibitem[{\citenamefont{Tachikawa and
  Yonekura}(2016{\natexlab{a}})}]{tachikawa2016a}
\bibinfo{author}{\bibfnamefont{Y.}~\bibnamefont{Tachikawa}} \bibnamefont{and}
  \bibinfo{author}{\bibfnamefont{K.}~\bibnamefont{Yonekura}}
  (\bibinfo{year}{2016}{\natexlab{a}}), \eprint{arXiv:1610.07010}.

\bibitem[{\citenamefont{Tachikawa and
  Yonekura}(2016{\natexlab{b}})}]{tachikawa2016b}
\bibinfo{author}{\bibfnamefont{Y.}~\bibnamefont{Tachikawa}} \bibnamefont{and}
  \bibinfo{author}{\bibfnamefont{K.}~\bibnamefont{Yonekura}}
  (\bibinfo{year}{2016}{\natexlab{b}}), \eprint{arXiv:1611.01601}.

\bibitem[{\citenamefont{Bhardwaj}(2016)}]{bhardwaj2016}
\bibinfo{author}{\bibfnamefont{L.}~\bibnamefont{Bhardwaj}}
  (\bibinfo{year}{2016}), \eprint{arXiv:1611.02728}.

\bibitem[{\citenamefont{Pollmann and
  Turner}(2012{\natexlab{b}})}]{PollmannTurner}
\bibinfo{author}{\bibfnamefont{F.}~\bibnamefont{Pollmann}} \bibnamefont{and}
  \bibinfo{author}{\bibfnamefont{A.~M.} \bibnamefont{Turner}},
  \bibinfo{journal}{Phys. Rev. B} \textbf{\bibinfo{volume}{86}},
  \bibinfo{pages}{125441} (\bibinfo{year}{2012}{\natexlab{b}}).

\bibitem[{\citenamefont{Perez-Garcia et~al.}(2007)\citenamefont{Perez-Garcia,
  Verstraete, Wolf, and Cirac}}]{MPS}
\bibinfo{author}{\bibfnamefont{D.}~\bibnamefont{Perez-Garcia}},
  \bibinfo{author}{\bibfnamefont{F.}~\bibnamefont{Verstraete}},
  \bibinfo{author}{\bibfnamefont{M.}~\bibnamefont{Wolf}}, \bibnamefont{and}
  \bibinfo{author}{\bibfnamefont{J.}~\bibnamefont{Cirac}},
  \bibinfo{journal}{Quantum Inf. Comput.} \textbf{\bibinfo{volume}{7}},
  \bibinfo{pages}{401} (\bibinfo{year}{2007}).

\bibitem[{\citenamefont{Tambara and Yamagami}(1998)}]{Tambara1998}
\bibinfo{author}{\bibfnamefont{D.}~\bibnamefont{Tambara}} \bibnamefont{and}
  \bibinfo{author}{\bibfnamefont{S.}~\bibnamefont{Yamagami}},
  \bibinfo{journal}{Journal of Algebra} \textbf{\bibinfo{volume}{209}},
  \bibinfo{pages}{692} (\bibinfo{year}{1998}).

\end{thebibliography}
